\documentclass[twoside,11pt]{article}
\usepackage{epsfig}
\usepackage{amsmath}
\usepackage{amssymb}
\usepackage[english]{babel}
\usepackage{fancyhdr}
\usepackage{psfrag}

\newcommand{\be}{\begin{eqnarray}}
\newcommand{\ee}{\end{eqnarray}}
\def\sss{\scriptscriptstyle}
\def\dbar{{\overline{D}}}
\def\units{{{\rm m}^{-2}{\rm s}^{-1}{\rm sr}^{-1} {\rm GeV}^{-1}}}
\def\unitsn{{{\rm m}^{-2}{\rm s}^{-1}{\rm sr}^{-1} {\rm (GeV/n)}^{-1}}}
\def\gsim{\gtrsim}
\newcommand{\met}{\not \!\! E_T}

\begin{document}

\begin{titlepage}
\vspace*{0.5cm}
\begin{center}
{\bf\huge Dark Matter and Collider Phenomenology\\[0.5cm] of Universal Extra Dimensions${}^*$}\\[1.8cm]

{\Large {\bf Dan Hooper${}^{(a)}$} and {\bf Stefano Profumo${}^{(b,c)}$}}\\[1.8cm]

{\em ${}^{(a)}$ Theoretical Astrophysics Group, Fermilab, Batavia, IL 60510, USA {\tt dhooper@fnal.gov}}\\[0.2cm]
{\em ${}^{(b)}$ Division of Physics, Mathematics and Astronomy, California Institute of Technology\\ Mail Code 106-38, Pasadena, CA 91125, USA {\tt profumo@caltech.edu}}\\[0.2cm]
\mbox{\hspace{-0.5cm}\em ${}^{(c)}$ Santa Cruz Institute for Particle Physics, University of California, Santa Cruz, CA 95064, USA}\\[1cm]
\end{center}
\rule{\textwidth}{.1pt}\\
{\bf Abstract}\\[0.3cm]
We review the phenomenology of models with flat, compactified extra dimensions where all of the Standard Model fields are allowed to propagate in the bulk, known as Universal Extra Dimensions (UED). UED make for an interesting TeV-scale physics scenario, featuring a tower of Kaluza-Klein (KK) states approximately degenerate in mass at the scale set by the inverse size of the compactification radius. KK parity, the four-dimensional remnant of momentum conservation in the extra dimensions, implies two basic consequences: (1) contributions to Standard Model observables arise only at loop level, and KK states can only be pair-produced at colliders, and (2) the lightest KK particle (LKP) is stable, providing a suitable particle dark matter candidate. After a theoretical overview on extra dimensional models, and on UED in particular, we introduce the model particle spectrum and the constraints from precision electroweak tests and current colliders data. We then give a detailed overview of the LKP dark matter phenomenology, including the LKP relic abundance, and direct and indirect searches. We then discuss the physics of UED at colliders, with particular emphasis on the signatures predicted for the Large Hadron Collider and at a future Linear Collider, as well as on the problem of discriminating between UED and other TeV-scale new physics scenarios, particularly supersymmetry. We propose a set of reference benchmark models, representative of different viable UED realizations. Finally, we collect in the Appendix all the relevant UED Feynman rules, the scattering cross sections for annihilation and coannihilation processes in the early universe and the production cross section for strongly interacting KK states at hadron colliders.\\
\rule{\textwidth}{.1pt}\\[1cm]

${}^*${\em Invited Review Article for Physics Reports}
\end{titlepage}

\small
\tableofcontents
\normalsize

\clearpage
\newpage
\section{Introduction}

In 1919, Theodor Kaluza extended general relativity to five dimensional space-time in an effort to unify gravity with the force of electromagnetism \cite{Kaluza:1921tu}. Seven years later, Oskar Klein proposed that this fifth dimension could be hidden from our observations by being compactified around a circle of very small radius \cite{klein}. Particles moving through the extra dimension would appear very massive, as a result of their extra-dimensional momentum being perceived as a contribution to its four-dimensional rest mass. These excited states of ordinary particles have become to be known as Kaluza-Klein modes.

The original proposal of Kaluza and Klein suffered from a number of problems and attracted only marginal interest for many decades. In the 1980s, however, the relevance of the work of Kaluza and Klein grew dramatically, bolstered by the conclusion that string theory could not be formulated in a four dimensional space-time.

More recently, interest in theories including extra spatial dimensions has undergone a resurgence. In particular, a great deal of attention has been given to two classes of extra dimensional theories over the past decade. First, scenarios featuring one or more large (millimeter-scale), flat extra dimensions were proposed by Arkani-Hamad, Dimopoulos and Dvali (ADD) in 1998~\cite{Arkani-Hamed:1998rs,Arkani-Hamed:1998nn}. In the ADD model, all of the fields of the Standard Model are confined to a four dimensional brane, leaving only gravity to propagate in the space of the extra dimensions (the bulk). As a result, the fundamental Planck scale as perceived in the bulk could be similar in magnitude to the electroweak scale, providing a potential connection between
those scales and the hierarchy problem.\footnote{For earlier work along similar lines, see Refs.~\cite{Antoniadis:1990ew,Antoniadis:1993jp,Dienes:1998vh}.}

In addition to the possibility of large, flat extra dimensions, Randall and Sundrum proposed that the fundamental Planck scale could be lowered to around the electroweak scale by introducing an additional small dimension of space with a large degree of spatial curvature~\cite{Randall:1999ee}. The presence of an exponential ``warp factor'' in the metric of the Randall-Sundrum model leads to the fundamental Planck scale (which is near the electroweak scale) to appear to observers on our brane to be much higher, thus addressing the hierarchy problem.

A further class of extra-dimensional models, which goes by the name of Universal Extra Dimensions (UED), was proposed in 2001 by Appelquist, Cheng and Dobrescu \cite{Appelquist:2000nn}, and is the focus of this review. By {\em universal}, it is meant that all of the Standard Model (SM) fields are free to propagate through all of the dimensions of space, rather than being confined to a brane as some or all of them are in the ADD and Randall-Sundrum models. Models featuring some or all of the SM gauge bosons and fermions in the bulk have appeared in the literature in various forms over the past decade, in particular in the work of Dienes, Dudas and Gherghetta~\cite{Dienes:1998vg} (see also Ref.~\cite{Dienes:1998hx}). Appelquist, Cheng and Dobrescu were the first, however, to recognize that the introduction of KK-parity considerably relaxes the constraints from electroweak precision observables, allowing for much lower scales of compactification.

For a number of reasons, which we will discuss later, extra dimensions of size $R \sim \rm{TeV}^{-1}$ are particularly well motivated within the context of UED, much smaller than those found in the ADD model.
In many ways the UED model is more similar to the original proposals of Kaluza and Klein than either the ADD or Randall-Sundrum scenarios. Although the motivations for UED are very different from those of Kaluza and Klein, each introduce extra flat and compact dimensions of space in which all types of particles are free to propagate.\footnote{For a recent review comparing and contrasting the ADD, Randall-Sundrum and UED models, see Ref.~\cite{Kribs:2006mq}.} 

One aspect of UED models which has received quite some attention is their ability to provide a natural candidate for the dark matter of our universe. In particular, the lightest Kaluza-Klein state can be stable in UED models and, as we will see, naturally colorless, electrically neutral and produced in the early universe with an abundance similar to that of the measured dark matter density. For these and other reasons, Kaluza-Klein dark matter has become a popular alternative to neutralinos \cite{Jungman:1995df} and other commonly studied dark matter candidates~\cite{Bertone:2004pz}.

In this review, we will discuss the dark matter and collider phenomenology of UED models. With the Large Hadron Collider soon to begin its operation, and numerous dark matter experiments being developed and carried out, we hope this article will be found useful in the coming years.

\clearpage
\newpage

\section{Universal Extra Dimensions}\label{ch:theory}

In this section, we introduce the basic UED model and discuss its motivations, the predicted mass spectrum of Kaluza-Klein states, and the current experimental constraints. 

The basic UED model is remarkably simple: a flat metric with one or more compact extra dimensions. There is one complication to this otherwise very simple picture, however. Fermions in a number of dimensions larger than four are non-chiral in the case of the simple compactification on a circle. If the extra dimensions are orbifolded, however, this problem can be avoided. In the five dimensional case, an $S^1/Z_2$ orbifold removes the unwanted fermionic degrees of freedom, allowing for the existence of chiral fermions. The orbifold compactification consists of projecting a circular extra dimension onto a line with two fixed points. In six or more dimensions, many orbifolding schemes become possible. The case of a $T^2/Z_2$ orbifold in six dimensions has particularly attractive features, for example.

Whenever possible, we will not commit ourselves to a particular number of extra dimensions, or choice of orbifold. In some cases, when conclusions can only be drawn in the case of specific scenarios, we will specify which class of UED models we will consider.

\subsection{Motivations for UED}\label{sec:ued}

A wide range of motivations for the UED model can be found in the literature. We summarize below a few of the arguments we find especially compelling.

\subsubsection*{Proton Stability}

In the Standard Model, there exist six-dimensional baryon and lepton number violating operators which cause the proton to decay with a lifetime on the order of
\begin{equation}
\tau_p \sim \frac{M^4}{m^5_p}
\end{equation}
where $M$ is the cutoff scale of the theory where quantum gravity effects set in. To evade the constraints on proton stability from Super-Kamionkande  ($\tau_p \gsim 10^{33}$ s)~\cite{Shiozawa:1998si}, the cutoff scale must be at least $\sim 10^{16}$ GeV. In many extra dimensional scenarios, the cutoff scale is well below this value. Although there have been several possible solutions proposed  \cite{Arkani-Hamed:1998rs,Arkani-Hamed:1999dc,Davoudiasl:2005ks}, in the original versions of the ADD and Randall-Sundrum models, protons are generically predicted to decay at a rate much higher than is experimentally acceptable.

In the case of UED, the conclusion which is reached is rather different \cite{Appelquist:2001mj}. The global symmetries in UED can limit the higher dimensional operators which lead to proton decay. Considering the case of UED in six dimensions, for example, invariance under standard model gauge transformations and under the SO(1,5) Lorentz symmetry implies that the lowest dimension operator which can lead to proton decay is 9-dimensional (in the 4-D effective theory). Higher dimensional operators can be shown to be generically suppressed \cite{Appelquist:2001mj}, and the leading proton decay mode to be such as $p \rightarrow e^- \pi^+ \pi^+ \nu \nu$, taking place with a lifetime of
\begin{equation}
\tau_p \sim 10^{35} \, \rm{years} \, \bigg(\frac{1/R}{500 \, \rm{GeV}}\bigg)^{12} \, \bigg(\frac{\Lambda R}{5}\bigg)^{22}
\end{equation}
where $R$ is the size of the extra dimensions and $\Lambda$ is the cutoff of the theory. In this particular six-dimensional case, the proton decay rate is well below the limits imposed by Super-Kamiokande. More generally speaking, many of the operators leading to rapid proton decay are forbidden or suppressed in UED models, thus alleviating or at least relaxing some of the problems faced by the ADD and Randall-Sundrum models.

Similar arguments can be made to show that interactions leading to neutron-anti-neutron oscillations, lepton number violating interactions leading to large Majorana neutrino masses, and interactions leading to large flavor changing neutral currents (FCNC), can each be highly suppressed in UED.

\subsubsection*{Anomaly Cancellation With Three Generations}

Motivation for UED has also been found from the requirements of anomaly cancellation. In particular, it was demonstrated in Ref.~\cite{Dobrescu:2001ae} that gauge anomalies can be avoided in UED models only if there exist three generations of fermions. In particular, in the case of six-dimensional UED, an $SU(2)_L$ global gauge anomaly exists unless the number of doublets with positive and negative chirality differ by an integer multiple of six. For each generation, this difference is either two or four, thus with three generations the global gauge anomaly is canceled in six-dimensional UED.

\subsubsection*{Dark Matter}

As we will discuss at length in this review, the UED model naturally encompasses a stable, electrically neutral, and non-colored state which can constitute a viable candidate for dark matter. Although this was not among the first motivations for UED, it is one which has later received a great deal of attention. In fact, the UED model is the extra-dimensional scenario which most naturally provides a suitable particle dark matter candidate, to-date one the most compelling pieces of evidence for particle content beyond the Standard Model.

\subsubsection*{Collider Searches}

Unlike other extra-dimensional scenarios, including models where only gauge bosons propagate in the bulk \cite{Masip:1999mk,Rizzo:1999br}, in UED KK parity suppresses the constraints on the size of the extra dimension(s) from electroweak precision tests, FCNC and other processes. As a result, KK states are generically light, making UED, in principle, a fully testable theory at present (Tevatron) and future (LHC, LC) colliders. This is especially true in the regions of parameter space favored by the abundance of KK dark matter produced thermally in the early universe.

\subsubsection*{``Bosonic Supersymmetry''}

The UED model shares striking similarities as well as several relatively subtle differences with supersymmetry, the by far most studied TeV-scale new physics scenario. This led some to dub the UED scenario ``bosonic supersymmetry'', referring to the spin of the heavier, yet-to-be-discovered states, which, unlike in supersymmetry, feature the same spin as their Standard Model counterparts. Both in the realm of dark matter detection and collider searches, this has triggered a lot of work on how to distinguish between different new physics scenarios. For instance, the cascade decays allowed by KK number conservation have led to a better understanding of the similar signals in supersymmetry, and have generated interest and a number of studies on the possibility of measuring spins at the LHC.

\subsubsection*{A Non-Confining Strongly Coupled Theory}

The UED model provides an interesting example of an effective field theory which behaves, at the cut-off scale, as a non-confining strongly coupled theory. In particular this may lead to a dynamical origin for the SM Higgs doublet, along the lines of the analysis of Ref.~\cite{Arkani-Hamed:2000hv}. The main ingredient to achieve this is the fact that fermionic fields propagate in the extra dimensions \cite{Cheng:1999bg}.

\clearpage
\newpage

\subsection{The UED Model}

This section is devoted to a theoretical overview of the UED model. We start in Sec.~\ref{sec:lagr} with an outline of the Lagrangian of the theory and of its four-dimensional counterpart. Sec.~\ref{sec:kkp} introduces Kaluza-Klein parity, a key feature for the model phenomenology. Finally, in Sec.~\ref{sec:spectrum}, we present the particle mass spectrum including the crucial ingredient of radiative corrections.

\subsubsection{The UED Lagrangian}\label{sec:lagr}

We consider the Standard Model (SM) in $4+D$ space-time dimensions, where all the SM fields are allowed to propagate in the extra (universal) dimensions, compactified at a scale $1/R$.

Following Ref.~\cite{Appelquist:2000nn}, we indicate the usual non-compact space-time coordinates with the notation $x^\mu$, $\mu=0,1,2,3$, and the coordinates of the extra-dimensions with $y^a$, $a=1,...,D$. Indexes running over all space-time dimensions are indicated with capitalized letters. The 4-dimensional Lagrangian density is obtained by dimensional reduction of the ($4+D$)-dimensional theory integrating over the compactified extra-dimensions as follows:
\be\label{eq:masterlagrangian}
\nonumber {\mathcal L}(x^\mu)&=&\int\ {\rm d}^Dy\Big\{-\sum_{i=1}^3\ \frac{1}{2\hat g^2_i}{\rm Tr}\Big[F_i^{AB}(x^\mu,y^a)F_{iAB}(x^\mu,y^a)\Big]+\\[0.2cm]
\nonumber &&+|(D_\mu+D_{3+a})H(x^\mu,y^a)|^2+\mu^2H^*(x^\mu,y^a)H(x^\mu,y^a)-\lambda\big[H^*(x^\mu,y^a)H(x^\mu,y^a)\big]^2+\\[0.2cm]
\nonumber &&+i\left(\overline{ Q},\overline{ u},\overline{ d},\overline{ L},  \overline{ e}\right)(x^\mu,y^a)\left(\Gamma^\mu D_\mu+\Gamma^{3+a}D_{3+a}\right)\left({ Q},{ u},{ d},{ L},{ e}\right)(x^\mu,y^a)+\\[0.2cm]
\nonumber &&\Big[ \overline{ Q}(x^\mu,y^a)\left(\hat\lambda_{ u}\ { u}(x^\mu,y^a)i\sigma_2H^*(x^\mu,y^a)+\hat\lambda_{ d}\ { d}(x^\mu,y^a)H(x^\mu,y^a)\right)+{\rm H.c.}\Big]+\\[0.2cm]
&&\Big[ \overline{ L}(x^\mu,y^a)\hat\lambda_{ e}\ { e}(x^\mu,y^a)H(x^\mu,y^a)+{\rm H.c.}\Big].
\ee

In Eq.~(\ref{eq:masterlagrangian}), the summation over fermion generations has been suppressed, and we indicate with $F_i^{AB}$ the $(4+D)$-dimensional gauge field strengths associated with the $SU(3)_c\times SU(2)_W\times U(1)_Y$ gauge group. $D_\mu=\partial/\partial x^\mu-{\mathcal A}_\mu$ and $D_{3+a}=\partial/\partial y^a-{\mathcal A}_{3+a}$ are the covariant derivatives, with ${\mathcal A}_A=-i\sum_{i=1}^3\hat g_i {\mathcal A}^r_{Ai}T^r_i$ being the $(4+D)$-dimensional gauge fields, and $\hat g_i$ the $(4+D)$-dimensional gauge couplings. The latter, as well as the Yukawa matrices, $\hat \lambda_{u,d,e}$, have dimension (mass)$^{-D/2}$.
The symbols $Q,u,d,L,e$ describe the $(4+D)$-dimensional fermions, whose zero modes correspond to the SM fermions. Capitalized letters indicate $SU(2)_W$ doublets, while lower case letters indicate $SU(2)_W$ singlets. The $(4+D)$-dimensional gamma matrices, $\Gamma^A$, are anticommuting $2^{K+2}\times2^{K+2}$ matrices, where $D=2K$ if $D$ is even, and $D=2K+1$ for odd $D$, satisfying the $(4+d)$-dimensional Clifford algebra $\{\Gamma^A,\Gamma^B\}=2g^{AB}$. In particular, for the case of $D=1$, one can set $\Gamma_\mu=\gamma_\mu$, and $\Gamma_4=i\gamma_5$.

The last step needed to extract the effective four-dimensional Lagrangian from Eq.~(\ref{eq:masterlagrangian}) is to specify the compactification of the extra-dimensions. The requirement of producing chiral fermions in four dimensions forces one, on general grounds, to consider orbifold compactifications \cite{Appelquist:2000nn}. We consider here an $[(S^1\times S^1)/Z_2]^K$ orbifold for $D$ even, and an  $[(S^1\times S^1)/Z_2]^K\times(S^1/Z_2)$ orbifold for $D$ odd. At this point one must specify how the fields transform under the orbifold projection. Choosing these transformation appropriately allows one to eliminate the extra, phenomenologically unwanted, degrees of freedom appearing at the zero modes level. For definiteness, we shall focus below on the case of $D=1$. 

Requiring that the ${\mathcal A}_\mu$ components of the gauge fields are even under the $y\rightarrow-y$ transformation, and that ${\mathcal A}_5$ is odd, will leave the zero modes corresponding to the physical SM gauge fields and will eliminate the extra zero modes (which would appear as massless scalar fields after dimensional reduction). Analogously, to retain the (zero mode) SM Higgs, one has to require that $H$ is even under the orbifold transformation. With these properties, we can derive the following decomposition of the gauge and scalar fields in KK modes:
\be
(H,{\mathcal A}_\mu)(x^\mu,y)&=&\frac{1}{\sqrt{\pi R}}\Big[(H_0,{\mathcal A}_{\mu,0})(x_\mu)+\sqrt{2}\sum_{n=1}^\infty H_n,{\mathcal A}_{\mu,n})(x_\mu)\cos\left(\frac{ny}{R}\right)\Big]\\
{\mathcal A}_5&=&\sqrt{\frac{2}{\pi R}}\sum_{n=1}^\infty {\mathcal A}_{5,n}(x_\mu)\sin\left(\frac{ny}{R}\right).
\ee
The normalization is chosen once we assume the range for the $y$ variable to be from 0 to $\pi R$. The case of the fermions requires taking into account a subtlety related to the problem of chiral fields in 5 dimensions. Since one cannot construct, in 5 dimensions, a matrix with the properties of $\gamma_5$ in 4 dimensions (i.e. anti-commuting with all $\Gamma^A$, and whose square is the identity), there is no chirality in 5 dimensions. Practically , this means that bilinears like $\psi\gamma^\mu\gamma_5\psi$ are not invariant under 5D Lorentz transformations, so they cannot appear in the 5D Lagrangian. Consequently, one cannot have the left and right-handed components of the zero components of the zero excitation, $\psi_0$, couple differently to the gauge fields, as in the SM. Therefore, one cannot get a chiral SM fermion using a single 5D fermion field, and is forced to introduce two 5D fermionic fields for each Dirac fermion field~\cite{Macesanu:2005jx}. Given a SM Dirac fermion field, $\psi^{\rm SM}$, one introduces two 5D fermion fields, $\psi_{L,R}$, each with the quantum numbers of left and right handed spinors, $\psi^{\rm SM}_{L,R}$. One can then set $\psi^{\rm SM}=P_L\psi_{L,0}+P_R\psi_{R,0}$, with $P_{L,R}=(1\mp\gamma_5)/2$ the 4-dimensional chiral projection operators. The generic KK decomposition will then be cast, for any given quark and lepton field, as
\be
\psi (x^\mu,y)=\frac{1}{\sqrt{\pi R}}\Big[\psi^{\rm SM}(x^\mu)+\sqrt{2}\sum_{n=1}^\infty P_L\ \psi_{L,n}(x^\mu)\cos\left(\frac{ny}{R}\right)+P_R\ \psi_{R,n}(x^\mu)\sin\left(\frac{ny}{R}\right)\Big].
\ee
As a result, the higher KK modes are 4-dimensional vector-like fermions, while the zero modes are chiral. This will induce a set of special Feynman rules, which we review and list in Appendix \ref{sec:feynrules}.

\subsubsection{Kaluza-Klein Parity}\label{sec:kkp}\label{sec:kkparity}

As the number of the KK level of a particle is a measure of its momentum in an extra dimension of space, one might expect KK-number to be a conserved quantity by the virtue of extra-dimensional momentum conservation. In UED models, however, the introduction of orbifold compactifications breaks this translational symmetry along the extra dimension and, through Noether's theorem, leads to KK-number violating interactions. Even after an orbifold is introduced, however, a subgroup of KK-number conservation known as KK-parity can remain present which insures the conservation of the ``evenness" or ``oddness" of KK number in an interaction \cite{Appelquist:2000nn}. For instance, in the case of one extra dimension compactified on an $S^1/Z_2$ orbifold, KK-parity is a $Z_2$ symmetry under which only KK modes with odd KK number are charged. 

Phenomenologically, KK-parity acts similarly to how R-parity conserves the ``evenness" or ``oddness" of the number of superpartners within the context of supersymmetry. In general, KK-parity is conserved in UED if no explicit KK-parity violating interactions are introduced on the orbifold fixed points. To determine whether this is the case, one would have to consider a specific compactification scheme and have a full UV completion of the theory. Although KK-parity appears to be fairly natural within the context of UED models, without a UV completion we are unable to address with certainty whether KK-parity will be conserved. Throughout this review, however, we will proceed under the assumption that KK-parity is conserved.

KK-parity can be written simply as $P = (-1)^n$, where $n$ denotes the $n$th KK mode. There are several important phenomenological consequences of KK-parity conservation. Most importantly for dark matter, KK-parity insures that the Lightest Kaluza-Klein Particle (the LKP, as it is often called) is stable \cite{Servant:2002aq,Cheng:2002ej}. Of consequence for collider experiments is that odd level KK states can only be pair produced \cite{Appelquist:2000nn,Rizzo:2001sd,Macesanu:2002db,Cheng:2002ab}. Finally, all direct couplings to even number KK modes are loop-suppressed.

\subsubsection{The Particle Spectrum}\label{sec:spectrum}

In this section, we describe the spectrum of Kaluza-Klein (KK) states in a 5-dimensional UED model, including radiative corrections and the effect of boundary terms. Beginning at tree level, all of the Standard Model fields appear as towers of Kaluza-Klein states with masses of:
\begin{equation}
m^2_{X^{(n)}} = \frac{n^2}{R^2} + m^2_{X^{(0)}},
\label{tree}
\end{equation}
where $X^{(n)}$ is the $n$th Kaluza-Klein excitation of the Standard Model field, $X$, and $R\sim \rm{TeV}^{-1}$ is the size of the extra dimension. $X^{(0)}$ denotes the ordinary Standard Model particle (known as the zero mode). Assuming that $R^{-1}$ is considerably larger than any of the Standard Model zero mode masses, this leads us to expect a highly degenerate spectrum of Kaluza-Klein states at each level, $n$. This picture is somewhat modified, however, when the effects of radiative corrections and boundary terms are considered.

Corrections to KK masses are generated by loop diagrams traversing around the extra dimension, called bulk loops, and by brane-localized kinetic terms which appear on the orbifold boundaries. Together, these contributions (at one loop) are given by \cite{Cheng:2002iz}:
\begin{eqnarray}\label{rspec}
\delta(m^2_{B^{(n)}}) &=& \frac{g'^2}{16 \pi^2 R^2} 
  \left( \frac{-39}{2} \frac{\zeta(3)}{\pi^2} -\frac{n^2}{3} \ln \, 
  \Lambda R \right) \nonumber \\
\delta(m^2_{W^{(n)}}) &=& \frac{g^2}{16 \pi^2 R^2} \left ( 
  \frac{-5}{2} \frac{\zeta(3)}{\pi^2} + 15 n^2 \ln \,
  \Lambda R \right )\nonumber \\
\delta(m^2_{g^{(n)}}) &=& \frac{ g_3^2}{16 \pi^2 R^2} \left ( 
  \frac{-3}{2} \frac{\zeta(3)}{\pi^2} + 23 n^2 \ln \,
  \Lambda R \right )\nonumber \\
\delta(m_{Q^{(n)}}) &=& \frac{n}{16 \pi^2 R} \left ( 6 g_3^2+ \frac{27}{8} 
g^2 + \frac{1}{8} g'^2 \right) \ln \, \Lambda R \nonumber \\
\delta(m_{u^{(n)}}) &=& \frac{n}{16 \pi^2 R} \left ( 6 g_3^2+  
 2 g'^2 \right) \ln \, \Lambda R \nonumber \\
\delta(m_{d^{(n)}}) &=& \frac{n}{16 \pi^2 R} \left ( 6 g_3^2+  \frac{1}{2} 
g'^2 \right) \ln \, \Lambda R \nonumber \\
\delta(m_{L^{(n)}}) &=& \frac{n}{16 \pi^2 R} \left ( \frac{27}{8} 
g^2 + \frac{9}{8} g'^2 \right) \ln \, \Lambda R \nonumber \\
\delta(m_{e^{(n)}}) &=& \frac{n}{16 \pi^2 R}\frac{9}{2} g'^2 \ln \, \Lambda R \; . 
\end{eqnarray}
Here $\zeta(3)\approx 1.2020...$ is the third zeta function, and $\Lambda$ is the cutoff scale of the theory, defined as the scale where the effective 5-dimensional theory breaks down, that is where the 5-dimensional couplings become strong and the theory is no longer perturbative. $\Lambda$ is a parameter of the theory; to estimate its largest possible value, one can consider the loop expansion parameters
\be
\varepsilon_i=N_i\frac{\alpha_i(\Lambda)}{4\pi}\left(\Lambda R\right)
\ee
where the $\alpha_i$ are the 4-dimensional standard model gauge couplings, the index $i=1,2,3$ labels the corresponding gauge groups, and $N_i$ indicates the corresponding number of colors. Finally, $\Lambda R$ counts the number of KK modes below $\Lambda$. Although the 4-dimensional $SU(3)$ coupling becomes more asymptotically free above each KK level, the 5-dimensional $SU(3)$ interaction becomes non-perturbative in the ultraviolet before the other gauge interactions. In particular, $\varepsilon_3$ becomes of order one for
values of the cutoff scale typically of the order of $\Lambda \sim 10\, R^{-1}$.

The brane kinetic term contributions in (\ref{rspec}) are those proportional to the logarithm of the cutoff. In the case of KK scalars and spin-1 fields, the $(n/R)^2$ term in Eq.~(\ref{tree}) gets replaced by $(n/R)^2+\delta(m^2)$, while in the case of KK fermions $(n/R)^2\rightarrow(n/R+\delta(m))^2$. KK gravitons do not receive appreciable radiative corrections, and their masses are simply given by $n/R$.

After the KK modes of the $W$ and $Z$ bosons acquire masses by eating the fifth components of the gauge fields and Higgs KK modes, four scalar states remain at each KK level. These modes have masses given by:
\begin{eqnarray}
m^2_{H^0_n} = \frac{n^2}{R^2} + m^2_h + \delta m^2_{H_n}  \nonumber \\
m^2_{H^{\pm}_n} = \frac{n^2}{R^2} + m^2_W + \delta m^2_{H_n}  \nonumber \\
m^2_{A^0_n} = \frac{n^2}{R^2} + m^2_Z + \delta m^2_{H_n},  
\end{eqnarray}
where the radiative and boundary term corrections are given by
\begin{eqnarray}
\delta m^2_{H_n} =  \frac{n^2}{16 \pi^2 R^2} \bigg(3g^2 + \frac{3}{2}g'^2 - 2\lambda_H\bigg) \ln \, \Lambda R\ +\ \bar{m}^2_H.
\label{higgscor}
\end{eqnarray}
Here, $\lambda_H$ is the Higgs quartic coupling and $\bar{m}_H$ is the boundary mass term for the Higgs mode.

\begin{figure}
\centering
\includegraphics[width=0.6\textwidth,clip=true]{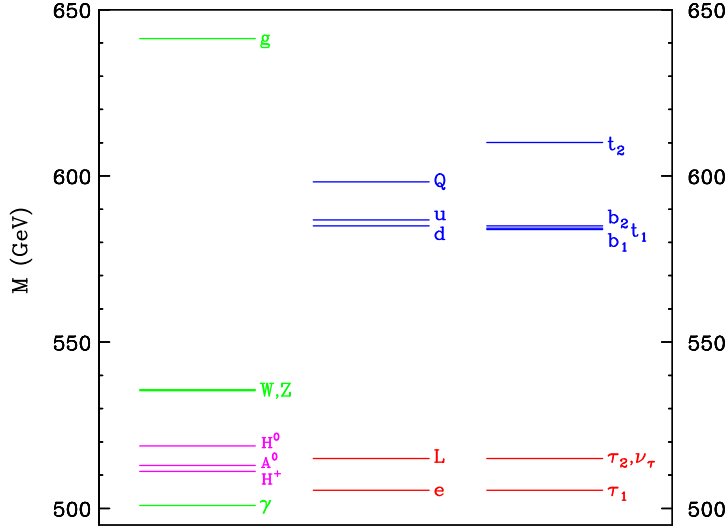} \hspace{1.0cm}
\caption{The spectrum of first level Kaluza-Klein states, including the effects of radiative corrections and boundary terms. A compactification radius of $R^{-1}=500$ GeV, Higgs mass of 120 GeV and cutoff of $\Lambda = 20\,R^{-1}$ have been used. From Ref.~\cite{Cheng:2002ab}.}
\label{radspec}
\end{figure}

Lastly, the additional contribution from the top quark Yukawa coupling yields:
\begin{eqnarray}
\delta_{h_t} (m_{Q^{(n)}_3}) = \frac{n}{R} \bigg(-\frac{3 h^2_t}{64 \pi^2}\ln \frac{\Lambda^2}{\mu^2}\bigg) \nonumber \\
\delta_{h_t} (m_{t^{(n)}}) =  \frac{n}{R} \bigg(-\frac{3 h^2_t}{32 \pi^2}\ln \frac{\Lambda^2}{\mu^2}\bigg).
\label{topyukawacor}
\end{eqnarray}
In Fig.~\ref{radspec}, an example of the first level Kaluza-Klein spectrum is shown. The choice of cutoff used ($\Lambda = 20\,R^{-1}$) leads to approximately 20\% corrections to the KK quark and KK gluon masses, and corrections of a few percent or less for the rest of the first level states.

The above expressions have been calculated under the assumption that the matching contributions from the brane-localized term cancel at the scale of the cutoff, $\Lambda$. Such contributions could, in principle, compete or even dominate over those corrections shown in Eqs.~(\ref{rspec}), (\ref{higgscor}) and (\ref{topyukawacor}). With this in mind, it is possible that the KK spectrum might not resemble the results described here. If such non-symmetric boundary term contributions are relatively small, however, the spectrum should closely reflect that described in this section~\cite{Cheng:2002iz,Cheng:2002ab}.

The observation that radiative corrections to the mass parameters of the UED theory are naturally expected to lie at the cutoff scale -- as defined in the usual sense of naive dimensional analysis as the scale for which loops are unsuppressed -- leads to a {\em little hierarchy problem} in the context of UED \cite{Burdman:2006jj}. For example, the zero mode Higgs field would naturally have mass at the cutoff $\Lambda\gg 1/R$. Possible solutions to this fine-tuning problem between the cutoff scale and the compactification scale arise in the context of extra-dimensional versions of Little Higgs, or Twin Higgs models, where a symmetry in the bulk forbids the potential for the Higgs field. The symmetry is then broken in such a way as to generate a potential leading to a mass of the order of the weak scale. In the Little Higgs model case, the phenomenology of the theory might differ significantly from that of the simplest UED model, in that the compactification scale approximately coincides  with the scale where the masses of the new states (the ``Little Higgs partners'') appear. The latter scale is instead considerably higher for the Twin Higgs model, and the expected phenomenology (unless both scales are significantly low) is similar to that of UED \cite{Burdman:2006jj}.

As we discussed in Sec.~\ref{sec:kkparity}, the lightest of the first level KK states can be stable in UED. Since we are particularly interested in dark matter within the context of these models, we would like to determine which of the first level KK states is the lightest and, therefore, potentially stable. The results of Ref.~\cite{Cheng:2002iz} would lead us to conclude that the KK photon will be the lightest KK particle (LKP), due to the small (and negative) correction to its mass. Things are complicated somewhat, however, by the fact that the KK photon is a mixture of the $B^{(1)}$ and $W^{3(1)}$ fields, just as the zero-mode of the photon is a mixture of $B^{(0)}$ and $W^{3(0)}$. The corresponding mass matrix, in the $B^{(n)}$, $W^{3(n)}$ basis, is as follows:
\begin{equation}
\left( \begin{array}{cc} 
\frac{n^2}{R^2} + \delta m^2_{B^{(n)}} + \frac{1}{4} g'^2 v^2 
  & \frac{1}{4} g'g v^2 \\
\frac{1}{4} g'g v^2 
  & \frac{n^2}{R^2} + \delta m^2_{W^{(n)}} + \frac{1}{4} g^2 v^2 
\end{array} \right).
\end{equation}
Here $v\approx 174$ GeV is the Higgs vacuum expectation value. In the zero-mode case ($n=0$), the off-diagonal terms are comparable in magnitude to the diagonal terms, leading to significant mixing between $B^{(0)}$ and $W^{3(0)}$ ($\sin^2\theta_W \approx 0.23$). At the first KK level, however, the diagonal terms are much larger than the off-diagonal terms, leading to only a small degree of mixing between the $B^{(1)}$ and $W^{3(1)}$ states. For $R^{-1}$=1 TeV, for example, the effective first KK level Weinberg angle is approximately $\sin^2 \theta_{W,1} \approx 10^{-3}$. Thus the mass eigenstate often called the ``KK photon'' is nearly identical to the state $B^{(1)}$.

If the corrections given by Eq.~(\ref{rspec}) do not fully describe the contributions to the KK spectrum, it is plausible that, instead of $B^{(1)}$, another of the first level KK states could be the LKP. Limiting our choices to electrically neutral and non-strongly interacting states, we are left with the possibilities of $\nu^{(1)}$, $W^{3(1)}$ or the first level KK graviton for the LKP. KK neutrinos were considered as a possible choice for the LKP in Ref.~\cite{Servant:2002aq}, but were found to generate unacceptably large rates in direct detection experiments in Ref.~\cite{Servant:2002hb}. As for the KK mode of the neutral SU(2) gauge boson, the SM contribution to its mass makes it difficult to envision a scenario where it is lighter than the U(1) KK gauge boson.

A latter possibility is that the LKP is the KK graviton (for a discussion of when this is the case in the context of the minimal UED model, see e.g. the analysis of Ref.~\cite{Cembranos:2006gt}). The cosmological consequences of a KK graviton LKP might turn out to be rather daunting. In particular, KK gravitons can be overproduced in the early universe, and can also distort the process of light elements nucleosynthesis, the cosmic microwave background, as well as the diffuse gamma-ray background. Even if the KK graviton is not the LKP, the presence of a KK graviton tower can potentially lead to similar problems. A range of reheating temperatures does exist, however, for which these problems can be avoided \cite{Shah:2006gs}. We discuss the physics of KK gravitons in some detail in Sec.~\ref{sec:kknlkp}.

Throughout the majority of this review, following most of the phenomenological analyses carried out so far in the literature, we will assume that the $B^{(1)}$ is the LKP.

\subsection{Electroweak Precision Constraints}\label{sec:ewprecision}

Very precise measurements have been made of a number of Standard Model observables, in particular in the electroweak sector. These measurements agree remarkably well with with the predictions of the Standard Model. Any model of physics beyond the Standard Model must be able to conform to this very restrictive body of data.

To illustrate how restrictive electroweak precision data can be, consider an extra-dimensional model in which only the gauge bosons are free to propagate in the bulk. In such a model, LEP data restricts these Kaluza-Klein gauge bosons to have masses far above the electroweak scale, at least as heavy as a few TeV \cite{Masip:1999mk,Rizzo:1999br}. 

Such constraints are weakened, however, in models in which all of the fields of the Standard Model are allowed to propagate in the bulk (UED models) \cite{Appelquist:2000nn,Appelquist:2002wb,Flacke:2005hb,Gogoladze:2006br}. In this class of models, the strongest constraints come from measurements at the $Z$-pole, including the relationship between the $Z$ and $W^{\pm}$ masses.

The full independent set of electroweak precision observables (EWPO) can be
defined by:
\begin{eqnarray}
\label{STUVXYWdef}
\hat{T} &\equiv& \frac{1}{m_W^2}\left(\Pi_{W_3W_3}(0)-\Pi_{W^+W^-}(0)\right)\nonumber\\
\hat{S} &\equiv& \frac{g}{g'}\Pi'_{W_3B}(0)\nonumber\\
\hat{U} &\equiv& \Pi'_{W^+W^-}(0)-\Pi'_{W_3W_3}(0)\nonumber\\
X &\equiv& \frac{m_W^2}{2}\Pi''_{W_3B}(0)\\
Y &\equiv& \frac{m_W^2}{2}\Pi''_{BB}(0)\nonumber\\
W &\equiv& \frac{m_W^2}{2}\Pi''_{W_3W_3}(0)\nonumber\\
V &\equiv& \frac{m_W^2}{2}\left(\Pi''_{W_3W_3}(0)-\Pi''_{W^+W^-}(0)\right)\nonumber
\end{eqnarray}
where $\Pi$ denote the new-physics contributions to the
transverse gauge boson vacuum polarization amplitudes,
with $\Pi'(0) = d\Pi(q^2)/dq^2 |_{q^2=0}$, etc. The parameters $\hat{S}$, $\hat{T}$ and $\hat{U}$ are related to the usual $S$, $T$ and $U$ by $S= 4 \sin^2 \theta_W \hat{S}/\alpha$, $T=\hat{T}/\alpha$ and $U=-4 \sin^2 \theta_W \hat{U}/\alpha$. Of these seven quantities, all but $V$ have been well determined by LEP data, and provide stringent constraints on new physics models.

Within the context of UED, the strongest constraints come from the parameter $\hat{T}$, and to a somewhat lesser extent from $\hat{S}$. In particular, these parameters acquire leading order contributions from the presence of KK modes given by:
\begin{eqnarray}
\hat{T}&=& \frac{g^2 m^4_t}{96 m^2_W m^2_{KK}} - \frac{5 g^2 \sin^2 \theta_W m^2_h}{1152 \cos^2\theta_W m^2_{KK}}, \nonumber \\
\hat{S}&=& \frac{g^2 m^2_t}{576 m^2_{KK}} + \frac{g^2 m^2_h}{2304 m^2_{KK}}.
\label{thatterms}
\end{eqnarray}
Comparing these two expressions, we see that the contributions to $\hat{T}$ are larger than those to $\hat{S}$ by a factor of $\sim 6 m^2_t/m^2_W$. Given that the experimental constraints on $\hat{T}$ and $\hat{S}$ are comparable, it is clear that the parameter $\hat{T}$ will dominate the electroweak precision constraints on UED models.

\begin{figure}
\psfrag{mH}[c]{$m_h$ [GeV]}
\psfrag{T}[c]{$\hat{T}_x$}
\psfrag{Rinv}[c]{$R^{-1}$[GeV]}
\centering
\mbox{\hspace*{-0.5cm}\includegraphics[width=0.47\textwidth,clip=true]{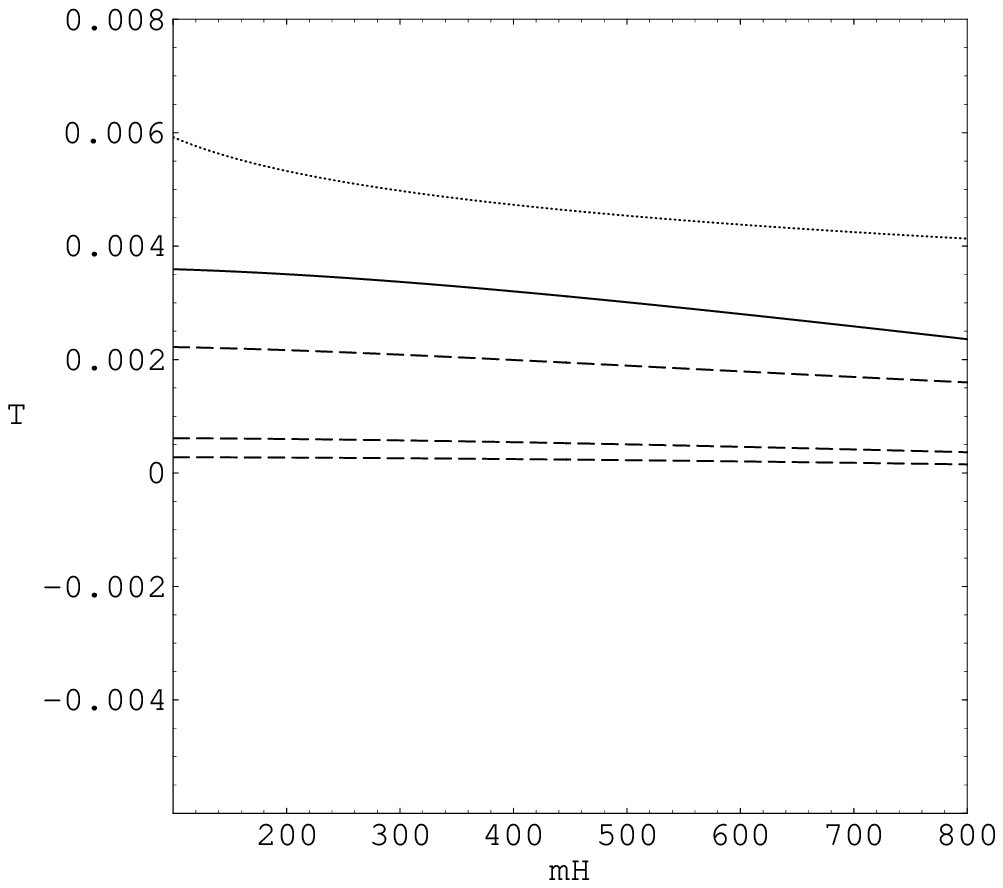}\qquad\qquad
\includegraphics[width=0.47\textwidth,clip=true]{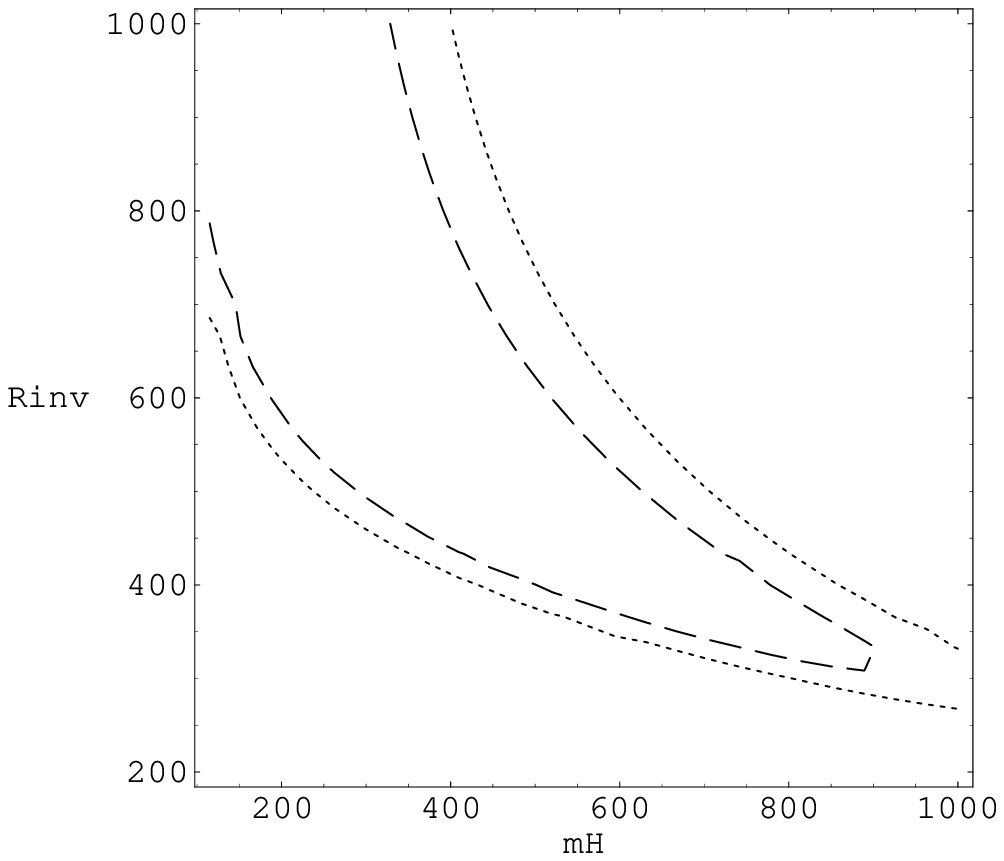}}
\caption{Left:  The magnitude of the contributions to the parameter $\hat{T}$ from each of the first three KK levels (dashed) for $m_{KK}=400$~GeV as a function of Higgs mass, as well as the sum over the first 10 KK modes (solid). The magnitude of the Higgs-dependent correction (the second term in the expression for $\hat{T}$ in Eq.~(\ref{thatterms})) is shown as the dotted line (first from above). Right: The 95\% (dashed) and 99\% (dotted) confidence limit exclusion regions, as a function of Higgs mass and mass $m_{KK}=1/R$. From Ref.~\cite{Flacke:2005hb}.}
\label{ewfig}
\end{figure}

In the left frame of Fig.~\ref{ewfig}, the size of the contributions to $\hat T$ are given for the first three KK modes for the case of $m_{KK}=400$ GeV, as a function of the Higgs mass. Also shown are the contributions from the zero modes alone, which depend significantly on the Higgs mass. The sign of these two classes of contributions are opposite enabling a cancellation to occur. Summing the contributions, the 95\% and 99\% confidence limit exclusion regions are shown in the $m_{KK}$-$m_h$ plane in the right frame of Fig.~\ref{ewfig}. For a light Higgs mass, limits as strong as $m_{KK} \gtrsim 700-800$ GeV can be obtained. For a heavy Higgs ($m_h\gtrsim 300$ GeV), electroweak precision observables constrain the viable range of $R^{-1}$ from below {\em and} from above: for instance, at 95\% C.L., $400\lesssim R^{-1}/{\rm GeV}\lesssim 600$ for $m_h=500$ GeV, and $300\lesssim R^{-1}/{\rm GeV}\lesssim 400$ for $m_h=800$ GeV~\cite{Flacke:2005hb}.

\subsection{Accelerator Searches: the Current Status}\label{sec:accconst}

Accelerator searches for extra dimensions have been performed for a variety of theoretical setups. The most recent and strongest constraints on models with large extra dimensions and warped extra dimensions were obtained at the Tevatron. Those results are summarized in Refs.~\cite{KaragozUnel:2004ti,Landsberg:2004mj}. The only existing analysis of direct searches for KK excitations in the UED scenario can be found in Ref.~\cite{Lin:2005ix}, and is based upon CDF Run IB data taken in the years 1994-1996 \cite{Abe:1998qm}. This analysis relies on the multi-lepton channel only, and is restricted to the case of one extra dimension. Also, it is assumed that the spectrum of KK particles follows the radiative corrections we summarized in Sec.~\ref{sec:spectrum}, with $\bar m_H^2$, the boundary mass term for the Higgs mode, set to zero\footnote{A non-zero value for $\bar m_H^2$ should not significantly affect the results of the analysis, though.}.

\begin{figure}
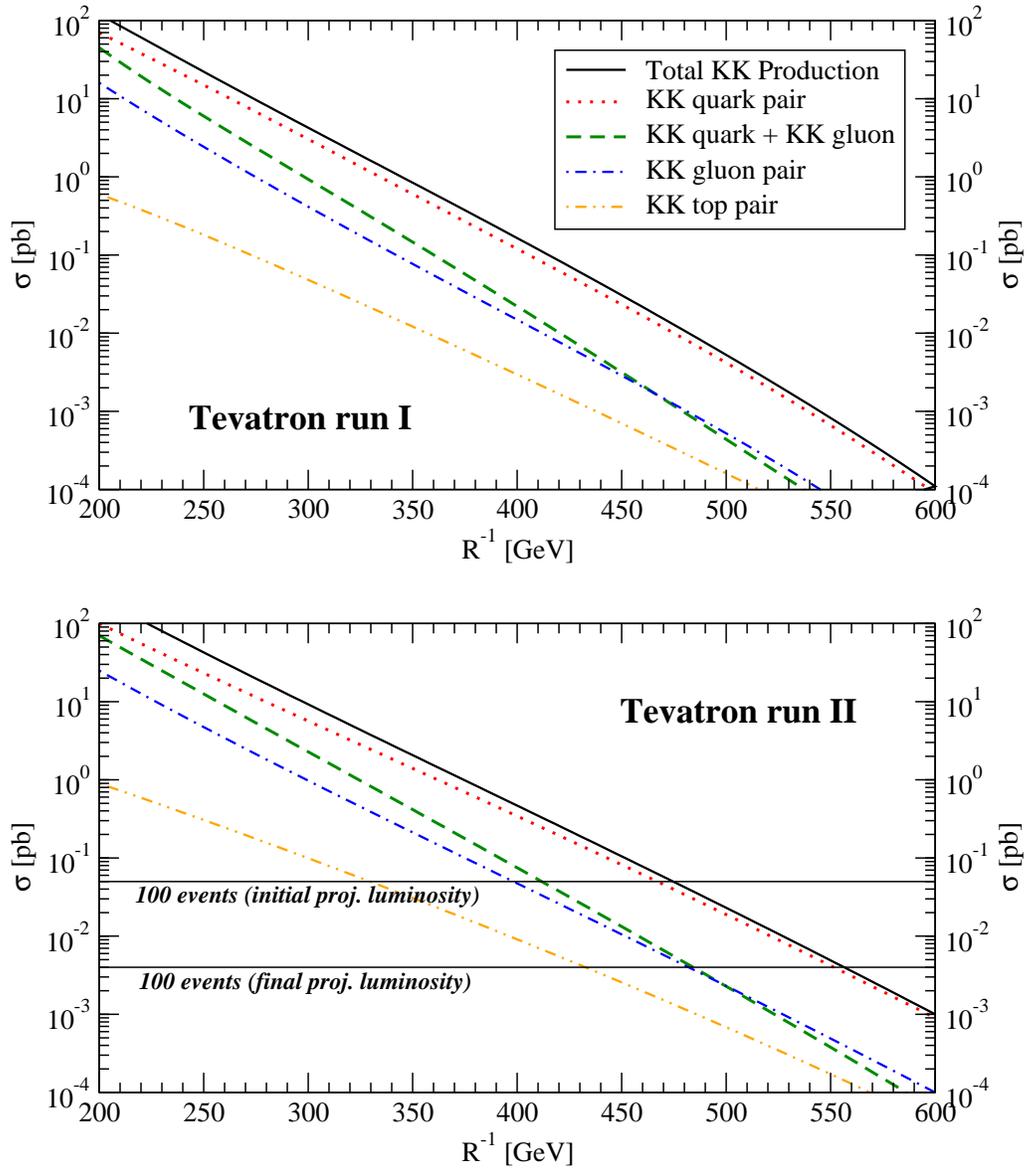

\centering
\includegraphics[width=0.85\textwidth,clip=true]{FIGURES/tevatronI.eps}\\[0.5cm]
 \includegraphics[width=0.85\textwidth,clip=true]{FIGURES/tevatronII.eps}
\caption{The production cross section of KK quarks and gluons at the Tevatron Run I (upper panel) and Run II (lower panel). The solid black curve represents the total production cross section, while the other lines show the separate contributions from KK quark pairs (red dotted line) KK quark-gluon (green dashed line) and KK gluon pairs (blue dot-dashed line). The KK top production cross section is indicated by the orange double-dot-dashed line. Adapted from Ref.~\cite{Macesanu:2002db}.}
\label{fig:prodxsectevatronone}
\end{figure}

\begin{figure}
\centering
\includegraphics[width=0.6\textwidth,clip=true]{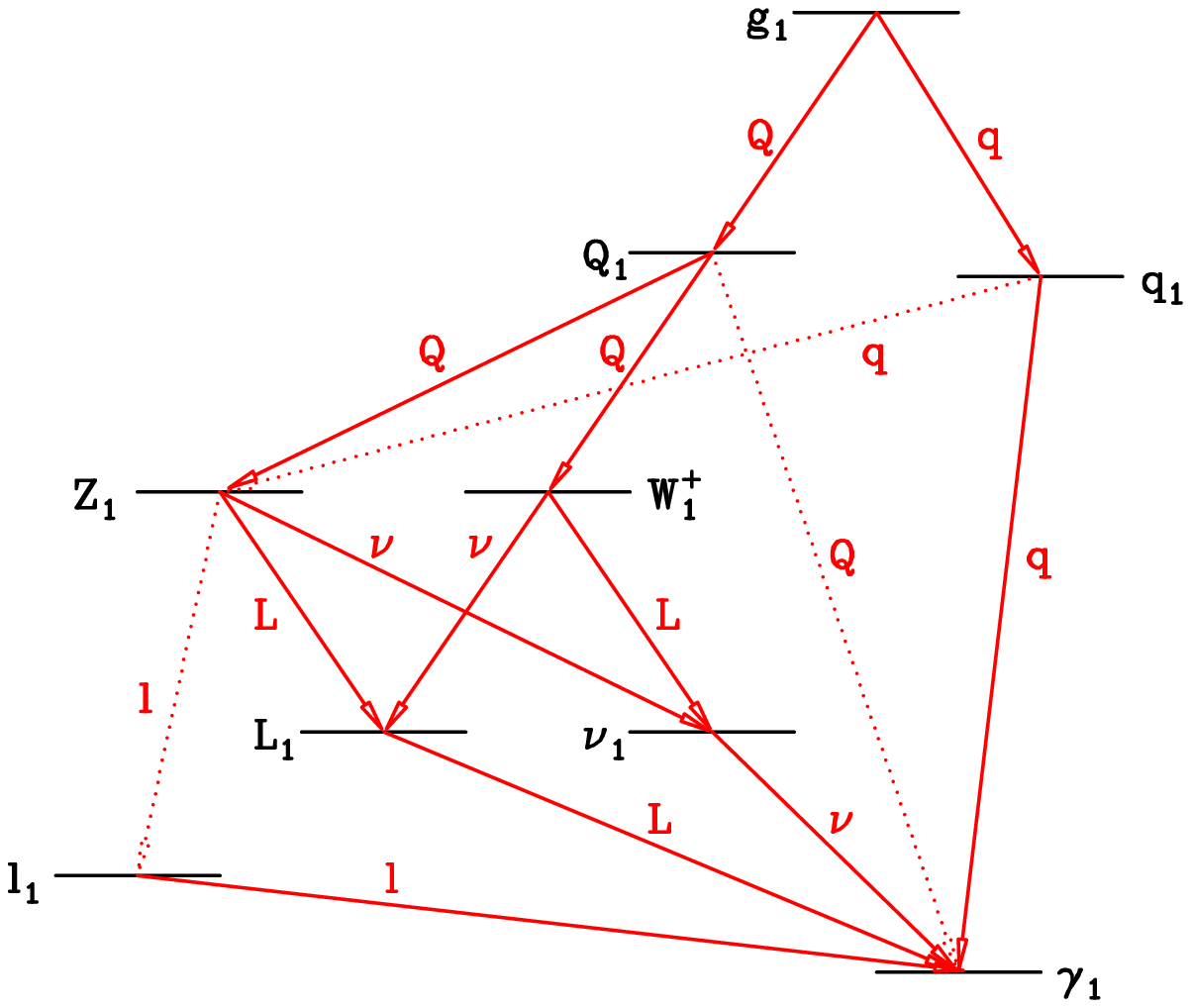} \hspace{1.0cm}
\caption{The $n=1$ KK decay chain, showing the dominant (solid) and sub-dominant (dotted) transitions and the resulting decay products. Adapted from Ref.~\cite{Cheng:2002ab}.}
\label{fig:decaychain}
\end{figure}

The heavier $n=1$ states cascade decay into the stable LKP, $B^{(1)}$, emitting soft SM particles. The LKP then escapes detection, leading to a missing energy signature. The KK modes which should be most abundantly produced at a hadron collider are KK quarks and gluons. KK parity dictates that KK states are pair produced (see Sec.~\ref{ch:darkmatter}). The production cross sections for KK modes at the Tevatron were first computed in Refs.~\cite{Rizzo:2001sd,Macesanu:2002db}\footnote{We collect in App.~\ref{sec:xsechad} the production cross sections for strongly interacting KK states at hadron colliders.}. We show the cross section for the production of a pair of KK quarks and/or gluons at the Tevatron Run I in Fig.~\ref{fig:prodxsectevatronone}. Once produced, the heavier strongly interacting KK states follow a homogeneous pattern in their decay chains (unlike, e.g., the wide range of possibilities in the case of generic supersymmetric spectra). We show a typical instance in Fig.~\ref{fig:decaychain}. In the figure, the dominant decay channels are represented by solid lines, and the sub-dominant by dotted lines. 

\begin{figure}
\centering
\includegraphics[width=0.6\textwidth,clip=true]{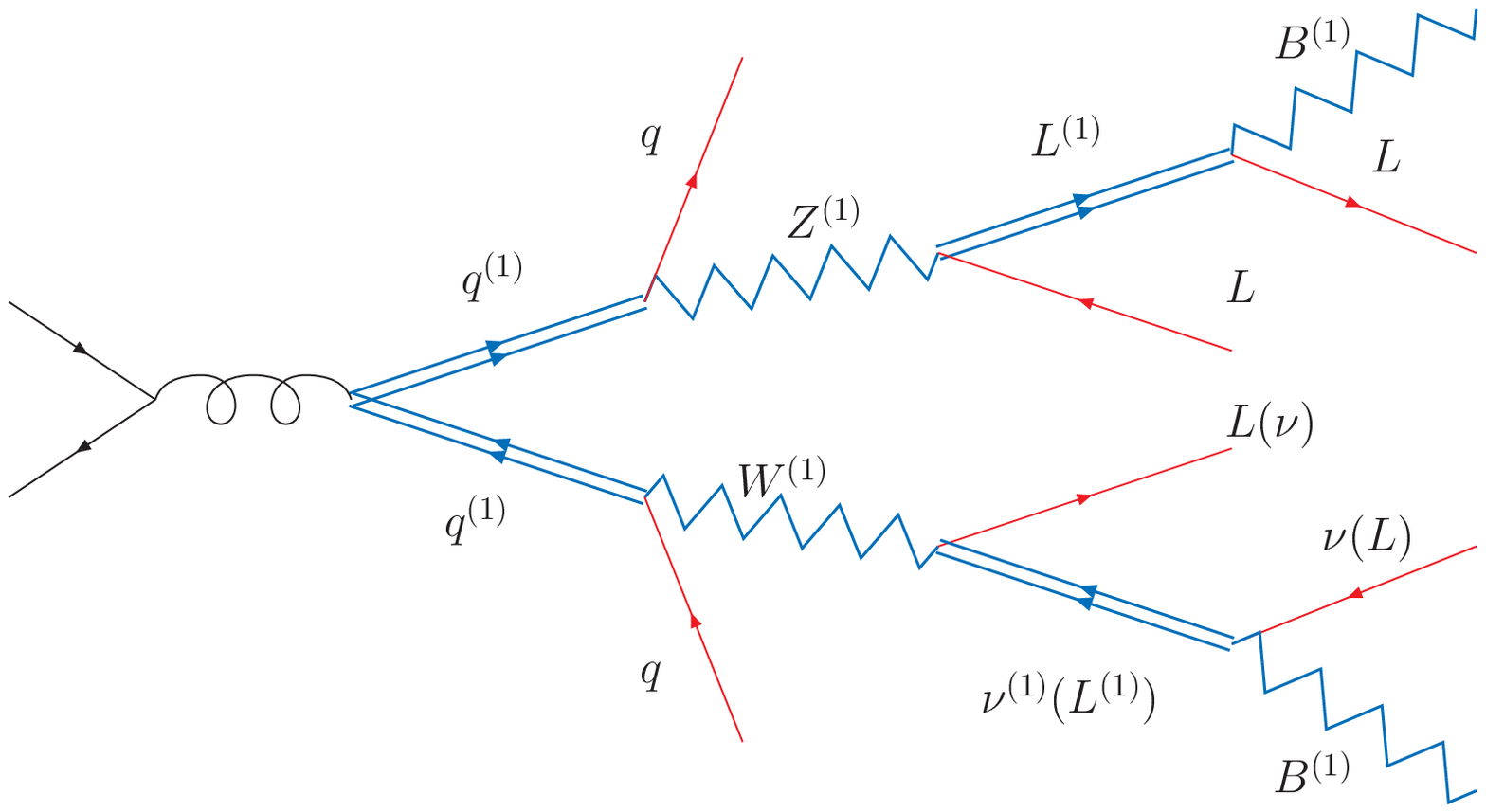} \hspace{1.0cm}
\caption{The Feynman diagram leading to the multi-lepton plus 2 jets plus missing energy signature in KK SU(2) double associated production.}
\label{fig:feyndecay}
\end{figure}

The heaviest KK mode, the KK gluon, decays with almost 100\% branching ratio (BR) to a KK quark and an ordinary quark. In particular, $BR(g^{(1)}\rightarrow Q^{(1)}Q)\simeq0.5$, and $BR(g^{(1)}\rightarrow u^{(1)}u)+BR(g^{(1)}\rightarrow d^{(1)}d)\simeq0.5$ \cite{Cheng:2002ab}. The SU(2) singlet KK quarks then decay into the zero-mode (ordinary) quarks and an LKP, $B^{(1)}$, resulting in a jets plus missing transverse energy signature. The SU(2) doublet KK quarks, $Q^{(1)}$, on the other hand, will mostly decay to the zero-mode quarks and a KK $W^{(1)}$ or a $Z^{(1)}$ gauge boson. The latter subsequently decay to KK leptons plus ordinary leptons. In their turn, KK leptons will decay to an ordinary lepton and a $B^{(1)}$. The resulting signature would then be 3 leptons plus 2 jets plus missing energy (from the $W^{(1)}Z^{(1)}$ channel), or 4 leptons plus 2 jets plus missing energy (from the $Z^{(1)}Z^{(1)}$ channel). Fig.~\ref{fig:feyndecay} illustrates the Feynman diagram leading to the multi-lepton plus 2 jets plus missing energy signature in KK SU(2) double associated production.

The analysis carried out in Ref.~\cite{Lin:2005ix} only considers the multi-lepton channels which contribute to the signal search in the Tevatron data, effectively disregarding final states which include taus. Furthermore, it was assumed in the Monte Carlo simulations for the signal that the relative branching ratio of KK SU(2) doublets into $W^{(1)}$ versus that into $Z^{(1)}$ is $4:1$. The resulting KK gauge bosons are then forced to decay into those channels included in the signal analysis (see Tab.~4.1 of Ref.~\cite{Lin:2005ix}).

Although the largest final state signature for UED at the Tevatron would be jets plus missing energy, the SM background suppression requires one to resort to the much cleaner multi-lepton plus missing energy signature. In this latter case, the main sources of background are $b\bar b/c\bar c$,  $t\bar t$, $WZ$ and $ZZ$ production, as well as Drell-Yan and $WW$ production with misidentified leptons. The blind analysis procedure followed in Ref.~\cite{Lin:2005ix} consists of identifying the data sample containing three leptons passing certain criteria in the CDF dataset, generating both the SM and the signal events through Monte Carlo simulation, and passing it through the CDF detector simulator. Selection criteria are then adopted to enhance the signal and to suppress the background, eventually obtaining an upper limit on the number of UED signal events at a given confidence level from the number of observed data events. Finally, this is translated into an upper limit on the production cross section, and on the size of the compactification radius.

The search for three or more charged leptons and missing transverse energy was performed with a total integrated luminosity of $87.5\ {\rm pb}^{-1}$ collected during CDF's Run IB period, resulting in a production cross section limit for SU(2) KK quark pairs at 95\% (90\%) Confidence Level (CL) of 3.3 pb (2.5 pb). This translates into a lower limit on the inverse extra-dimensional radius of $1/R>$270 GeV (280 GeV). A limit for the total KK quark pair plus KK gluon-quark plus KK gluon pair production cross section of 7.9 pb (6.0 pb) was found, which translates into a limit of $1/R>$280 GeV (290 GeV).

\subsection{Other Phenomenological Constraints}

Physics beyond the SM can manifest itself not only via direct production of new particles at colliders, but also indirectly, e.g.~contributing to precision electroweak data, as discussed in Sec.~\ref{sec:ewprecision}, or to other processes or observables such as rare decays, flavor physics, or the anomalous magnetic moment of the muon\footnote{See e.g. the recent review \cite{Ramsey-Musolf:2006vr} on low energy precision tests of supersymmetric models.}. In this section, we summarize the effect this latter class of phenomenological constraints has on the UED model.

\begin{figure}
\centering
\psfragscanon

  \psfrag{bsgammabsgammabsg}{ $Br(B\rightarrow X_s \gamma )\times 10^{4}$}
  \psfrag{rinvrinv}[][]{ \shortstack{\\  $R^{-1}$ [GeV] }}
  \includegraphics[scale=1]{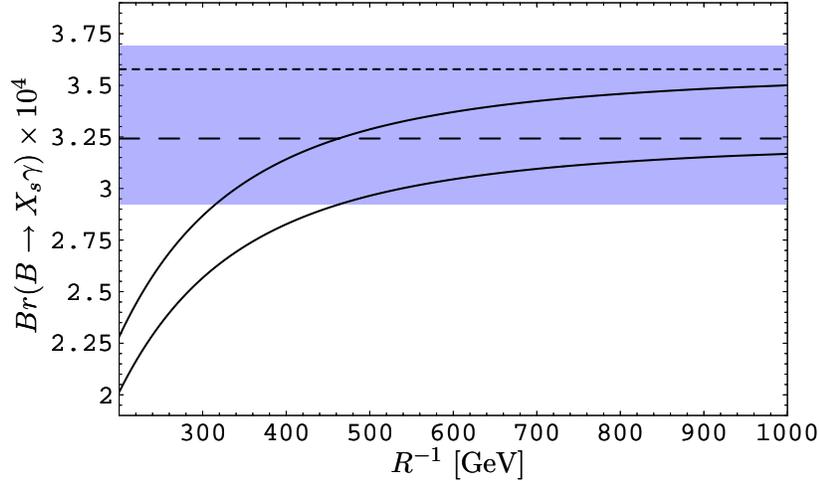}
\caption{The branching ratio for $b\rightarrow s\gamma$ and $E_\gamma>1.6$ GeV as a function of the inverse compactification radius, $1/R$. The grey band represents the experimental result, the dashed lines the SM values, and the solid lines the UED results. The upper (lower) curves make use of a value for the ratio $m_c/m_b=0.22$ ($m_c/m_b=0.29$), and all of the curves feature an estimated theoretical uncertainty of order 10\%. From Ref.~\cite{Buras:2003mk}.}
\label{fig:bsg}
\end{figure}

The inclusive $b\rightarrow s\gamma$ branching ratio is known to provide extremely stringent bounds on physics beyond the SM, for instance low energy supersymmetry \cite{Bertolini:1990if}. The contribution of UED KK states to $b\rightarrow s\gamma$ was first evaluated in Refs.~\cite{Agashe:2001ra,Agashe:2001xt}. The effective Hamiltonian for $\Delta S=1$ $B$ meson decays is given by
\be
{\cal H}_{\rm eff}=\frac{4G_F}{\sqrt{2}}V_{tb}V_{ts}^*\sum_{j=1}^{8}C_j(\mu){\cal O}_j
\ee
where, at leading order, the operator relevant for the transition $b\rightarrow s\gamma$ is
\be
{\cal O}_7=\frac{e}{16\pi^2}m_b\bar s_{L\alpha}\sigma^{\mu\nu}b_{R\alpha}F_{\mu\nu}.
\ee
In UED, the contributions to the relevant coefficient $C_7$ come both from loops involving the KK top quark and (1) the KK $W^{(n)}$ tower and (2) the KK states of charged (physical) would-be-Goldstone bosons. Contribution (2) is dominant, and gives
\be
C_7^{\rm UED}\sum_n\approx\frac{m_t^2}{m_t^2+(n/R)^2}\left(B\left(\frac{m_t^2+(n/R)^2}{(n/R)^2}\right)-\frac{1}{6}A\left(\frac{m_t^2+(n/R)^2}{(n/R)^2}\right)\right),
\ee
where the loop functions are given by
\be
A(x)&=&x\left(\frac{\frac{2}{3}x^2+\frac{5}{12}x-\frac{7}{12}}{(x-1)^3}-\frac{\left(\frac{3}{2}x^2-x\right)\ln x}{(x-1)^4}\right),\\
B(x)&=&\frac{x}{2}\left(\frac{\frac{5}{6}x-\frac{1}{2}}{(x-1)^2}-\frac{\left(x-\frac{2}{3}\right)\ln x}{(x-1)^3}\right).
\ee
Combining theory and experimental 2$\sigma$ errors in quadrature, the 95\% CL constraint on the contribution of KK states, with $\mu_b\rightarrow m_b$, can be cast as
\be
\big|\big[C_7^{\rm total}(\mu_b)\big]^2/\big[C_7^{\rm SM}(\mu_b)\big]^2-1\big|\lesssim0.36
\ee
In the limit of $m_t\ll (1/R)$, Ref.~\cite{Agashe:2001xt} finds the approximate expression
\be
\sum_n m_t/(m_t+(n/R)^2)\lesssim 0.5.
\ee
Using the exact formul\ae, the resulting bound on the size of the inverse compactification radius from $b\rightarrow s\gamma$ quoted in Ref.~\cite{Agashe:2001xt} reads $1/R\gtrsim 280$ GeV. A refined analysis of the UED contributions to the processes was then carried out in Ref.~\cite{Buras:2003mk}. The latter includes a next-to-leading order (NLO) treatment of the QCD corrections, which has two important consequences: first, depending on the scale $\mu_b$ the enhancement of the NLO result with respect to the LO one can be larger than a factor 2; and, secondly, and perhaps most importantly, the dependence upon the $\mu_b$ scale is vastly suppressed with respect to the LO result, and amounts, for reasonable values of $\mu_b$, to not more than 1.5\%. The final result of the analysis is shown in Fig.~\ref{fig:bsg}, where the grey band represents the experimental result, the dashed lines the SM result, and the solid lines the UED result; the upper (lower) curves make use of a value for the ratio $m_c/m_b=0.22$ ($m_c/m_b=0.29$), and all of the curves feature an estimated theoretical uncertainty of order 10\%. As can be deduced from the figure, taking into account the theory uncertainty, one can conclude that consistency with $b\rightarrow s\gamma$ implies $1/R\gtrsim 250$ GeV \cite{Buras:2003mk}, a value slightly more conservative than the one quoted in Ref.~\cite{Agashe:2001xt}. The more recent analysis carried out in \cite{Colangelo:2006vm}, making use of the exclusive branching ratio $B\rightarrow K^*\gamma$, confirmed the importance of the detailed knowledge of the relevant form factors, and shows that the experimental constraint, under conservative assumptions, can be translated into a bound of $1/R\gtrsim250$ GeV.

Another precision electroweak observable which is sensitive to physics beyond the SM is the $Z b\bar b$ vertex \cite{Akhundov:1985fc,Bernabeu:1987me,Bernabeu:1990ws}. In the context of UED, corrections to the $Z b\bar b$ vertex were first considered in the seminal paper of Ref.~\cite{Appelquist:2000nn}, and then re-examined in Refs.~\cite{Oliver:2002up} and \cite{Buras:2002ej}. Shifts in the $Z b\bar b$ coupling due to radiative corrections, either from the SM or from new physics, affect observables such as the branching ratio $R_b\equiv\Gamma_b/\Gamma_h$, where $\Gamma_b\equiv\Gamma(Z\rightarrow b\bar b)$ and $\Gamma_h\equiv\Gamma(Z\rightarrow{\rm hadrons})$, or the left-right asymmetry, $A_b$, defined as
\be
A_b\equiv\frac{g_L^2-g_R^2}{g_L^2+g_R^2},
\ee
where, in turn, the couplings
\be 
g_L&=&-\frac{1}{2}+\frac{s_w^2}{3}+\delta g_L^{\rm SM}+\delta g_L^{\rm NP} \\
g_R&=&\frac{s_w^2}{3}+\delta g_R^{\rm SM}+\delta g_R^{\rm NP}
\ee
enter in the $Z b\bar b$ vertex as
\be
\frac{g}{c_w}\bar b\gamma^\mu\left(g_LP_L+g_RP_R\right)bZ_\mu.
\ee
Both within the SM and within most of its extensions, the largest corrections, scaling with the square of the top quark mass, affect the $g_L$ coupling. A shift $\delta g_L^{\rm NP}$ then entails a shift in $R_b$ given by
\be
\delta R_b=2R_b(1-R_b)\frac{4g_R^2g_L}{g_L^2+g_R^2}\delta g_L^{\rm NP},
\ee
and a shift in $A_b$ of
\be
\delta A_b=\frac{4g_R^2g_L}{\left(g_L^2+g_R^2\right)^2}\delta g_L^{\rm NP}.
\ee
The dominant contributions to $\delta g_L^{\rm NP}$ within UED come from loops containing charged KK Higgses and top quarks. Summing over all possible loop corrections yields the result~\cite{Oliver:2002up}
\be\label{eq:rb}
\delta g_L^{\rm UED}\approx\frac{\sqrt{2}G_Fm_t^4R^2}{192}.
\ee
Using the SM and experimental input values for $R_b$, one gets from Eq.~(\ref{eq:rb}) a bound of $1/R\gtrsim230$ GeV from $R_b$, and a much weaker bound from $A_b$; however, sub-leading contributions in $m^2_W/m^2_t$, stemming from loops involving $W^{\pm(n)}_\mu$ and $W^{\pm(n)}_5$, give an extra contribution which pushes the limit on the compactification inverse radius from $R_b$ up to $1/R\gtrsim300$ GeV, according to the estimate of Ref.~\cite{Oliver:2002up}.

Loops of KK quarks and KK gauge and Higgs bosons also contribute to observables related to the unitarity triangle, as well as to rare $K$ and $B$ decays. The impact of UED on the $B^0_d-\bar B^0_d$ mixing was first addressed in Ref.~\cite{Chakraverty:2002qk}, and re-evaluated, together with the UED corrections to other quantities relevant to the unitarity triangle (including e.g. $\epsilon_K$, $\Delta M_s$, $\Delta M_K$), in Ref.~\cite{Buras:2002ej}. The conclusion of the analysis of Ref.~\cite{Chakraverty:2002qk} was that with the current $1\sigma$ experimental range on $\Delta M_d$, the mass difference between the $B^0_d$ mesons, one can derive a constraint of $1/R\gtrsim165$ GeV. If future measurements of the $B$ meson decay constant, $f_B$, and the QCD correction parameter, $B_B$, were able to reduce the errors on these quantities by one third (and yielded the same central value for the combination $f_B\sqrt{B_B}$), this could lead to a $1\sigma$ bound on the inverse compactification radius as large as $1/R\gtrsim740$ GeV. Ref.~\cite{Buras:2002ej}, however, pointed out that analyzing all UED contributions to the unitarity triangle (UT), values of $1/R\sim200$ GeV are perfectly compatible with the present fits of the UT, and that an improvement in the accuracy of the determination of $f_B\sqrt{B_B}$ is not enough to derive more stringent bounds on $1/R$. Rather, an improvement on both $f_B\sqrt{B_B}$ {\em and} $|V_{td}|$, if determined through $\Delta M_d/\Delta M_s$, can yield an improved maximal or minimal bound on $1/R$ sensitively depending upon the maximal or minimal value of $f_B\sqrt{B_B}$.

Ref.~\cite{Buras:2002ej} pointed out that the short distance one-loop functions relevant for the rare decay modes $K^+\rightarrow\pi^+\nu\bar\nu$, $K_L\rightarrow\pi^0\nu\bar\nu$ and $B\rightarrow X_{s,d}\nu\bar\nu$ are larger than the SM value by around 10\%, when the compactification scale is set to $1/R\approx300$ GeV, due to KK contributions to the $Z$ penguin diagrams. For the decay modes $K^+\rightarrow\pi^+\nu\bar\nu$, $K_L\rightarrow\pi^0\nu\bar\nu$ and $B\rightarrow X_{d}\nu\bar\nu$ this enhancement is partly compensated by the fact that $|V_{td}|_{\rm UED}<|V_{td}|_{\rm SM}$, and that $\bar\eta_{\rm UED}<\bar\eta_{\rm SM}$, resulting in the enhancement of the branching ratios for the aforemenitoned rare decays with respect to the SM expectation of 9\%, 10\% and 12\% respectively, for $1/R\approx300$ GeV~\cite{Buras:2002ej}. As $B\rightarrow X_{s}\nu\bar\nu$ is governed by the CKMM matrix element $|V_{ts}|$, that is common to the SM and UED, the projected enhancement of the BR over the SM prediction is around 21\% ~\cite{Buras:2002ej}. 

Analogously, the short distance one-loop function relevant for the $B_{d,s}\rightarrow\mu^+\mu^-$ and $K_L\rightarrow\mu^+\mu^-$ decay is enhanced by around 15\% in UED. This results in a 33\% enhancement of BR$(B_{s}\rightarrow\mu^+\mu^-)$, 23\% of BR$(B_{d}\rightarrow\mu^+\mu^-)$ and 20 \% of BR$(K_L\rightarrow\mu^+\mu^-)$ for the UED case over the SM. The analysis of the exclusive decay modes $B\rightarrow Kl^+l^-$ and $B\rightarrow K^*l^+l^-$ carried out in Ref.~\cite{Colangelo:2006vm} confirms the enhancement with respect to the SM, and point out that in general no constraints can be drawn on $1/R$ from those decay modes. The subsequent analysis of Ref.~\cite{Colangelo:2006gv} considered the branching fractions and the lepton polarization asymmetries for the rare decay modes $B\rightarrow X_s\tau^+\tau^-$ and $B\rightarrow K^{(*)}\tau^+\tau^-$, focusing in particular on the transverse asymmetries. They find that for the exclusive transitions the uncertainties due to the hadronic matrix elements are small, and in the large energy limit for the light hadron in the final state, the polarization asymmetries are basically free of hadronic uncertainties.

The branching ratio $B\rightarrow X_s\mu^+\mu^-$ has also been investigated in Ref~\cite{Buras:2003mk}. Interestingly enough, UED enhance the SM result by a factor of up to 12\% (for $1/R\simeq300$ GeV), which goes in the right direction, as the data from BELLE \cite{Kaneko:2002mr} indicate that the measured branching ratio is larger than the SM expectation. In view of the current experimental and theoretical uncertainties, however, this is certainly not enough to draw any significant constraint on the UED model. However, it was pointed out in Ref.~\cite{Buras:2003mk} that a quantity
of particular interest for the discrimination of a signature from UED is
the forward-backward asymmetry in the $B\rightarrow X_s\mu^+\mu^-$ decay,
defined as
\be
A_{\rm FB}(\hat s)\equiv\frac{1}{\Gamma(b\rightarrow ce\bar\nu)}\int_1^1{\rm d}\cos\theta_l\frac{{\rm d}^2\Gamma(b\rightarrow s\mu^+\mu^-)}{{\rm d}\hat s{\rm d}\cos\theta_l}{\rm sgn}(\cos\theta_l),
\ee
where 
\be
\hat s=\frac{\left(p_{\mu^+}+p_{\mu^-}\right)^2}{m_b^2}
\ee
and $\theta_l$ is the angle between the $\mu^+$ and $B$ meson momenta in the center of mass frame. The location of $\hat s_0$, the zero of $A_{\rm FB}(\hat s)$ 
is a particularly good observable, as it is sensitive to short distance physics, and it is subject to very small non-perturbative uncertainties. This motivated the authors of Ref.~\cite{Buras:2003mk} to state that future experimental results on $A_{\rm FB}(\hat s)$ could lead to particularly strong tests of the UED model. This point of view has been confirmed by the recent analysis of Ref.~\cite{Colangelo:2006vm}. Furthermore, the experimental results of the BELLE collaboration indicate that the relative signs of the Wilson coefficients $C_7$ and $C_9$ is negative, implying that $A_{\rm FB}$ should have a zero~\cite{Ishikawa:2006fh}. An analogous analysis focusing on the exclusive $\Lambda$ decay was carried out in Ref.~\cite{Aliev:2006xd}, where it was shown that for $1/R\sim300$ GeV the branching ratio is enhanced by around 20\% with respect to the SM result, and where the importance of the forward-backward asymmetry was also emphasized. Ref.~\cite{Aliev:2006gv} studied double-lepton polarization asymmetries in $\Lambda_b\rightarrow \Lambda l^+ l^-$ decays. Ref.~\cite{Aliev:2006gv} showed that, unlike the case of single-lepton polarization, where a discrimination between the SM and UED does not seem feasible, various double-lepton polarization asymmetries in the above mentioned decays can be useful tools for establishing new physics signals as predicted in the UED model, as the predictions for the SM and for UED differ substantially, and to a level that might be experimentally accessible in the foreseeable future.

The branching ratio for the rare decay mode $K_L\rightarrow\pi^0e^+e^-$ is enhanced in UED up to 10\% more than its SM value~\cite{Buras:2003mk}; the latter is however still more than 2 orders of magnitude smaller than the current experimental sensitivity~\cite{Alavi-Harati:2000sk}. More promisingly, the $\epsilon^\prime/\epsilon$ ratio, which parameterizes the size of the direct $CP$ violation with respect to the indirect $CP$ violation in $K_L\rightarrow\pi\pi$ decays,  is typically of the same order, or larger, than the current experimental uncertainty for $1/R\lesssim 400$ GeV. However, Ref.~\cite{Buras:2003mk} pointed out that, even fixing a specific value for $m_s(m_c)$, the uncertainties on the main non-perturbative parameters $B_6^{1/2}$ and $B_8^{3/2}$ is such that no constraints can be drawn model-independently from the experimental results and the UED expectations.

An interesting by-product of the analysis of Ref.~\cite{Buras:2002ej} is that $B$ and $K$ physics can, in principle, distinguish between UED and supersymmetry. For instance, the expectation for the MSSM in the large $\tan\beta$ regime is a {\em suppression} of $\Delta M_s$ with respect to the SM prediction, while a generic prediction of UED is an {\em enhancement}: the ratio $\Delta M_s^{\rm exp}/\Delta M_s^{\rm SM}$ can thus indicate a preference for one option rather than the other. Similarly, in the regime of small $\tan\beta$, the MSSM predicts a suppression of the branching ratios for $K^+\rightarrow\pi^+\nu\bar\nu$, $K_L\rightarrow\pi^0\nu\bar\nu$, $B\rightarrow X_{d}\nu\bar\nu$ and $B_{d}\rightarrow\mu^+\mu^-$, while an enhancement is predicted in the UED scenario, as pointed out above.

Among other flavor changing neutral current effects considered in Ref~\cite{Buras:2003mk}, the UED predictions for the process $b\rightarrow s g$ show very significant departures from the SM result as well as, once again, a strong enhancement due to NLO QCD corrections \cite{Greub:2000an}. However, both the SM and the UED results strongly depend upon $\mu_b$ and on the ratio $m_c/m_b$. In principle, a good determination of the two quantities could allow for a discrimination of UED effects up to $1/R\sim600$ GeV. Unfortunately, however, a significant reduction of the uncertainties related to $\mu_b$ and to $m_c/m_b$ looks problematic, and the extraction of the branching ratio from the experimental data also appears very difficult, if not impossible~\cite{Greub:2000an,Lenz:1997aa}.

The role of the muon $(g-2)$ for UED was first considered in Refs.~\cite{Agashe:2001ra}, \cite{Appelquist:2001jz} and \cite{Rizzo:2001sd}. The contributions to the anomalous magnetic moment of the muon in UED arise from loops involving KK gauge bosons or KK scalars and KK muons or muon neutrinos. The contribution of the $n$-th KK level to the muon $(g-2)$  reads
\be
(g-2)_n=\frac{5e^2m_\mu^2}{128\pi^2(n/R)^2}-\frac{3e^2(2s_w^2-1)m_\mu^2}{32\pi^2c_w^2(n/R)^2}\int_0^1{\rm d}x\int_0^{1-x}{\rm d}y\frac{6x^2+12xy-11x}{1+(x+y)\left(\frac{m_Z}{n/R}\right)^2},
\ee
leading to the approximate result, expanding in powers of the small parameter $\epsilon\equiv (m_ZR)^2$, of \cite{Rizzo:2001sd}
\be
-439\times10^{-11}\left(\epsilon-0.23\epsilon^2+...\right).
\ee
The first thing to notice is that the overall sign is opposite to the potential difference between the experimental data and the SM prediction. Secondly, the absolute value of the contribution of UED is at most $(g-2)_{\rm UED}\simeq-4\times10^{-10}$ which is only one quarter as large as the SM electroweak contribution, and is smaller than the final expected sensitivity of the muon $(g-2)$ experiment at BNL \cite{Brown:2001mg}. In addition, the independent assessment of $(g-2)_{\rm UED}$ carried out in Ref.~\cite{Appelquist:2001jz} finds a final result that is smaller  by a factor of approximately 3 than that found in Ref.~\cite{Rizzo:2001sd}, corroborating the conclusion that the muon $(g-2)$ does not yet provide any useful constraint on the UED scenario.

As a final remark, we wish to comment on the interesting recent result of Ref.~\cite{Bigi:2006vc}, where it was pointed out that the charged lepton flavor violating (CLFV) decays $l_i\rightarrow l_j\gamma$ in UED are suppressed relative to the SM result by factors of $(m_W/m_W^{(n)})^8$, making them an absolutely hopeless channel to observe effects of the presence of UED. Five-dimensional UED can therefore produce observable lepton flavor violation only if explicitly flavor-violating boundary terms are assumed to be present. Finally, CLFV decays in the context of UED with more than one extra-dimension were discussed in Ref.~\cite{He:2002pv}.

We give a summary of the constraints on the size of the extra-dimensional radius from flavor physics and other low-energy probes of new physics discussed in this section in Tab.~\ref{tab:flavourconstraints}.

\begin{table}[!h]
\begin{center}
\begin{tabular}{ c |c| c}
 & Limit on $1/R$ & $\quad$ Ref. $\quad$  \\[0.3cm]
\hline
&&\\[-0.1cm]
BR($b\rightarrow s\gamma$) & $\gtrsim 250$ GeV & \cite{Buras:2003mk}\\[0.3cm]
BR($B\rightarrow K^*\gamma$) & $\gtrsim 250$ GeV & \cite{Colangelo:2006vm}\\[0.3cm]
$R_b\quad\left(\frac{\Gamma(Z\rightarrow b\bar b)}{\Gamma(Z\rightarrow{\rm hadrons})}\right)$ & $\gtrsim 230$ GeV & \cite{Oliver:2002up}\\[0.5cm]
$R_b\quad$(plus sub-lead. contrib.) & $\gtrsim 300$ GeV & \cite{Oliver:2002up}\\[0.3cm]
$\Delta M_d\quad \left(m_{B^0_d}-m_{\bar B^0_d}\right)$ & $\gtrsim 165$ GeV & \cite{Chakraverty:2002qk}\\[0.3cm]
\hline

\end{tabular}
\vspace*{1cm}
\caption{\label{tab:flavourconstraints} A summary of the constraints on the 5-dimensional UED model from rare decays and flavor physics.}
\end{center}
\end{table}

\clearpage
\newpage
\section{Kaluza-Klein Dark Matter}\label{ch:darkmatter}


Particle dark matter candidates must fulfill an array of phenomenological constraints pertaining to their interactions with ordinary matter, dictated by the results of dark matter search experiments and other considerations. In particular, in the range of particle masses of relevance here (above $\sim300$ GeV and below or around the TeV scale) a dark matter candidate which is charged under strong or electromagnetic interactions would become bound with ordinary matter, forming anomalously heavy isotopes. Negative results of searches for anomalously heavy isotopes lead to the exclusion of any strong or electromagnetically interacting dark matter particles in the mass range of relevance here \cite{Rich:1987jd,Hemmick:1989ns}. This rules out the KK gluon, all KK quarks and the charged KK fermions, as well as the charged KK Higgs and gauge bosons, as the LKP.

Neutral KK SU(2) gauge bosons and Higgs bosons are unlikely to be the lightest KK particle since their masses squared are shifted by their large zero mode masses with respect to KK neutrinos and the KK photon (see Sec.~\ref{sec:spectrum}). Although departures from the spectrum described in Sec.~\ref{sec:spectrum} could potentially change this conclusion, it is somewhat difficult to envision a theoretically motivated setup where the bulk corrections and the boundary terms overturn this natural hierarchy. As shown in Ref.~\cite{Cembranos:2006gt}, one such scenario is possible if the SM Higgs mass is assumed to be very heavy ($m_h\gtrsim230-250$ GeV). In this case, depending upon the value of the compactification radius, and assuming, again, a minimal UED setup for the radiative mass corrections, the LKP can be the charged Higgs boson, $H^{\pm(1)}$. This latter option, as explained above, is phenomenologically excluded, however, unless special, very low reheating temperature ranges, $T_{\rm RH}\lesssim 1$ GeV, and large compactification radii, $R\gtrsim1$ TeV, are assumed \cite{Cembranos:2006gt}. In any case, the $H^{\pm(1)}$ relic density would be insufficient to be a significant contributor to the overall dark matter budget.

The only KK fermion which could play the role of a dark matter particle is the KK neutrino. Relaxing the restriction that SU(2) charged fields are heavier than their singlet counterparts (as it is the case in the minimal setup \cite{Cheng:2002iz}), one can consider the possibility that one of the KK neutrinos is the LKP \cite{Servant:2002aq}. The relic abundance of KK neutrinos was computed in Ref.~\cite{Servant:2002aq} in four different setups: with one and three neutrino flavors, and with or without coannihilations with the left handed KK leptons, taken to be degenerate in mass with the KK neutrinos. While most coannihilations play a relatively small role here, it turns out that the number of flavors is a critical input, as extra $Z^{(1)}$ $t$-channel exchanges can occur between neutrinos with different flavors. As a result, the overall effective cross section is increased, and the range of masses compatible with the observed cold dark matter abundance is shifted toward larger values. Namely, the acceptable mass range for one KK neutrino flavor is between 0.8 and 1.0 TeV, while with three flavors the favored mass range lies between 1.1 and 1.3 TeV.

As discussed in Sec.~\ref{sec:spectrum}, the constraints on a KK neutrino LKP from the negative results of elastic scattering experiments such as CDMS and EDELWEISS are particularly strong. Quantitatively, we find that the current CDMS limits \cite{cdmssi} and the results of Ref.~\cite{Servant:2002hb} on the KK neutrino-nucleon scattering cross section imply that a KK neutrino must be heavier than around 1200 TeV if it is to be a viable particle candidate to make up the dark matter of our universe. This is clearly at odds with the range favored by thermal relic abundance considerations. As is well known, thermal production in a standard cosmological setup is not the only mechanism available to produce the dark matter in the early universe. Alternatives include non-thermal production and late time entropy injection. Such alternatives can in principle enable a very heavy relic, which would be produced in the standard scenario in excessive amounts, to be brought into accord with the desired cold dark matter abundance (for example, see the recent analysis of Ref.~\cite{Gelmini:2006pq}). Nevertheless, an inverse compactification radius of ${\mathcal O}(10^3)$ TeV, although phenomenologically viable, is certainly not theoretically favored, nor welcome from the point of view of detecting KK states at colliders or at dark matter searches other than elastic scattering experiments. We therefore conclude that a KK neutrino LKP is at present not a favored option.

The choice of LKP which has received the greatest interest is by all means the $B^{(1)}$, sometimes called the KK photon. In this section, we discuss at length the phenomenology of a $B^{(1)}$ LKP from the point of view of its implications for cosmology and dark matter searches. It is worth bearing in mind, however, that another type of LKP may also constitute a viable dark matter candidate, the KK graviton. We devote Sec.~\ref{sec:kkgraviton} to a discussion of the cosmology and phenomenology of KK gravitons\footnote{Another possibility for dark matter within the context of extra dimensional models in which scalar fields are allowed to propagate in the bulk is that of soliton-type states. The existence and stability of such states was demonstrated in Ref.~\cite{Stojkovic:2001qi}. The lightest such state might feature a mass around a TeV, be electrically neutral, stable, and in principle be a viable dark matter candidate.}.

\subsection{Relic Abundance}\label{sec:relicab}

In UED models, KK dark matter (KKDM) states are abundant in the early universe ($T \gg R^{-1}\sim \, \rm{TeV}$), being freely created and annihilated in pairs. As the universe expands and the temperature drops below that needed to produce such states in chemical equilibrium, however, the number density becomes rapidly suppressed. A certain density of stable KK states (here, the $B^{{(1)}}$s) freezes out, and remains in the form of a thermal relic of the universe's hot youth. In this section, we review the calculation which is performed to determine the thermal relic abundance of KK states in the universe today~\cite{kolbturner}.
\begin{figure}[t]
\centering
\includegraphics[width=0.4\textwidth,clip=true]{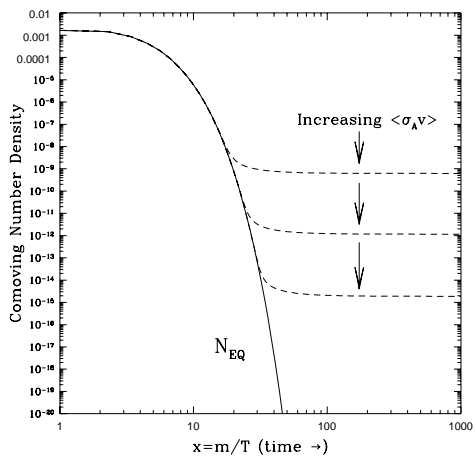}
\caption{The process of the thermal freeze-out of a stable, weakly interacting particle in the early universe. As the temperature drops below the particle's mass ($x \gtrsim 1$), the number density of such particles becomes Boltzmann suppressed. As the universe expands further, eventually these particles encounter no others of their species with which to self-annihilate, leading their density to ``freeze-out''.}
\label{freeze}
\end{figure}

The number density of $B^{(1)}$s evolves according to the Boltzmann equation:
\begin{equation}
\frac{dn_{B^{(1)}}}{dt} + 3 H n_{B^{(1)}}
= -\langle\sigma v\rangle \bigg[(n_{B^{(1)}})^2 - (n^{\rm{eq}}_{B^{(1)}})^2 \bigg], 
\end{equation}
where $H=\sqrt{8\pi\rho/3M_{\rm{pl}}}$ is the Hubble rate and $\langle\sigma v\rangle$ is the $B^{(1)}$s self-annihilation cross section. The equilibrium number density of $B^{(1)}$s is given by:
\begin{equation}
n^{\rm{eq}}_{B^{(1)}} = g \bigg(\frac{m_{B^{(1)}}T}{2\pi}\bigg)^{3/2} \, \rm{exp}\bigg(\frac{-m_{B^{(1)}}}{T}\bigg),
\end{equation}
where $g=3$ is the number of degrees of freedom of the LKP. At $T \gtrsim m_{B^{(1)}}$, the number density of $B^{(1)}$s was very close to its thermal equilibrium value. As the temperature dropped below $m_{B^{(1)}}$, the number density became exponentially suppressed, until eventually the annihilation rate was overcome by the effects of Hubble expansion~(see Fig.~\ref{freeze}).

Numerical solutions of the Boltzmann equation yield a relic density of \cite{kolbturner}: 
\begin{equation}
\Omega_{B^{(1)}} h^2 \approx \frac{1.04 \times 10^9 x_F}{M_{\rm{Pl}} \sqrt{g^*} (a+3b/x_F)},
\label{sol}
\end{equation}
where $x_F = m_{B^{(1)}}/T_F$, $T_F$ is the temperature at freeze-out, $g^*$ is
the number of relativistic degrees of freedom available at freeze-out ($g^* \approx 92$
for the case at hand), and $a$ and $b$ are terms in the partial wave expansion of the
LKP annihilation cross section, $\sigma v = a + bv^2 +\vartheta(v^4)$. Evaluation of $x_F$ leads to:
\begin{equation}
x_F = \ln\bigg[c(c+2) \sqrt{\frac{45}{8}}
\frac{g \, m_{B^{(1)}} M_{\rm{Pl}}  (a+6b/x_F)}{ 2 \pi^3 \sqrt{g^* x_F}}\bigg],
\label{xf}
\end{equation}
where $c$ is an order 1 parameter determined numerically. Note that since $x_F$ appears in the logarithm as
well as on the left hand side of the equation, this expression must be solved by
iteration. WIMPs with masses near the electroweak scale generically freeze-out at temperatures in the range of approximately
$x_F\approx$ 20 to 30.

The $B^{(1)}$ self-annihilation cross section is roughly constant as a function of temperature, and is approximately given by (see Appendix~\ref{sec:ann}):
\begin{equation}
\langle\sigma v\rangle \approx \frac{95 g^4_1}{324 \pi m^2_{B^{(1)}}}.
\label{annlkp}
\end{equation}
This is unlike what is generically expected in the case of supersymmetry: when the lightest neutralino is bino-like, the pair-annihilation into gauge or higgs bosons pairs final states is suppressed, and the main annihilation channel is into fermion pairs through sfermion $t$-channel exchange. As a consequence of the Majorana nature of neutralinos, fermion pair ($f\bar f$) final states feature an $s$-wave amplitude for the pair annihilation cross section suppressed by a factor $m^2_f/m^2_\chi$, where $m_\chi$ is the neutralino mass \cite{Jungman:1995df}. Therefore one expects a sizable variation of the thermally averaged pair annihilation cross section with temperature. In addition, the mentioned suppression affects the pattern of pair annihilation final states branching ratios at $T=0$, relevant for indirect detection: namely, independently of the composition of the neutralino, prompt annihilation in lepton pairs is always greatly suppressed. This is unlike the case of $B^{(1)}$ LKP, where $f\bar f$ final states are not helicity suppressed, but, rather, feature a sizable branching ratio. In the approximation that all heavier level one KK modes have the same mass, the relative annihilation fraction can be determined from simply the hypercharge of the final state fermions. We collect the branching ratios for $B^{(1)}$ pair annihilations in Tab.~\ref{tab:br}: in the first column, $\Delta_i=0$, we assume a completely degenerate KK mass spectrum, while in the second one, $\Delta_{q^{(1)}}=0.14$ the relative mass splitting between the LKP and KK quarks is assumed to be 0.14. Clearly the relative annihilation fractions are not particularly sensitive to the details of the spectrum so long as KK leptons are the same mass or lighter than KK quarks as the one-loop radiative corrections suggest.
\begin{table}
\begin{center}
\renewcommand{\arraystretch}{1.2}\small\normalsize
\begin{tabular}{rcl|ll} \hline
\multicolumn{3}{c}{process} & \multicolumn{2}{|c}{annihilation fraction} \\
& & & $\Delta_i = 0$ & $\Delta_{q^{(1)}} = 0.14$ \\ \hline
$B^{(1)}B^{(1)}$ & $\rightarrow$ & $\nu_e \overline{\nu}_e$, $\nu_\mu \overline{\nu}_\mu$,
                 $\nu_\tau \overline{\nu}_\tau$ & $0.012$ & $0.014$ \\
       & $\rightarrow$ & $e^+e^-$, $\mu^+\mu^-$, 
                 $\tau^+\tau^-$ & $0.20$ & $0.23$ \\
       & $\rightarrow$ & $u\overline{u}$, $c\overline{c}$, 
                 $t\overline{t}$ & $0.11$ & $0.077$ \\
       & $\rightarrow$ & $d\overline{d}$, $s\overline{s}$, 
                 $b\overline{b}$ & $0.007$ & $0.005$ \\ 
       & $\rightarrow$ & $\phi \phi^*$
                 & $0.023$ & $0.027$ \\ \hline
\end{tabular}
\end{center}
\caption{The relative annihilation fraction into various 
final states.  The numbers shown are not summed over generations,
and the Higgs mass was assumed to be lighter than $m_{B^{(1)}}/2$.}
\label{tab:br}
\end{table}

\begin{figure}[t]
\centering
\mbox{\hspace*{-0.5cm}\includegraphics[width=0.52\textwidth,clip=true]{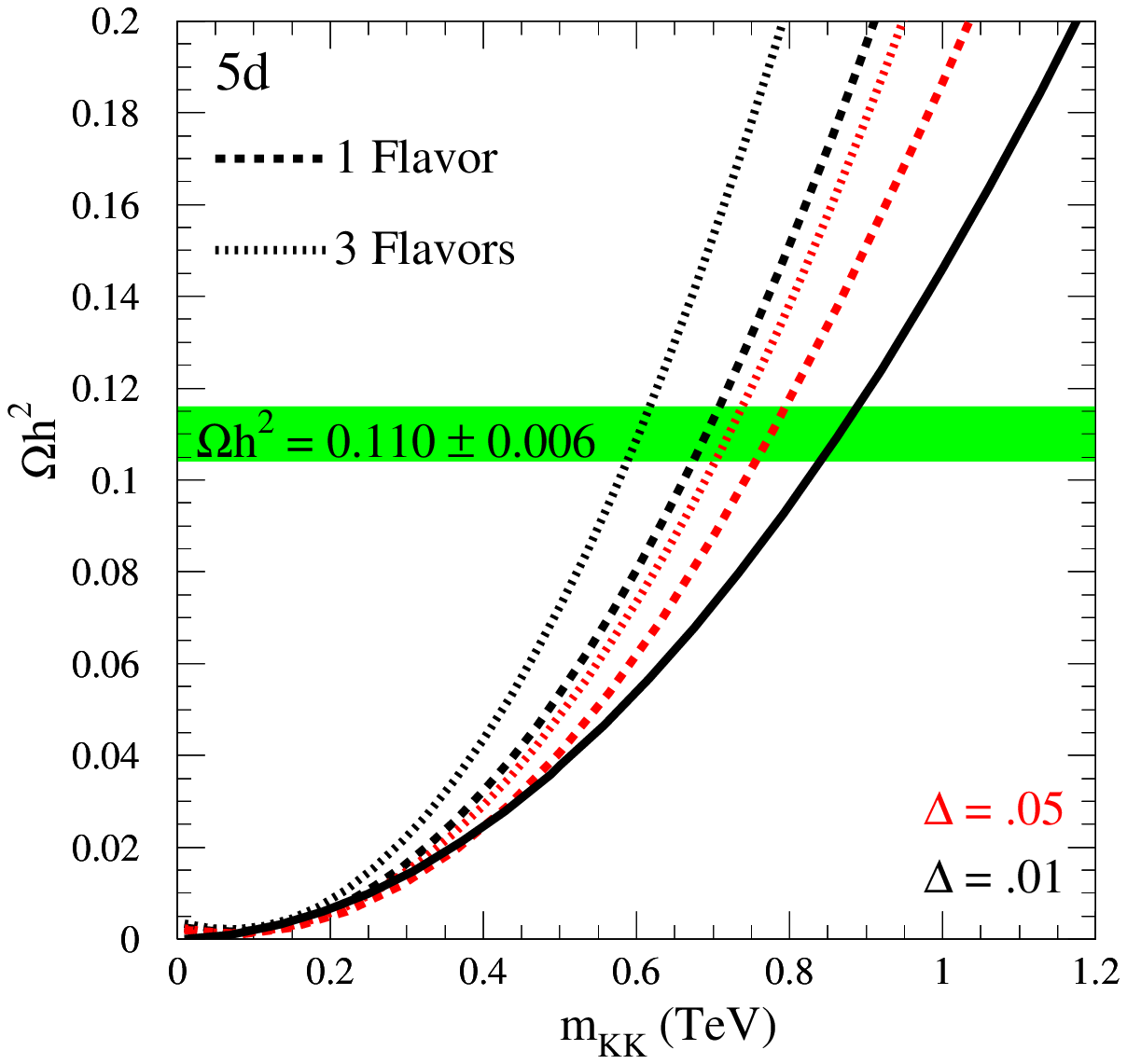}\includegraphics[width=0.52\textwidth,clip=true]{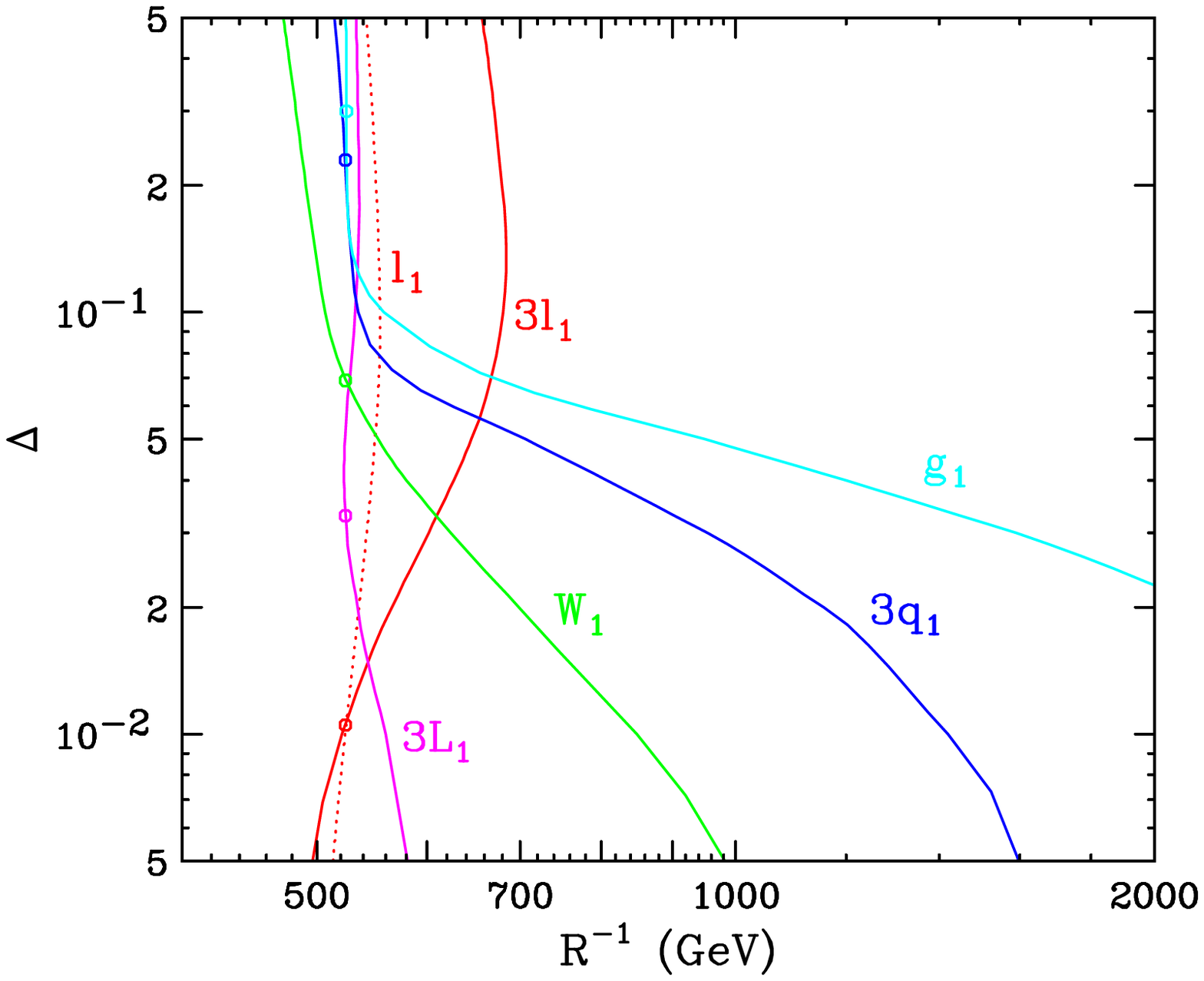}}
\caption{Left: The thermal relic abundance of $B^{(1)}$s without the effects from any other KK species (solid line) and including the effects of KK leptons 5\% and 1\% heavier than the LKP (dashed and dotted lines). Shown as a horizontal band is the measured dark matter abundance from the WMAP experiment \cite{wmap}. Adapted from Ref.~\cite{Servant:2002aq}. Right: The change in the cosmologically preferred value for $R^{-1}$ as a result of varying the different KK masses away from their nominal values in the minimal UED scenario (indicated by the circles on each line). Adapted from Ref.~\cite{Kong:2005hn}.}
\label{relic1}
\end{figure}

When the cross section in Eq.~(\ref{annlkp}) is inserted into Eqs.(\ref{sol}) and~(\ref{xf}),
a relic abundance within the range of the cold dark matter density measured by WMAP
($0.095 < \Omega h^2 < 0.129$) \cite{wmap} can be obtained for $m_{B^{(1)}}$ approximately
in the range of 850 to 900 GeV. Lighter values lead to the LKP over-annihilating and thus under-producing the abundance of dark matter. Heavier values generate more dark matter than is observed (see the solid line in Fig.~\ref{relic1}).

This conclusion can be substantially modified if other KK
modes have masses quasi-degenerate to the LKP and thus contribute to the freeze-out process \cite{Griest:1990kh}. This is expected to be especially important in UED models due to the quasi-degenerate nature of the KK spectrum, leading to many particle species being present at the time of freeze-out. To include the effects of other KK modes in the freeze-out process, we adopt the formalism outlined in Ref.~\cite{Griest:1990kh} to account for co-annihilation processes in the process of thermal freeze-out of species. In Eqs.(\ref{sol}) and~(\ref{xf}), we replace the cross section
($\sigma$, denoting the appropriate combinations of $a$ and $b$) with an effective
quantity which accounts for all particle species involved:
\begin{equation}
\sigma_{\rm{eff}} = \sum_{i,j} \sigma_{i,j} \frac{g_i g_j}{g^2_{\rm{eff}}}
\, (1+\Delta_i)^{3/2} \, (1+\Delta_j)^{3/2} \, e^{-x(\Delta_i+\Delta_j)}. 
\end{equation}
Similarly, we replace the number of degrees of freedom, $g$, with the effective quantity:
\begin{equation}
g_{\rm{eff}} = \sum_{i} g_i \, (1+\Delta_i)^{3/2} e^{-x \Delta_i}.
\end{equation}
In these expressions, the sums are over KK species, $\sigma_{i,j}$ denotes the
coannihilation cross section between species $i$ and $j$, and, finally, the $\Delta$s
denote the fractional mass splitting between that state and the LKP.

To illustrate how the presence of multiple KK species can affect the freeze-out
process, we describe below two illustrative cases. First, consider a case in which the
coannihilation cross section between the two species, $\sigma_{1,2}$, is large
compared to the LKP's self-annihilation cross section, $\sigma_{1,1}$. If the
second state is not much heavier than the LKP ($\Delta_2$ is small), then
$\sigma_{\rm{eff}}$ may be considerably larger than $\sigma_{1,1}$, and thus
the residual relic density of the LKP will be reduced. Physically, this case
represents a second particle species depleting the WIMP's density through
coannihilations. This effect is often found in the case of supersymmetry models
in which coannihilations between the lightest neutralino and another superpartner,
such as a chargino, stau, stop, gluino or heavier neutralino, can substantially
reduce the abundance of neutralino dark matter \cite{Profumo:2004wk}.

The second illustrative case is quite different. If $\sigma_{1,2}$ is comparatively
small, then the effective cross section for two species will tend toward $\sigma_{\rm{eff}} \approx
\sigma_{1,1}\, g^2_1/(g_1+g_2)^2 + \sigma_{2,2}\, g^2_2/(g_1+g_2)^2$. If $\sigma_{2,2}$
is not too large, $\sigma_{\rm{eff}}$ may be smaller than the LKP's self-annihilation
cross section alone. Physically, this scenario corresponds to two species freezing out
quasi-independently, followed by the heavier species decaying into the LKP, thus
enhancing its relic density. An example of this behavior was found in Ref.~\cite{Servant:2002aq}, in which the impact of KK leptons only slightly heavier than the LKP were considered in the freeze-out process. The resulting enhancement to the relic abundance of $B^{(1)}$s is shown in Fig.~\ref{relic1}. Notice that a similar situation can occur in the context of supersymmetry, for instance with slepton coannihilations when the neutralino is higgsino- or wino-like, see e.g. Ref.~\cite{Profumo:2006bx}.

Recently, the LKP freeze-out calculation, including all
coannihilation channels, has been performed by two independent groups~\cite{Burnell:2005hm,Kong:2005hn}. We will briefly summarize their conclusions here. 

As expected, the effects of coannihilations on the LKP relic abundance depend critically
on the KK spectrum considered. If strongly interacting KK states are less than roughly
$\sim 10\%$ heavier than the LKP, the effective LKP annihilation cross section
can be considerably enhanced, thus reducing the relic abundance. KK quarks which are
between 5\% and 1\% heavier than the LKP lead to a $B^{(1)}$ with the measured dark matter abundance over the range of masses of $m_{B^{(1)}} \approx $ 1500 to 2000 GeV (instead of 850-900 GeV for the case of the $B^{(1)}$ alone).
If KK gluons are also present with similar masses, $m_{B^{(1)}}$ as heavy as 2100 to
2700 GeV is required to generate the observed relic abundance.

On the other hand, if the strongly interacting KK modes are
considerably heavier than the LKP, other KK states may still
affect the LKP's relic abundance. If, for example, all three families of KK leptons are
each 1\% more massive than the LKP, then the observed relic abundance is generated for
$m_{B^{(1)}}$ between approximately 550 and 650 GeV. If the KK leptons are instead 5\% more massive than the LKP, the
observed abundance is found for $m_{B^{(1)}}\approx$ 670 to 730 GeV (see Fig.~\ref{relic1}). 

The discussion above is quantitatively illustrated in the right panel of fig.~\ref{relic1}. The plane indicates the value of the compactification radius $R^{-1}$ and $\Delta_i$, the relative mass splitting between the $B^{(1)}$ LKP and a given class of coannihilating KK particles: one (red dotted) or three (red solid) generations of $SU(2)_W$-singlet
KK leptons; three generations of $SU(2)_W$-doublet leptons (magenta);
three generations of $SU(2)_W$-singlet quarks (blue)
(the result for three generations of $SU(2)_W$-doublet quarks
is almost identical); KK gluons (cyan) and
electroweak KK gauge bosons (green). Each line indicates the the value of $R^{-1}$ needed for a given mass splitting $\Delta_i$ to get a $B^{(1)}$ relic abundance $\Omega_{B^{(1)}} h^2=0.1$.

\begin{figure}[t]
\centering
\includegraphics[width=0.48\textwidth,clip=true]{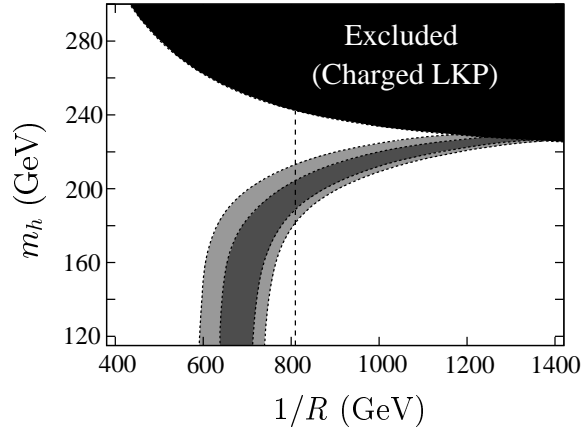}
\caption{Regions, in the $(R^{-1},m_h)$ plane, where the $B^{(1)}$ relic abundance is consistent with the CDM abundance as inferred from the WMAP results, at the 1 and 2-$\sigma$ level (dark and light grey, respectively). The relic abundance computation takes here into account the effect of the second KK level resonances. Adapted from Ref.~\cite{Kakizaki:2006dz}.}
\label{relic2}
\end{figure}
Additionally, it was pointed out in Refs.~\cite{Kakizaki:2005en,Kakizaki:2005uy} that second level KK modes can also play a significant role in the freeze-out process, in particular in the case that $m_{h^{(2)}}$ is very close to $2 m_{B^{(1)}}$. For $m_{\rm{KK}}=$1 TeV, the cross section given in Eq.~(\ref{annlkp}) can underestimate the full value by up to $\sim$15\%, thus increasing the mass range which generates the observed dark matter abundance. If  $(m_{h^{(2)}}-2 m_{B^{(1)}})/2 m_{B^{(1)}}$ is larger than a few percent, however, second KK level particles have a negligible effect on the relic abundance calculation.

Finally, assuming a minimal UED model where the matching conditions from brane-localized terms cancel at the cutoff scale, $\Lambda$, barring second-level KK modes resonances and setting the Higgs mass $m_h=120$ GeV and the cutoff scale $\Lambda R=20$, Ref.~\cite{Kong:2005hn} finds that the preferred range for the compactification scale (and the LKP mass) is $500\lesssim R^{-1}/{\rm GeV}\lesssim600$. In this range, the full computation of the thermal relic density of $B^{(1)}$'s, including all coannihilating particles, falls in the measured range of the cold dark matter abundance. Ref.~\cite{Matsumoto:2005uh} showed, again for a minimal UED setup, that for large Higgs masses, as the mass splitting between the $B^{(1)}$ and the KK Higgses becomes small, coannihilation effects shift the viable region of compactification radii to larger values. For instance, at $m_h=230$ GeV, Ref.~\cite{Matsumoto:2005uh} quotes the range $800\lesssim R^{-1}/{\rm GeV}\lesssim1200$. Including the effect of the second KK level resonances, taking into account all coannihilation processes and varying the Higgs mass (but fixing $\Lambda R=20$), according to the analysis of Ref.~\cite{Kakizaki:2006dz}, the preferred range of $R^{-1}$ for the minimal UED model consistent with the observed cold dark matter abundance is $600\lesssim R^{-1}/{\rm GeV}\lesssim1400$, as illustrated in Fig.~\ref{relic2}.

\subsection{KKDM Protohalos}\label{sec:protohalos}

While freeze-out signals the departure of WIMPs from
{\em chemical} equilibrium, it does \emph{not} signal the end of WIMP
interactions in the early universe.  Elastic and inelastic scattering processes of the
form $X f \rightarrow X f$ or $X f \rightarrow X^{\prime}
f^\prime$ keep the dark matter particle, $X$, in \emph{kinetic}
equilibrium until later times (lower temperatures)
\cite{Boehm,Chen:2001jz,GreenWIMPy}. Here $f$ and $f^\prime$ are
SM particles in the thermal bath (leptons, quarks, gauge bosons) and
$X^{\prime}$ is an unstable particle that carries the same
conserved quantum number as $X$.  The temperature of {\em
kinetic} decoupling, $T_{\rm kd}$, sets the
distance scale at which linear density perturbations in the
dark-matter distribution get washed out --- the small-scale 
cutoff in the matter power spectrum.  In turn, this small-scale
cutoff sets the mass, $M_c$, of the smallest protohalos that
form when these very small-scales go nonlinear at a redshift $z \sim 70$.
There may be implications of this small-scale cutoff for direct
\cite{Diemand:2005vz} and indirect \cite{Ando:2005xg} dark-matter detection (see also Ref.~\cite{Goerdt:2006hp}), although this depends critically upon the degree at which these structures are tidally disrupted through encounters with stars~\cite{Zhao:2005mb,Green:2006hh,Goerdt:2006hp}. For instance, the clumpiness of dark matter on small scales enhances the rates for indirect detection by a factor that can be approximately cast as $f\cdot \delta$ \cite{Bergstrom:1998jj}, where $f$ is the fraction of dark matter mass in substructures, and $\delta$ is the dark matter density contrast,
\be
\delta=\frac{\int{\rm d}^3r_{\rm cl}\left(\rho_{\rm cl}(\vec{r_{\rm cl}})\right)^2}{\rho_0\int{\rm d}^3r_{\rm cl}\ \rho_{\rm cl}(\vec{r_{\rm cl}})}.
\ee
At present, however, no clear-cut, first-principle predictions are available for the size of the {\em boost factor} resulting from a given small-scale cutoff in the dark matter power spectrum. Sufficiently large substructures can be in principle also directly detected with gamma-ray telescopes, as we discuss in Sec.~\ref{sec:gr}.

Ref.~\cite{Profumo:2006bv} showed that the WIMP's kinetic decoupling temperatures can vary over a large range, from several MeV to a few GeV. According to the estimate of Ref.~\cite{Loeb:2005pm},
\begin{equation}\label{eq:minihalos}
     M_c \simeq 30\left(T_{\rm kd}/10\ {\rm
     MeV}\right)^{-3}~{\rm M}_{\oplus},
\end{equation}
which leads to a range of masses for the smallest protohalos (in generic WIMP models thermally producing the right amount of relic cold WIMPs) from $10^{-6}~{\rm M}_{\oplus}$ to  $10^{2}~{\rm M}_{\oplus}$ (${\rm M}_{\oplus}$ denotes the mass of the Earth). Eq.~(\ref{eq:minihalos}) accounts for both the acoustic oscillations imprinted on
the power spectrum by the coupling between the dark
matter and the relativistic particles in the primordial plasma
prior to kinetic decoupling and the cutoff due to free-streaming
of dark matter after kinetic decoupling. In particular, Ref.~\cite{Profumo:2006bv} addressed, together with supersymmetric models, the case of KK dark matter within the context of UED models.

The kinetic decoupling temperature $T_{\rm kd}$ is defined by $\tau_r(T_{\rm kd})=H^{-1}(T_{\rm kd})$
\cite{GreenWIMPy}, where $H(T)$ is the Hubble expansion rate,
and the relaxation time $\tau_r$ is given by
\begin{equation}
    \tau_r^{-1}\equiv \sum_l
    n_{l}(T,m_l)\sigma_{lX}(T)(T/m_X). 
\end{equation}
 Here, $n_{l}(T,m_l)\sim T^3$ is the equilibrium number density
of the relativistic particle species $l$ (the true mass
dependence can be crucial here for some of the species under
consideration such as the $\mu$ and $\tau$ leptons),
$\sigma_{lX}(T)$ is the thermally averaged scattering cross
section of the WIMP $X$ off $l$'s, and the factor $(m_X/T)^{-1}$
counts the number of scatters needed to
keep the WIMPs in kinetic equilibrium. Here, we consider $l \in \{ \nu_{e,\mu,\tau},\ e,\ \mu,\  \tau,\ u,\ d,\ s,\ c\}$ and neglect the scattering off
light quarks below $\Lambda_{\rm QCD}\simeq 150$ MeV (our results are insensitive to the precise value taken for $\Lambda_{\rm QCD}$ and the detailed nature of the QCD phase transition). We neglect the scattering of WIMPs off mesons and baryons below $\Lambda_{\rm QCD}$ because this process is suppressed with respect to their scattering off light leptons by the relative abundance of the species in the thermal bath and by hadronic form factors. 

\begin{figure}
\centerline{\epsfig{file=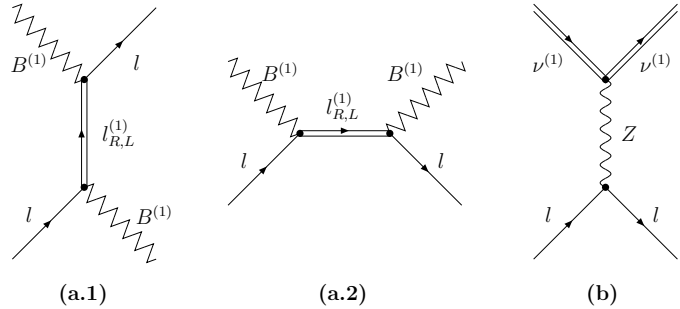,width=3.5in,angle=0}}
\caption{Feynman diagrams contributing to the scattering of $B^{(1)}$ (a.1 and a.2) and of $\nu^{(1)}$ (b) off leptons.} 
\label{fig:feyn}
\end{figure}
Within the context of UED, the scattering cross section of $B^{(1)}$ off light fermions proceeds through the Feynman diagrams shown in Fig.~\ref{fig:feyn}(a). Unlike the case of neutralino dark matter~\cite{Chen:2001jz}, the intrinsically degenerate nature of the KK spectrum, where $m_{B^{(1)}}\approx m_{L^{(1)}}$, clearly enforces a resonant enhancement. Ref.~\cite{Profumo:2006bv} finds, to leading order in $E_l/m_{X}$, and in
the relativistic limit for $l$ and non-relativistic limit for
the LKP particle, that
\begin{eqnarray}
     \sigma_{B^{(1)}l}\simeq 
     \frac{E^2_l}{2\pi}\sum_{R,L}\frac{\left(g_1
     Y_{R,L}\right)^4}{\left(m_{B^{(1)}}^2-m_{l^{(1)}_{R,L}}^2\right)^2},
\end{eqnarray}
where $Y_{R,L}$ denote the hypercharge
quantum number. In analogy with the case of neutralino dark matter, Ref.~\cite{Profumo:2006bv} pointed out that the
$\sigma_{Xl}\propto E^2_l$ scaling~\cite{Chen:2001jz} holds for a $B^{(1)}$ KK dark matter WIMP as well. The case of a KK neutrino was also studied in Ref.~\cite{Profumo:2006bv}, even though the KK neutrino is strongly disfavored as a dark matter particle by direct detection searches. KK neutrinos belong to a category of dark matter candidates which feature the same couplings to the $Z$ gauge bosons as ordinary heavy neutrinos. In this case, one gets
\begin{eqnarray}
     \sigma_{\nu^{(1)}l}&\simeq
     &\frac{\left|g_{\sss\nu^{(1)}\nu^{(1)}Z}\right|^2}{4\pi
     m_Z^4}\left(g_L^2+g_R^2\right)\ E_l^2,
\end{eqnarray}
where $g_{R,L}$ stand for the couplings of the left and right handed lepton, $l$, to the gauge boson, $Z$, and
$g_{\sss\nu^{(1)}\nu^{(1)}Z}=e/(\sin2\theta_W)$. Again, the $\sigma_{Xl}\propto E^2_l$ scaling holds, and, remarkably, $\sigma_{\nu^{(1)}l}$ does not depend on the LKP mass.

\begin{figure}
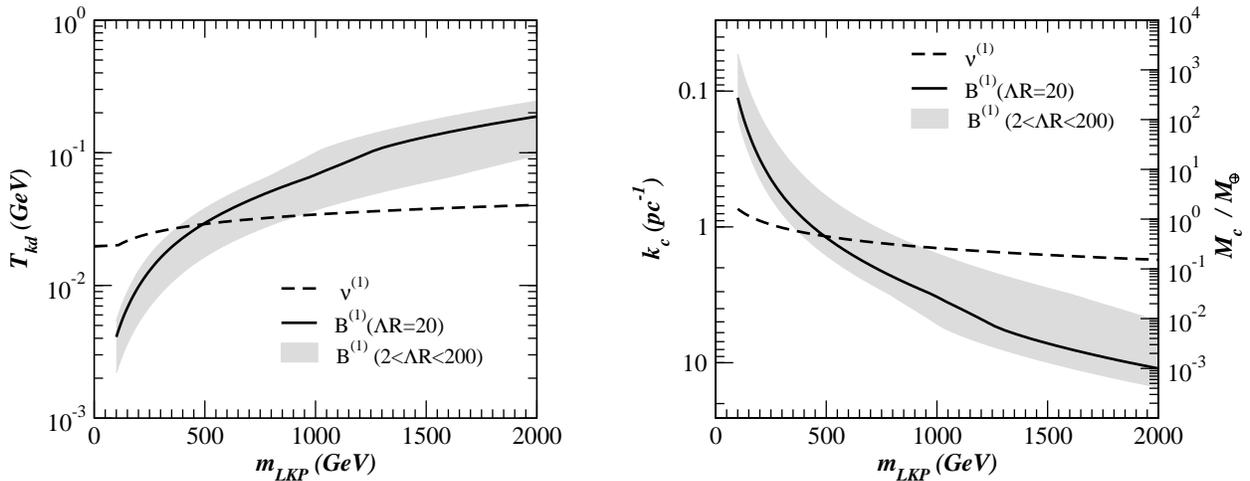

\centering
\mbox{\includegraphics[height=0.4\textwidth,clip=true]{FIGURES/ued_minihalos_2.eps}
\qquad \includegraphics[height=0.4\textwidth,clip=true]{FIGURES/ued_minihalos_1.eps}}
\caption{Left: The kinetic-decoupling temperature, $T_{\rm kd}$, as a function of the LKP mass for minimal UED models featuring a
     $B^{(1)}$ (with $2<\Lambda R<200$, grey shaded region, and for $\Lambda R=20$, solid black line) and a $\nu^{(1)}$ LKP. Right: The WIMP protohalo characteristic comoving wavenumber, $k_c$, (left axis) and mass, $M_c$, (right axis) as a function of the LKP mass.}
\label{fig:ued_minihalos}
\end{figure}
In Fig.~\ref{fig:ued_minihalos} we show the results for $T_{\rm kd}$ (left)
and for $M_c$ (right), respectively, versus the LKP mass. We set the KK spectrum according to the minimal UED prescription for radiative corrections to the KK
masses \cite{Cheng:2002iz} (see Sec.~\ref{sec:spectrum}), setting the cutoff scale $2<\Lambda R<200$, and showing the $\Lambda R=20$ case with a
black solid line. 

We conclude that KK dark matter protohalos feature a kinetic decoupling temperature lying between 10 MeV and a few hundred MeV. The resulting size of the smallest KK dark matter protohalos varies between $10^{-3}$ and $10^3$ Earth masses. Recently, Ref.~\cite{Bringmann:2006mu} also addressed the computation of kinetic decoupling of WIMPs in the early universe, including the case of the LKP (in particular, see their App.~C for a complete derivation of the LKP scattering amplitude off light SM fermions). In the relevant range of masses, the resulting estimates for the kinetic decoupling temperatures agree with those of Ref.~\cite{Profumo:2006bv}.

\subsection{Direct KKDM Searches}\label{sec:direct}

A large number of experiments have been designed and developed in the hope of observing the elastic scattering of dark matter particles with nuclei. Some of these experiments include CDMS \cite{cdmssi,cdmssd}, Edelweiss \cite{edelweiss}, ZEPLIN \cite{zeplin}, CRESST~\cite{cresst}, CLEAN, COUPP, DEAP, DRIFT, EURECA, SIGN, XENON~\cite{xenon}, WARP \cite{warp}, KIAS, NaIAD \cite{naiad}, Picasso \cite{picasso}, Majorana \cite{majorana}, DUSEL, IGEX \cite{igex}, ROSEBUD~\cite{rosebud}, ANAIS, KIMS, Genius~\cite{genius}, DAMA~\cite{dama} and LIBRA. This collection of experimental programs is collectively known as direct detection.

\subsubsection{Spin-Independent Scattering}

The elastic scattering of a WIMP with nuclei can be separated into spin-independent and spin-dependent contributions. Spin-independent scattering can take place coherently with all of the nucleons in a nucleus, leading to a cross section proportional to the square of the nuclei mass. As a result, the current direct detection constraints on spin-independent scattering are considerably stronger than for the spin-dependent component. 

The spin-independent $B^{(1)}$-nuclei elastic scattering cross section is given by:
\begin{equation}
\sigma_{B^{(1)}N,\rm{SI}} \approx \frac{4 m^2_{B^{(1)}} m^2_{N}}{\pi (m_{B^{(1)}}+m_N)^2} [Z f_p + (A-Z) f_n]^2,
\end{equation}
where $m_N$, $Z$ and $A$ are the mass, atomic number and atomic mass of a target nucleus.  $f_p$ and $f_n$ are the $B^{(1)}$'s couplings to protons and neutrons, given by:
\begin{equation}
f_{p,n}=\sum_{q=u,d,s} f^{(p,n)}_{T_q} a_q \frac{m_{p,n}}{m_q} + \frac{2}{27} f^{(p,n)}_{TG} \sum_{q=c,b,t} a_q  \frac{m_{p,n}}{m_q},
\label{feqn}
\end{equation}
where $a_q$ are the $B^{(1)}$-quark couplings and $f^{(p)}_{T_u} \approx 0.020\pm0.004$,  
$f^{(p)}_{T_d} \approx 0.026\pm0.005$,  $f^{(p)}_{T_s} \approx 0.118\pm0.062$,  
$f^{(n)}_{T_u} \approx 0.014\pm0.003$,  $f^{(n)}_{T_d} \approx 0.036\pm0.008$ and 
$f^{(n)}_{T_s} \approx 0.118\pm0.062$ \cite{nuc1,nuc2,nuc3}. The first term in Eq.~(\ref{feqn}) 
corresponds to interactions with the quarks in a target nucleon, whereas the second term corresponds to interactions with gluons in the target through a quark/KK-quark 
loop diagram. $f^{(p)}_{TG}$ is given by $1 -f^{(p)}_{T_u}-f^{(p)}_{T_d}-f^{(p)}_{T_s} 
\approx 0.84$, and analogously, $f^{(n)}_{TG} \approx 0.83$.
 
\begin{figure}
\centering
\includegraphics[width=0.6\textwidth,clip=true]{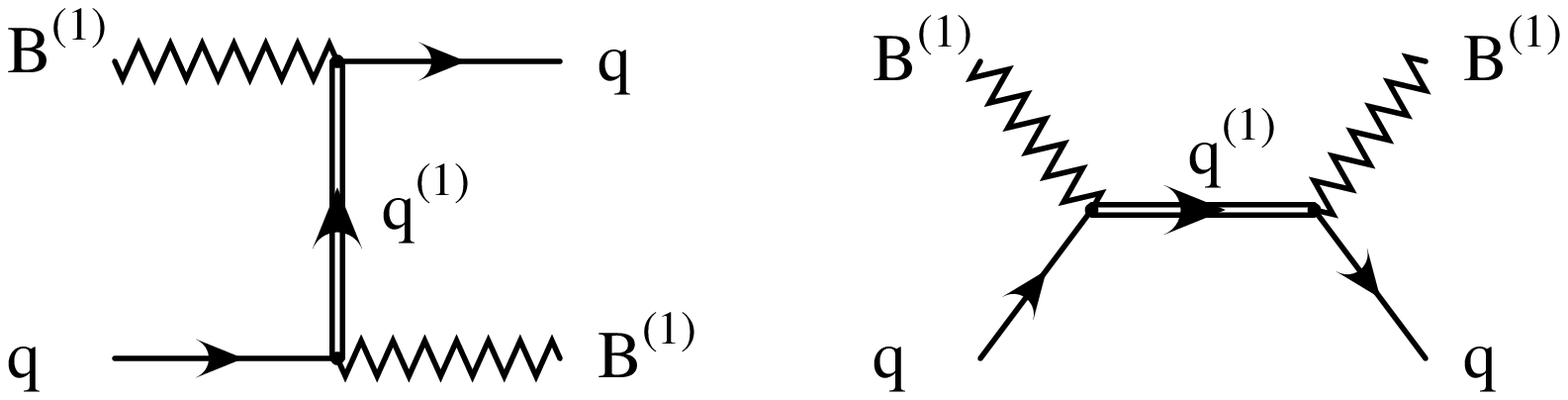}
\includegraphics[width=0.3\textwidth,clip=true]{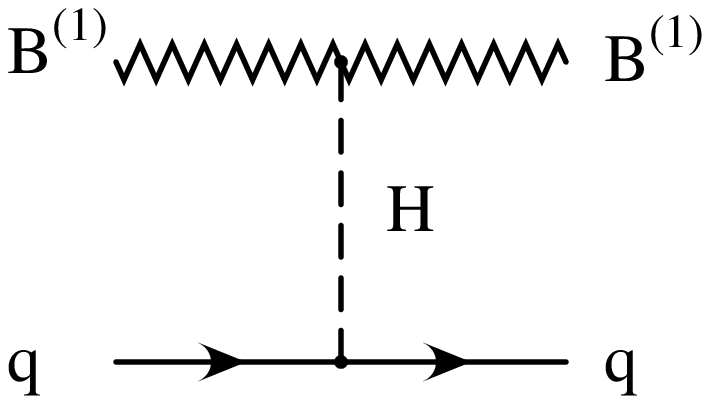}
\caption{Leading Feynman diagrams for $B^{(1)}$-quark elastic scattering. $B^{(1)}$-gluon scattering can also occur through analogous diagrams with a quark/KK-quark loop. From Ref.~\cite{Servant:2002hb}.}
\label{elasticdiagrams}
\end{figure}

\begin{figure}
\centering
\includegraphics[width=0.45\textwidth,clip=true]{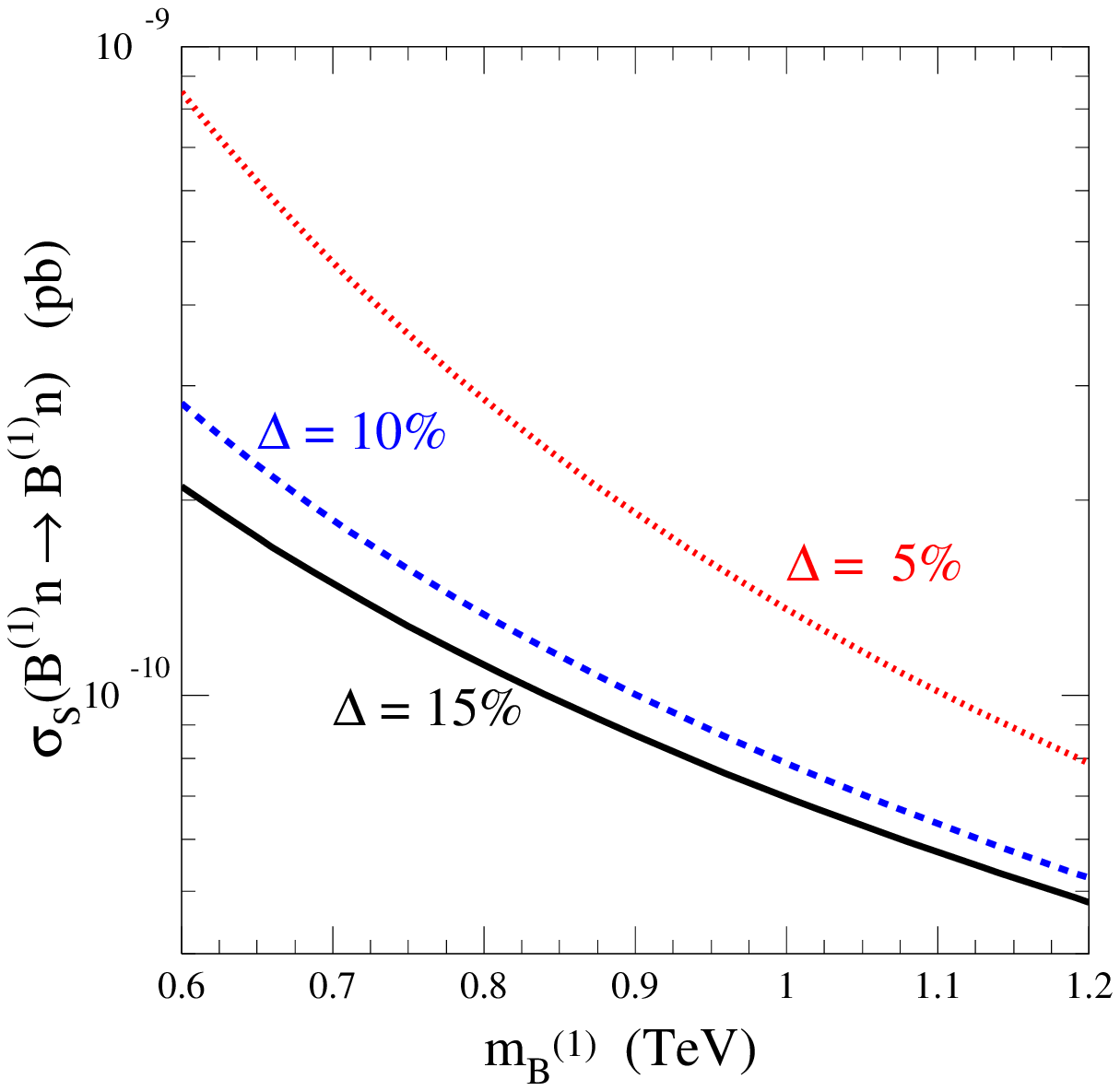}
\includegraphics[width=0.45\textwidth,clip=true]{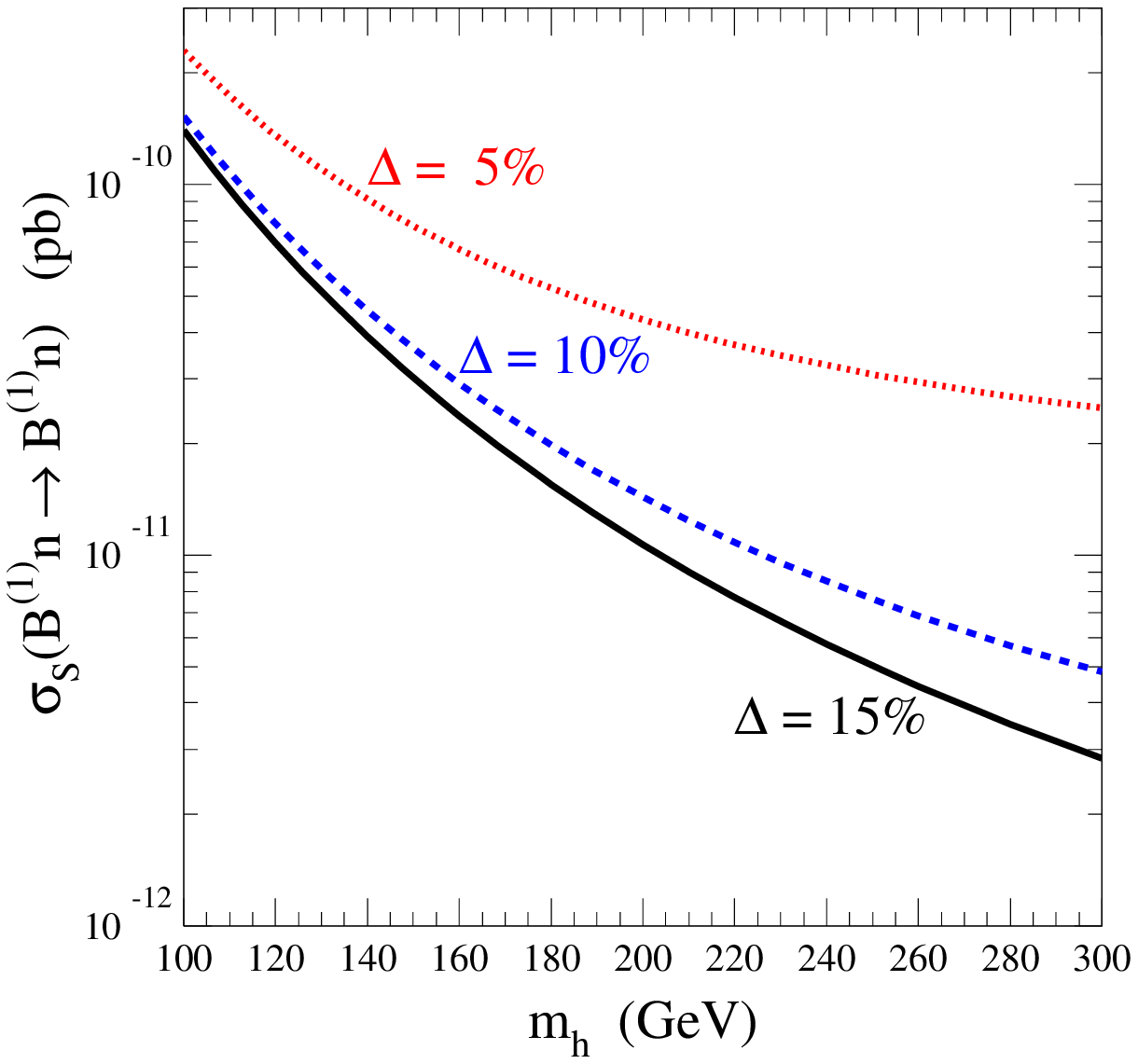}
\caption{The spin-independent $B^{(1)}$-nucleon elastic scattering cross section. In the left frame, $m_h$ has been fixed to 120 GeV. In the right frame, $m_{B^{(1)}}$ has been fixed to 1 TeV. The quantity $\Delta$ is defined as $\Delta \equiv (m_{q^{(1)}}-m_{B^{(1)}})/m_{B^{(1)}}$. From Ref.~\cite{Servant:2002hb}.}
\label{elasticsigma}
\end{figure}

The coupling $a_q$ receives contributions from two classes of diagrams: the s-channel exchange of KK quarks and t-channel Higgs boson exchange (see Fig.~\ref{elasticdiagrams}). This leads to \cite{Cheng:2002ej,Servant:2002hb}:
\begin{eqnarray}
\label{aq}
a_q  =  \frac{m_q \, g^2_1 \, (Y^2_{q_R}+Y^2_{q_L}) \,  ( m^{2}_{B^{(1)}} + m^{2}_{q^{(1)}})}{4 m_{B^{(1)}}(m^{2}_{B^{(1)}} - m^{2}_{q^{(1)}})^2}  + \frac{m_q \, g^2_1}{8 m_{B^{(1)}}\, m^2_h}, 
\end{eqnarray}
The first term in this expression should only be included for the light quarks ($q=u, d, s$), while the second term contributes to both light and heavy species. This leads to a contribution to the cross section from Higgs exchange which is proportional to $1/(m^2_{B^{(1)}}\,m^4_h)$ and a contribution from KK-quark exchange which is approximately proportional to $1/(m^6_{B^{(1)}}\, \Delta^4)$, where $\Delta=(m_{q^{(1)}}-m_{B^{(1)}})/m_{B^{(1)}}$.
Numerically, this leads to a $B^{(1)}$-nucleon cross section approximately given by:
\begin{equation}
\sigma_{B^{(1)}n,\rm{SI}}  \approx  1.2 \times 10^{-10} \, {\rm pb}\,\bigg(\frac{1\,\rm{TeV}}{m_{B^{(1)}}}\bigg)^2 \, \bigg[\bigg(\frac{100\, \rm{GeV}}{m_h}\bigg)^2 + 0.09 \, \bigg(\frac{1\,\rm{TeV}}{m_{B^{(1)}}}\bigg)^2 \bigg(\frac{0.1}{\Delta}\bigg)^2\bigg]^2.
\end{equation}
In Fig.~\ref{elasticsigma}, this cross section is plotted as a function of $m_{B^{(1)}}$ and $m_h$.

The range of elastic scattering cross sections predicted in this model are well below the current experimental limits. The strongest constraint on spin-independent scattering has been reported by the CDMS collaboration \cite{cdmssi}, which finds roughly $\sigma_{\rm{SI}} \gtrsim 10^{-6}$ pb in the mass range of interest here. A number of other experiments have placed limits which are only mildly less restrictive, including ZEPLIN, Edelweiss and CRESST. 


%
CDMS and other direct detection experiments are currently working toward improving their sensitivity. Although the current constraints are around three orders of magnitude or more below the cross sections predicted for KKDM, the advanced phases of Super-CDMS should be able to reach the needed sensitivity to directly observe dark matter in these models. In addition to Super-CDMS, ton-scale experiments using liquid elements such as Argon or Xenon appear very promising. Although such noble liquid detectors may very well reach the level of sensitivity needed to observe KKDM, these technologies are currently less well understood than the cryogenic Germanium and Silicon detectors used by CDMS and Edelweiss. 

\begin{figure}[!t]
\centering
\includegraphics[width=1.0\textwidth,clip=true]{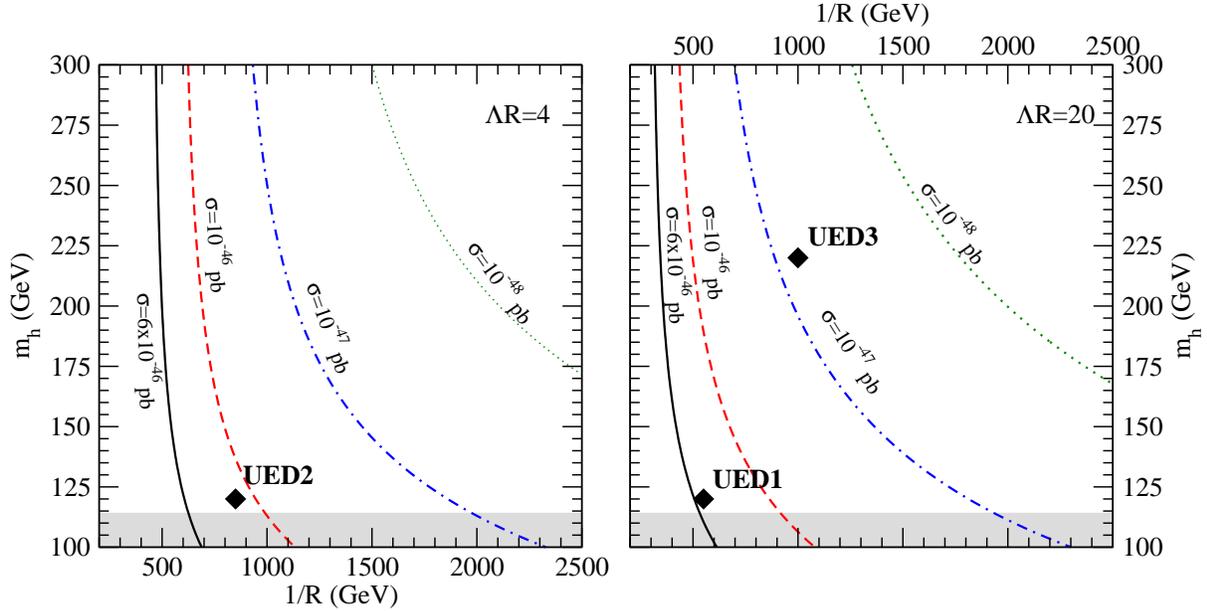}
\caption{Contours of constant spin-independent $B^{(1)}$-proton scattering cross sections in the ($1/R,m_h$) plane, for two choices of $\Lambda R=4$ and 20. The reach of the future direct detection experiments ``Xenon-1 ton'' and ``Super-CDMS C'' approximately correspond to the black solid line and to the red dashed line. We also indicate the location of three of the benchmark models of Appendix~\ref{ch:benchmarks}.}
\label{fig:bench_si1}
\end{figure}

\begin{figure}[!]
\centering
\includegraphics[width=1.0\textwidth,clip=true]{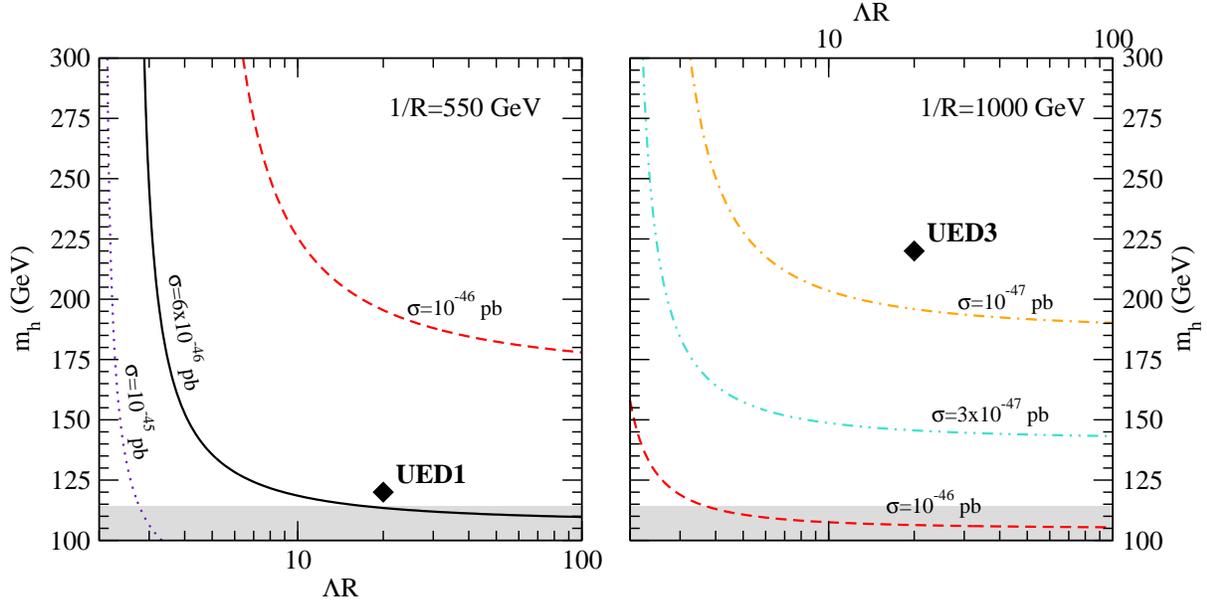}
\caption{As in Fig.~\protect{\ref{fig:bench_si1}}, but in the ($\Lambda R,m_h$) plane.}
\label{fig:bench_si2}
\end{figure}

We illustrate the prospects for direct KKDM detection in future direct detection experiments in Figs.~\ref{fig:bench_si1} and \ref{fig:bench_si2}, focusing on three minimal UED benchmark models, and exploring various slices of the minimal UED parameter space (for a detailed description of the motivations and of the particle spectra for these models see Appendix~\ref{ch:benchmarks}). For all benchmark models, a vanishing Higgs boundary mass term Model is assumed, and the models are therefore fully defined (through the radiative mass corrections outlined in Sec.~\ref{sec:spectrum}) by the values of $1/R$, $\Lambda R$ and $m_h$. All benchmark models feature a $B^{(1)}$ LKP with a thermal relic abundance in accord with the cold dark matter content of the universe. Model {\bf UED1} has $1/R=550$, $\Lambda R=20$ and $m_h=120$ GeV, model {\bf UED2} has $1/R=850$, $\Lambda R=4$ and $m_h=120$ GeV, and model {\bf UED3} $1/R=1000$, $\Lambda R=20$ and $m_h=220$ GeV.

Specifically, we explore the $B^{(1)}$-proton scattering cross section in Fig.~\ref{fig:bench_si1}, where we scan the ($1/R,m_h$) plane, for two representative values of $\Lambda R=4,\ 20$ (which determine the mass splitting, $\Delta$). The grey shaded regions are excluded by the negative results of the LEP2 Higgs searches. We also plot curves representing the reach of representative future direct dark matter search experiments (the projected sensitivity at a WIMP mass of 1 TeV is that indicated on each curve). The black diamonds locate the benchmark models of Appendix~\ref{ch:benchmarks}. The approximate sensitivity reach of the future Xenon-1 ton and Super-CDMS C experiments approximately correspond to the solid black and to the dashed red curves, respectively. While models {\bf UED1} and {\bf UED2} will both escape detection at Xenon-1t, they lie within reach of the planned sensitivity for the phase C of Super-CDMS. Model {\bf UED3}, on the other hand, will not be detectable with foreseeable direct detection devices. Fig.~\ref{fig:bench_si2} shows the same curves, and two of the benchmark models, this time in the ($\Lambda R,m_h$) plane. 

These two figures demonstrate how the three parameters defining the minimal UED setup affect direct detection rates. We notice that in the low-$1/R$ regime the role of $m_h$ is less critical than with a heavier KK spectrum. Also, the parameter $\Lambda R$ plays a role (affecting the LKP-KK quark spectrum) only when it takes low values $\lesssim 10$; at larger $\Lambda R$, the direct detection rates become largely independent of $\Lambda R$.

%

\subsubsection{Spin-Dependent Scattering}
\label{subsec:spindep}

In addition to spin-independent elastic scattering, axial-vector couplings can lead to $B^{(1)}$-nuclei spin-dependent scattering. As we mentioned earlier, the current experimental constraints on spin-dependent scattering are far weaker than for spin-independent processes. New experimental technologies, however, such as those being developed by the COUPP collaboration, may enable greater sensitivity to spin-dependent couplings. Spin-dependent scattering can also play an important role in the capture of WIMPs in the Sun (see Section~\ref{sec:ne}).

The cross section for spin-dependent scattering is given by:
\begin{equation}
\sigma_{B^{(1)}N,\rm{SD}} \approx \frac{2 g^4_1 \, m^2_{B^{(1)}} m^2_{N}}{3 \pi (m_{B^{(1)}}+m_N)^2} \frac{\Lambda^2 \, J(J+1)}{(m^2_{q^{(1)}}-m^2_{B^{(1)}})^2},
\end{equation}
where $J$ is the spin of the nuclei species and the quantity $\Lambda$ is given by:
\begin{equation}
\Lambda \equiv \frac{a_p  \langle S_p\rangle + a_n  \langle S_n\rangle}{J},
\end{equation}
where
\begin{eqnarray}
a_p &=&  \frac{1}{4}(Y^2_{u_R}+Y^2_{u_L})\Delta u +  \frac{1}{4}(Y^2_{d_R}+Y^2_{d_L})\Delta d +  \frac{1}{4}(Y^2_{d_R}+Y^2_{d_L})\Delta s \\ \nonumber
&=& \frac{17}{36} \Delta u +  \frac{5}{36} \Delta d +  \frac{5}{36} \Delta s, \\ \nonumber
a_n &=&  \frac{1}{4}(Y^2_{u_R}+Y^2_{u_L})\Delta d +  \frac{1}{4}(Y^2_{d_R}+Y^2_{d_L})\Delta u +  \frac{1}{4}(Y^2_{d_R}+Y^2_{d_L})\Delta s \\ \nonumber
&=&\frac{5}{36} \Delta u +  \frac{17}{36} \Delta d +  \frac{5}{36} \Delta s.
\end{eqnarray}
Here the $\Delta q$s parameterize the fraction of spin carried in a proton by each quark species: $\Delta u = 0.78 \pm 0.02$, $\Delta d = -0.48 \pm 0.02$, and $\Delta s = -0.15 \pm 0.07$~\cite{spinfraction1,spinfraction2}.

For spin-dependent scattering with a proton, the cross section is well approximated by:
\begin{eqnarray}
\sigma_{B^{(1)}p, \rm{SD}} &=& \frac{g_1^4 m_p^2}{8 \pi m_{B^{(1)}}^2 
(m_{q^{(1)}_R} - m_{B^{(1)}})^2} \,
\left(  (Y^2_{u_R}+Y^2_{u_L}) \Delta u +  (Y^2_{d_R}+Y^2_{d_L}) (\Delta d + \Delta s) \right)^2 \\ \nonumber 
&\approx& 1.8 \times 10^{-6} \, {\rm pb}\, \bigg(\frac{1 \,{\rm TeV}}{m_{B^{(1)}}}\bigg)^4 \bigg(\frac{0.1}{\Delta}\bigg)^2.
\end{eqnarray}
And with a neutron:
\begin{eqnarray}
\sigma_{B^{(1)}n, \rm{SD}} &=& \frac{g_1^4 m_n^2}{8 \pi m_{B^{(1)}}^2 
(m_{q^{(1)}_R} - m_{B^{(1)}})^2} \,
\left(  (Y^2_{u_R}+Y^2_{u_L}) \Delta d +  (Y^2_{d_R}+Y^2_{d_L}) (\Delta u + \Delta s) \right)^2 \\ \nonumber 
&\approx& 0.4 \times 10^{-6} \, {\rm pb}\, \bigg(\frac{1 \,{\rm TeV}}{m_{B^{(1)}}}\bigg)^4 \bigg(\frac{0.1}{\Delta}\bigg)^2.
\end{eqnarray}

The strongest experimental limits on spin-dependent scattering in the mass range we are interested in have been set by the CDMS \cite{cdmssd} and NaIAD \cite{naiad} experiments, for scattering with neutrons and protons, respectively. These experiments only constrain $\sigma_{\rm{SD}} \lesssim 0.1-1\,$pb, however, and thus do not have the sensitivity needed to test KKDM models.

\subsection{Gamma Ray Searches}\label{sec:gr}

In addition to detecting WIMPs directly, efforts are underway to observe the stable products of dark matter annihilations, including gamma rays, energetic neutrinos and antimatter. The relevant particle model input quantities for indirect detection are the pair annihilation cross section and the relative frequency (branching ratio) of a given pair-annihilation final state. We outlined both in Sec.~\ref{sec:relicab} and in Tab.~\ref{tab:br} above.

In the case of KKDM, gamma rays can be produced as continuum photons from final state radiation and the cascades of other annihilation products, and as line emission from loop-diagrams to $\gamma \gamma$, $\gamma Z$ or $\gamma h$ final states. 

The locations most likely to produce an observable flux of gamma rays from dark matter annihilation are those which contain a very high density of dark matter and are relatively nearby. The center of our Galaxy has long been considered one of the most promising regions of the sky for observing dark matter annihilation radiation~\cite{bousasi,Stecker:dz,ber}.

The flux of gamma rays from the Galactic center is found by integrating the density squared along the line-of-sight to the observer \cite{Bergstrom:1997fj}:
\begin{equation}
\label{fluxgamma2}
\Phi_{\gamma}(\psi,E_{\gamma})=\sigma v \frac{dN_{\gamma}}{dE_{\gamma}} \frac{1}{4 \pi m_{B^{(1)}}^2}
\int_{\mbox{\small{line-of-sight}}}{\rm d}\,s
\rho^2\left(r(s,\psi)\right),
\end{equation}
where $\sigma v$ is the WIMP's annihilation cross section, $dN_{\gamma}/dE_{\gamma}$ is the gamma ray spectrum produced per annihilation, and the coordinate $s$ runs along the line-of-sight, in a
direction making an angle, $\psi$, from the direction
of the Galactic center. $\rho(r)$ is the density of dark matter at a distance $r$ from the Galactic center.

This expression can be separated into factors which depend on the dark matter distribution, and which depend on the characteristics of the dark matter species. The first of these is described by the quantity $J(\psi)$:
\begin{equation}
J\left(\psi\right) = \frac{1} {8.5\, \rm{kpc}} 
\left(\frac{1}{0.3\, \mbox{\small{GeV/cm}}^3}\right)^2
\int_{\mbox{\small{line-of-sight}}}d\,s\rho^2\left(r(s,\psi)\right)\,.
\label{gei}
\end{equation}
We further define $\overline{J}(\Delta\Omega)$ as the average of $J(\psi)$ over
the solid angle, $\Delta\Omega$, centered on $\psi=0$.

We can now express the flux of gamma rays over a solid angle $\Delta \Omega$ around the Galactic center:
\begin{equation}
\Phi_{\gamma}(\Delta\Omega, E_{\gamma})\approx 2.8\times10^{-12}\,\frac{dN_{\gamma}}{dE_{\gamma}} 
\left( \frac{\sigma v}
{3\times 10^{-26}\,\rm{cm}^3/\rm{s}}\right)\left(\frac{1\rm{TeV}} 
{m_{B^{(1)}}} \right)^2 \overline{J}\left(\Delta\Omega\right) 
 \; \Delta\Omega\,\,\rm{cm}^{-2} \rm{s}^{-1}.
\label{final}
\end{equation}
\begin{center}
\begin{table}
\centering
\begin{tabular}{|c|ccccc|}
\hline 
&$\alpha$&$\beta$&$\gamma$&R (kpc)&$\overline{J}\left( 10^{-3} \, \rm{str}\right)$ \\
\hline 
NFW& 1.0& 3.0& 1.0& 20& $1.352 \times 10^3$\\
Moore& 1.5& 3.0& 1.5& 28.0 &$ 1.544 \times 10^5$ \\
Iso& 2.0& 2.0& 0& 3.5&$2.868 \times 10^1$\\ 
Kra& 2.0& 3.0&0.4 & 10.0 &$ 2.166 \times 10^1$ \\
\hline 
\end{tabular}
\label{profiles}
\caption{Parameters, and values of $J$ averaged over $10^{-3}$ steradians, for some commonly used galactic halo profiles, as used in Eq.~(\ref{profile}). The models shown are those of Navarro, Frenk and White~(NFW)~\cite{Navarro:1995iw}, Moore et al.~(Moore)~\cite{Moore:1999gc}, modified isothermal~(Iso)~\cite{Bergstrom:1997fj} and Kravtsov et al.~(Kra)~\cite{kra}.}
\end{table}
\end{center}
There are considerable uncertainties involved in the distribution of dark matter in the Galactic center region. N-body simulations suggest that there exists a universal dark matter profile, with the same shape
for all masses, epochs and input power spectra \cite{Navarro:1995iw}. 
Such a profile is often parameterized as
\begin{equation}
  \rho(r)= \frac{\rho_0}{(r/R)^{\gamma}
  [1+(r/R)^{\alpha}]^{(\beta-\gamma)/\alpha}} \;\;.
\label{profile} 
\end{equation}
In Table~\ref{profiles}, the parameters used in the parameterization of Eq.~(\ref{profile}) are given for several representative halo profiles. Of these profiles, NFW~\cite{Navarro:1995iw} and Moore et al.~\cite{Moore:1999gc} each predict very dense concentrations of dark matter in the inner regions of the Galactic halo ($\rho(r) \propto r^{-1}$ or $r^{-1.5}$). Such a cusped halo profile is typically required if gamma rays from dark matter annihilations are to be observed from the Galactic center.

None of above profiles derived from N-body simulations account for the effects of baryonic matter, however. As the baryonic density dominates over that of dark matter in the inner regions of the Milky Way, it is possible that N-body simulations will not accurately predict the dark matter distribution in this region. The effects of baryons on the dark matter profile are, unfortunately, not easily predicted. One possibility is that the effects of adiabatic compression on the dark matter profile near the Galactic center could play an important role, potentially enhancing the dark matter density by up to an order of magnitude in the inner parsecs of the Milky Way \cite{prada}. Baryonic matter could also flatten the inner cusp of the dark matter distribution.

At the center of the Milky Way is a $2.6 \times 10^6$ solar mass black hole. It is possible that the adiabatic accretion of dark matter onto this object could lead to extremely high densities of dark matter~\cite{peebles72}. An initial density profile with a slope, $\rho(r) \propto r^{-\gamma}$, would lead to a density ``spike'' with a slope of $\gamma_{\rm{sp}}=(9-2\gamma)/(4-\gamma)$ \cite{Gondolo:1999ef}. Such a feature would dramatically increase the rate of dark matter annihilations near the Galactic center.

A spectrum of gamma rays extending from at least 200 GeV to 10 TeV (and above) has recently been observed from the Galactic center by the HESS \cite{hessgc} and MAGIC \cite{magicgc} Atmospheric Cerenkov Telescopes (ACTs). While initially it was thought to be possible that this signal was the result of dark matter annihilations~\cite{gcdark1,gcdark2,gcdark3,Bergstrom:2004cy}, the spectrum now appears incompatible with that predicted from dark matter and is likely to be of an astrophysical nature. This spectrum represents a challenging background to future searches for dark matter in the Galactic center~\cite{gabi}.

With this potential obstacle involving searches for dark matter near the Galactic center in mind, other regions are beginning to appear more attractive. In particular, a class of companion galaxies to the Milky Way, called dwarf spheroidal galaxies, may provide an observable flux of dark matter annihilation radiation from low background regions of the sky \cite{dwarf1,dwarf2,dwarf3,dwarf4}. Galaxies external to the Milky Way have also been considered \cite{external}. Alternatively, if density spikes form around Galactic intermediate mass black holes, a number of point sources of dark matter annihilation radiation may become observable in the near future~\cite{imbh,Bertone:2006kr}.

\subsubsection{The Gamma Ray Spectrum}

The majority of KKDM annihilations produce pairs of quarks or leptons. Many of these particles then proceed to decay and fragment, producing a continuous spectrum of gamma rays \cite{Bertone:2002ms}. At modest and high energies, this spectrum is dominated by photons produced in the semi-leptonic decays of tau leptons and by radiative processes (bremsstrahlung) associated to $\mu^+\mu^-$ and $e^+e^-$ final states. All these charged lepton pairs final states are produced in approximately 20\% of $B^{(1)}$ annihilations \cite{Bergstrom:2004cy} (see Tab.~\ref{tab:br}). This is clearly in contrast with the case of the lightest neutralino, where, as explained in Sec.~\ref{sec:relicab}, leptonic final states are suppressed by factors of the order $m_f^2/m_\chi^2$, and the high energy end of the gamma-ray spectrum is typically dominated by gauge bosons pair final states, when these are open. The latter, here, do not, instead, contribute appreciably to shape the gamma-ray spectrum.

\begin{figure}
\psfrag{l+}[][][1]{\mbox{\footnotesize $\ell^+$}}
\psfrag{l-}[][][1]{\mbox{\footnotesize $\ell^-$}}
\psfrag{l1}[r][r][1]{\mbox{\footnotesize $\ell^{\text{{\tiny(1)}}}$}}
\psfrag{b}[][][1]{\mbox{\footnotesize $B^{\text{{\tiny(1)}}}$}}
\psfrag{g}[][][1]{\mbox{\footnotesize $\gamma$}}
\centering
\includegraphics[width=0.7\columnwidth]{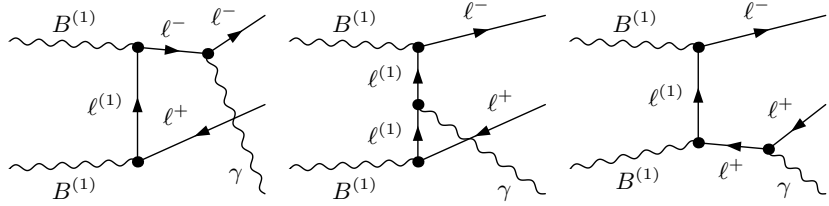}
\caption{Contributions to $B^{(1)} B^{(1)} \rightarrow
\ell^+ \ell^- \gamma$. From Ref.~\cite{Bergstrom:2004cy}.}
\label{feynfsr}
\end{figure}

\begin{figure}[t]
\psfrag{y}[][][1.0]{$x^2 \text{d}N_\gamma^\text{eff}/\text{d}x$}
\psfrag{x}[][][0.95]{$x=E_\gamma / m_{B^{(1)}}$}
\psfrag{1}[][][0.8]{\textsf{1}}
\psfrag{0.1}[][][0.8]{\textsf{0.1}}
\psfrag{0.03}[][][0.8]{\textsf{0.03}}
\psfrag{0.01}[][][0.8]{\textsf{0.01}}
\centering
\includegraphics[width=0.6\columnwidth]{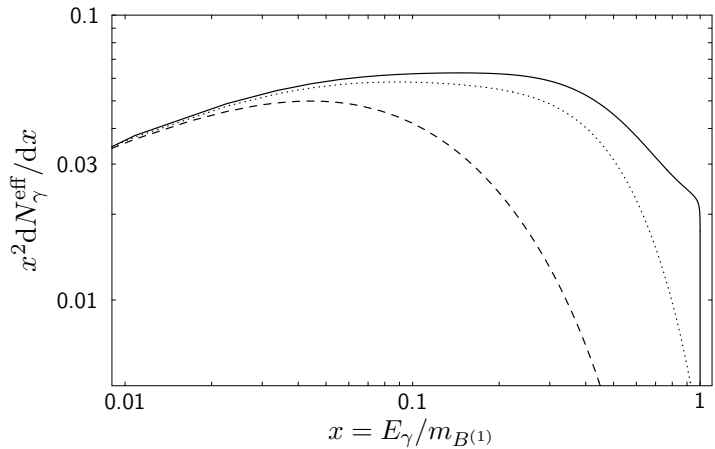}
\caption{The spectrum of photons per $B^{(1)}B^{(1)}$ annihilation (solid
line). Shown as a dashed line is the contribution from quark
fragmentation alone. The dotted line is the quark fragmentation contribution plus the contribution from $\tau$ leptons. From Ref.~\cite{Bergstrom:2004cy}.}
\label{gammaspectrum}
\end{figure}

Final state radiation from the process $B^{(1)} B^{(1)} \rightarrow l^+ l^- \gamma$ (see Fig.~\ref{feynfsr}) plays, in the context of KKDM, a particularly important role. The spectrum of photons produced through this process (for a given lepton species) is given by:
\begin{equation}
\frac{dN_{\gamma}}{dx} \equiv \frac{d (\sigma_{l^+l^-\gamma} v)/dx}{\sigma_{l^+l^-}v} = \frac{\alpha}{\pi}\frac{(x^2-2x+2)}{x} \ln\bigg[\frac{m^2_{B^{(1)}}}{m^2_l}(1-x)\bigg],
\end{equation}
where $x=E_{\gamma}/m_{B^{(1)}}$. Final state radiation is very important in determining the gamma ray spectrum from KKDM largely because of the mentioned large fraction of annihilations that generate light lepton pairs (approximately 20\% to each of $e^+ e^-$ and $\mu^+ \mu^-$). 
In Fig.~\ref{gammaspectrum}, the contributions from quark fragmentation, tau decays, and final state radiation are compared. At the highest energies, final state radiation dominates. At more modest energies, semi-leptonic tau decays dominate. At low energies, most of the gamma rays are produced through quark fragmentation.

In addition to continuum gamma rays, the processes $B^{(1)} B^{(1)} \rightarrow \gamma \gamma$,  $B^{(1)} B^{(1)} \rightarrow \gamma Z$ and  $B^{(1)} B^{(1)} \rightarrow \gamma h$ result in mono-energetic gamma ray lines with energies of $E_{\gamma} = m_{B^{(1)}}$, $m_{B^{(1)}}\, (1-m^2_Z/4m^2_{B^{(1)}})$ and $m_{B^{(1)}}\, (1-m^2_h/4m^2_{B^{(1)}})$, respectively. Such gamma ray lines would be a ``smoking gun" for dark matter annihilations if they were to be observed.

The cross section for $B^{(1)} B^{(1)} \rightarrow \gamma \gamma$ involves contributions from twelve Feynman diagrams containing a fermion/KK-fermion loop and 22 diagrams involving the scalar sector of the theory~\cite{Bergstrom:2004nr}. The fermion diagrams dominate the cross section, which was found to be approximated by $\sigma_{B^{(1)} B^{(1)} \rightarrow \gamma \gamma}v \sim 2 \times 10^{-30}\, \rm{cm}^3/\rm{s} \,\, (1\,\rm{TeV}/m_{B^{(1)}})^2$. The cross section also grows as the mass of the KK-fermions is moved closer to the LKP mass. Varying the $(m_{f^{(1)}}-m_{B^{(1)}})/m_{B^{(1)}}$ between 0.3 and 0.01 increases the cross section by a factor of roughly 3~\cite{Bergstrom:2004nr}.

Calculations of the cross sections of $B^{(1)} B^{(1)} \rightarrow \gamma Z$ and  $B^{(1)} B^{(1)} \rightarrow \gamma h$ have not yet appeared in the literature, although preliminary work indicates that they are not expected to be as favorable channels as $B^{(1)} B^{(1)} \rightarrow \gamma \gamma$ for the monochromatic gamma ray line emission \cite{bergpriv}.

\subsubsection{Observational Prospects}

Both ground and space-based gamma ray detectors are potentially sensitive to dark matter annihilation radiation. A number of Atmospheric Cerenkov Telescopes (ACTs), including HESS \cite{hess}, MAGIC \cite{magic} and VERITAS \cite{veritas} are currently operational. These experiments are sensitive to gamma rays above 50--500 GeV, depending on the telescope and the location of the source in the sky (its zenith angle). Additionally, the GLAST satellite is scheduled to begin its mission in 2007 \cite{glast1,glast2}. GLAST will be sensitive to gamma rays between sub-GeV energies and about 300 GeV. Unlike ACTs, GLAST will be able to observe large regions of the sky at a time, although with a much smaller effective area (approximately one square meter) than ACTs.

The prospects for the detection of gamma rays from KKDM annihilations depend critically on the distribution of dark matter in the center of our Galaxy, in dwarf spheroidals, in dark matter clumps and in other such regions. In most cases, these distributions are very poorly known, making the prospects for future gamma ray dark matter searches difficult to predict.

For the case of the Galactic center, considering an NFW halo profile and $m_{B^{(1)}}\sim 1$ TeV, we expect GLAST to detect $\sim 10$ photons above a few GeV over its first three years (assuming the Galactic center is in its field-of-view about one third of the time). This rate could potentially represent a detection of dark matter if it were not for the backgrounds which have been observed by HESS and MAGIC \cite{gabi}. A rate roughly an order of magnitude larger would be needed to overcome these backgrounds, assuming that the power-law spectrum observed by HESS and MAGIC extends to lower energies \cite{gabi}. A lighter dark matter particle would be somewhat less difficult to detect with GLAST.

The annihilation rate and corresponding gamma ray flux from dwarf spheroidals is generally expected to be somewhat lower than is predicted from cusped halo models of the Galactic center, although predictions can vary considerably. The prospects for detecting dark matter annihilation radiation from such objects is difficult to assess at this time. The prospects for the detection of a signal from dark matter mini-halos or clumps (see Sec.~\ref{sec:protohalos}) strongly depends on the survival probability of such objects and on their specific dark matter distribution. Potentially GLAST might be able to detect gamma-rays from annihilations of KKDM in such structures~\cite{Baltz:2006sv}, or even the proper motion of the sources themselves~\cite{Koushiappas:2006qq}.

\subsection{Neutrino Searches}\label{sec:ne}

As WIMPs annihilate in the halo, they do not only produce gamma rays, but also other particles including neutrinos. Neutrinos are far more difficult to detect than gamma rays, however. For this reason, another approach must be taken if neutrinos from dark matter annihilations are to be observed. In particular, WIMPs can be captured in the Sun, where they annihilate efficiently. Of the various annihilation products, only neutrinos can escape the solar medium and potentially be observed.

WIMPs become captured in the gravitational potential of the Sun at a rate given by \cite{capture}:
\begin{equation} 
C_{\odot} \approx 3.35 \times 10^{18} \, \mathrm{s}^{-1} 
\left( \frac{\rho_{\mathrm{local}}}{0.3\, \mathrm{GeV}/\mathrm{cm}^3} \right) 
\left( \frac{270\, \mathrm{km/s}}{\bar{v}_{\mathrm{local}}} \right)^3  
\left( \frac{\sigma_{\mathrm{H, SD}} + 2.6 \, \sigma_{\mathrm{H, SI}}
+ 0.175 \, \sigma_{\mathrm{He, SI}}}{10^{-6} \, \mathrm{pb}} \right) 
\left( \frac{1 \, \mathrm{TeV}}{m_{B^{(1)}}} \right)^2 
\label{c-eq}
\end{equation} 
where $\rho_{\mathrm{local}}$ is the local dark matter density and $\bar{v}_{\mathrm{local}}$ is the local RMS velocity of halo dark matter particles. $\sigma_{\mathrm{H,SD}}$, $\sigma_{\mathrm{H, SI}}$ and $\sigma_{\mathrm{He, SI}}$  are the spin-dependent, $B^{(1)}$-proton (hydrogen), spin-independent, $B^{(1)}$-proton and spin-independent, $B^{(1)}$-helium elastic scattering cross sections, respectively. The factors of $2.6$ and $0.175$ include information on the solar abundances of elements, dynamical factors and form factor suppression.

As shown in Sec.~\ref{sec:direct}, $B^{(1)}$s elastically scatter with protons far more efficiently through spin-dependent than spin-independent couplings (note that this is overturn by coherence effects once one considers the scattering off nuclei rather than nucleons). The dominant elastic scattering cross section for capture in the Sun is thus $\sigma_{B^{(1)}p, \rm{SD}} \approx 1.8 \times 10^{-6}\, {\rm pb}\, (m_{B^{(1)}}/\rm{TeV})^{-4} (\Delta/0.1)^{-2}$.

The evolution of the number of $B^{(1)}$s in the Sun, $N$, is given by:
\begin{equation}
\dot{N} = C_{\odot} - A_{\odot} N^2 \; ,
\end{equation}
where $A_{\odot}$ is the 
annihilation cross section multiplied by the relative WIMP velocity per unit volume, given by:
\begin{equation}
A_{\odot} = \frac{\langle \sigma v \rangle}{V_{\mathrm{eff}}} \;.
\end{equation}
Here, $V_{\mathrm{eff}}$ is the effective volume of the Sun's core, which is
determined by matching the core temperature to
a WIMP's gravitational potential energy at the core
radius.  This was found to be \cite{equ1,equ2}
\begin{equation}
V_{\rm eff} = 1.8 \times 10^{26} \, \mathrm{cm}^3 
\left( \frac{1 \, \mathrm{TeV}}{m_{B^{(1)}}} \right)^{3/2} \; .
\end{equation}
This leads to a present annihilation rate of $B^{(1)}$s in the Sun of:
\begin{equation} 
\Gamma = \frac{1}{2} A_{\odot} N^2 = \frac{1}{2} \, C_{\odot} \, 
\tanh^2 \left( \sqrt{C_{\odot} A_{\odot}} \, t_{\odot} \right) \; 
\end{equation}
where $t_{\odot} \simeq 4.5$ billion years is the age of the solar system.
If $\sqrt{C_{\odot} A_{\odot}} t_{\odot} \gg 1$, then the annihilation and capture rates reach equilibrium, maximizing the resulting neutrino flux. Capture-annihilation equilibrium is reached if the following condition is met:
\begin{equation}
3.35 \times \, \bigg(\frac{\langle \sigma v \rangle}{3 \times 10^{-26}\,\rm{cm}^3/\rm{s}}\bigg)^{1/2} \,\left( \frac{\sigma_{\mathrm{H, SD}} + 2.6 \, \sigma_{\mathrm{H, SI}}
+ 0.175 \, \sigma_{\mathrm{He, SI}}}{10^{-6} \, \mathrm{pb}} \right)^{1/2} \, \bigg(\frac{1\,\rm{TeV}}{m_X}\bigg)^{1/4} \gg 1.
\end{equation}
Inserting the appropriate annihilation and elastic scattering cross sections, the left-hand side of this condition reduces to:
\begin{equation}
\sqrt{C_{\odot} A_{\odot}} t_{\odot} \approx 3.4\, \bigg(\frac{1 \, \rm{TeV}}{m_{B^{(1)}}}\bigg)^{13/4} \bigg(\frac{0.1}{\Delta}\bigg),
\end{equation}
leading us to conclude that capture-annihilation equilibrium will generally be reached in the case of KKDM.

Annihilations of $B^{(1)}$s then proceed to generate neutrinos either directly or through the cascades of other annihilation products. The resulting muon neutrino spectrum at Earth is given by:
\begin{equation}
\frac{dN_{\nu}}{dE_{\nu}} =\frac{C_{\odot} F_{\rm{Eq}}}{4 \pi D^2_{\rm{ES}}} \bigg(\frac{dN_{\nu}}{dE_{\nu}}\bigg)^{\rm{Inj}}.
\label{nuspec}
\end{equation}
Here, $F_{\rm{Eq}} \equiv \tanh^2 (\sqrt{C_{\odot} A_{\odot}} t_{\odot})$ is the non-equilibrium suppression factor ($\approx 1$ for capture-annihilation equilibrium), $D_{\rm{ES}}$ is the Earth-Sun distance, and $(\frac{dN_{\nu}}{dE_{\nu}})^{\rm{Inj}}$ is the neutrino spectrum injected in the Sun per annihilating $B^{(1)}$. Most of the energetic neutrinos produced come from annihilations to either neutrino or tau pairs. For the 2\% to 3\% of $B^{(1)}$ annihilations which produce tau neutrino or muon neutrino pairs, this injected spectrum is simply a delta function at $E_{\nu}=m_{B^{(1)}}$ (neglecting the effects of neutrino scattering in the solar medium). More than 20\% of $B^{(1)}$ annihilations produce tau pairs, which generate a hard spectrum of neutrinos through their leptonic decays $\tau^- \rightarrow \mu^- \nu_{\tau} \bar{\nu}_{\mu}$ and a softer component through charged pions from hadronic cascades.

The spectrum of Eq.~(\ref{nuspec}) can be modified due to a number of factors. Firstly, energetic neutrinos can scatter with nucleons in the Sun. The probability of a neutrino escaping from the Sun without interacting is approximately given by~\cite{crotty,edsjo}:
\begin{equation} 
P = e^{-E_{\nu}/E_k}
\end{equation}
where $E_k \simeq$ $130$, $160$, $200$, $230$ GeV for $\nu_\mu$, $\nu_\tau$, $\overline{\nu}_\mu$, $\overline{\nu}_\tau$, respectively. Most of these interactions are through charged current, and therefore result in the absorption of the neutrino. Charged current interactions of tau neutrinos, however, generate a tau-lepton which can decay before losing the majority of its energy to the solar medium. This leads to an effect known as tau-regeneration. Finally, neutrino oscillations must also be taken into account. Muon and tau neutrinos mix into equal flavors as they travel from the Sun to the Earth.

\begin{figure}
\centering
\mbox{\includegraphics[width=0.45\textwidth,clip=true]{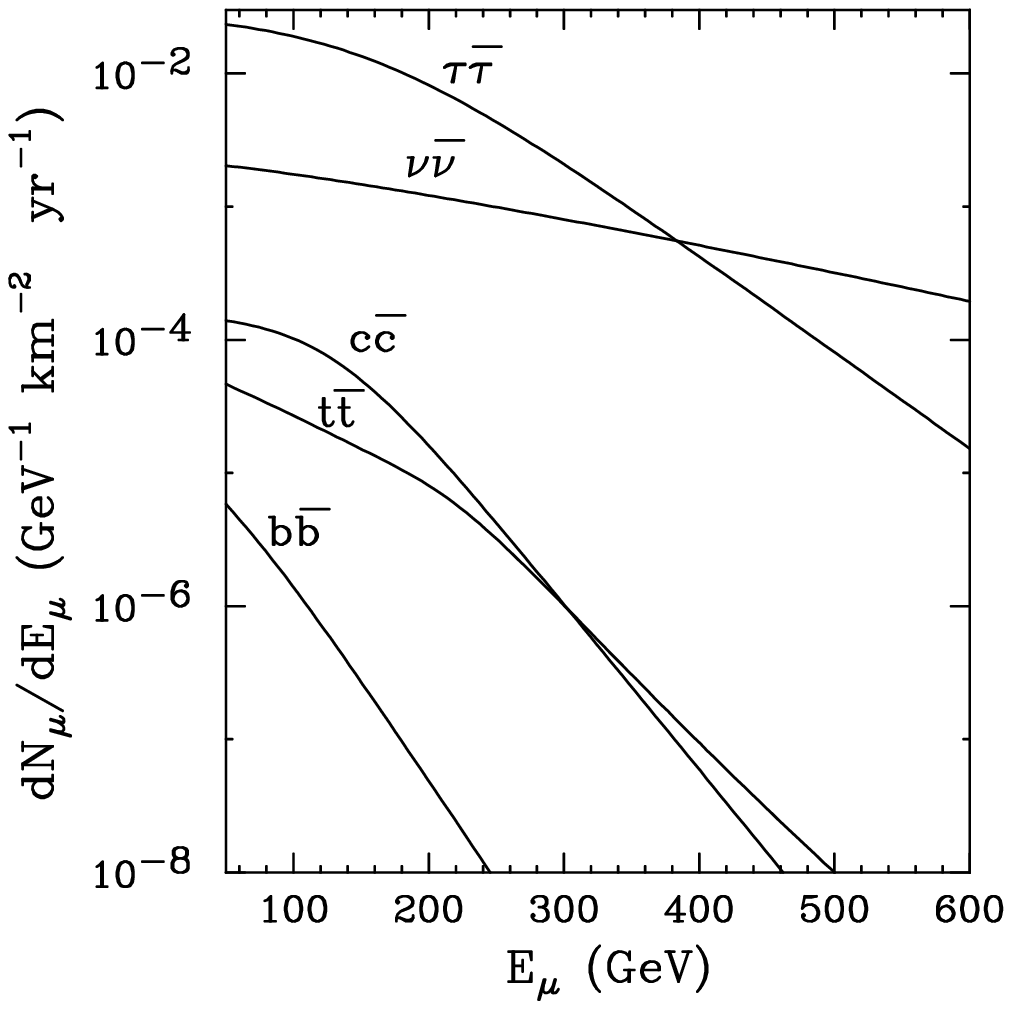}\qquad
\includegraphics[width=0.45\textwidth,clip=true]{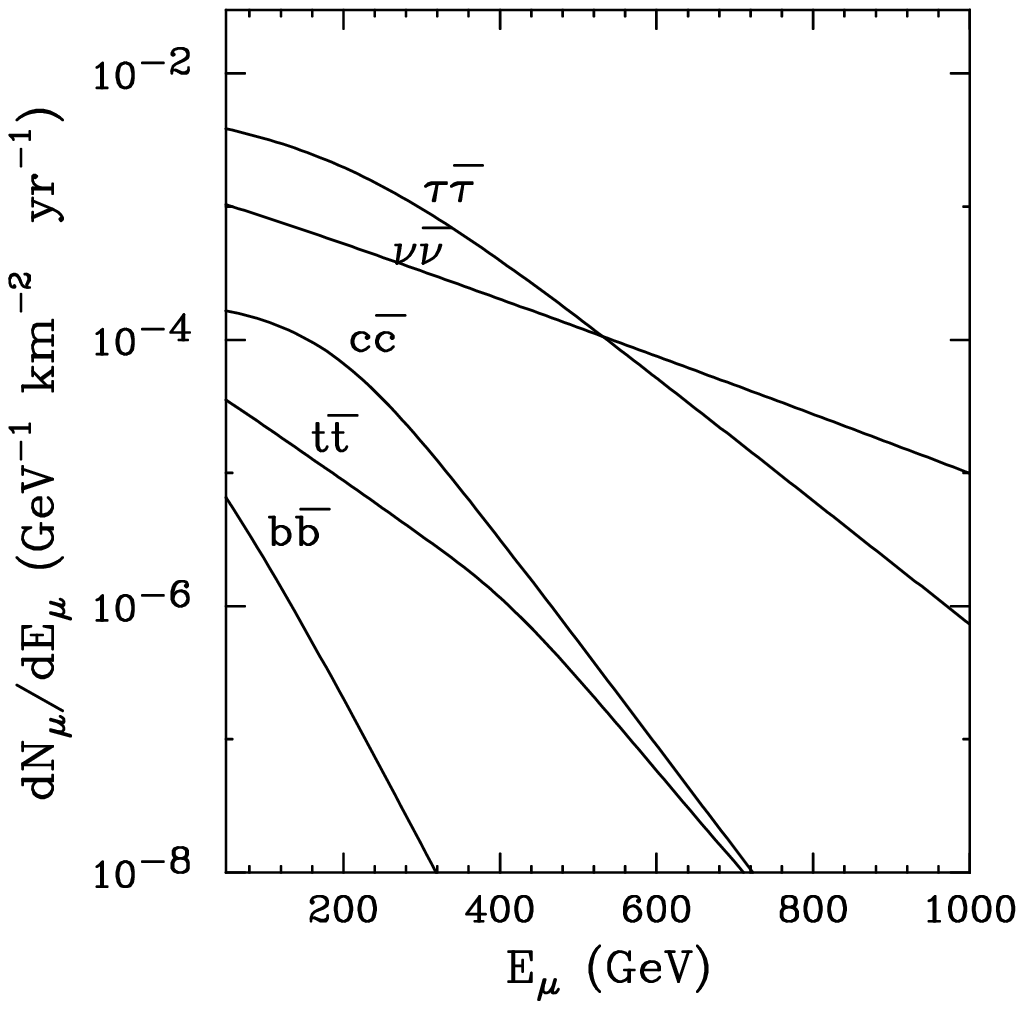}}
\caption{The spectrum of muons generated through charged current neutrino interactions at Earth, for the various annihilation modes of the $B^{(1)}$. In the left and right frames, $m_{B^{(1)}}=$ 600 GeV and 1000 GeV, respectively. The spectrum is dominated by annihilations to neutrino and tau pairs. From Ref.~\cite{Hooper:2002gs}.}
\label{muonspec}
\end{figure}

\begin{figure}
\centering
\includegraphics[width=0.55\textwidth,clip=true]{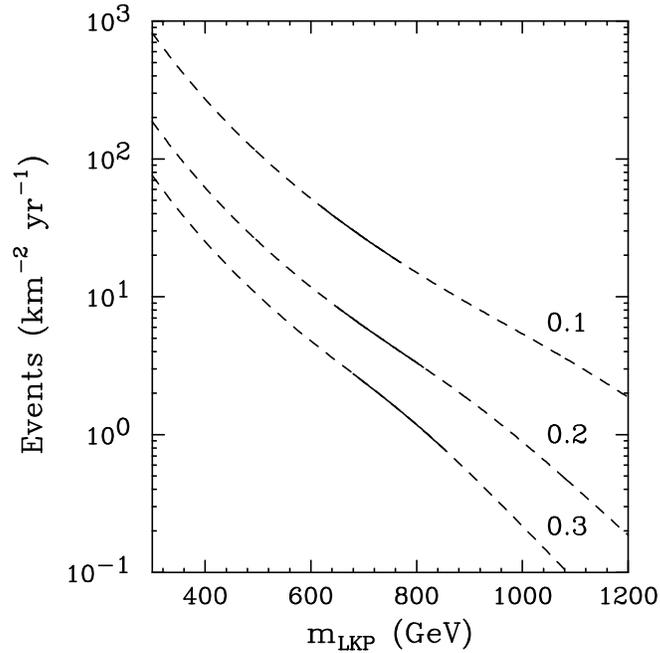}
\caption{The event rate in a kilometer-scale neutrino telescope resulting from $B^{(1)}$ annihilations in the Sun. Values of 0.1, 0.2 and 0.3 for the quantity $\Delta \equiv (m_{q^{(1)}}-m_{B^{(1)}})/m_{B^{(1)}}$ have been used. From Ref.~\cite{Hooper:2002gs}.}
\label{neurate}
\end{figure}

Upon reaching the Earth, a small fraction of the muon neutrinos interact via charged current in the ice or water near a neutrino telescope, generating energetic muon tracks at a rate of:
\begin{equation}
N_{\rm{events}}\simeq \int \int \frac{dN_{\nu_{{\mu}}}}{dN_{\nu_{{\mu}}}} \frac{d\sigma_{\nu}}{dy}(E_{\nu_{\mu}},y)\,R_{\mu}((1-y)E_{\nu})\,A_{\rm{eff}}\, dE_{\nu_{\mu}} dy,
\end{equation}
where $R_{\mu}(E_{\mu})$ is the distance a muon travels below falling below the energy threshold of the detector, called the muon range, and $A_{\rm{eff}}$ is the effective area of the detector. In Fig.~\ref{muonspec}, we show the spectrum of muons produced in this way. These muons lead to an event rate in a kilometer-scale neutrino telescope such as IceCube \cite{icecube1,icecube2} or KM3 \cite{km3}, which is shown in Fig.~\ref{neurate}.

For these neutrinos to be identified, they must overcome the background of atmospheric neutrinos. Above 100 GeV, this corresponds to roughly 80 background muons in the Sun's angular window per square kilometer, per year. Over a decade of observation, a 3$\sigma$ detection at a kilometer-scale experiment would, therefore, require a rate of $\sim 3 \sqrt{80/10} \sim 8$ per square kilometer, per year.

We see from Fig.~\ref{neurate} that observable event rates can potentially be generated for a combination of a light $B^{(1)}$ and small mass splittings between the LKP and KK-quarks. For KK-quarks 10\% heavier than the LKP, an observable rate is expected for $m_{B^{(1)}} \lesssim 900$ GeV, for example.


Currently, constraints on the energetic neutrino flux from the Sun have been placed by the Super-Kamiokande \cite{superklimit}, BAKSAN \cite{baksanlimit}, AMANDA \cite{amandalimit} and MACRO \cite{macrolimit} experiments. These experiments only limit the rate to less than $\sim$2000 events per square kilometer, per year, however, so do not yet constrain these models. Future experiments with considerably larger effective areas will improve upon these limits considerably, potentially reaching the level of sensitivity needed to observe on the order of ten events per square kilometer, per year.

Comparing the case of KKDM to neutralinos in supersymmetric models, we point out that larger neutrino rates are typically expected for KKDM for a number of reasons. Firstly, the flux of neutrinos produced is enhanced by the relatively large fraction of $B^{(1)}$ annihilations which produce $\tau^+ \tau^-$ (20--25\%) and $\nu_{\mu} \bar{\nu}_{\mu}$ or $\nu_{\tau} \bar{\nu}_{\tau}$ (2--3\%). Neutralinos, in contrast, annihilate largely to heavy quarks and gauge or Higgs bosons, which produce fewer energetic neutrinos in their decays. Secondly, $B^{(1)}$s elastically scatter in the Sun primarily through spin-dependent couplings, and thus are not limited by the results of CDMS and other direct detection experiments. The elastic scattering of neutralinos in the Sun can be dominated by either spin-dependent or spin-independent contributions. Spin-independent scatterings typically lead to either a direct detection rate in excess of the CDMS limit, or a neutrino rate below the reach of kilometer-scale neutrino telescopes \cite{halzen}.

\subsection{Anti-Matter Searches}\label{sec:anti}

The possibility of revealing the presence of an exotic particle population in our Galaxy through cosmic-ray searches has been considered for a long time~\cite{Zeldovich:1980st}. It was soon realized that if the ``missing matter'' was made up of WIMPs, one could hope to indirectly detect it through the detection of its stable annihilation products. The relative under-abundance of antimatter in cosmic rays, and the fact that WIMP annihilations typically yield as much matter as antimatter, motivated the possibility of disentangling an exotic signature originating from WIMP annihilations in the Galactic halo from the secondary and tertiary cosmic-ray background. 

The possibility of using positrons \cite{Silk:1984zy,Stecker:1985jc,Tylka:1989xj} and low-energy antiprotons \cite{Silk:1984zy} in this way has been discussed for some time. Although early studies mainly focused on the possibility of tracing anomalies in the observed antimatter spectra back to an exotic contribution from neutralino annihilations in the Galactic halo \cite{Tylka:1989xj,Kane:2001fz,Baltz:2001ir}, the possibility of {\em constraining} dark matter models (particularly neutralino dark matter scenarios) through their antimatter yields was also outlined \cite{Baltz:1998xv,Bergstrom:1999qv,Donato:2003xg}. Following the subsequent release of experimental data \cite{positronsexp,antiprotonsexp}, it became clear that a standard cosmic ray background could in most cases provide an excellent fit to the available antimatter fluxes \cite{Bergstrom:1999qv}. What is sometimes dubbed ``HEAT-anomaly'' \cite{heatanomaly}, an excess in the high energy end of the positron fraction (the ratio of positrons to positrons plus electrons), is a possible exception, although the measured positron spectrum itself (as opposed to the positron fraction) can be satisfactorily fitted with a pure astrophysical background, and the associated statistics are rather poor. The possible interpretation of this excess in terms of positrons produced through Galactic dark matter annihilations can be found, for example, in Refs.~\cite{Baltz:2001ir,Hooper:2003ad,Hooper:2004xn}.

Antimatter searches are entering an exciting era, where space based experiments such as PAMELA and AMS-02 \cite{Circella:2004gu,Spillantini:2004dp} will collect an unprecedented wealth of data covering a significantly larger energy range than has been explored thus far by balloon-borne experiments \cite{positronsexp,antiprotonsexp}. Furthermore, these new experiments will help in constraining and appropriately modeling the cosmic-ray background as well. Recent analysis addressing the question of whether these next-generation experiments will have the ability to statistically discriminate a WIMP-annihilation-induced positron and antiproton component for various particle physics setups and dark matter halo profiles can be found in Refs.~\cite{Donato:2003xg,Hooper:2004bq,Profumo:2004ty}.

The possibility of detecting a dark matter annihilation signature in the antideuteron flux was envisioned only much more recently \cite{Donato:1999gy,Mori:2001dv}. Compared to the flux of antiprotons produced in WIMP annihilations, the number of antideuterons is significantly suppressed. Since the threshold energy for the production of an antideuteron in proton-proton collisions is rather large, however, the impinging proton must feature a large kinetic energy. Consequently, the center-of-mass velocity of the proton-proton system is typically large, and the resulting antideuteron kinetic energy sizable. Therefore, the background of low kinetic energy antideuterons is expected to be highly suppressed. The exciting experimental possibility of selectively capturing slow antideuterons, and of identifying them with a remarkably high accuracy through atoms X-ray de-excitations, has lead to the design of a dedicated device, the general antimatter particle spectrometer, GAPS \cite{Hailey:2005yx}. In some setups, even the detection of a single low-energy antideuteron could constitute a signature of exotic physics, including that of dark matter annihilations in the Galactic halo \cite{Baer:2005tw}.

In the following sub-section, we discuss the latter case of KKDM detection by low-energy antideuteron search experiments. The case of antiproton and positron searches are discussed in Sec.~\ref{sec:pb} and \ref{sec:ep}, respectively. In Sec.~\ref{sec:spaceam}, we discuss the prospects for KKDM detection by space-based antimatter exploration experiments.

\subsubsection{Antideuterons}\label{sec:db} 

Antideuteron searches can be performed either with magnetic spectrometers mounted on balloon-borne (BESS/BESS-Polar) or space-borne (AMS) missions, 
or through GAPS-like devices, based on the radiative emissions of antiparticles captured into exotic atoms. 
The latter can be installed either on balloons or on satellites, and are specifically designed to look for low-energy antiparticles.

For both antideuterons and other antimatter species (see the following two sections) the effects of solar modulation can be accounted for within the framework of the Gleeson-Axford analytical force-field approximation~\cite{GleesonAxford}, 
where the interstellar flux at the heliospheric boundary, ${\rm d}\Phi_b/{\rm d}T_b$, and at the Earth, ${\rm d}\Phi_\oplus/{\rm d}T_\oplus$, are related by
\begin{equation}
\frac{{\rm d}\Phi_\oplus}{{\rm d}T_\oplus}(T_\oplus)=\frac{p^2_\oplus}{p^2_b}\ \frac{{\rm d}\Phi_b}{{\rm d}T_b}(T_b),
\end{equation}
where the energy at the heliospheric boundary is given by $E_b=E_\oplus+|Ze|\phi_F$, and $p_b$ and $p_\oplus$ stand for the momenta at the boundary and at the Earth, 
and $\phi_F$ is the above mentioned solar modulation Fisk~\cite{fisk} parameter, which is taken to be approximately charge--sign independent. Physically, the Fisk parameter corresponds to the effective electric potential associated to the presence of solar wind.

The flux of low-energy antideuterons is currently constrained by the recent results of Ref.~\cite{Fuke:2005it}. The BESS experiment looked for low-energy antideuterons during four flights (1997, 1998, 1999, and 2000) in the kinetic energy interval $0.17\div 1.15$ GeV/n. The upper and lower kinetic energy limits come, respectively, 
from the particle identification procedure and from the decrease of the geometrical acceptance and mean free path through the detector. 
Without assumptions on the $\dbar$ spectral shape, the BESS collaboration, by combining all four missions, 
derived an upper limit on the $\dbar$ flux, at 95\% C.L.,  of
\begin{equation}\label{eq:besslimit}
\phi^{\rm BESS}_{\dbar}\ < 1.9\times 10^{-4}\ \unitsn.
\end{equation}
The Fisk solar modulation parameter $\Phi_F$ was derived from the $p$ data from the same experiment, 
and is set to 500, 610, 648, 1334 MV for the four years of data-taking.

The future space-based spectrometer AMS-02 will be sensitive to low-energy antideuterons~\cite{ams}. For a total data-taking time of 3 years, and for an antideuteron kinetic energy band extending 
from the AMS threshold of 100 MeV/n to $\sim2.7$ GeV/n, the inferred acceptance reads $5.5\times 10^7\ {\rm m}^2\ {\rm s}\ {\rm sr}\ {\rm GeV}$ \cite{Donato:1999gy}. 

The GAPS experiment has recently undergone a rich phase of R\&D, carried out at the KEK accelerator in Japan \cite{Hailey:2005yx}. It has been realized that solid and liquid targets can greatly simplify the needed payload mass (thanks to the removal of the dead mass of the gas handling system) and the complexity of the apparatus, 
yielding an increased background rejection capability enabling the capture of more than 3 X-rays, as initially conceived. 
Furthermore, pion showers ($\pi^*$) and nuclear X-rays from the antiparticle annihilation in the target nuclei, neglected in the original sensitivity calculations~\cite{Mori:2001dv}, 
have been shown to significantly increase the antiparticle identification capability. 
The GAPS collaboration plans to test the finalized payload with a prototype as early as 2009, and to achieve a long duration balloon (LDB) 
flight from Antarctica, or an ultra-LDB (ULDB) flight from Australia as early as 2011 \cite{Hailey:2005yx,hailey_privcom}. 
A preliminary evaluation of the two balloon-borne options (with the LDB sensitivity based on 3 flights) found projected sensitivities of~\cite{hailey_privcom}
\begin{equation}
\phi^{\rm LDB}_{\dbar}\ \approx 1.5\times 10^{-7}\quad \units \ {\rm and}\quad \phi^{\rm ULDB}_{\dbar}\ \approx 3.0\times 10^{-8}\ \units
\end{equation}
over a bandwidth of $0.1<T_{\dbar}/({\rm GeV/n})<0.25$. The sensitivity of the GAPS prototype mounted on a satellite was assessed in Ref.~\cite{Mori:2001dv}. For a 3 year satellite mission, the projected sensitivity is
\begin{equation}
\phi^{\rm GAPS/S}_{\dbar}\ \approx 2.6\times 10^{-9}\quad \units,\quad\quad {\rm for} \quad\quad  0.1<T_{\dbar}/({\rm GeV/n})<0.4 .
\end{equation}
A last, very optimistic option, mentioned in Ref.~\cite{Mori:2001dv}, is to send GAPS on a deep-space probe, where eventually solar modulation 
effects can be significantly reduced. Depending on the spectral shape of the differential $\dbar$ yield from WIMP pair annihilation, solar modulation can deplete the low energy antideuteron flux, hence this interplanetary GAPS setup might represent the ultimate probe for dark matter searches via the detection of low energy $\dbar$s~\cite{Profumo:2004ty}.

The computation of the differential flux of $\dbar$s per kinetic energy per nucleon interval induced by WIMP pair-annihilations involves a number of steps, 
which we briefly review below, referring the reader to Refs.~\cite{Donato:1999gy,Profumo:2004ty,Baer:2005tw} for more details. 

The computation of the source spectrum for the primary antideuteron flux originating from WIMP pair annihilation is based on three hypothesis: 
(1) correlations in the associated production of two antinucleons are neglected, {\em i.e.} the probability of producing a pair of antinucleons is assumed to be equal to the product of the probabilities of producing two single antinucleons ({\em factorization}) 
(2) the antineutron production cross section is equal to the antiproton production cross section ({\em isospin invariance}) and 
(3) the formation of an antideuteron can be described by the {\em coalescence model}. 
We refer the reader to Refs.\cite{Chardonnet:1997dv,Donato:1999gy,Donato:2001sr,Duperray:2005si} for a through discussion of the validity of these assumptions. 
The factorization assumption is conservative, in that the probability of pair producing antinucleons in the same jet 
is presumably not factorized, since their momenta will not be isotropically distributed. 
The main idea of hypothesis (3) is that whenever the difference of the momenta of an antiproton and an antineutron produced in a jet 
resulting from a WIMP pair annihilation is less than a phenomenologically given value $2p_0$, where $p_0$ indicates the {\em coalescence momentum}, 
then an antideuteron is formed. 

The differential energy spectrum of primary antideuterons produced in the pair annihilation of KKDM
can then be expressed by \cite{Donato:1999gy}
\begin{equation}
\frac{{\rm d}N_{\dbar}}{{\rm d}E_{\dbar}}=\left(\frac{4\ p_0^3}{3\ k_{\dbar}}\right)\ \left(\frac{m_{\dbar}}{m_{\bar p}\ m_{\bar n}}\right)\ 
\sum_{f}BR(B^{(1)} B^{(1)}\rightarrow f)\times\left(\frac{{\rm d}N^{(f)}_{\bar p}}{{\rm d}E_{\bar p}}\left(E_{\bar p}=E_{\dbar}/2\right)\right)^2 ,
\end{equation}
where $E_{\dbar}^2=m_{\dbar}^2+k_{\dbar}^2$, $f$ indicates any final state of the WIMP pair annihilation process occurring with a branching ratio 
$BR(B^{(1)} B^{(1)}\rightarrow f)$, and ${\rm d}N^{(f)}_{\bar p}/{\rm d}E_{\bar p}$ is the antiproton differential yield for the final state $f$. 
The latter can be computed using the {\tt Pythia} Monte Carlo event generator \cite{pythia,pythia2}, as implemented in the {\tt DarkSUSY} package~\cite{Gondolo:2004sc}, for example. The reference value we assume for the coalescence momentum is 58 MeV, the same choice as in Refs.~\cite{Chardonnet:1997dv} and \cite{Donato:1999gy}, 
not too far from what is expected from the antideuteron binding energy, $\sqrt{m_p\ B}\approx 46$ MeV. For a discussion of the effects of the propagation of antideuterons through the Galactic magnetic fields, we refer the reader to Ref.~\cite{Baer:2005tw}.

The source spectrum of antideuterons is specified at every point in the Galactic halo once the shape of the dark matter halo itself is given. 
We take here as an (optimistic) reference model the adiabatic contraction \cite{blumental} of the N03 halo profile \cite{n03} (see Ref.~\cite{pierohalos} for details), 
which closely resembles the profile proposed by Moore et al., \cite{Moore:1999nt}. We consider here a smooth halo profile, but we discuss in Sec.~\ref{sec:spaceam} the potential effects of dark matter halo substructures.

\begin{figure}[!t]
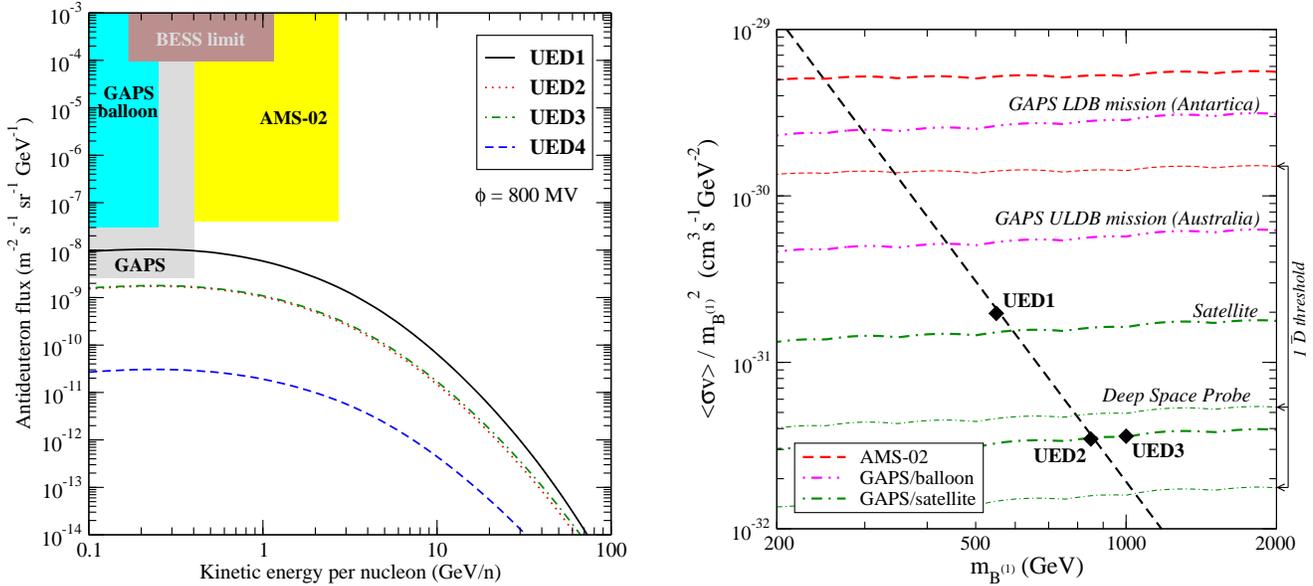

\centering
\mbox{\hspace*{-0.05\textwidth}\includegraphics[width=0.52\textwidth,clip=true]{FIGURES/dbar_dnde.eps}\qquad\includegraphics[width=0.52\textwidth,clip=true]{FIGURES/dbar_all.eps}}
\caption{Left: The differential antideuteron flux from the four benchmark models ({\bf UED1}-{\bf 4}) of Appendix~\ref{ch:benchmarks} (the results for {\bf UED2} and {\bf UED3} fall almost on top of each other). The shaded regions correspond to the sensitivities of various existing and proposed experiments featuring antideuteron searches. Right: The sensitivity reach of antideuteron search experiments for the UED model, with a $B^{(1)}$ LKP (dashed black lines) in the $(m_{B^{(1)}},\langle\sigma v\rangle_0/m_{B^{(1)}}^2)$ plane. We also indicate the location of three of the benchmark models of Appendix~\ref{ch:benchmarks}.}
\label{fig:dbar}
\end{figure}
We collect in the left frame of Fig.~\ref{fig:dbar} the results of the computation of the antideuteron flux at Earth for the four benchmark models found in Appendix~\ref{ch:benchmarks}. We assume here a Fisk parameter $\phi_F=800$ MV. In the figure, the shaded regions correspond to the sensitivities and kinetic energy range of various existing and proposed experiments featuring antideuteron searches, as outlined above. Of the benchmark models we consider here, models {\bf UED1}, {\bf UED2} and {\bf UED3} are likely to generate at least one low-energy $\dbar$ at a space-based GAPS experiment (grey-shaded area). Model {\bf UED1} might even produce some events at a balloon-borne GAPS experiment or at AMS-02. 

In the case of highly sensitive space-based antideuteron search experiments, it has been pointed out in Ref.~\cite{Baer:2005tw} that the role of the cosmic ray background cannot be neglected (unlike the case of, {\em e.g.} balloon-borne setups).  Previously neglected antideuteron production and energy loss processes, 
including secondary antideuteron production from antiproton scattering off the ISM, and a tertiary antideuteron component originating from the previously neglected non-annihilating inelastic scattering processes, 
have been shown to largely populate the low-energy end of the antideuteron spectrum~\cite{Duperray:2005si}. As a result, the detection of a single low-energy antideuteron would not be sufficient to claim an exotic signal, as it would more likely originate from the standard cosmic ray background \cite{Baer:2005tw}. A statistical approach showed that the number of low-energy $\dbar$ events needed to claim, at the 95\% confidence level, a $\dbar$ component of exotic origin corresponds to one event for ULDB GAPS/balloon, five for GAPS/satellite, three for a GAPS detector mounted on an interplanetary probe, and six for AMS.

In the right frame of Fig.~\ref{fig:dbar}, the 95\% confidence level sensitivies of the various $\dbar$ search experiments are shown in the  $(m_{B^{(1)}},\langle\sigma v\rangle_0/m_{B^{(1)}}^2)$ plane. The dashed black line in this figure indicates the standard value of the pair annihilation cross section of KKDM neglecting second-level KK mode resonances (notice that model {\bf UED3} does not lie on the dashed black line precisely because $n=2$ resonances do contribute). UED will be probed at balloon-borne GAPS missions for KKDM masses below $\approx 400$ GeV in the present halo model setup. Analogously, satellite--based experiments extend the reach in the $\dbar$ channel up to masses around $\approx 600$ GeV, and an interplanetary probe up to $m_{B^{(1)}}\approx 800$ GeV.

\subsubsection{Antiprotons}\label{sec:pb}

\begin{figure}[!t]
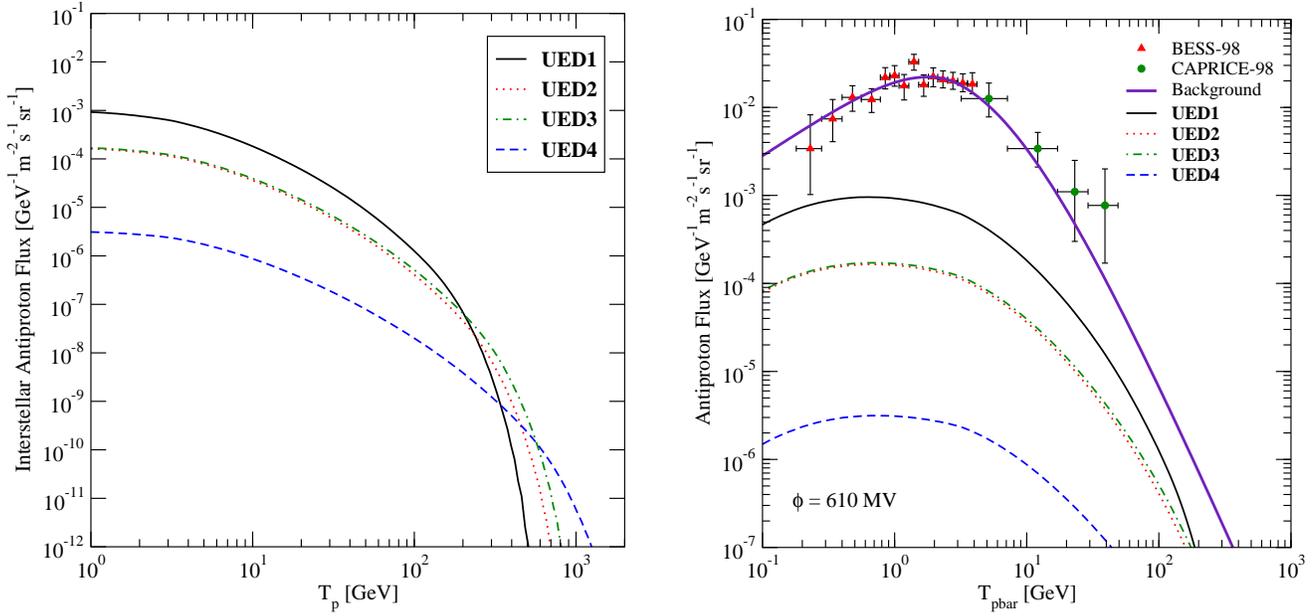

\centering
\mbox{\hspace*{-0.05\textwidth}\includegraphics[width=0.52\textwidth,clip=true]{FIGURES/pb_IS.eps}\qquad\includegraphics[width=0.52\textwidth,clip=true]{FIGURES/pb_SM.eps}}
\caption{The local interstellar (left) and solar-modulated (right) antiproton flux, as a function of the antiproton kinetic energy, for the four benchmark models ({\bf UED1}-{\bf 4}) of Appendix~\ref{ch:benchmarks} (the results for {\bf UED2} and {\bf UED3} fall almost on top of each other). In the right frame, we also include an estimate of the cosmic ray antiproton background \cite{Profumo:2004ty} and the current antiproton flux measurements \cite{antiprotonsexp}}
\label{fig:pb}
\end{figure}
The number of antiprotons (the case of positron is completely analogous) per unit time, energy and volume element produced in WIMP annihilations is called the source function, $Q$, and is given by~\cite{Bergstrom:1999jc}
\begin{equation}
Q(T,\vec{x})=\langle\sigma v\rangle_0\left(\frac{\rho_{B^{(1)}}(\vec{x}\,)}{m_{B^{(1)}}}\right)^{2}
  \sum_{f}^{}\frac{dN^{(f)}}{dT}BR(B^{(1)} B^{(1)}\rightarrow f)
\end{equation}
where $T$ stands for the antiproton (positron) kinetic energy, and for a given annihilation channel final state, $f$, $dN^{(f)}/dT$ is the fragmentation function into antiprotons (positrons). The effects of antiproton propagation through the Galaxy is described in Ref.~\cite{Bergstrom:1999jc} (see also Ref.~\cite{Profumo:2004ty}). The effects of solar modulation are as described in the previous section for the case of antideuterons.

In Fig.~\ref{fig:pb}, the spectra of antiprotons expected for the four benchmark models ({\bf UED1}-{\bf 4}) of Appendix~\ref{ch:benchmarks} are shown. In the left frame, the spectrum does not include the effects of solar modulation, while in the right frame, those effects are included. Also shown in the right frame is the expected cosmic ray background \cite{Profumo:2004ty} and the spectrum of antiprotons as measured by balloon-borne experiments~\cite{antiprotonsexp}. In the right frame, for consistency with the time of experimental data-taking for the measured fluxes shown, we make use of an averaged Fisk parameter $\phi_F=610$ MV. 

The spectrum of antiprotons in UED is rather soft and mostly originates from quark-antiquark final states. Compared to expectations for neutralino dark matter, the large branching fraction of $B^{(1)}$ pair annihilations in leptonic states suppresses the antiproton yield. Ref.~\cite{Bringmann:2005pp} also computed the contribution from the radiative pair annihilation channel into two gluons, and found that the resulting branching ratio is smaller than 0.5\%. We notice however that the ratio of the expected signal over background can be large, especially at very large energies ($T_{\bar p}\gtrsim50$ GeV), which will be probed at future space-based antimatter exploration experiments (see Sec.~\ref{sec:spaceam}). The low energy region ($0.1\lesssim T_{\bar p}/{\rm GeV}\lesssim10$ GeV), instead, is not only plagued with large uncertainties in the background computation and in the effects of solar modulation, but also typically features a very suppressed signal-to-background ratio. With smooth dark matter halos, currently available data on antiproton spectra (that are limited to relatively low antiproton kinetic energies) do not provide significant constraints on the UED model, even in the case of a relatively optimistic halo model setup (such as the one used here). 

Ref.~\cite{Bringmann:2005pp}, where the antiproton yield from KKDM was analyzed in detail, gets to similar conclusions to those sketched above, pointing out that the most promising energy range for the detection of a signal over the secondary background in the antiproton spectra to be measured in forthcoming space-based experiments lies between 10 GeV and a few 100 GeV. The presence of clumpiness in the dark matter distribution in the galactic halo can significantly enhance the predicted flux of antiprotons \cite{Bringmann:2005pp}. The resulting increase in the signal depends upon the fraction of dark matter mass assumed to lie in clumps, the assumed density contrast, and the antiproton energy \cite{Bringmann:2005pp}, and might be crucial for the detection of a WIMP annihilation signature, possibly to be correlated with a signal in the positron channel as well (see the following Sec.~\ref{sec:ep}). Ref.~\cite{Barrau:2005au}, which also addressed the flux of primary antiprotons in $B^{(1)}$ annihilations in the galactic halo, arrives at similar conclusions, emphasizing that the best energy range to detect a signal appears to be the high-energy tail, and the necessity of an increase in the antiproton flux from a clumped dark matter distribution. The uncertainties in the secondary antiproton background in that energy range were analyzed in Ref.~\cite{Bringmann:2006im} and compared to the expected primary antiproton fluxes from WIMP pair annihilations (including the case of $B^{(1)}$ dark matter). 

\subsubsection{Positrons}\label{sec:ep}
\begin{figure}[!t]
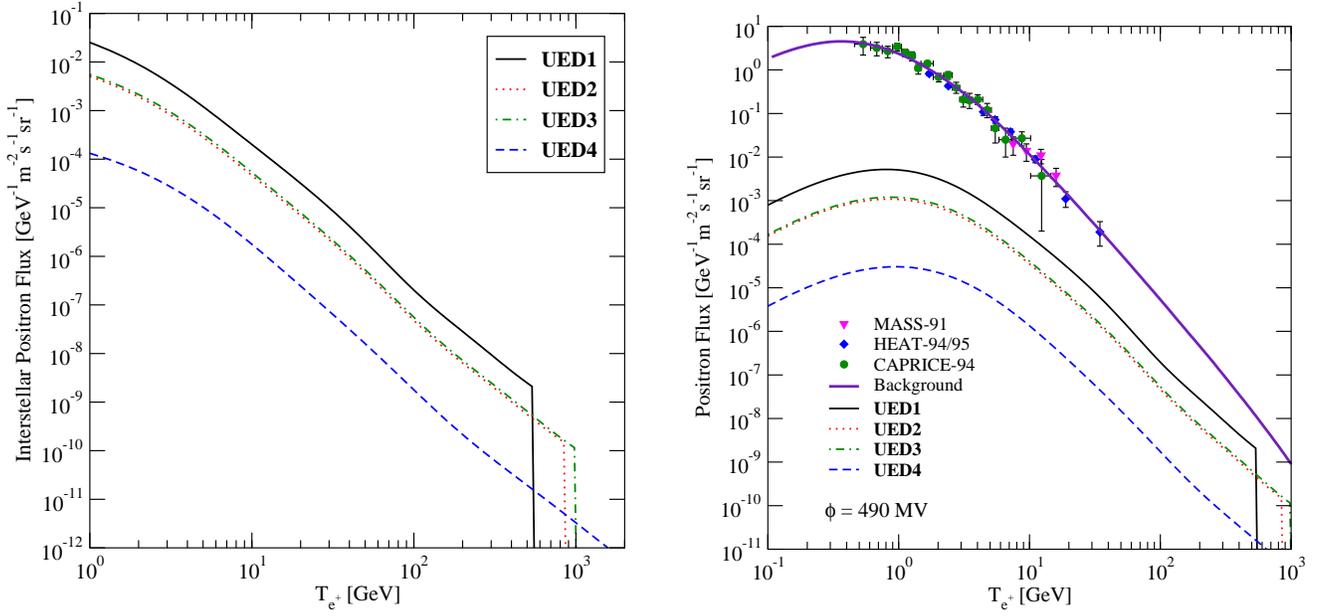

\centering
\mbox{\hspace*{-0.05\textwidth}\includegraphics[width=0.52\textwidth,clip=true]{FIGURES/ep_IS.eps}\qquad\includegraphics[width=0.52\textwidth,clip=true]{FIGURES/ep_SM.eps}}
\caption{The local interstellar (left) and solar-modulated (right) positron flux, as a function of the positron kinetic energy, for the four benchmark models ({\bf UED1}-{\bf 4}) of Appendix~\ref{ch:benchmarks} (the results for {\bf UED2} and {\bf UED3} fall almost on top of each other). In the right frame, we also include an estimate of the cosmic ray positron background \cite{Profumo:2004ty} and the current positron flux measurements \cite{positronsexp}.}
\label{fig:ep}
\end{figure}

$B^{(1)}$ pairs annihilate very frequently (around 20\% of the time) into monoenergetic electron-positron pairs with energies at production corresponding to the $B^{(1)}$ mass. An additional 40\% of time they annihilate into either muon or tau pairs which, again, yield energetic electrons at production. As pointed out in Refs.~\cite{Cheng:2002ej,Hooper:2004xn}, this implies a sizable positron flux from KKDM pair annihilations in the local Galactic halo.

Shown in the left frame of Fig.~\ref{fig:ep} is the interstellar positron flux expected from the four benchmark models ({\bf UED1}-{\bf 4}) found in Appendix~\ref{ch:benchmarks}. These spectra exhibit, especially at large energies, a spectral shape close to a power law, following (for this particular choice of a halo model and propagation setup) the approximate scaling
\begin{equation}
\frac{dN_{e^+}}{dE_{e^+}}\approx3\times 10^7\left(\frac{\rm 1\ GeV}{m_{B^{(1)}}}\right)^6\left(\frac{m_{B^{(1)}}}{E_{e^+}}\right)^3\ {\rm GeV}^{-1}\ {\rm cm}^{-2}\ {\rm s}^{-1}\ {\rm sr}^{-1}.
\end{equation}
The naive scaling with mass of the combination $\langle\sigma v\rangle/m_{B^{(1)}}^2\propto m_{B^{(1)}}^{-4}$ is compensated somewhat by the differential number of positrons produced per LKP annihilation, which grows approximately linearly with the $B^{(1)}$ mass.

The effects of solar modulation on the interstellar positron fluxes are included in the right frame of Fig.~\ref{fig:ep}. Also shown is the expected background \cite{Profumo:2004ty} and the current experimental data on the positron spectrum. We use an averaged solar modulation parameter $\phi\simeq490$ MV, appropriate for the period when the data shown were taken. We notice that the positron background has a significantly softer spectrum than the signal from KKDM pair annihilations, particularly at large energies. At high energies, the signal-to-background can be very large, even close to unity. Unfortunately, this occurs at energies which will probably be beyond the sensitivity of even space--based antimatter experiments.

To minimize the effects of solar modulation, a useful quantity to consider is the so-called positron fraction, {\em i.e.} the ratio, at a given energy, of the positron flux to the sum of the positron and electron fluxes. In this context, the background is given by the standard expectation for the positron fraction, and the dark matter contribution shows up as an extra component, enhancing the positron fraction.

In 1994 and 1995, the High-Energy Antimatter Telescope (HEAT) reported 
an excess in cosmic positron fraction, peaking in the range of 7-10 GeV, 
and continuing to higher energies \cite{heatanomaly}.  In 2000, 
an additional HEAT flight confirmed this observation \cite{heat2000}. 
Many previous experiments, although less precise, have also recorded a larger 
than expected positron flux above about 10 GeV (see Ref.~\cite{heatanomaly} 
and references therein). The excess of positrons at large energies has also been recently confirmed by AMS-01 data, although with a rather poor statistics \cite{Gast:2006hb}. The study of the astrophysical production 
of positrons \cite{Protheroe} has been thoroughly investigated \cite{secbg}, 
with the conclusion that the ratio of cosmic positrons to electrons 
above about 10 GeV is higher than is suggested by secondary production 
in a model of a diffusive halo.

\begin{figure}[!t]
\centering
\mbox{\includegraphics[width=0.48\textwidth,clip=true]{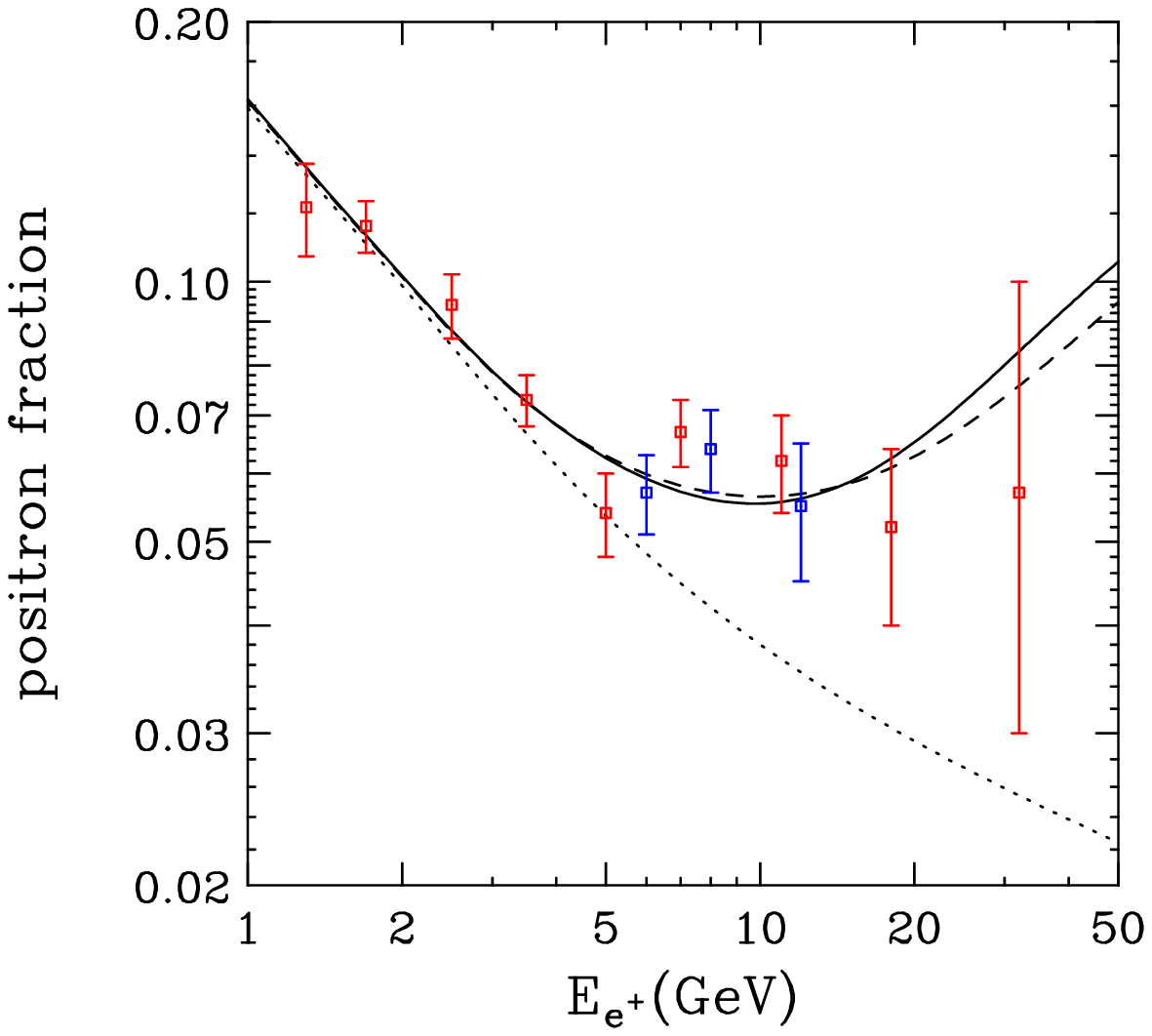}\qquad\includegraphics[width=0.48\textwidth,clip=true]{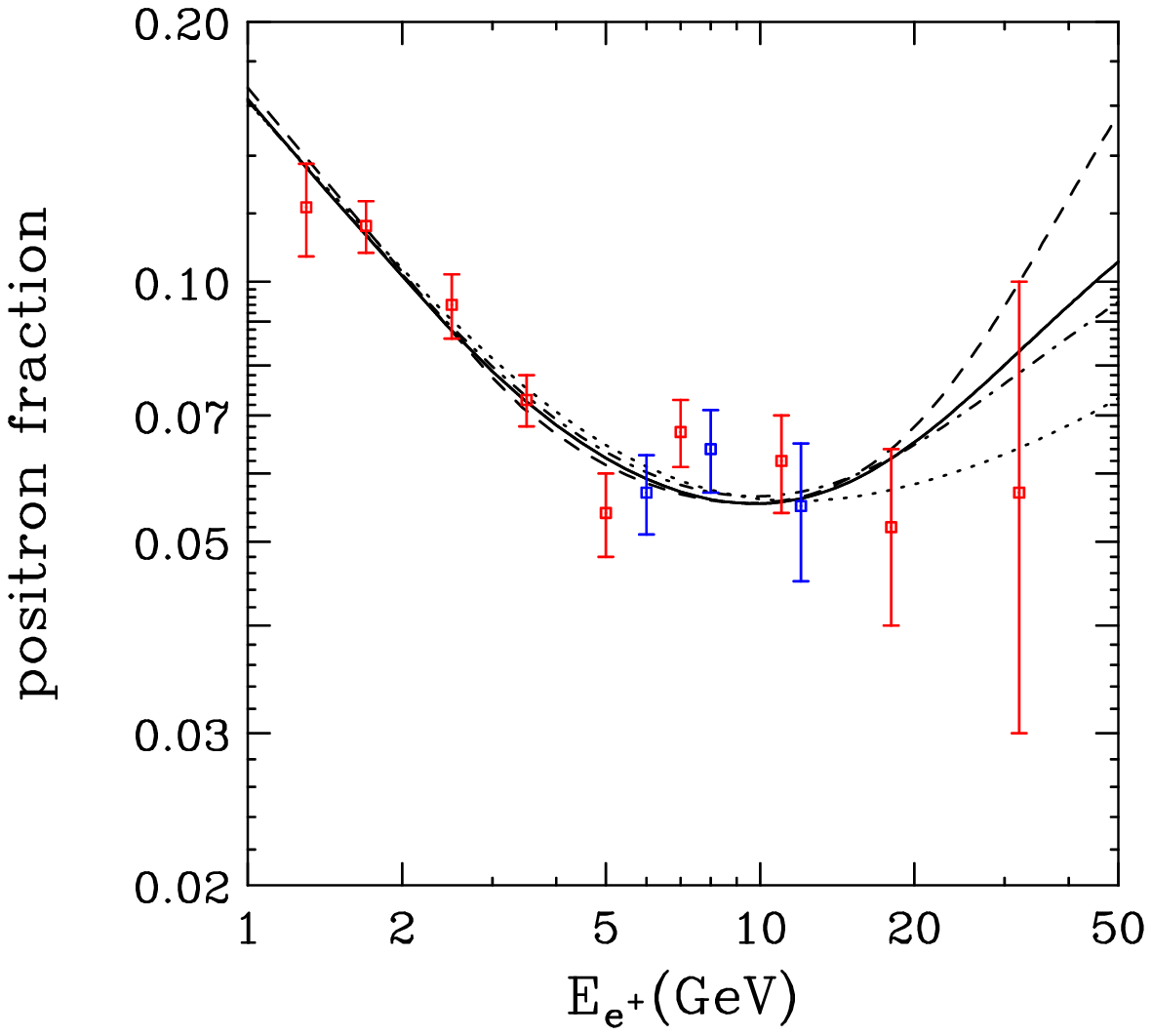}}
\caption{Left: The positron fraction from KKDM annihilations as 
a function of positron energy.  The solid and dashed lines represent 
300 and 600 GeV $B^{(1)}$s, respectively. The annihilation rate was treated 
as a free parameter, used for normalization. The dotted line represents 
the background predicted with no contribution from dark matter annihilation. 
The error bars shown are from the 1994-95 and 2000 HEAT flights. Right: The positron fraction from annihilation of KKDM for several 
choices of propagation parameters. The solid line represents the model 
with the same propagation parameters as in the left frame.  
The dashed line is for a model with an energy loss rate smaller by a 
factor of two.  The models represented by dot-dashed and dotted lines 
use the full energy loss rate but diffusion constants that are 
$80\%$ and $50\%$, respectively, of the value used in the left frame.  Lastly, 
the spectrum with {\em both} half the diffusion constant and half the 
energy loss rate falls on top of the solid line. 
In the right frame, $m_{B^{(1)}} = 300$ GeV.}
\label{fig:posfrac}
\end{figure}
This excess can be, and indeed was, tentatively interpreted as originating from the production of positrons in Galactic dark matter annihilations (see {\em e.g.} Refs.~\cite{Baltz:2001ir,Hooper:2003ad}). Ref.~\cite{Hooper:2004xn} explored this interpretation in the context of KKDM. Shown in Fig.~\ref{fig:posfrac} is the HEAT data compared to the expectation from $B^{(1)}$ pair annihilations. In the left frame, two values for the $B^{(1)}$ mass are shown (300 and 600 GeV). In each case, the normalization of the annihilation rate was chosen to obtain the best possible fit to the HEAT data. This procedure implies annihilate rates larger than those expected from a smooth dark matter halo distribution. To fit the HEAT data with KKDM, annihilation rates boosted by a factor on the order of $10^2$ are typically required.

The spectrum shown in Fig.~\ref{fig:posfrac} can be modified somewhat through the choice of diffusion parameters which are adopted. In the left frame, the following diffusion parameters were used: a diffusion constant of $K(E_{e^{+}}) = 3.3 \times 10^{28} (E_{e^{+}}/1\,\rm{GeV})^{0.47} \, 
\rm{cm}^2/\rm{s}$ \cite{strong02}, an energy loss rate of 
$b(E_{e^{+}}) = 10^{-16} (E_{e^{+}}/1\,\rm{GeV})^2 \,\, \rm{GeV/s}$ and a $2 L =8\,$ kpc thick slab for the diffusion zone \cite{strong02}.   In the right frame of Fig.~\ref{fig:posfrac}, these parameters are varied somewhat to illustrate the uncertainties involved.

Space-based antimatter search experiments, discussed in the following section, will collect far better statistics and extend the explored range of positron energies considerably. It is reasonable to expect that PAMELA and AMS-02 will definitively confirm, or rule out, the above mentioned HEAT positron excess and its interpretation in terms of dark matter annihilation.

\subsubsection{Prospects For Space-Based Antimatter Experiments}\label{sec:spaceam}

Upcoming space-based antimatter experiments will tremendously enhance the resolution and accuracy of positron and antiproton spectra measurements in comparison to existing balloon borne results. The PAMELA experiment~\cite{Picozza:2006nm}, on board the Resurs DK1 satellite, was launched from the Baikonur cosmodrome on June 15, 2006.  Also, AMS-02~\cite{ams} is expected to be launched and installed on board the International Space Station within the next few years. With the purpose of assessing the capabilities of these experimental facilities to identify signatures of dark matter annihilations, we follow Ref.~\cite{Profumo:2004ty}. To this extent, we will implement a statistical $\chi^2$ analysis to compare the case of a pure background measurement to that of a signal from dark matter annihilations. 

The relevant experimental parameters entering the estimate are given by:
\begin{itemize}
\item[-] The {\em geometrical factor} of the experimental facility, {\it i.e}. its effective area, $A$
\item[-] The time of data acquisition, $T$
\item[-] The energy coverage of experiment, as defined by the number of and size of the energy bins
\end{itemize}

We will declare that a given model is {\em discriminable} at a certain future experiment at given confidence level, $X\%$, if the $\chi^2$ induced by the dark matter model is larger than the $(\chi^2)^{X\%}_{n_{\rm b}}$ corresponding to $n_{\rm b}$ degrees of freedom. Letting $N_i^P=N_i^S+N_i^B$ be the number of projected signal plus backgroud events in a given bin, $i$, and $N_i^O$ be the number of observed events, with a standard deviation, $\Delta_{N_i^O}$, the $\chi^2$ is defined as 
\begin{equation}\label{eq:chin}
\chi^2=\sum_{i=1}^{n_{\rm b}}\frac{\left(N_i^P-N_i^O\right)^2}{\left(\Delta_{N_i^O}\right)^2},\, \, \, \, N_i^P=N_i^S+N_i^B. 
\end{equation}
Supposing that the standard deviation has a Gaussian distribution, one finds
\begin{equation}
\Delta_{N_i^O}\simeq \sqrt{N_i^O}.
\end{equation}
We are interested in finding the limiting cases, {\it i.e}. those cases for which the signal is a small component with respect to the background, as is the case for all KKDM models considered here (see Figs.~\ref{fig:pb} and \ref{fig:ep}). We therefore make the assumption that:
\begin{equation}\label{eq:approximation}
N_i^S\ \ll N_i^B\;,\;\;\;\;{\rm or}\;\;\;\;N_i^O\ \simeq\ N_i^B\;.
\end{equation}
Eq.~(\ref{eq:chin}) then reduces to
\begin{equation}\label{eq:chinbis}
\chi^2=\sum_{i=1}^{n_{\rm b}}\frac{\left(N_i^S\right)^2}{\left(\sqrt{N_i^B}\right)^2}.
\end{equation}
Since the number of events in an energy bin $\Delta E$ is given, as a function of the flux of particles $\phi$, by
\begin{equation}
N=(\Delta E)\cdot\phi\cdot A\cdot T,
\end{equation}
and indicating with $\phi_s$ and $\phi_b$ the signal and background fluxes, respectively, Eq.~(\ref{eq:chinbis}) can be recast as
\begin{equation}\label{eq:chisquareddef}
\chi^2=\sum_{i=1}^{n_{\rm b}}\ \frac{\left(\phi_s^i\right)^2}{\phi_b}\cdot(\Delta E)_i\cdot A\cdot T.
\end{equation}
The quantity given in Eq.~(\ref{eq:chisquareddef}) is what we use to asses the future sensitivity at antimatter experiments. We declare that a model is {\em within discrimination capabilities} of a given future experiment at the X\% confidence level if it satisfies the relation
\begin{equation}
\chi^2\ >\ (\chi^2)^{X\%}_{n_{\rm b}}.
\end{equation}
We focus here, for definiteness, on the case of the PAMELA detector~\cite{Picozza:2006nm}, and compute the reduced $\chi^2$  for an effective area of 24.5 ${\rm cm}^2{\rm sr}$, an exposure time of 3 years, and resorting to a trial energy binning as sketched in Ref.~\cite{pamelasens}. The results we show are in the limit of ``known background'', {\em i.e.} in the (optimistic) scenario in which degeneracies in the parameters used to model the propagation of charged cosmic rays in the Galaxy are resolved, for example, by precision measurements of ratios of secondaries to primaries for several light cosmic-ray nuclei species. As a rule of thumb, the sensitivity of AMS-02 is expected to be one order of magnitude better than that of PAMELA \cite{Feng:2000zu}. 

\begin{figure}[!t]
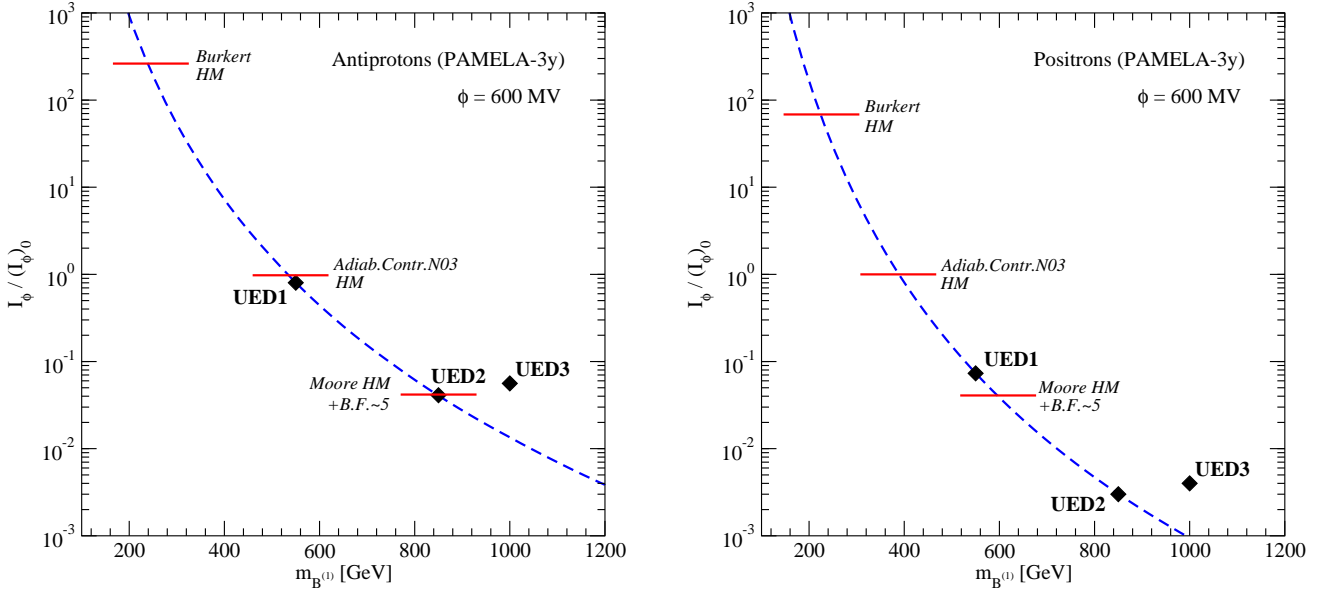

\centering
\mbox{\hspace*{-0.05\textwidth}\includegraphics[width=0.52\textwidth,clip=true]{FIGURES/pb_all.eps}\qquad\includegraphics[width=0.52\textwidth,clip=true]{FIGURES/ep_all.eps}}
\caption{The projected ``Visibility Ratio'' for the PAMELA experiment, defined as the ratio $I_\phi/(I_\phi)_0$ (see Eq.~(\ref{eq:IPHIdiscrim})) as a function of the LKP mass, $m_{B^{(1)}}$, for antiprotons (left) and positrons (right). The visibility ratio was computed using a reference halo model (the adiabatically contracted N03 profile), and two other halo models (the Burkert profile, and the Moore {\it et al.} profile with a boost factor of 5 resulting from dark matter substructure). The reach of the PAMELA experiment is indicated in each case by a horizontal red line. We also indicate the location of three of the benchmark models of Appendix~\ref{ch:benchmarks}.}
\label{fig:am}
\end{figure}
Eq.~(\ref{eq:chisquareddef}) can be recast in the limit of a large number of energy bins approximating the discrete sum with an integral. The region of integration will be given by the lowest and the largest energies, $T_{\rm min}$ and $T_{\rm max}$, accessible by the experiment. We henceforth define the quantity
\begin{equation}
I_\phi\equiv\ \int_{T_{\rm min}}^{T_{\rm max}} \frac{\left(\phi_s\right)^2}{\phi_b}{\rm d}E.
\end{equation}
The visibility condition, which reflects the continuum version of Eq.~(\ref{eq:chisquareddef}), is then given by
\begin{equation}\label{eq:IPHIdiscrim}
I_\phi\ >\ \frac{(\chi^2)^{X\%}_{n_{\rm b}}}{A\cdot T}.
\end{equation}
In the parameter $I_\phi$, the dependence on the extreme of integration is small as long as the peak of the signal-to-background ratio falls within them (which is always the case for the models we are considering and for the values expected for PAMELA and AMS).
The effective area is assumed to be independent of energy. If this is not a good approximation, a weighting function should be introduced accordingly. For the PAMELA setup already exploited, supposing a number of bins given by $n_b=55\div60$, and taking $(\chi^2)^{95\%}_{n_{\rm b}}\approx 75$, we find the following critical values for $I_\phi$, after one and three years of data taking:
\begin{equation}\label{eq:Iphis}
I_\phi^{1{\rm yr}}=9.7\cdot 10^{-8}\, \, {\rm cm}^{-2}{\rm sr}^{-1}{\rm s}^{-1}, \qquad I_\phi^{3{\rm yr}}=3.2\cdot 10^{-8}\, \, {\rm cm}^{-2}{\rm sr}^{-1}{\rm s}^{-1}.
\end{equation}

In Fig.~\ref{fig:am}, the reach of PAMELA (after 3 years of data) to antiprotons (left) and positrons (right) from KKDM annihilations is shown. The solar modulation Fisk parameter has been set to $\phi_F=600$ MV, a value appropriate for the projected solar activity over PAMELA's data taking time. The reference dark matter halo model is, again, the adiabatically contracted N03 halo profile described in Sec.~\ref{sec:db}. However, also shown are the results for two other halo setups, namely a (more conservative) cored Burkert profile~\cite{burkert}, and a (more optimistic) Moore {\it et al.} profile featuring a degree of dark matter substructures responsible for a boost factor of 5 in the antimatter fluxes.

We find that it will be modestly more easy to disentangle an exotic signature in the antiproton spectrum than in the positron channel. With antiprotons, KKDM models with masses $m_{B^{(1)}}\lesssim550$ GeV will give a significant signal for the case of the reference halo model (adiabatically contracted N03). In the case of positrons, this figure is reduced down to $m_{B^{(1)}}\lesssim400$ GeV. The comparison between these two channels is similar when other halo profiles are considered.

\subsection{KK Gravitons}\label{sec:kkgraviton}

The inclusion of the gravitational sector within the framework of a model with universal extra dimensions yields the existence of a KK graviton tower. Depending upon the KK spectrum adopted, the lightest KK graviton, which we shall hereafter indicate with the symbol $G^{(1)}$ (not to be confused with the KK gluon, $g^{(1)}$), could possibly constitute the LKP. If the LKP is the first-level KK graviton, the resulting dark matter phenomenology will be that of a super-Weakly Interacting Massive Particle (a ``SuperWIMP", see {\em e.g.} Refs.~\cite{Feng:2003xh,Feng:2003uy}). If instead KK gravitons are not the LKP, they will be meta-stable, in which case the production of these relatively long-lived states can significantly affect the cosmology of the early universe, including the relic abundance of LKPs. In Sec.~\ref{sec:kkglkp} we consider the phenomenology of models where the $G^{(1)}$ is the LKP, while in Sec.~\ref{sec:kknlkp} we address the consequences of the existence of a tower of (unstable) KK gravitons in scenarios where the LKP is not the $G^{(1)}$.

\subsubsection{The KK Graviton as the LKP}\label{sec:kkglkp}

As in the case of supersymmetry, where the LSP need not be a SM partner, within the context of UED the LKP could in principle be the first KK excitation of the graviton. Since radiative corrections to the KK masses \cite{Cheng:2002iz} are typically positive, and they are negligible for the $G^{(1)}$ (being Planck-scale suppressed), this is a fairly natural option. Assuming the boundary conditions found in the minimal UED model (see Sec.~\ref{sec:spectrum}), and assuming $m_{G^{(1)}}=R^{-1}$, the mass splitting between the $B^{(1)}$ and the $G^{(1)}$ can be approximated, neglecting electroweak mixing between the KK neutral gauge bosons, as~\cite{Shah:2006gs}
\begin{equation}
m^2_{B^{(1)}}-m^2_{G^{(1)}}\simeq-
\left(\frac{39}{2}\frac{\alpha_1\zeta(3)}{4\pi^3}+\frac{1}{6}\frac{\alpha_1}{4\pi}\ln(\Lambda R)^2\right)\left(\frac{1}{R}\right)^2+\frac{g_1^2v^2}{4}.
\end{equation}
Making use of the full expression, the mass splitting is positive (hence the $G^{(1)}$ is the LKP) for $R^{-1}\lesssim 809.1$ GeV, and negative elsewhere (hence the $B^{(1)}$ is the LKP for $R^{-1}\gtrsim 809.1$ GeV). Also, again within the context of minimal UED, one finds that the maximal mass splitting, for $R^{-1}\lesssim 809.1$ GeV, amounts to only $(m_{B^{(1)}}-m_{G^{(1)}})\lesssim1.7$ GeV. We stress, however, that the minimal UED spectrum can be affected by further contributions to the KK masses, {\em e.g.} from large boundary terms at the cutoff (relaxing the assumption that the boundary kinetic terms vanish at the cutoff scale, $\Lambda$).

The production of KK gravitons in the early universe proceeds in general through two mechanisms: (1) non-thermal production through NLKP decays, and (2) production at reheating after the end of the inflationary era. In the first case, the NLKP undergoes the usual thermal freeze-out when its pair annihilation rate drops below the Hubble expansion rate, and subsequently decays into KK gravitons. As the NLKP-LKP interactions are suppressed, the NLKP freeze-out occurs at much earlier times than the NLKP decays into the KK graviton. In this scenario, since every NLKP decays into one KK graviton, the KK graviton relic abundance is simply given by
\begin{equation}
\Omega_{G^{(1)}}\ =\ \Omega_{\rm NLKP}\ \frac{m_{G^{(1)}}}{m_{\rm NLKP}}.
\end{equation}
Since the KK graviton mass and the NLKP mass are generically predicted to be rather close to one another, if the NLKP relic abundance is similar to the CDM abundance the KK gravitons will inherit the correct relic abundance through NLKP decays. As long as the reheating temperature is low enough to avoid regenerating large numbers of KK gravitons, this is the main production mechanism for LKP KK gravitons. 

In any case, KK gravitons are also produced at reheating. If reheating occurs at a temperature, $T_{\rm RH}$, a population of KK gravitons is produced through the collisions of particles in the thermal bath. The evolution of the number density of $n$th level KK gravitons, $n_{G^{(n)}}$, follows the Boltzmann equation,
\begin{equation}
\frac{dn_{G^{(n)}}}{dt}+3Hn_{G^{(n)}}=C_{G^{(n)}},
\end{equation}
where $C_{G^n}$ is the collision operator and can be parameterized
as follows:
\begin{equation}
\label{eq:CG} C_{G^{(n)}}=C\sigma(g_*(T)n_0)^2,
\end{equation}
where
\begin{equation}
\label{eq:sigma_G} \sigma=\frac{\alpha_3}{4\pi M^2_4}, \quad M_4^2\equiv(16\pi G_N)^{-1}
\end{equation}
and where $\alpha_3$ is the strong coupling constant. In the equation above, $C$ is the graviton production parameter that can be
understood as the fraction of all possible collisions which will
interact strongly to produce gravitons, $g_*(T)\simeq g_*^{KK} T R$ is the number of effective degrees of freedom ($g_*^{KK}$ is a model dependent quantity that for the case of one extra-dimension compactified on the $S^1/Z_2$ orbifold equals approximately 197.5), and $n_0=\zeta(3)T^3/\pi^2$. The cumulative effect of all KK levels up to a KK mass of the same order of $T_{\rm RH}$ gives, again for the $S^1/Z_2$ UED model and for $m_{G^{(1)}}=R^{-1}$~\cite{Feng:2003nr},
\begin{equation}
\Omega_G^{(1)}\approx 0.84 \times C\times\left(\frac{g_*^{KK}}{200}\right)^{1/2}\left(\frac{1\ {\rm TeV}}{m_{G^{(1)}}}\right)^{2}\left(\frac{T_{\rm RH}}{m_{G^{(1)}}}\right)^{7/2}.
\end{equation}
The value of $C$ is subject to some controversy in the literature. Ref.~\cite{Feng:2003nr} estimates $C\sim{\cal O}(0.01)$ based upon the analogy with the supersymmetric gravitino case worked out in detail in Refs.~\cite{Bolz:1998ek,Bolz:2000fu}. The independent estimate of Ref.~\cite{Shah:2006gs}, in contrast, yields $C\sim{\cal O}(1)$. As a rule of thumb, when $T_{\rm RH}\gtrsim10^2$ TeV, LKP production from the KK graviton tower is expected to be the dominant production mechanism over the $G^{(1)}$ production from long-lived NLKP decays, and vice-versa. A lower limit on $T_{\rm RH}$, in the case of $G^{(1)}$ production from long-lived NLKP decays, stems from the requirement that the universe is not reheated after thermal freeze-out, {\em i.e.} $T_{\rm RH}\gtrsim R^{-1}/25$.

\begin{figure}[!t]
\centering
\includegraphics[width=0.9\textwidth,clip=true]{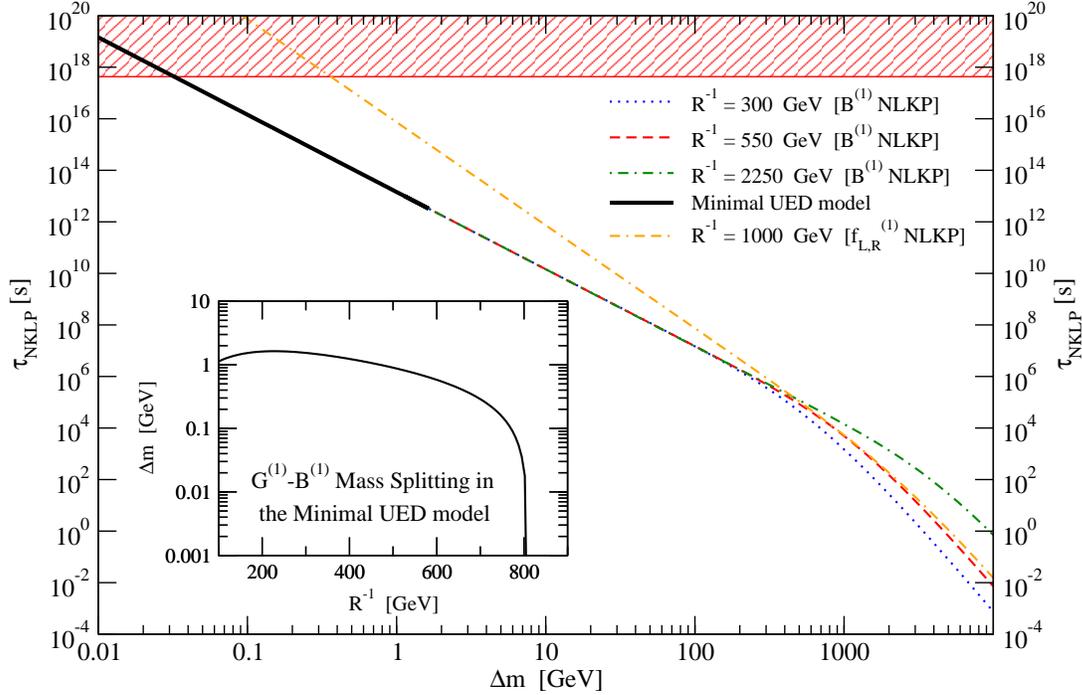}
\caption{The lifetime of NLKP in the KK graviton LKP scenario, as a function of the mass splitting $\Delta m$ between the NLKP and the $G^{(1)}$. In the inset we show the relation between $R^{-1}$ and $\Delta m$ in the minimal UED scenario.}
\label{fig:lifetimeskkgrav}
\end{figure}
Constraints on the KK graviton LKP scenario come from limits on the amount of energy injected through NLKP decays into $G^{(1)}$s. NLKP decays at early times  can affect the predictions of Big Bang Nucleosynthesis (BBN), the Cosmic Microwave Background (CMB) or conflict with observations of the diffuse photon background~\cite{earlydecays}. The first necessary step to determine these constraints is to compute the lifetime of the NLKP \cite{Feng:2003xh,Feng:2003nr}. If the NLKP is the $B^{(1)}$, the resulting decay width is given by
\begin{equation}
 \Gamma(B^{(1)} \to G^{(1)} \gamma)
\ =\  \frac{\cos^2\theta_W}{144\pi M_4^2}
\frac{m_{B^{(1)}}^7}{m_{G^{(1)}}^4}
\left(1 - \frac{m_{G^{(1)}}^2}{m_{B^{(1)}}^2} \right)^3 \times \left(1 + 3 \frac{m_{G^{(1)}}^2}{m_{B^{(1)}}^2}
 + 6 \frac{m_{G^{(1)}}^4}{m_{B^{(1)}}^4} \right),
\end{equation}
while if the NLKP is a KK chiral fermion, $f_{L,R}^{(1)}$, one has \cite{Feng:2003nr}
\begin{equation}
 \Gamma(f_{L,R}^{(1)} \to G^{(1)} f_{L,R})
\ =\  \frac{1}{96\pi M_4^2}
\frac{m_{f^{(1)}}^7}{m_{G^{(1)}}^4}
\left(1 - \frac{m_{G^{(1)}}^2}{m_{f^{(1)}}^2} \right)^4 \times \left(2 + 3 \frac{m_{G^{(1)}}^2}{m_{f^{(1)}}^2}\right).
\end{equation}
We show in Fig.~\ref{fig:lifetimeskkgrav} the predictions for the NLKP lifetime as a function of the NLKP-$G^{(1)}$ mass difference, $\Delta m$, in the case of a $B^{(1)}$ NLKP, at three different values for the $G^{(1)}$ mass (two of them feature the same value of $R^{-1}$ as that of models {\bf UED1} and {\bf UED4}), and for $R^{-1}=1$ TeV for the case of a chiral fermion NLKP. We also show the prediction for the $B^{(1)}$ lifetime in the case of minimal UED, where the $B^{(1)}$-$G^{(1)}$ mass splitting, as a function of $R^{-1}$ is given in the inset. The hatched region near the top of the plot contains an NLKP with a lifetime larger than the age of the universe (hence the possibility of a mixed $G^{(1)}$-$B^{(1)}$ dark matter scenario. A KK fermion NLKP in this region is ruled out because either a fraction of the dark matter would be charged or, for a KK neutrino NLKP, would be excluded by the negative results from WIMP direct detection searches).

The most relevant bounds on the KK graviton scenario depend upon the total energy release into photons in $B^{(1)}$ NLKP decays, {\em i.e.} on the quantity $\varepsilon_\gamma Y_\gamma$, where $\varepsilon_\gamma$ is the energy of the photons when created and $Y_\gamma = n_{\gamma}/n_{\gamma}^{\text{BG}}$ is the number density of photons from NLKP decay normalized to the number density of background photons $n_{\gamma}^{\text{BG}} = 2 \zeta(3) T^3/\pi^2$, where $T$ is the temperature during NLKP decay. Since NLKPs decay essentially at rest, $\varepsilon_\gamma = (m_{\rm NLKP} - m_{G^{(1)}}^2)/(2m_{\rm NLKP})$, and since one KK graviton is produced in association with each photon, $Y_\gamma = Y_{G^{(1)}}$, the photon abundance is given by~\cite{Feng:2003xh}
\begin{equation}
Y_{G^{(1)}} \simeq 3.0 \times 10^{-12}
\left(\frac{1\ {\rm TeV}}{m_{G^{(1)}}}\right)
\left(\frac{\Omega_{G^{(1)}}}{0.23}\right) .
\end{equation}

\begin{figure}[!t]
\centering
\includegraphics[width=0.7\textwidth,clip=true]{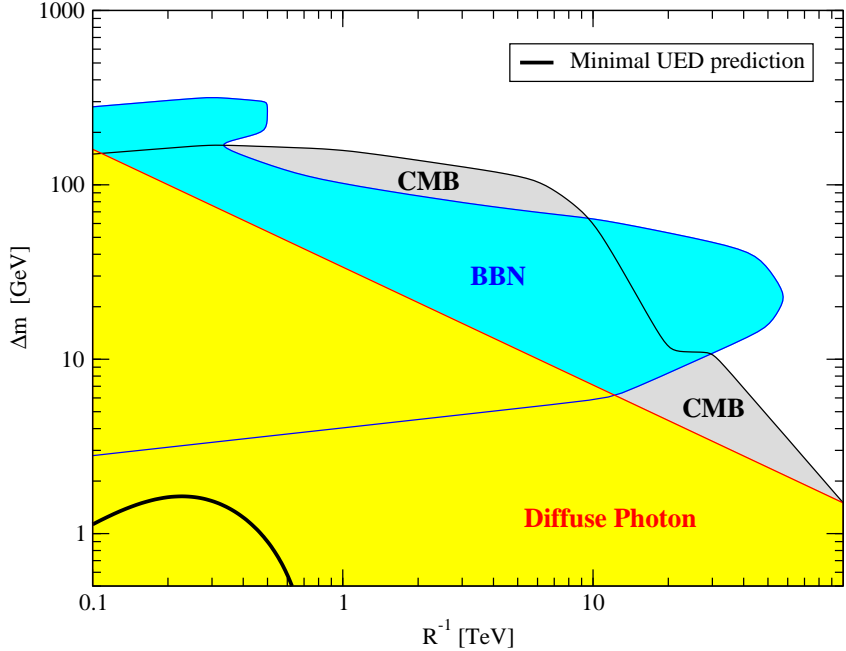}
\caption{The constraints on the $R^{-1}, \Delta m$ parameter space of the KK graviton LKP mass versus the $B^{(1)}$-$G^{(1)}$ mass splitting from BBN (the excluded region is shaded in light blue), CMB (light grey) and the diffuse photon background (yellow). The black line indicates the minimal UED prediction for the $B^{(1)}$-$G^{(1)}$ mass splitting as a function of $R^{-1}$.}
\label{fig:kkgrav}
\end{figure}
Fig.~\ref{fig:lifetimeskkgrav} shows that NLKP decays can occur before and during BBN, hence they can in principle destroy the successful BBN predictions for the abundances of light elements. Photons produced in NLKP decays rapidly thermalize through scattering off background photons, $\gamma \gamma_{\text{BG}} \to e^+e^-$, as well as through Compton scattering. The BBN constraints therefore very weakly depend upon the initial energy distribution of the injected photons. We show in Fig.~\ref{fig:kkgrav} (adapted from Ref.~\cite{Feng:2003xh}) the region of the $R^{-1},\Delta m$ parameter space ruled out by the BBN constraint~\cite{Cyburt:2002uv} with a light blue shading. In the figure, we assume $\Omega_{G^{(1)}}\simeq\Omega_{\rm CDM}$.

The injection of photons can also distort the CMB spectrum from the observed blackbody distribution as a result of elastic Compton scattering, bremsstrahlung and double Compton scattering, $e^- \gamma \to e^- \gamma \gamma$. Before redshifts $z\sim10^5$, the bounds come from corrections to the chemical potential $\mu$ of the Bose-Einstein distribution to which the photon spectrum relaxes. At redshifts smaller than  $z\sim10^5$, deviations from the black-body spectrum can be parameterized through the Sunyaev-Zeldovich $y$ parameter~\cite{Yao:2006px}. The resulting bounds~\cite{Feng:2003xh} exclude the region shaded in light grey in Fig.~\ref{fig:kkgrav}.

The non-thermal production of soft gamma rays in NLKP late decays contributes to the diffuse photon spectrum, with a differential photon flux peaking at an energy
\begin{equation}
E^{\rm max}_\gamma=490\ {\rm keV}\ \left(\frac{1\ {\rm GeV}}{\Delta m}\right)
\end{equation}
and giving a maximal differential photon flux of~\cite{Feng:2003xh}
\begin{equation}
\frac{d\Phi_\gamma}{dE_{\gamma}} (E_\gamma^{\text{max}})
\ \simeq \ 2.1\times \left( \frac{1\ {\rm TeV}}{m_{G^{(1)}}} \right)
\left( \frac{\Delta m}{1\ {\rm GeV}} \right)
\left( \frac{\Omega_{G^{(1)}}}{0.23} \right)\ \ {\rm cm}^{-2}~{\rm s}^{-1}~{\rm sr}^{-1}~{\rm MeV}^{-1}. 
\end{equation}
Constraints from the observatories HEAO, OSSE and COMPTEL \cite{dpb} translate into the excluded region shaded in yellow in Fig.~\ref{fig:kkgrav}. 

In the figure, we also show the prediction for the $B^{(1)}$-$G^{(1)}$ mass splitting as a function of $R^{-1}$ in the minimal UED model (where, recall, the $G^{(1)}$ can be the LKP for $m_{G^{(1)}}\lesssim 809.1$ GeV). We conclude that the a graviton LKP is not a viable option within the context of minimal UED, as it badly violates both the diffuse photon background and the CMB bounds. A possible way out was proposed in Ref.~\cite{Matsumoto:2006bf}, through the introduction of right-handed neutrinos to the model. The $B^{(1)}$ then dominantly decays into neutrinos, and does not emit photons. Within this context, models with sub-dominant decay modes producing photons have been shown to feature branching ratios so small that the CMB and diffuse photon background spectra are not distorted at observable levels~\cite{Matsumoto:2006bf}.

As a last remark, we wish to shortly comment on the possibility of having detectable signatures from the KK graviton dark matter scenario. In the case that the NLKP is neutral (which appears to be the most natural option), the presence of KK graviton dark matter might reveal itself only in small perturbations to the quantities listed above (CMB, BBN) or through spectral features in the diffuse photon background. If the NLKP is instead charged, other detection methods based on the production of the meta-stable charged NLKP at colliders or in high energy neutrino-nucleon collisions in the atmosphere could lead to detectable signatures, at least in principle. Ref.~\cite{Feng:2004yi} proposed to build water tanks surrounding the LHC to trap long-lived charged particles (the charged NLKP), which could afterwords be studied in their decays in an appropriate low-background environment. High energy neutrinos colliding with nucleons in the atmosphere can also produce long-lived charged NLKPs, which can in principle travel through the Earth long enough to be eventually detected at neutrino telescopes \cite{Albuquerque:2003mi,Albuquerque:2006am}. The tracks resulting from charged NLKPs can be disentangled from the di-muon background on the basis of both their spatial separation (muons must be created close to the detector, unlike charged NLKP) and on their energy spectrum (charged NLKPs feature a significantly more energetic spectrum)~\cite{Albuquerque:2006am}.

\subsubsection{KK Gravitons and the Early Universe}\label{sec:kknlkp}

\begin{figure}[!t]
\centering
\mbox{\hspace*{-1cm}\includegraphics[width=0.58\textwidth,clip=true]{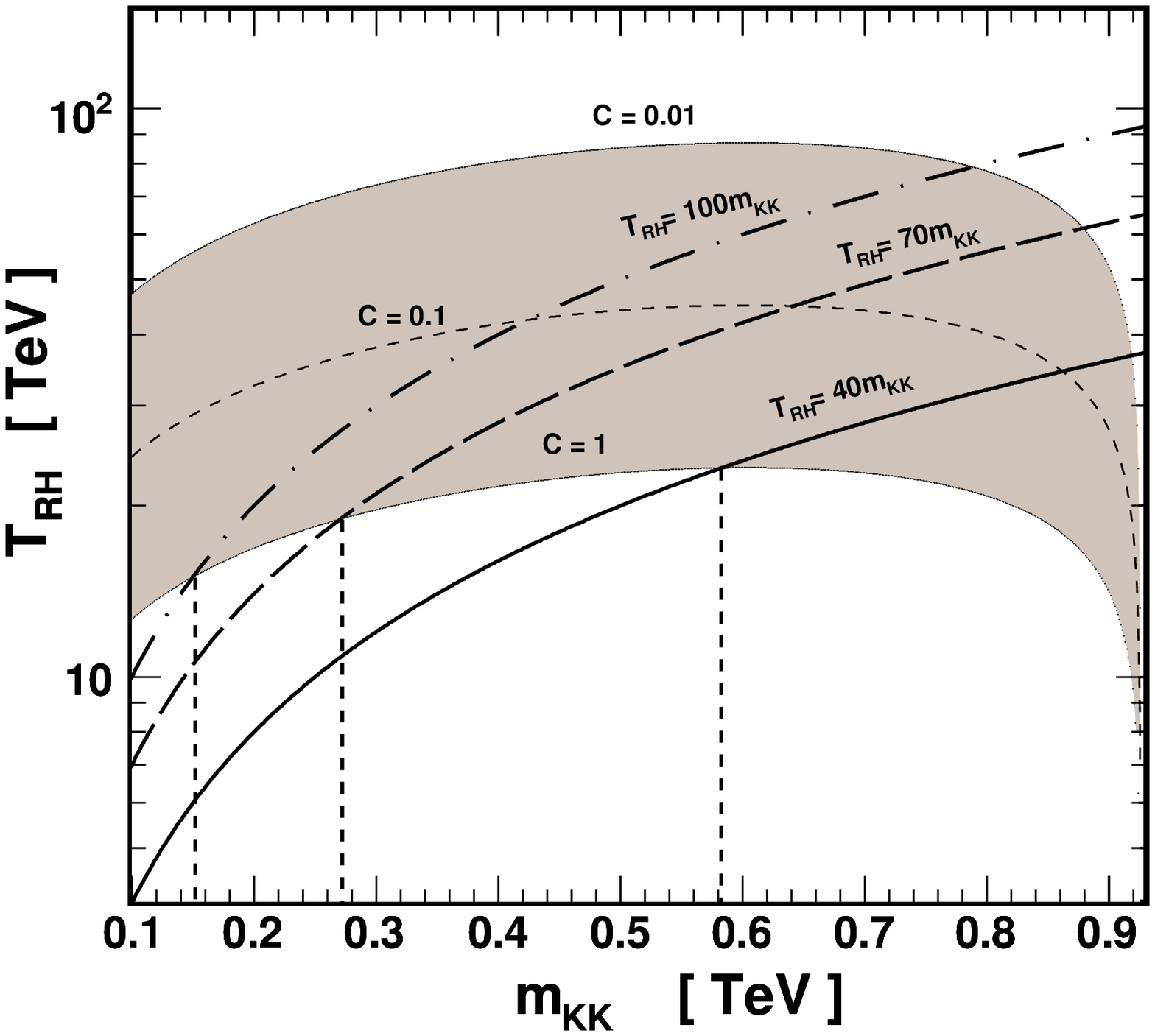}\hspace*{-0.5cm}\includegraphics[width=0.58\textwidth,clip=true]{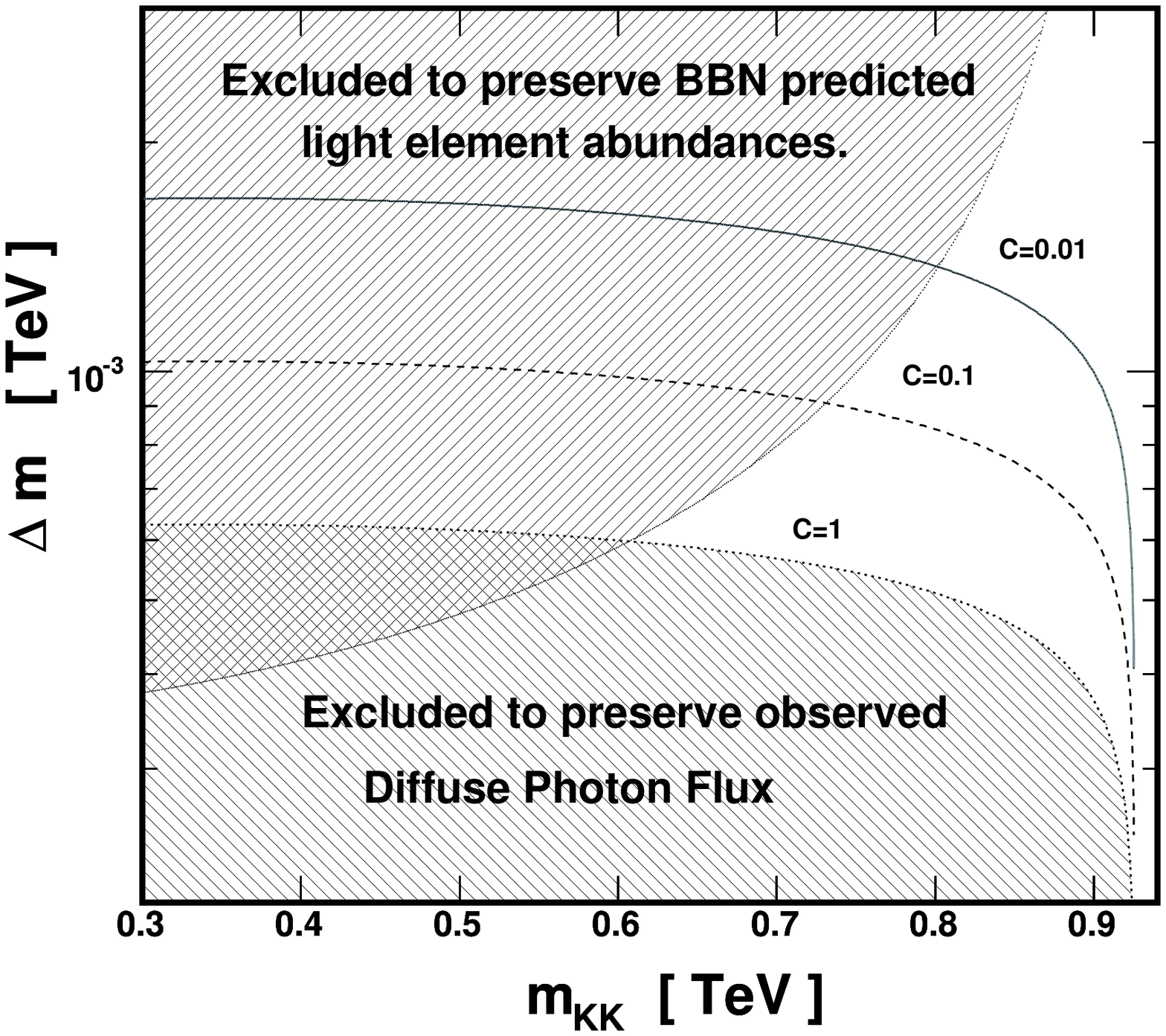}}
\caption{Left: Values of
the reheating temperature, $T_{\rm RH}$, obtained by demanding a proper dark
matter density, $\Omega_{B^{(1)}}=\Omega_{\rm DM} \approx 0.23$, for
different values of the graviton production parameter: $C=1$, $0.1$,
$0.01$. Also shown are lines of constant ratios
of the reheating temperature to the lightest KK mass: $T_{\rm RH}=40 \;
m_{\rm KK}$, $70 \; m_{\rm KK}$ and $100 \; m_{\rm KK}$. Right: Constraints on the value of the mass difference, $\Delta m$, between $G^{(1)}$ and
the LKP assuming $\Omega_{B^{(1)}}=\Omega_{\rm DM} \approx 0.23$ and $C=1$, $0.1$, $0.01$. Possible coannihilation and second KK level effects have been neglected. Figures adapted from Ref.~\cite{Shah:2006gs}.}
\label{fig:kkgrav_wagner}
\end{figure}

Even when the LKP corresponds to the $B^{(1)}$, its relic abundance will be, in general, affected by $B^{(1)}$ production through the decays of the KK tower of gravitons. Indicating with $Y_\infty$ the abundance of LKPs without the inclusion of gravitons, the total relic abundance can be expressed as\cite{Shah:2006gs}
\begin{eqnarray}
Y_{B^{(1)}}&=&Y_\infty+Y_G\\
n_{B^{(1)}}&=&s_0\ Y_{B^{(1)}}\\
\Omega_{B^{(1)}}&=&\frac{m_{B^{(1)}}\ n_{B^{(1)}}}{\rho_c}
\end{eqnarray}
where $\rho_c=5.3\times10^{-9}\mbox{ TeV/}\mbox{cm}^3$, and 
\begin{eqnarray}\label{eq:YG}
Y_G&=&\int_0^{T_{\rm RH}R}nY_{G^{(n)}}\,{\rm d}n\nonumber\\
   &=&\frac{45\ \sqrt{5}\ \zeta^2(3)}{7\ \pi^8}\ \frac{\alpha_3\  C\ \sqrt{g_*^{KK}}}{M_4\  R}\left(T_{\rm RH}\ R \right)^{7/2}.
\end{eqnarray}
Requiring that $\Omega_{B^{(1)}}\approx 0.23$, one gets a prediction for the reheating temperature of
\begin{equation}\label{eq:Tr_const}
T_{\rm RH} \simeq R^{-1}\times\left(\frac{\Omega_{B^{(1)}}\ \rho_c-s_0\ R^{-1}\ Y_\infty}{1.8\times 10^{-18}\  C\  s_0\ R^{-2}}\right)^{2/7},
\end{equation}
where $s_0$ is the entropy density today. The left frame of Fig.~\ref{fig:kkgrav_wagner} illustrates, as a function of $m_{\rm KK}\equiv R^{-1}$, the reheating temperature needed to produce  $\Omega_{B^{(1)}}\approx 0.23$ for different values of the graviton production parameter $C$.

The constraints from the diffuse photon background and from BBN mentioned above apply also in the case of a meta-stable KK graviton, $G^{(1)}$, decaying into the $B^{(1)}$ LKP. In particular, since photons from $n>1$ KK gravitons are produced before matter domination, they don't affect the diffuse photon background. On the other hand, a line of reasoning analogous to that presented above leads to a constraint on the mass difference between the lightest KK graviton and the LKP, $\Delta m$, which reads
\begin{equation}
\Delta m\;\gsim\left(2.48\times10^{-3}\left(\left(T_{\rm RH} R\right)^{3/2}-1\right)\left(\frac{R^{-1}}{1\ \mbox{TeV}}\right)\right)^{1/2}\;\mbox{GeV},\label{eq:kkgravdpb}
\end{equation}
where it was assumed that $\Omega_{B^{(1)}}=\Omega_{\rm DM} \approx 0.23$. 

Constraints on BBN come from electromagnetic energy released by the decay of the $G^{(1)}\to B^{(1)}\gamma$, and from the hadronic decays of the whole KK graviton tower (the $G^{(n)}$s, $n>1$, decay right after BBN). The latter bounds depend strongly on the branching ratio of the decay of gravitons into hadrons. Indicating with ${\rm B}_{\rm EM/Had}$ the electromagnetic/hadronic branching ratio, one obtains the following constraint on $\Delta m$ (again assuming $\Omega_{B^{(1)}}=\Omega_{\rm DM} \approx 0.23$)~\cite{Shah:2006gs}:
\begin{equation}
\Delta m<\frac{n_\gamma \xi}{{\rm B}_{\rm EM/Had}\left(\rho_c\ \Omega_{B^{(1)}}\ -\ R^{-1}\ s_0\ Y_\infty\right)},\label{eq:kkgravbbn}
\end{equation}
where $\xi_B$ is the bound on the energy released per background photon. Conservatively, Ref.~\cite{Shah:2006gs} assumed $\xi_B<10^{-15}$ TeV.

The bounds quoted above are shown in the right frame of Fig.~\ref{fig:kkgrav_wagner} for the same model considered in the left frame, for three values of the graviton production parameter, $C$. As pointed out in Ref.~\cite{Shah:2006gs}, the prediction of the minimal UED model for $\Delta m$ can be, in principle, compatible with the bounds in Eq.~(\ref{eq:kkgravdpb}-\ref{eq:kkgravbbn}) within small ranges of the inverse compactification radius.

Setups that mimic the features of a UED scenario without KK gravitons are the so-called ``deconstruction models'', where KK states are merely the manifestation of a chain of gauge groups~\cite{Arkani-Hamed:2001ca,Hill:2000mu}. For instance, a simple two-site model could resemble the first KK level of UED without the introduction of a KK graviton. 

Scenarios where the UED model is encompassed in a higher dimensional setup, and where all SM particles are confined to the five-dimensional space-time but not the gravitons (which therefore can be significantly heavier than the rest of the KK spectrum), also effectively get rid of the implications of KK gravitons discussed above. One example was given in Ref.~\cite{Dienes:2001wu} in the context of shape moduli associated to large extra dimensions.

The cosmology of KK gravitons was also discussed in Ref.~\cite{Kolb:2003mm}, with similar results to those outlined above. In that analysis, UED and other models featuring flat extra-dimensions have been studied in a cosmological context addressing, in particular, the issue of the cosmology of gravi-scalars, or ``radions'', i.e. the role of the scalar field associated to the geometrical modulus that determines the size of the compactified dimensions. The main parameters entering radion cosmology are the compactification radius $R^{-1}$ and the scale $M_I\equiv V_I^{1/4}$ of the inflaton potential. Different cosmological scenarios are then discussed, depending on whether inflation took place before, after or around the same time of the compactification of the extra spatial dimensions. Since, on rather model independent grounds, the radion decay width reads
\begin{equation}
\Gamma=\tau^{-1}\simeq\frac{\sqrt{3}(R^{-1})^6}{64\pi m_{\rm Pl}^5},
\end{equation}
the radion is effectively stable (i.e. its lifetime is larger than $H_0^{-1}\sim4\times 10^{17}$ s) if $R^{-1}\lesssim 7\times 10^8$ GeV. This implies a stable radion for the UED model discussed here, and, consequently, a generic over-closure problem, as the radion can easily dominate the energy density of the universe at late times. The radion energy density can be sufficiently damped if the inflation scale $M_I$ lies around the TeV scale. As the radion is stable, reheating would have to come from a different field, which would also imprint its fluctuations into the cosmic microwave radiation. Moreover, Ref.~\cite{Kolb:2003mm} points out that there is a narrow window of parameter space where the radion can be the dark matter as well. In general, inflation model building for $R^{-1}\sim$ TeV is problematic. Specifically, a low reheating temperature, as needed to account for the dilution of the radion energy density, necessitates a very efficient reheating, and almost certainly a preheating phase as well~\cite{Kolb:2003mm}. 

Along similar lines, Ref.~\cite{Mazumdar:2003vg} investigated the case of $n$ universal, ``small'' extra-dimensions and $p$ larger spatial dimensions where gravity also propagates. This setup is motivated by the idea of explaining the weakness of gravity by the presence of large extra-dimensions orthogonal to the $4+n$ dimensions of the Standard Model brane. The model features two radions, that can be, in turn, identified with a ``shape'' and a ``volume'' mode, respectively. Ref.~\cite{Mazumdar:2003vg} notices that models where a radion plays the role of an inflaton, or where the inflaton is a brane scalar field, generically present problems; however, it is possible to find setups where bulk scalar fields drive successful inflation, although the success of radion cosmology depends upon a more complete understanding of the radion potential arising from the fundamental theory governing the low energy action.

\clearpage
\newpage

\section{Universal Extra Dimensions and Colliders}\label{ch:colliders}

The KK states of UED can be produced in high energy colliders such as the Tevatron, the Large Hadron Collider (LHC), or a future International Linear Collider (ILC). At hadron colliders, first level KK modes can be abundantly pair-produced, depending upon the value of the compactification scale. The colored KK modes then cascade decay according to decay chains similar to those shown in our Fig.~\ref{fig:decaychain} of Sec.~\ref{sec:accconst}. Since the $B^{(1)}$ is stable and weakly interacting, it will escape the detector, leading to the typical signature $l^+l^-l^\pm+2\,{\rm jets}+{\rm missing\ energy}$. Given the approximate degeneracy of the KK modes, the resulting jets will be rather soft and challenging to distinguish from the background. The leptons will also be soft, but usually they can pass some reasonably chosen cuts. In the following section (Sec.~\ref{sec:lhc}), we review the prospects for the discovery of UED at the LHC. 

The advantage of a linear collider featuring $\sqrt{s}\gtrsim 1/R$ over an hadronic collider would be enormous, and would entail the possibility of a very accurate reconstruction of the KK particle spectrum, as well as the unambiguous identification of the spin of the KK modes. We devote Sec.~\ref{sec:ilc} to the discussion of the collider physics related to UED that could be performed at a future ILC. We remind the reader that the current constraints from the Tevatron and the discovery potential of Run II were already reviewed in Sec.~\ref{sec:accconst}. 

A crucial issue in the assessment of the potential of collider experiments in exploring various beyond-the-Standard-Model scenarios is how to differentiate among them. As testified by the catchy title of the seminal paper, ``{\em Bosonic supersymmetry? Getting fooled at the LHC}\,''~\cite{Cheng:2002ab}, if new physics signals are detected at colliders, a basic point will be how to discriminate between supersymmetry and UED (or other TeV-scale physics scenarios). In a nutshell, although the expected collider signals from the two setups are strikingly similar, there are three basic handles to potentially discriminate between the two: (1) KK first level states in UED have the same spin as their SM counterparts, while SUSY partners have opposite spin, (2) the Higgs sector of UED carries a different KK parity assignment than the heavy Higgs bosons of the MSSM ($H$, $A$, $H^{\pm}$), making them more similar to the SUSY {\em higgsinos} than to the SUSY Higgs sector (although the two Higgs sectors in SUSY and UED share exactly the same gauge quantum numbers), and (3) UED feature higher level KK modes, unlike the case of supersymmetry. We devote the following Sec.~\ref{sec:distinguish} for a thorough discussion of these points, as well as of the role, in the discrimination between supersymmetry and UED, of dark matter search experiments.

\subsection{The Large Hadron Collider}\label{sec:lhc}

The signatures of the production of new heavy particles at collider experiments crucially depend upon those particles' interactions and mass spectrum. Here, we focus on the particular case of minimal UED~\cite{Cheng:2002iz}, where all boundary couplings are assumed to be flavor-conserving, and all the boundary terms are assumed to vanish at the cutoff scale, $\Lambda$ (following Ref.~\cite{Cheng:2002ab}, we shall adopt here $\Lambda R=20$). Once the radiative corrections are taken into account, the KK mass degeneracy is lifted, and the heavier KK states promptly decay. The resulting collider phenomenology is similar to the case of supersymmetric models where all superpartners lie relatively close in mass. As in the case of supersymmetry, the decay cascades will terminate in UED with the stable lightest KK particle leaving the detector undetected, resulting in generic missing energy signatures.

\begin{figure}
\centering
\includegraphics[width=0.85\textwidth,clip=true]{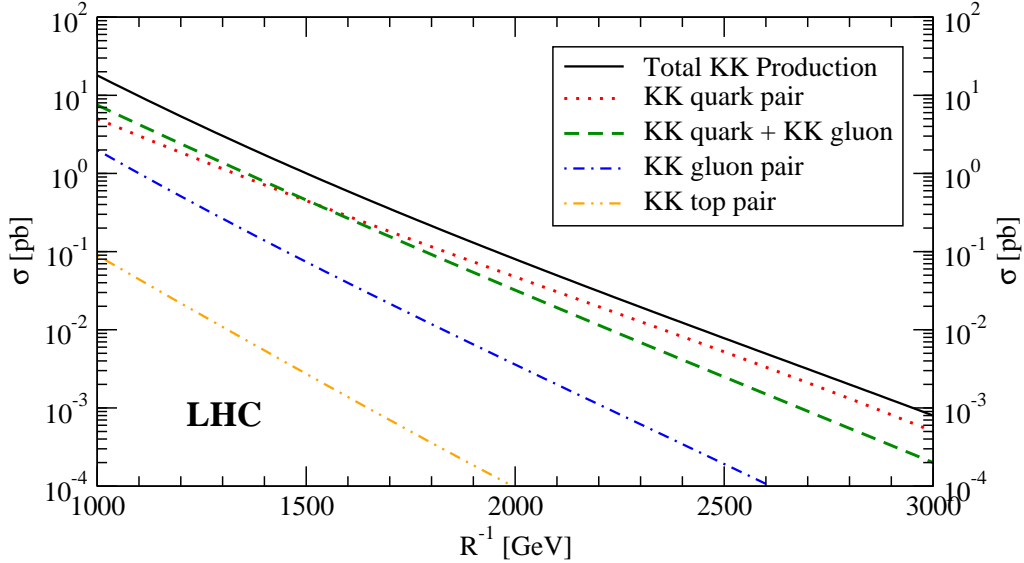}
\caption{The production cross section of KK quarks and KK gluons at the LHC. The solid curve represents the total production cross section, while the remaining curves show the separate contributions from the production of KK quark pairs, KK quark + KK gluon, KK gluon pairs and KK top pairs. Adapted from Ref.~\cite{Macesanu:2002db}.}
\label{fig:lhcprodxsec}
\end{figure}

At hadron colliders, the by far most abundantly produced KK particle species will be KK gluons and KK quarks, {\em i.e.} the strongly interacting states. Fig.~\ref{fig:lhcprodxsec} shows the production cross sections for strongly interacting first level KK states at the LHC ($\sqrt{s}=14$ TeV). As a first step to assess the capability of extracting a signature of UED at the LHC, one needs to consider the allowed decays of of the first level KK particles and estimate the relevant branching fractions. 

The heaviest KK particle in the present setup is the KK gluon, $g^{(1)}$. The two-body decays to the KK quarks $Q^{(1)},\ u^{(1)},\ d^{(1)}$ are always open and have comparable branching ratios,
\begin{equation}
{\rm BR}(g^{(1)}\to Q^{(1)}Q^{(0)})\approx{\rm BR}(g^{(1)}\to q^{(1)}q^{(0)})\simeq0.5,\quad {\rm where}\ q=u,d.
\end{equation}
The $SU(2)$-singlet KK quarks, $q^{(1)}$, can only decay to the hypercharge gauge boson, $B^{(1)}$, their branching ratio into $Z^{(1)}\equiv W^{3(1)}$ being suppressed by the level-1 Weinberg angle (recall that $\theta^{(1)}_W\ll\theta_W$, leading to ${\rm BR}(q^{(1)}\to Z^{(1)} q^{(0)})\approx \sin\theta^{(1)}_W\approx10^{-2}-10^{-3}$, while ${\rm BR}(q^{(1)}\to B^{(1)} q^{(0)})\approx 1$). As a consequence, $q^{(1)}$ production yields jets plus missing energy. An exception to this situation occurs for the KK top quark, $t^{(1)}$, which can decay to $W^{+(1)}b^{(0)}$ and $H^{+(1)}b^{(0)}$ (as noticed in Ref.~\cite{Cheng:2002ab}, this is also the dominant mechanism for the production of $H^{\pm(1)}$ at hadron colliders).

The decay chain of $SU(2)$ doublet quarks, $Q^{(1)}$, is somewhat less trivial, as they can decay into a SM quark plus $B^{(1)}$, $Z^{(1)}$ or $W^{(1)}$. In the $\sin\theta^{(1)}_W\ll1$ limit, $SU(2)$ symmetry dictates
\begin{equation}
{\rm BR}(Q^{(1)}\to W^{(1)}Q^{\prime(0)})\approx 2 \times {\rm BR}(Q^{(1)}\to Z^{(1)}Q^{(0)}).
\end{equation}
For $m_{Q^{(0)}}\ll1/R$ one also has
\begin{equation}
\frac{{\rm BR}(Q^{(1)}\to Z^{(1)}Q^{(0)})}{{\rm BR}(Q^{(1)}\to B^{(1)}Q^{(0)})}\approx\frac{g_2^2T_{3Q}^2(m^2_{Q^{(1)}}-m^2_{Z^{(1)}})}{g_1^2Y_{Q}^2(m^2_{Q^{(1)}}-m^2_{B^{(1)}})},
\end{equation}
where $T_3$ and $Y$ stand for the weak isospin and hypercharge of the quark. The KK quark decays into $SU(2)$ gauge bosons, although suppressed by phase space, are enhanced by the ratio of the couplings squared as well as by the quantum numbers. Typically, one numerically obtains
\begin{equation}
{\rm BR}(Q^{(1)}\to W^{(1)}Q^{\prime(0)})\approx65\%,\ {\rm BR}(Q^{(1)}\to Z^{(1)}Q^{(0)})\approx33\%,\ {\rm BR}(Q^{(1)}\to B^{(1)}Q^{(0)})\approx2\%.
\end{equation}
Once produced, the heavy KK gauge bosons cannot decay hadronically for kinematic reasons, and will democratically decay into all lepton flavors (we indicate here with $L$ the charged $SU(2)$ doublet lepton of any generation, and with $l$ the $SU(2)$ singlets),
\begin{eqnarray}
{\rm BR}(W^{\pm(1)}\to \nu^{(1)}L^{\pm(0)})&\approx&{\rm BR}(W^{\pm(1)}\to L^{\pm(1)}\nu^{(0)})\approx\frac{1}{6}\\
{\rm BR}(Z^{(1)}\to \nu^{(1)}\bar\nu^{(0)})&\approx&{\rm BR}(Z^{(1)}\to L^{\pm(1)}\bar L^{\mp(0)})\approx\frac{1}{6}
\end{eqnarray}
for each generation. The branching fraction for $Z^{(1)} \to l^{(1)} \bar l^{(0)}$ is suppressed by $\sin\theta^{(1)}_W\ll1$.

The resulting level-1 KK leptons will directly decay into $B^{(1)}$ and the corresponding SM lepton. As a result, the heavy KK gauge bosons will almost always decay as
\begin{equation}
W^{\pm(1)}\to B^{(1)}L^{\pm(0)}\nu^{(0)}\quad {\rm and} \quad Z^{\pm(1)}\to B^{(1)}L^{\pm(0)}L^{\mp(0)}\quad {\rm and} \quad Z^{\pm(1)}\to B^{(1)}\nu^{(0)}\bar\nu^{(0)}
\end{equation}
resulting in large $e$ and $\mu$ yields. Lastly, the KK Higgs will decay according to the available channels, dictated by the mass of the SM Higgs. If the KK Higgs is heavier than the KK $W$, $Z$, $t$ and/or $b$, it will decay into those states. Otherwise, the tree level two-body decays will be suppressed, leading it to decay to $B^{(1)}$ and a virtual zero-mode Higgs boson, or through a loop to $B^{(1)}$ and a photon. In any case, the production cross section for the KK Higgs is negligible with respect to the strongly interacting KK states.

At the LHC, the signature with the largest overall rate is $E_T^{\rm miss}+(N\geq2)\ {\rm jets}$, analogous to the well-known traditional squark and gluino searches \cite{Abbott:1999xc}. Roughly one quarter of the total production cross section of KK states at the LHC gives rise (directly or indirectly) to $q^{(1)}$ pairs which produce the aforementioned signature. However, even though the missing mass in these events is rather large, the measured missing energy is rather small, as it is correlated with the energy of the relatively soft recoiling jets. The estimate of the LHC reach in this channel is around $1/R\lesssim$1.2 TeV, as can be inferred from the analysis carried out in Ref.~\cite{Bityukov:1999am}.

A cleaner channel for UED discovery at the LHC is that of multilepton final states arising from associated heavy KK gauge boson production. The inclusive $Q^{(1)}$ pair production cross section amounts to roughly one quarter of the total production cross section. The subsequent decays of the $Q^{(1)}$s yield $W^{(1)}W^{(1)}$, $W^{(1)}Z^{(1)}$ and $Z^{(1)}Z^{(1)}$ pairs in the approximate proportion of 4:4:1. The subsequent heavy KK gauge boson decays yield final states with missing energy and up to four leptons, each of which offer the possibility of discovery. Following Ref.~\cite{Cheng:2002ab}, we consider here the gold-plated $4l+E_T^{\rm miss}$ signature.

Ref.~\cite{Cheng:2002ab} studied the reach of the LHC in this channel. They conservatively neglected direct KK gauge boson production and $Q^{(1)}W^{(1)}$ and $Q^{(1)}Z^{(1)}$ associated production processes and used for the single lepton trigger the relatively soft cuts of $p_T>\{35,\ 20,\ 15,\ 10\}$ GeV with $|\eta(l)<2.5|$, and, in addition, $E_T^{\rm miss}>50$ GeV. The SM background from $ZZ\to l^\pm l^\mp\tau^+\tau^-\to4l+E_T^{\rm miss}$, where $Z$ stands for a real or virtual $Z$ or $\gamma$ \cite{Matchev:1999yn}, can be reduced by invariant mass cuts for any pair of opposite sign, same flavor leptons. To this extent, Ref.~\cite{Cheng:2002ab} uses $|m_{ll}-M_Z|>10$ GeV and $m_{ll}>10$ GeV. Additional sources of background, including multiple gauge bosons and/or top quark production \cite{Matchev:1999yn}, fakes, and leptons from $b$-jets, are estimated conservatively to give a background of the order of 50 events after cuts per 100 ${\rm fb}^{-1}$.

\begin{figure}
\centering
\includegraphics[width=0.65\textwidth,clip=true]{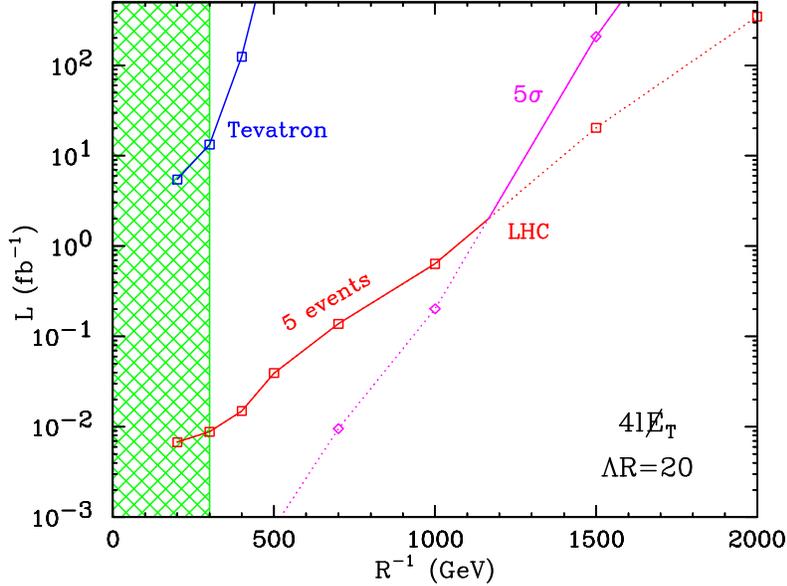}
\caption{Discovery reach for the minimal UED scenario at
the Tevatron (blue) and at the LHC (red), in the $4\ell + \met$ channel. 
The reach contours correspond to a $5\sigma$ excess or the observation of 5 signal events. The $y$-axis corresponds to the required total integrated
luminosity per experiment (in ${\rm fb}^{-1}$) as a function
of $R^{-1}$, for $\Lambda R=20$. Both the $5\sigma$ excess and the 5 signal events refer to single LHC experiments, and not to the combination of the two). From Ref.~\cite{Cheng:2002ab}.}
\label{fig:lhcreachcheng}
\end{figure}

The estimate of the LHC reach in the $(1/R,L)$ plane (where $L$ stands for the total integrated luminosity) is shown in Fig.~\ref{fig:lhcreachcheng}, where the two requirements of 5 signal events and of a 5$\sigma$ significance over background are shown separately. The Tevatron Run II performance is also estimated, making use of similar cuts and assumptions \cite{Cheng:2002ab}. The final LHC reach in the $4l+E_T^{\rm miss}$ channel extends up to inverse compactification radii just below roughly 1.5 TeV.

Other leptonic channels such as two or three leptons plus $E_T^{\rm miss}$ are affected by larger backgrounds, but take advantage of the larger branching fraction for $Q^{(1)}\to W^{(1)}Q^{\prime(0)}$ and can therefore offer higher statistics which can be useful, for instance, to enhance the reach of Tevatron Run II.

It is important to realize that the results presented above are somewhat model-dependent, in that if the cutoff scale was smaller than the assumed value, $\Lambda R=20$, the mass splitting between the first level KK excitations would be reduced, along with the energy of the final state leptons, making it harder to differentiate the signal from the background \cite{Macesanu:2005jx}. Moreover, should the boundary mass terms be non-zero at the cutoff scale, the phenomenology might also be significantly affected \cite{Macesanu:2005jx}. For example, if the KK leptons are heavier that the heavy KK gauge bosons, then the cascade decays will be significantly modified from the minimal UED case, as will the KK gauge boson decay widths. In such a scenario, new competing decay modes, such as $W^{(1)}\to W^{(0)} B^{(1)}\to l\nu B^{(1)}$, though suppressed by $\sin\theta^{(1)}_W\ll1$, would be competitive with the off-shell decay process through a KK lepton.

\subsection{The International Linear Collider}\label{sec:ilc}

An International Linear Collider (ILC) with a center of mass energy smaller than 3 TeV would not be, within the context of UED, a discovery machine. As pointed out before, a signal from UED is in fact expected at the LHC when $R^{-1}\lesssim1.5$ TeV, at least in the multi-lepton plus missing energy channel. However, given the cleanness of the environment in $e^+e^-$ collisions, an ILC would play an invaluable role in accurately measuring the properties of the particles under consideration, including their mass, couplings and spin. Such determinations will be critical for differentiating between UED and other beyond-the-SM scenarios (see Sec.~\ref{sec:distinguish}). 

As alluded to above, a crucial test for the discrimination of UED from other scenarios is the possibility of producing KK level-2 modes. If the ILC is energetic enough to pair produce first level KK modes, than single production of second level KK states should also be possible. As single production of KK level-1 modes is not allowed by KK parity, the ILC will face a none-or-both situation. We devote Sec.~\ref{sec:ilcn1} to the discussion of the production and detection of level-1 KK modes at the ILC \cite{Bhattacharyya:2005vm}, while Sec.~\ref{sec:ilcn2} deals with the physics of singly-produced level-2 KK modes~\cite{Bhattacherjee:2005qe}.

If $R^{-1}$ is large and the mass splitting between the heavier KK modes and the LKP is small, a serious problem a future high energy ILC will face is that of the huge background from two photon processes for a signal consisting of very soft final state leptons or jets~\cite{Cheng:2002rn}. The latter are produced by the collisions of two soft photons from initial state radiation, while the $e^+$ and $e^-$ go down the pipe, resulting in large missing energies. The leptons coming from the decays of KK leptons can thus potentially be buried by the two photon background. Larger mass splittings than those predicted in the minimal UED model, or the production of KK quarks, can alleviate this problem. Another way out is to use $e^-e^-$ collisions \cite{Cheng:2002rn}. In this case the signal would be two soft {\em same} charge fermions plus large missing energy, and the opposite charge soft fermion background from soft initial state radiation would not be as problematic as with $e^+e^-$ collisions, making it possible to identify and study in detail the properties of KK electrons, as well as, in principle, to test the possibility of lepton flavor violation in UED \cite{Cheng:2002rn}.

\subsubsection{KK $n=1$ Pair Production and Detection at the ILC}\label{sec:ilcn1}

\begin{figure}
\centering
\mbox{\includegraphics[height=0.52\textwidth,clip=true,angle=-90]{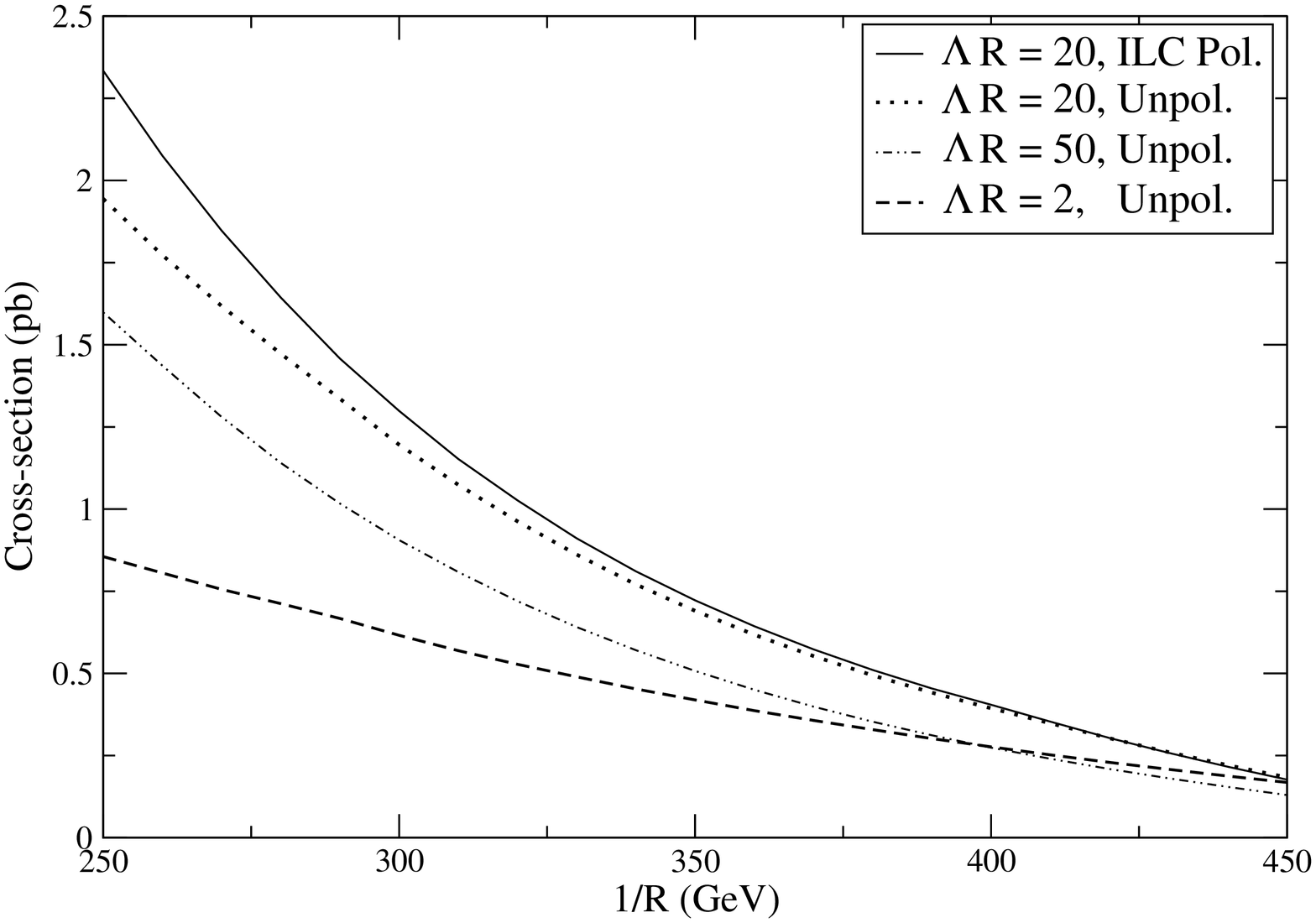}\includegraphics[height=0.52\textwidth,clip=true,angle=-90]{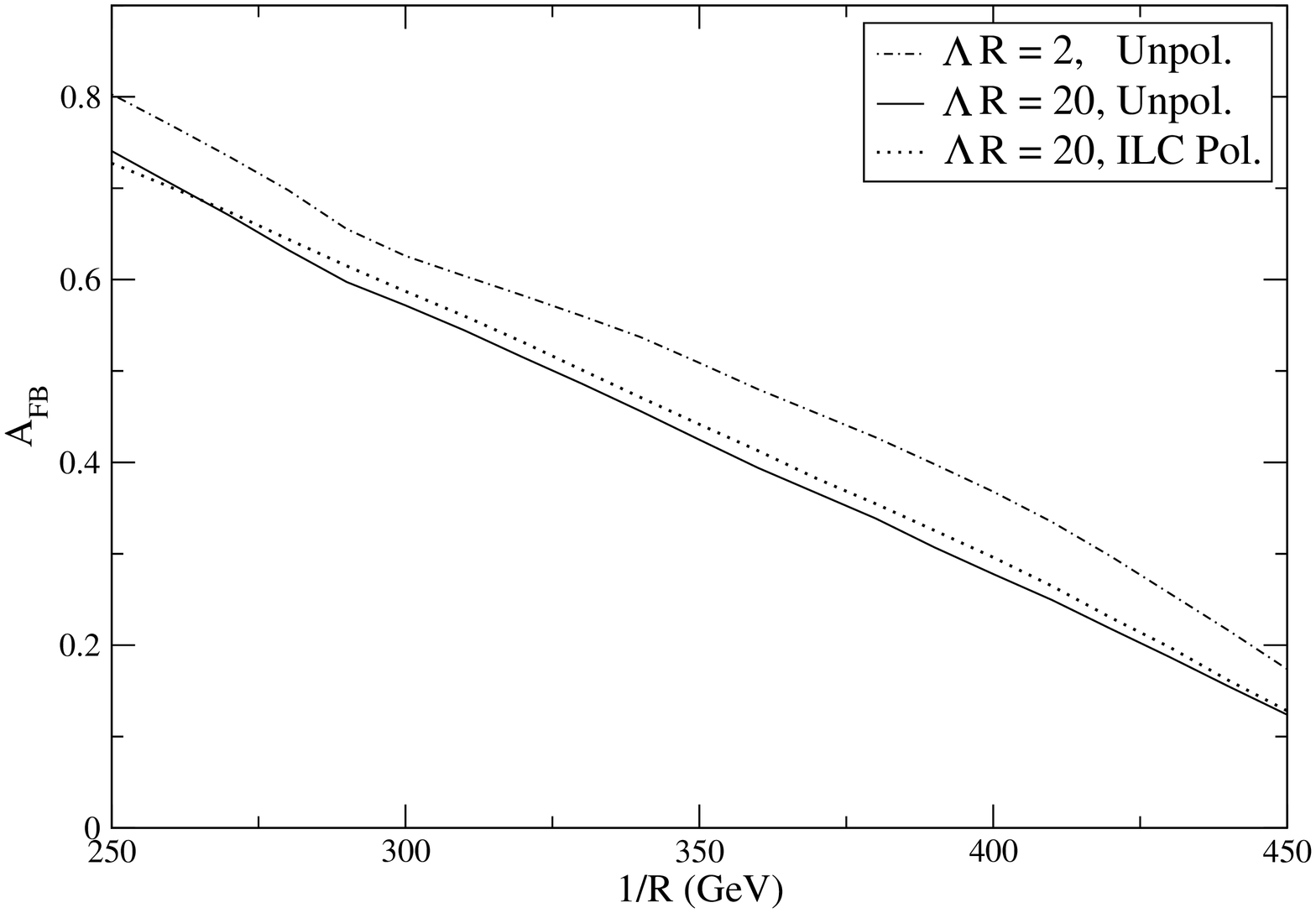}}
\caption{Left: Cross section versus $1/R$ for the process
$e^+e^-\to e^+e^- + $ missing energy, in minimal UED for various polarization choices and values of $\Lambda R$. Right: The forward-backward asymmetry, $A_{FB}$, versus $1/R$ for the same process (from Ref.~\cite{Bhattacharyya:2005vm}).}
\label{fig:ilcn1}
\end{figure}

Following the analysis presented in Ref.~\cite{Bhattacharyya:2005vm}, we assume an ILC featuring a center-of-mass energy of $\sqrt{s}=1$ TeV, and consider the range 250 GeV $<R^{-1}<$ 450 GeV within the minimal UED framework. The production cross section for SU(2) doublet and singlet KK electrons is shown in the left frame of Fig.~\ref{fig:ilcn1} for different polarizations of the incident beams (including the optimal ILC polarization of 80\% for $e^-$ and 50\% for $e^+$) and for different values of the cutoff parameter, $\Lambda R$. The collider parameters employed in the simulation are specified below. The $e^+e^-\to E^{+(1)}_{L,R}E^{-(1)}_{L,R}$ process proceeds through $s$ and $t$ channels, the first mediated by $\gamma$ and $Z$ exchange, and the latter by $B^{(1)},Z^{(1)}$ and $A^{(1)}$ exchange. The KK electrons promptly decay into an electron and an LKP. The mass splitting is large enough for the decay to occur well within the detector with a nearly 100\% branching ratio. The observation of a displaced vertex might, in principle, be possible for very small mass splittings. The final state signature will then be $e^++e^-+2B^{(1)}$, the latter resulting in missing energy. 

The same final state can be obtained from $e^+e^-\to W^{+(1)}W^{-(1)}$ as well. The $W^{(1)}$ can then decay into a KK neutrino and a lepton, or a KK lepton and a neutrino. Since all three lepton flavors are open, and assuming that only the final state electrons are tagged, the $e^+e^-\to W^{+(1)}W^{-(1)}$ reaction is suppressed by a branching fraction of 1/9 to the final state $e^++e^-+$ missing energy and, therefore, provides a subdominant contribution. An even smaller contribution comes from the charged KK scalars, $A^{\pm(1)}$.

\begin{figure}[!t]
\centering
\includegraphics[width=0.3\textwidth,clip=true]{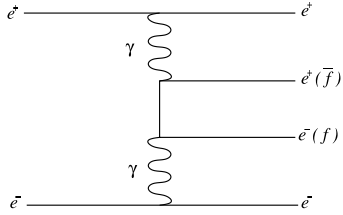}
\caption{The initial state radiation background from the 2 photon processes (from Ref.~\cite{Cheng:2002rn}).}
\label{fig:2gamma}
\end{figure}

The dominant SM background, as explained above, comes from initial state radiation photon-photon to $e^+e^-$ events \cite{Cheng:2002rn,Bhattacharyya:2005vm}. We illustrate the process in Fig.~\ref{fig:2gamma}. The $\gamma^*\gamma^*\to e^+e^-$ production cross section at a 1 TeV ILC is expected to be around $10^4$ pb, with about half of those events resulting in visible particles. The background $e^+e^-$ pairs are usually quite soft and coplanar with the beam axis~\cite{Bhattacharyya:2005vm}. Therefore, an acoplanarity cut significantly helps in removing this background, while not significantly reducing the signal \cite{Bhattacharyya:2005vm}. Ref.~\cite{Bhattacharyya:2005vm} quotes that, for instance, excluding events which deviate from coplanarity within 40 mrad reduces the signal by only 7\%. Another strategy to estimate and eliminate the aforementioned background would be to derive it from observations of $\mu^+\mu^-$ events, whose production cross section is suppressed by a factor 20 with respect to the $e^+e^-$ case (due to the $s$-channel suppression). Once measured, this could be then be used to normalize the actual background. Ref.~\cite{Bhattacharyya:2005vm} estimates that, for an integrated luminosity of 300 fb$^{-1}$, the signal events divided by the square root of the number of background events would still be around 10 for the range of $R^{-1}$ considered. Less significant backgrounds originate from $e^+e^-\to W^+W^-,e\nu W$ and $e^+e^-Z$, followed by the appropriate leptonic decays of the gauge bosons. The cross section shown in the left frame of Fig.~\ref{fig:ilcn1} also assumes kinematic cuts on the lower and upper energies of the final state charged leptons of 0.5 and 20 GeV, respectively. The lower cut is the minimum energy resolution for identification, while the upper cut eliminates most of the SM backgrounds. A rapidity cut is also assumed, admitting only those final state electrons away from the beam pipe by more than 15 degrees.

A further handle to deal with the SM background is to make use of the forward-backward (FB) asymmetry, defined as $A_{FB}\equiv(\sigma_F-\sigma_B)/(\sigma_F+\sigma_B)$. For the SM background, $A_{FB}=0$, while the result for KK electrons, as a function of $R^{-1}$ for various values of $\Lambda R$, is shown in the right frame of Fig.\ref{fig:ilcn1}. Beam polarization does not seem to help in enhancing $A_{FB}$. However, it is important to bear in mind that the main background (electron-positron pairs from two initial radiation photons, see Fig.~\ref{fig:2gamma}), is forward-backward symmetric, and therefore does not pollute the new physics signal from the forward-backward asymmetry.

The phenomenology of the KK Higgs sector at the ILC was investigated in Ref.~\cite{Bhattacherjee:2006jb}, where both the importance of soft $\tau$ leptons detection and the role of backgrounds from pair production of KK gauge bosons and KK taus were emphasized. The most promising channel appears to be charged scalar pair production. Ref.~\cite{Bhattacherjee:2006jb} also concludes that, in the context of the minimal UED model, the determination of the value of the parameter $\bar m_h$ from collider data will be particularly challenging.

\subsubsection{KK $n=2$ Pair Production and Detection at the ILC}\label{sec:ilcn2}

The production and observation of $n=2$ KK states would constitute a highly distinctive signature for many extra-dimensional scenarios, particularly for the case of UED \cite{Cheng:2002ab}. While pair production of such states would be challenging at the energy of the LHC (and is surely beyond the reach of an ILC), the resonant production of a single $B^{(2)}$ or $Z^{(2)}$ would appear as two narrow, closely spaced peaks~\cite{Battaglia:2005zf,Riemann:2005es}. Although this will likely not be resolvable by the LHC, particularly because, as we show below, the $B^{(2)}$ decay pattern is dominated by two jets plus no missing energy (see, however, the discussion in Sec.~\ref{sec:distinguish} for prospects with other final states), an ILC in the $\sqrt{s}$ scan mode is potentially the ideal environment to produce and study $n=2$ KK modes. Due to the structure of the KK spectrum, the production of $n=2$ KK states at an ILC stands on the same grounds as the pair production of first level KK states. As single production of $n=1$ modes is forbidden, the ILC will face a none-or-both situation. 

The decay of an $n=2$ KK state to two $n=0$ (zero mode) states is allowed by KK parity conservation, but suppressed by the boundary-to-bulk ratio, as it is KK number forbidden. However, there is no phase space suppression, not even if the final state is a $t\bar{t}$ pair. On the other hand, KK number conserving decay modes ($2\to 2+0,1+1$) have large couplings, but are kinematically suppressed, if kinematically open at all.

The couplings of $n=2$ KK gauge bosons to fermion-antifermion pairs are given by \cite{Cheng:2002iz}:
\be\label{eq:level2}
\left(-ig_{1,2}\gamma^\mu T_a P_+\right) \frac{\sqrt{2}}{2} \left(
\frac{\bar\delta(m_{B^{(2)},Z^{(2)}}^2)}{(2R^{-1})^2}-2\frac{\bar\delta(m_{f_2})}{2R^{-1}}\right),
\ee
where $g_{1,2}$ stand for the U(1)and SU(2) gauge couplings, respectively, $T_a$ is the group generator (hypercharge and third component of isospin, respectively) and $P_+$ is the $Z_2$-even projection operator (which is $P_L=(1-\gamma_5)/2$ for $Z^{(2)}$, but can be both $P_L$ or $P_R$ for $B^{(2)}$). The expressions for the boundary corrections, $\bar\delta$, are given by the part of the expressions in Eq.~(\ref{rspec}) proportional to $\ln(\Lambda R)$. Eq.~(\ref{eq:level2}) includes the sum of four contributions: (1) one-loop vertices, (2) $n=2$(external)-$n=0$ gauge bosons kinetic mixing, (3) $(n=2)$-$(n=0)$ gauge bosons mass mixing and (4) $(n=2)$-$(n=0)$ fermion mass mixing (see Fig.~11 and 12 and App.~C in Ref.~\cite{Cheng:2002iz} for more details). 

\begin{figure}
\centering
\mbox{\includegraphics[height=0.47\textwidth,clip=true,angle=-90]{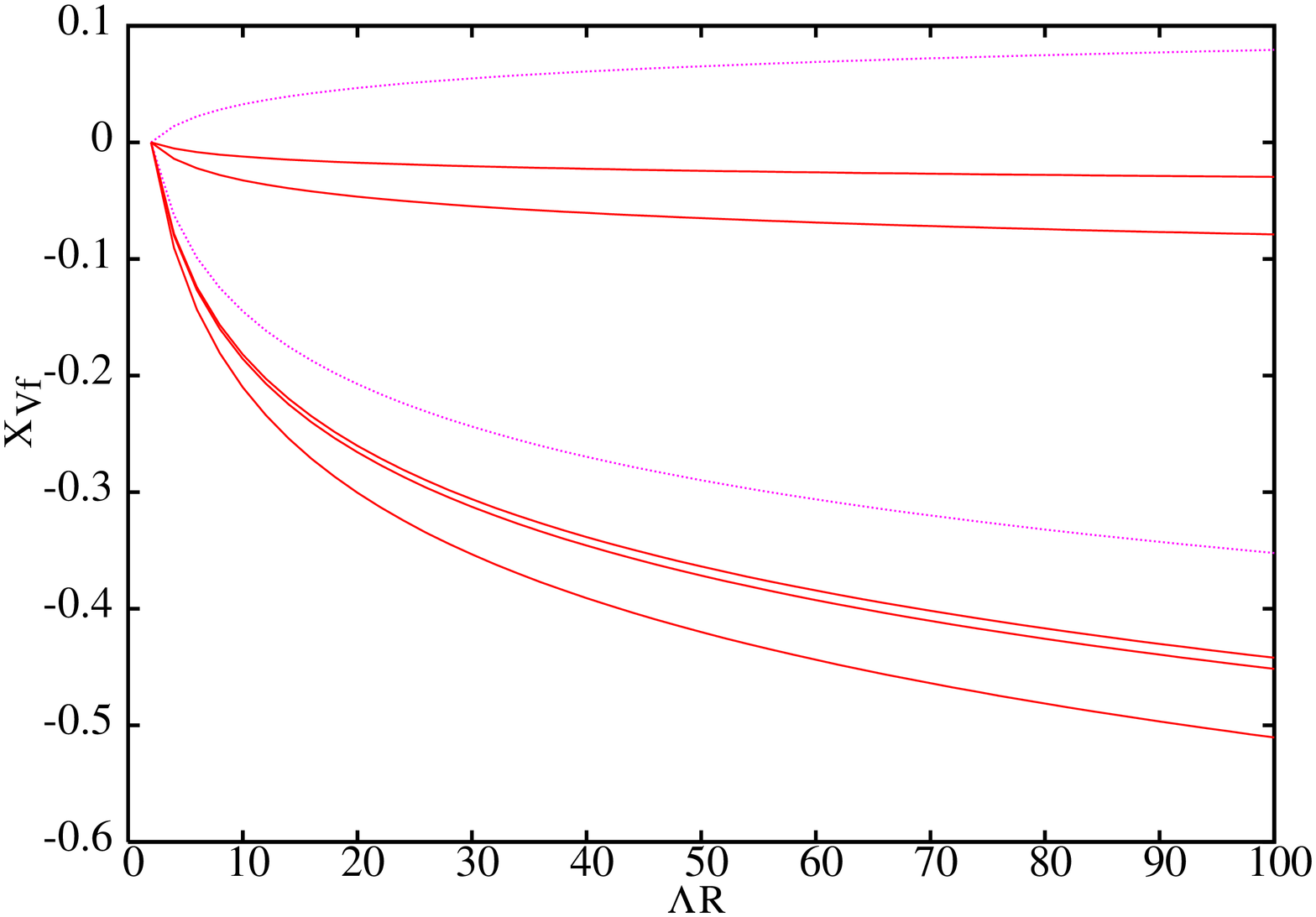}\qquad\includegraphics[height=0.47\textwidth,clip=true,angle=-90]{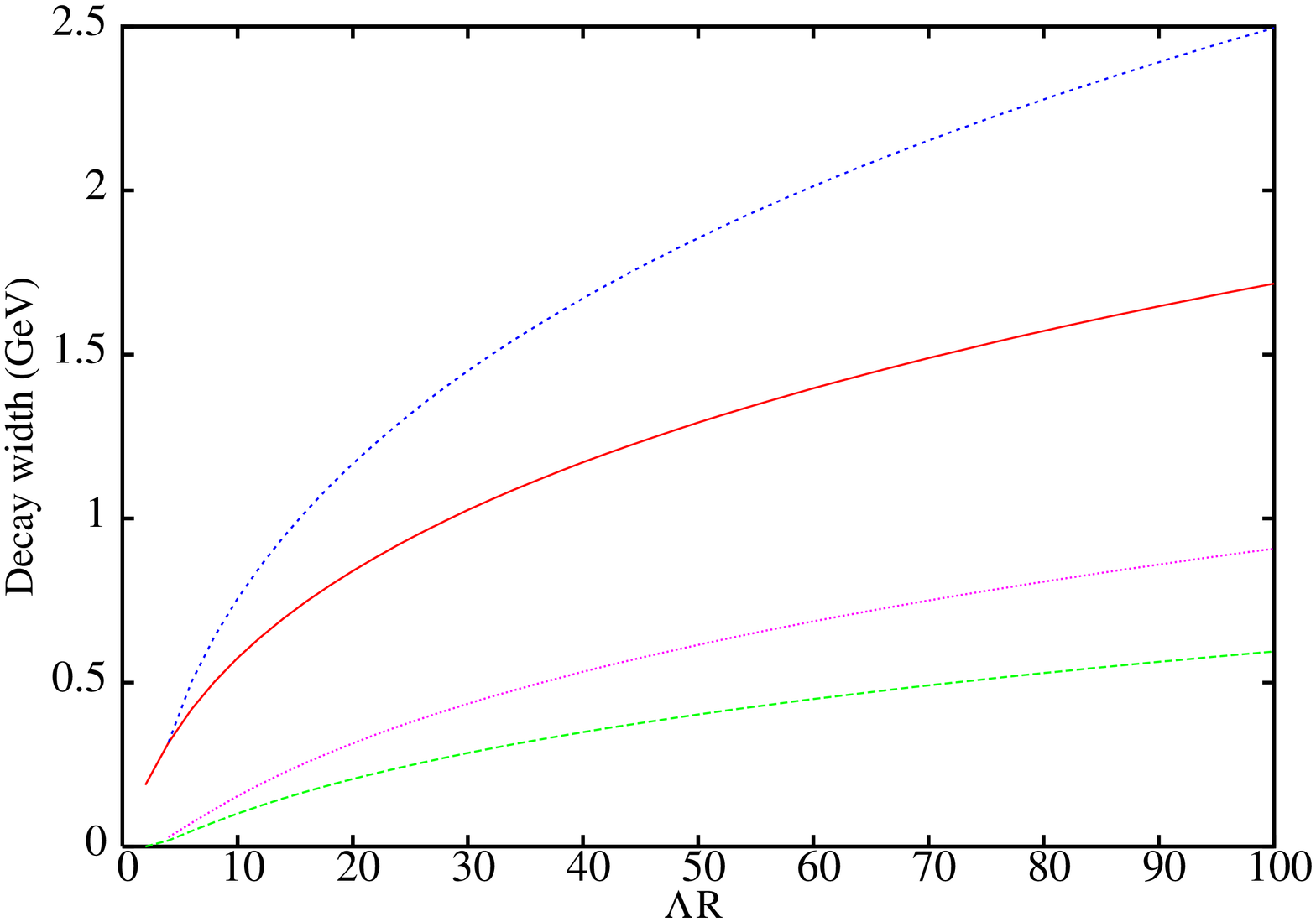}}
\caption{Left: $X_{Vf}$, the KK number violating couplings, as
a function of $\Lambda R$, for $R^{-1}=300$ GeV (the values are independent
of $R$). From top to bottom, the curves correspond to $X_{ZL}$, $X_{\gamma e}$,
$X_{\gamma L}$, $X_{ZQ}$, $X_{\gamma d}$, $X_{\gamma u}$, and $X_{\gamma Q}$, respectively.
Right: Decay widths of $Z^{(2)}$ (upper pair) and $B^{(2)}$ (lower pair)
as a function of $\Lambda R$, for $R^{-1}=450$ GeV and 300 GeV (upper and 
lower curves in each pair, respectively). From Ref.~\cite{Bhattacherjee:2005qe}.}
\label{fig:ilc2}
\end{figure}

For any level, $B^{(n)}$ is the lightest KK state. Thus, $B^{(2)}$ cannot decay into other $n=2$ states plus SM particles. It turns out that the decay into $n=1$ pairs is also kinematically forbidden for $B^{(2)}$, leaving as the only possibility the KK number violating decay into zero-mode fermion-antifermion pairs. The left frame of Fig.~\ref{fig:ilc2} shows the coupling 
\be
X_{Vf}\equiv\frac{\sqrt{2}}{2} \left(
\frac{\bar\delta(m_{B^{(2)},Z^{(2)}}^2)}{(2R^{-1})^2}-2\frac{\bar\delta(m_{f_2})}{2R^{-1}}\right),
\ee
as a function of $\Lambda R$ (notice that for $\Lambda R=2$, $\bar\delta=0$, see Eq.~(\ref{rspec})). Here, $V$ denotes $B^{(2)}$ or $Z^{(2)}$. As can be seen from the figure, the $B^{(2)}$ decays almost entirely into $q\bar{q}$ pairs, giving as a signature two jets with no missing energy. As pointed out in Ref.~\cite{Bhattacherjee:2005qe}, $B^{(2)}$ cannot decay into KK-number conserving three or four-body channels, {\em e.g}, $B^{(2)}\to  E^{(1)}E^{(1)*} \to E^{(1)} e B^{(1)}$, at least for the spectrum of the minimal UED type. The reason is that KK-number conserving decays must result in two LKPs in the final state, and $2m_{B^{(1)}} > m_{B^{(2)}}$ over the entire parameter space.

The decay pattern of $Z^{(2)}$ is more complicated \cite{Bhattacherjee:2005qe}. The $Z^{(2)}$ is an almost pure $W$-like state and, therefore, couples only to left-handed doublet fermions. The $Z^{(2)}$ decay to an $n=1$ pair of doublet leptons ($Z_2$-even) is kinematically allowed, except for very low values of $\Lambda R<3$. The $n=1$ states then decay to the corresponding $n=0$ leptons, plus the LKP $B^{(1)}$, so that the resulting collider signature is a pair of soft leptons (for charged lepton channels) plus a huge amount of missing energy (excited neutrinos, of course, will go undetected). These final state soft leptons should be detectable \cite{Bhattacharyya:2005vm,Battaglia:2005zf}. Similarly, the $Z_2$ can also decay into $n=2$ and $n=0$ doublet leptons. Both of these modes are KK-number conserving, but there is an important difference: while the coupling is the usual $g$ for the latter channel, it is $g/\sqrt{2}$ for the former ones \cite{Bhattacherjee:2005qe}. 

The $Z^{(2)}$ also features KK-number violating decay modes, but only to left-handed particles. Since the lower limit on $R^{-1}$ is about 300 GeV, both of these gauge bosons can decay even to a (zero mode) $t\bar{t}$ pair. 

In the minimal UED model with ${\bar{m}_h}^2=0$, the $Z^{(2)}$ cannot decay through the Bjorken channel to $Z^{(1)} H^{(1)}$, as it is kinematically forbidden. Moreover, the three-body channels, with a virtual $Z^{(1)}$ or $H^{(1)}$, will be even more suppressed. However, if ${\bar{m}_h}^2 < 0$, all of the Higgs masses will be lowered, and the production of a neutral CP-even Higgs will be possible. The decay pattern of the $H^{(1)}$ is dominantly into right-handed $\tau$ pairs (assuming the mixing in the $n=1$ level is small) plus an LKP. If the taus are soft enough, they may escape detection, leading to an invisible $H^{(1)}$ decay mode.

The second level KK gauge bosons are produced as $s$-channel resonances in $e^+e^-$ collision through KK-number violating couplings. This suppression brings down the peak cross-section from an otherwise expected nanobarn level to about 35-45 pb for $Z^{(2)}$ and to about 63 pb for $B^{(2)}$ (for $R^{-1}=300$ GeV). For $R^{-1}=450$ GeV, these figures drop down to 16-21 pb and 28 pb, respectively. The reason for the larger $B^{(2)}$ production cross-section is its narrower width compared to that of the $Z^{(2)}$. Shown in the right frame of Fig.~\ref{fig:ilc2} are the decay widths of the $Z^{(2)}$ and $B^{(2)}$ as a function of $\Lambda R$.

Since the $B^{(2)}$ decays almost entirely to two jets, it will be very challenging to detect at the LHC, as the signal would be swamped by the very large QCD background, and because the resonance itself is quite narrow. The case of the $Z^{(2)}$ is instead more promising, as it features various hadronically quiet decay modes, and soft leptons with energy greater than 2 GeV might be, in principle, detectable. For a precision study of these resonances, the ILC would be an ideal machine. In particular, on the $Z^{(2)}$ peak, the ratio, ${\cal R}$, of $e^+e^-$ to two jets over $e^+e^-$ to $\mu^+\mu^-$ would show a sharp dip. The latter can be even more marked including the missing energy events. In fact, the $Z^{(2)}$-width is dominated by decays into $n=1$ lepton pairs, and quarks can appear only from KK-number violating interactions. On the other hand, ${\cal R}$ should show a sharp peak corresponding to the $B^{(2)}$ resonance. Furthermore, the total $e^+e^-$ cross-section would show a kink between the two peaks, corresponding to the value of $\sqrt{s}$ where the KK-number conserving channels open up.

With polarized beams, the behavior of the two peaks will also be quite different. Since $Z^{(2)}$ couples only to the left-handed fermions, with suitable polarization the peak may entirely vanish, or it may get enhanced by a factor of 3 (assuming 80\% $e^-$ polarization and 60-70\% $e^+$ polarization) \cite{Bhattacherjee:2005qe}. The $B^{(2)}$ peak will get enhanced by about a factor of 2 with a left-polarized $e^-$ beam, but will never vanish altogether, as the hypercharge gauge boson, $B$, couples to both right and left-handed fermions.

Although in the minimal UED model the $B^{(2)}$ does not have any KK-number conserving decay channels, in non-minimal models there are two possibilities for such channels. First, non-universal boundary terms can enhance the $B^{(2)}$ mass, possibly kinematically opening KK-number conserving decay channels. Second, asymmetric boundary terms (different for $y=0$ and $y=\pi R$) would break KK-parity, resulting in decay modes like $B^{(2)}\to E^{(1)}+e$. A precision study at the ILC would in all cases allow for the discrimination between such different realizations of the UED model. As a concluding remark, we point out that the expected background level from the continuum at the ILC is less than 10 pb for $\sqrt{s}=600$-900 GeV \cite{Abe:2001gc}, and may be further reduced by suitable cuts \cite{Bhattacherjee:2005qe}. 
\begin{figure}
\centering
\includegraphics[width=0.4\textwidth,clip=true]{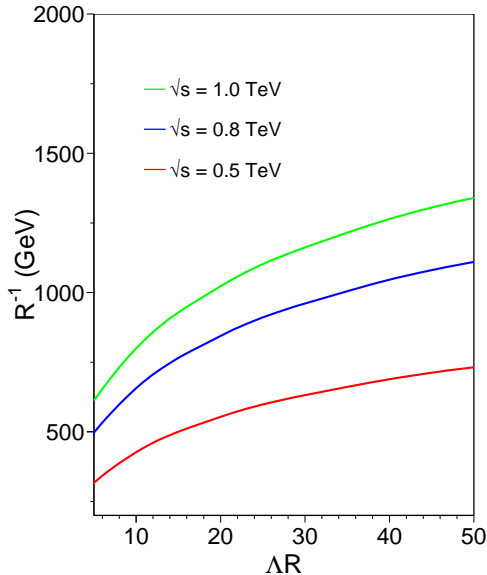}
\caption{Projected 95\% confidence level exclusion limits for minimal UED in the $\Lambda R,R^{-1}$ plane from combined measurements of leptonic and hadronic final states at an ILC. Deviations from the SM expectations can be observed up to the values given. From Ref.~\cite{Riemann:2005es}.}
\label{fig:ilcreach}
\end{figure}

The region of the minimal UED parameter space that can be {\em ruled out} at an ILC (at the 95\% confidence level, in the $\Lambda R,R^{-1}$ plane) is shown in Fig.~\ref{fig:ilcreach}. The assumed ILC integrated luminosity corresponds to 1~ab$^{-1}$ with $\delta L_{int}=$0.1\%. Uncertainties due to the identification of leptonic and hadronic final states are assumed to result in systematic errors of 0.1\%. In the analysis of Ref.~\cite{Riemann:2005es} a polarization of 80\% for the electron beam and 60\% for the positron beam, and an error of 0.1\% for the polarisation measurement were considered. Also, Ref.~\cite{Riemann:2005es} assumes that 50\% of the luminosity is spent for each sign combination, (+ , -) and  (- , +), of the beam polarization.

\clearpage
\newpage

\section{Distinguishing Between UED and Supersymmetry}\label{sec:distinguish}

If signals of new physics are detected at the LHC in the form of the production and subsequent decay of new heavy particles, a crucial question will be whether it is possible to discriminate between different beyond the Standard Model scenarios. As we pointed out before, the phenomenology of low energy supersymmetry and of UED at colliders are strikingly similar. In both models, the lightest new state is neutral and stable, leading to collider signatures including missing transverse momentum (energy) plus a number of jets and leptons. Furthermore, both supersymmetry and UED predict the same couplings for the SM particles and their heavier counterparts. Distinguishing between supersymmetry, UED and other beyond the SM scenarios will likely be a challenging task for the LHC.

On general grounds, there are a few handles one has which could be used to distinguish between supersymmetry and UED. Provided that the LHC measures the mass spectrum of the new heavy particles, a quasi-degenerate mass pattern would admittedly favor a UED-like scenario, while a split spectrum with sizable mass differences between the heavy states masses would certainly be more suggestive of a supersymmetric scenario. Theoretically, in fact, the assumption of universality in the soft supersymmetry breaking parameters is thought to hold at some high energy scale, e.g. the grand unification scale. As a result, degeneracy at the weak scale in supersymmetry is generically rather unnatural. In contrast, the KK structure of the UED spectrum predicts an almost completely degenerate spectrum, up to radiative corrections.

A second distinguishing feature concerns the Higgs structure of the two theories. The particle content of the minimal UED setup does not include an analogue of the heavy Higgs bosons of the MSSM. More precisely, even though the level-1 KK modes of the Higgs/gauge bosons in minimal UED have exactly the same quantum numbers as the MSSM Higgs bosons $H^0,\ A^0$ and $H^\pm$, the latter do not carry $R$-parity. The UED states $H^{(1)},\ A^{1)}$ and $H^{(1)\pm}$, in contrast, carry KK parity, and are therefore are more analogous to the higgsino sector of supersymmetry. However, the occurrence of additional Higgs bosons is not a robust discriminatory feature of supersymmetry, at least at the LHC, as such particles are likely to escape detection over large portions of the supersymmetric parameter space.

One is, therefore, left with the two most striking differences between supersymmetry and UED. Firstly, UED predicts the existence of a tower of KK states with almost degenerate masses lying around $m_n\approx n/R$ for the $n$-th KK level, rather than the single heavier ``copy'' of the SM particle content found in supersymmetry. Secondly, in UED the spins of the KK states are the same as that of their SM counterparts, while in supersymmetry the spins of the superpartners differ from their SM counterparts by $1/2$.

In this section, we review the feasibility of discriminating between UED and supersymmetry at hadronic (Sec.~\ref{sec:discrhadro}) and leptonic (Sec.~\ref{sec:discrlepto}) colliders, as well as with astrophysical experiments (Sec.~\ref{sec:discrastro}). The punchline of this section is that at the LHC, if $R^{-1}\lesssim 1$ TeV,  signals of $n=2$ KK states and various spin correlation studies can provide marginal evidence favoring either UED or supersymmetry. At a multi-TeV ILC, the two models can be accurately distinguished through several processes, including angular distributions of final state leptons and total cross section measurements. The particle masses of a UED setup will also be measured to an impressive degree of precision at such a machine. Finally, direct and indirect dark matter detection experiment can, in many cases, provide signatures which would also allow for a discrimination between neutralino and KK photon dark matter, or complement and confirm results from collider data.

\subsection{Discrimination of UED and Supersymmetry at Hadron Colliders}\label{sec:discrhadro}

As alluded in the previous section, the detection of second level KK modes would be a provide strong evidence in favor of a UED scenario over an accidentally degenerate supersymmetric model. The production cross sections and decay patterns of $n=2$ KK states at the LHC were first addressed in Ref.~\cite{Datta:2005zs}. It was pointed out that the production cross section of strongly interacting $n=2$ KK states at the LHC is enhanced, at the same mass scale, with respect to the supersymmetric case. This is the case for several reasons. Firstly, the particle content in UED is ``duplicated'' in the KK modes. For example, there are both left-handed and right-handed SU(2) doublet KK fermions, while in supersymmetry there are only left-handed-doublet squarks. Secondly, the different angular distributions for fermions ($1+\cos^2\theta$) versus scalars ($1-\cos^2\theta$), when integrated over all angles, account for an extra factor 2. Furthermore, as heavy states at the LHC are produced close to threshold, the strong suppression of the cross section for the production of scalars in supersymmetry ($\sim\beta^3$) is replaced by the milder suppression for KK quarks ($\sim\beta$). However, Ref.~\cite{Datta:2005zs} points out that any signal from $n=2$ KK quarks and leptons decaying into $n=1$ modes would be likely swamped by the direct production of $n=1$ KK states. A possible exception might be that of decay chains leading to $Z^{(1)}, W^{(1)}$ pairs, subsequently decaying into leptons, and in principle giving rise to a very clean and potentially observable multilepton signature ($Nl+$ missing energy, $N\ge5$). The latter would, unfortunately, feature very limited statistics, and looks very challenging at the LHC~\cite{Datta:2005zs}. 

The best prospects for detecting second level KK modes appear to be connected with the gauge boson sector. In particular, decays of $n=2$ gauge bosons into fermion-antifermion pairs can be looked for as a bump in the invariant mass distribution of the decay products, in analogy with $Z^\prime$ searches. As discussed in Sec.~\ref{sec:ilcn2}, the decay widths of KK $n=2$ gauge bosons are always very suppressed, and much smaller than the typical width of a $Z^\prime$ with SM couplings. This implies, experimentally, that the width of the resonance will then be determined by the experimental resolution, rather than the intrinsic particle width.

The production of $n=2$ KK gauge bosons proceeds through three main channels: (1) single production through KK-number violating operators, (2) indirect production through decays of strongly interacting $n=2$ states and (3) direct pair production. The first channel was discussed in Sec.~\ref{sec:ilcn2} and is completely analogous for the case of charged gauge bosons. The production of electroweak KK modes in the decays of heavier $n=2$ particles is analogous to the dominant production of electroweak superpartners in supersymmetry. In particular, the large branching fractions of SU(2) doublet, $n=2$ KK quarks into $W^{(2)}$ and $Z^{(2)}$ and of SU(2) singlet, $n=2$ KK quarks into $B^{(2)}$ entail that this channel is a very significant source for the production of $n=2$ gauge bosons at the LHC. Direct pair production through KK-number conserving interactions, instead, is kinematically suppressed, since {\em two} heavy particles must be produced in the final state, and can be neglected for electroweak gauge bosons \cite{Datta:2005zs}.

In analogy to the case of $n=2$ KK quarks mentioned above, KK-number conserving decays are not very distinctive, since they simply contribute to the inclusive $n=1$ sample which is dominated by direct $n=1$ production. The decays of $n=1$ particles will then give relatively soft objects, and most of the energy will be lost in the LKP mass. Signatures of second level KK modes based on purely KK-number conserving decays are not very promising experimentally as the suppressed production cross section for the heavy $n=2$ particles is not compensated by the benefit of the large mass, since most of the energy is carried away by the invisible LKP. KK-number violating channels, in which the $n=2$ KK gauge boson decays are fully visible, are instead much more promising \cite{Datta:2005zs}. In particular, since KK-number conserving decays dominate the decays of the $n=2$ gluon, searches for the electroweak gauge bosons in leptonic channels is predicted to be the best search strategy, even though the branching fraction of $Z^{(2)},\ B^{(2)}\to l^+l^-$ is always rather suppressed ($\sim$2\%).

Ref.~\cite{Datta:2005zs} studied the inclusive production of $Z^{(2)},\ B^{(2)}$ and looked for a dilepton resonance in both the $e^+e^-$ and $\mu^+\mu^-$ channels. As mentioned above, a crucial parameter of the search is the width of the reconstructed resonance which, in turn, determines the size of the invariant mass window selected by the cuts. Since the intrinsic width of the $Z^{(2)}$ and $B^{(2)}$ resonances is so small, the mass window is entirely determined by the mass resolution in the dimuon and dielectron channels. For electrons, the resolution in CMS is approximately constant, on the order of $\Delta m_{ee}/m_{ee}\approx 1\%$, in the region of interest~\cite{Datta:2005zs}. On the other hand, the dimuon mass resolution is energy dependent and, in preliminary studies based on a full simulation of the CMS detector, has been parametrized as~\cite{muonres}
$$\frac{\Delta m_{\mu\mu}}{m_{\mu\mu}}=
0.0215+0.0128\left(\frac{m_{\mu\mu}}{1\ {\rm TeV}}\right)\ .$$
The analysis of Ref.~\cite{Datta:2005zs} used the following cuts: (1) lower cuts on the lepton transverse momenta, $p_T(\ell)>20$ GeV, (2) a central rapidity cut on the leptons, $|\eta(\ell)|<2.4$, and (3) dilepton invariant mass cuts for electrons of $2R^{-1}-2\Delta m_{ee}<m_{ee}<m_{V_2}+2\Delta m_{ee}$ and, for muons, $2R^{-1}-2\Delta m_{\mu\mu}<m_{\mu\mu}<m_{V_2}+2\Delta m_{\mu\mu}$. The main SM background to the signal is Drell-Yan.

\begin{figure}
\centering
\mbox{\includegraphics[width=0.47\textwidth,clip=true]{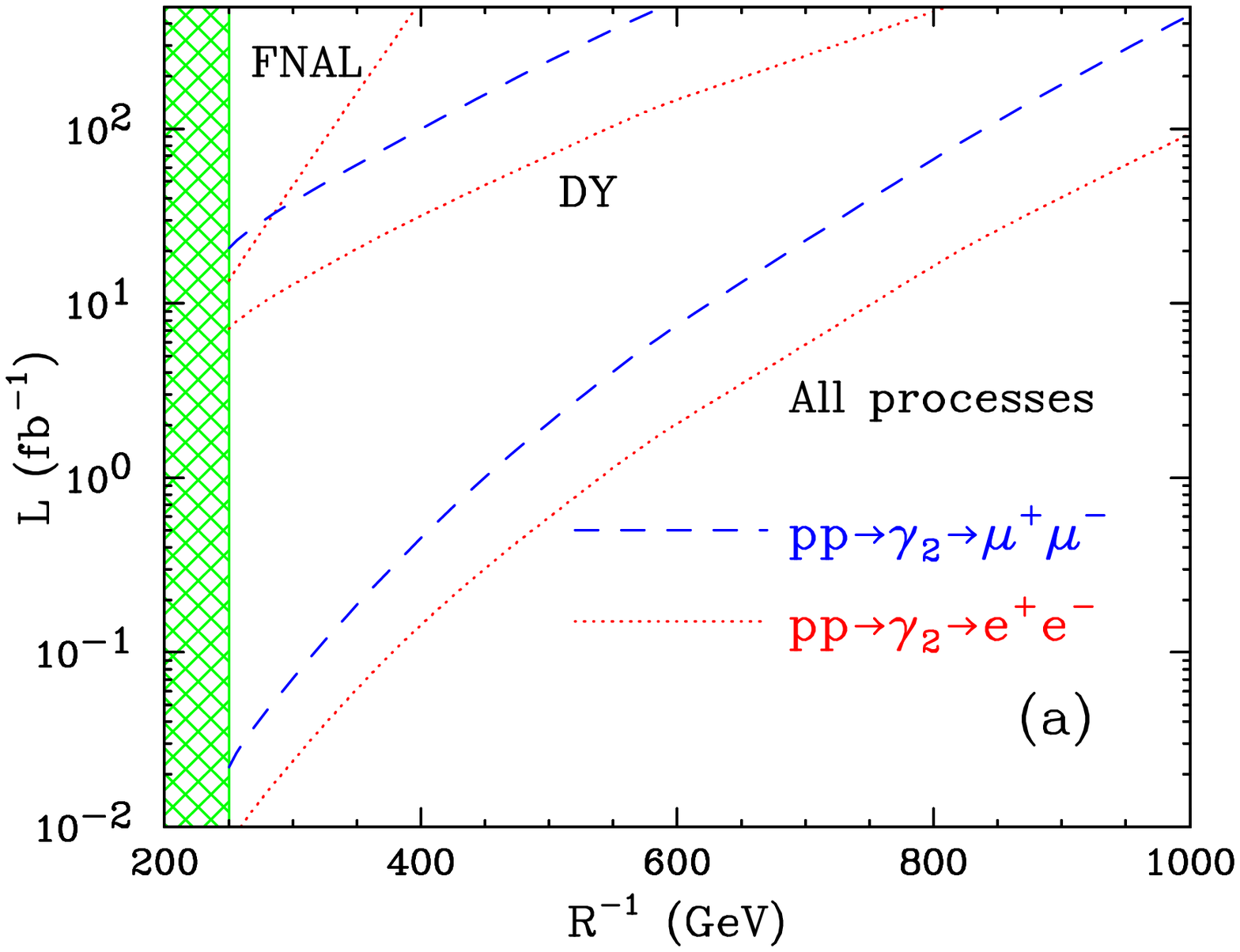}\qquad\includegraphics[width=0.47\textwidth,clip=true]{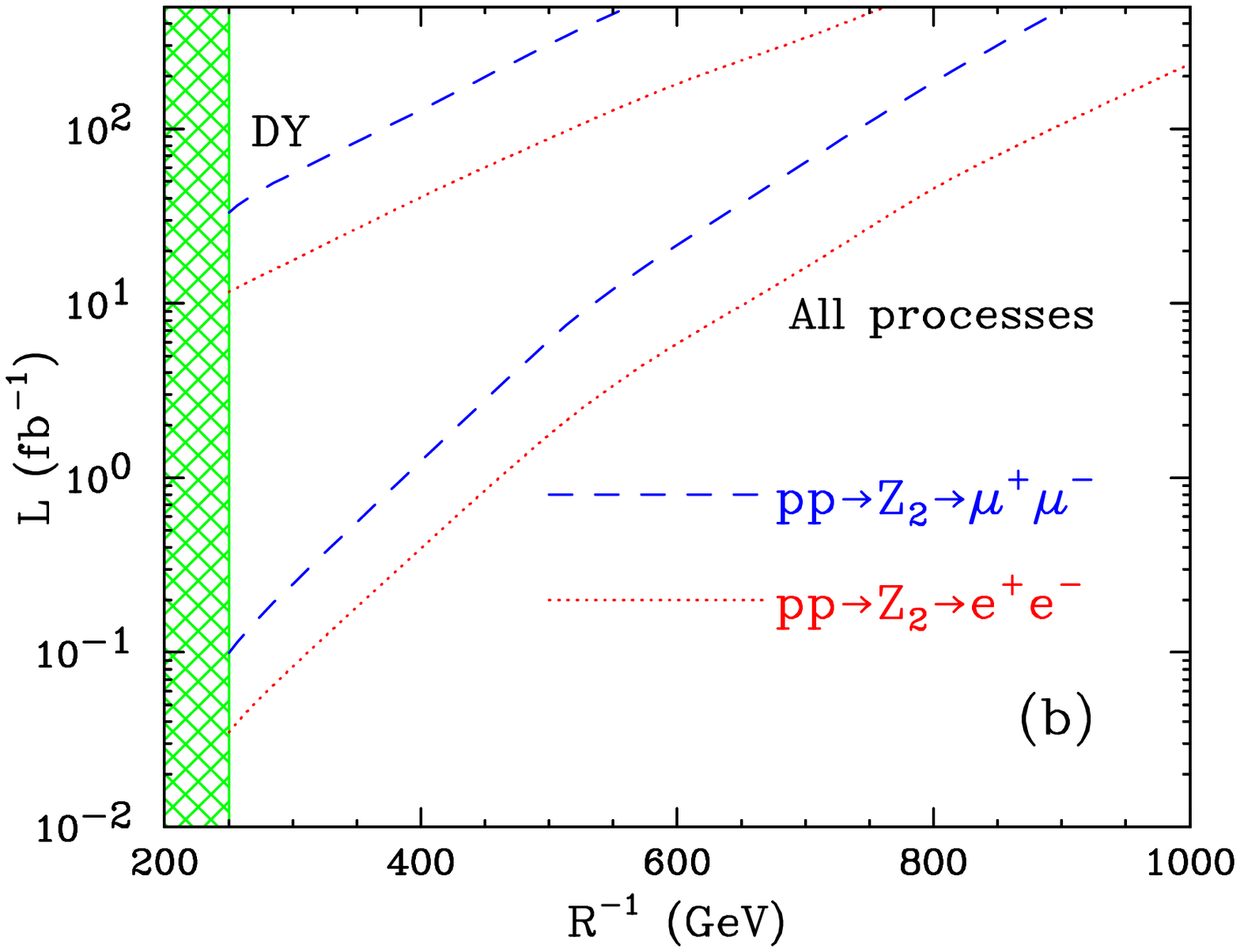}}
\caption{$5\sigma$ discovery reach for $B^{(2)}$ (left) and $Z^{(2)}$ (right).
The y-axis indicates the total integrated luminosity ${\rm L}$ (in ${\rm fb}^{-1}$) required for a $5\sigma$ excess of signal over background in the dielectron (red, dotted) or dimuon (blue, dashed) channel, as a function of $R^{-1}$. In each plot, the upper set of lines labeled ``DY'' makes use of the single production only, while the lower set of lines (labeled ``All processes'') includes indirect production from $n=2$ KK quark decays. The red dotted line marked ``FNAL'' in the upper left corner of the left panel reflects the expectations for a $\gamma_2\to e^+e^-$ discovery at the Tevatron in Run II. The shaded area below $R^{-1}=250$ GeV indicates the region disfavored by precision electroweak data~\cite{Appelquist:2002wb}. From Ref.~\cite{Datta:2005zs}.}
\label{fig:lhcreachn2}
\end{figure}

The discovery reach of the LHC and the Tevatron for the $B^{(2)}$ and $Z^{(2)}$ resonances, as assessed in Ref.~\cite{Datta:2005zs}, is shown in Fig.~\ref{fig:lhcreachn2}. The electron energy resolution assumed for the Tevatron was $\Delta E/E=0.01\oplus0.16/\sqrt{E}$ \cite{Blair:1996kx}. The Tevatron reach in the dimuon channel is worse due to the poorer resolution, and the reach for the $Z^{(2)}$ is worse still since $m_{Z^{(2)}}>m_{B^{(2)}}$ for a fixed $R^{-1}$. Even after one year of a low luminosity run, the LHC should be able to test for the existence of $n=2$ KK states for values of $R^{-1}\lesssim 750$ GeV. Eventually, the LHC should be able to probe $R^{-1}\lesssim 1$ TeV.

\begin{figure}
\centering
\mbox{\includegraphics[width=0.47\textwidth,clip=true]{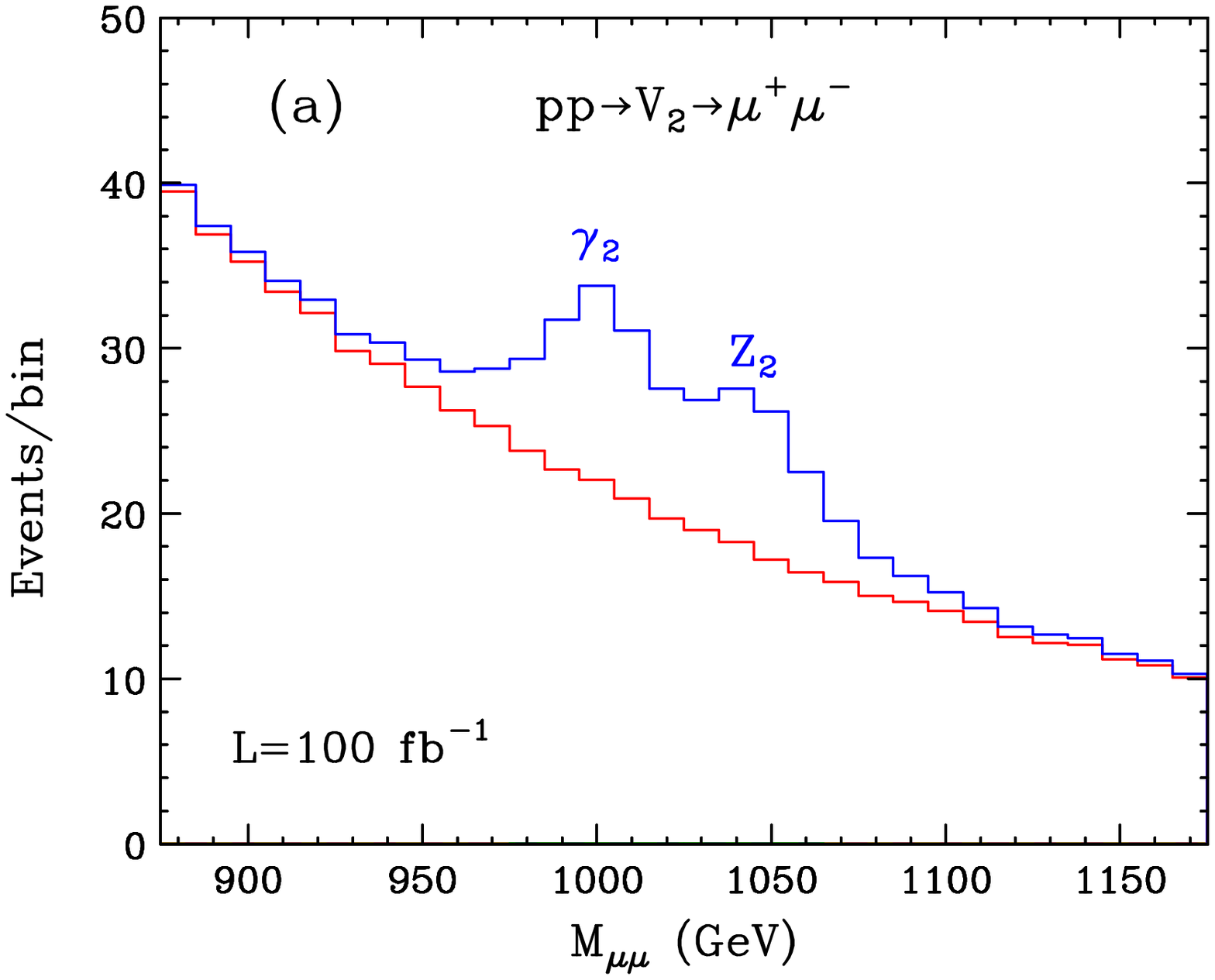}\qquad\includegraphics[width=0.47\textwidth,clip=true]{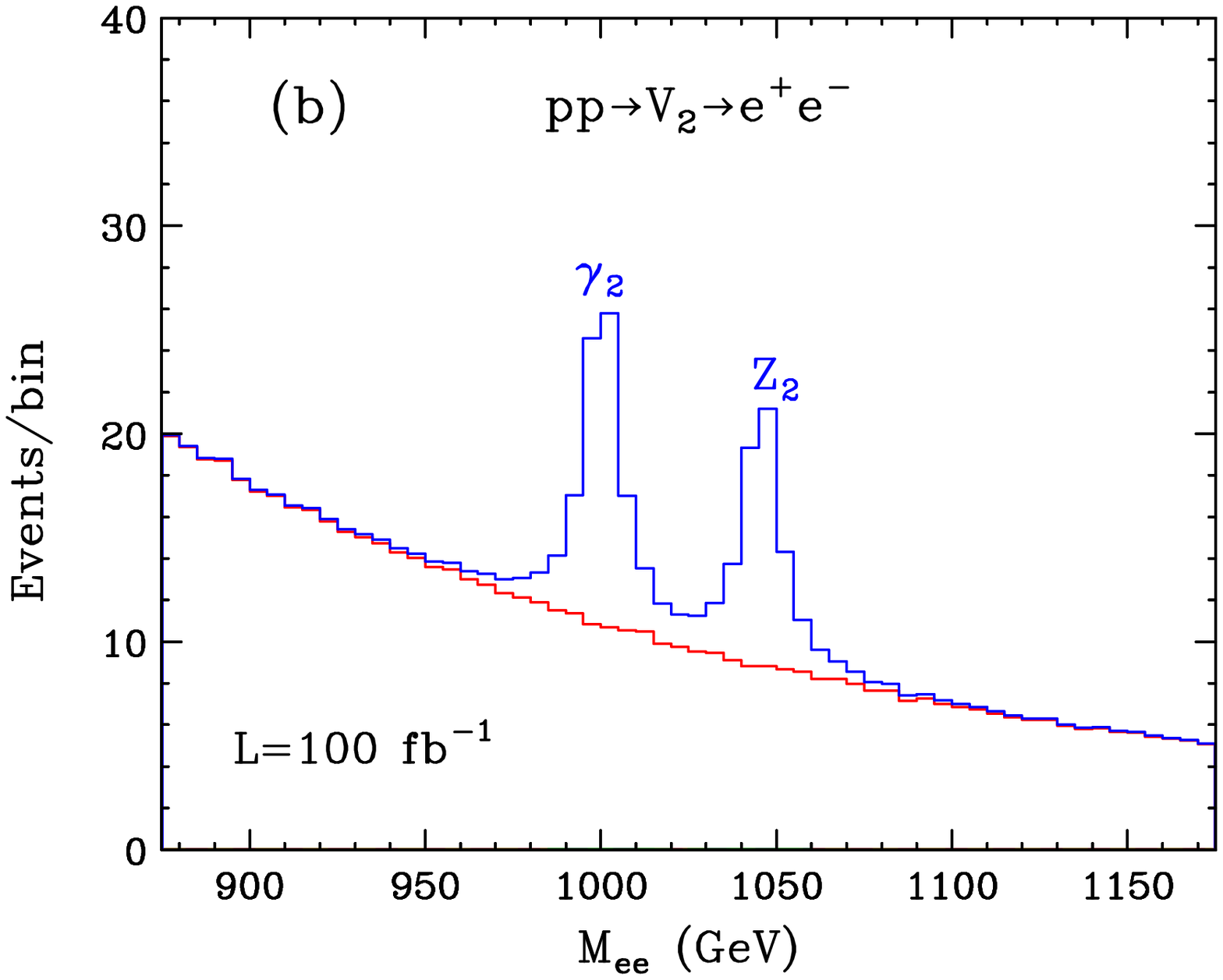}}
\caption{The $B^{(2)}-Z^{(2)}$ diresonance structure with $R^{-1}=500$ GeV, for the dimuon (left) and the dielectron (right) channels at the LHC with ${\rm L}=100\ {\rm fb}^{-1}$. The SM background is shown with the (red) continuous underlying histogram. From Ref.~\cite{Datta:2005zs}.}
\label{fig:dileptons}
\end{figure}

As pointed out in Ref.~\cite{Datta:2005zs}, the discovery of a resonance, while suggestive of a UED setup, might be confused with the effect of a $Z^\prime$ gauge boson. A much cleaner signal would be provided by multiple quasi-degenerate KK gauge boson resonances, as their is no compelling motivation for multiple $Z^\prime$s with approximate mass degeneracy. A double peak structure would be rather challenging to observe in the dijet channel due to the relatively poor LHC jet energy resolution. Again, the dilepton channel from $B^{(2)}$ and $Z^{(2)}$ is the most promising channel. The results for the invariant mass distribution of dimuons and dileptons, for $R^{-1}=500$ GeV and ${\rm L}=100\ {\rm fb}^{-1}$, and the same cuts described above, is shown in fig.~\ref{fig:dileptons}. The better mass resolution makes the diresonance structure easier to detect in the dielectron channel, although the structure is beginning to emerge in the dimuon channel as well.

The second strong handle on the discrimination between UED and supersymmetry we mentioned at the beginning of this section pertains the determination of the spin of particles at the LHC. Spin determinations at hadron colliders appear to be extremely challenging, as the center of mass energy of the underlying event is unknown. Furthermore, the momenta of the two undetected neutral stable particles in the event are also unknown. 

\begin{figure}
\centering
\mbox{\includegraphics[width=0.47\textwidth,clip=true]{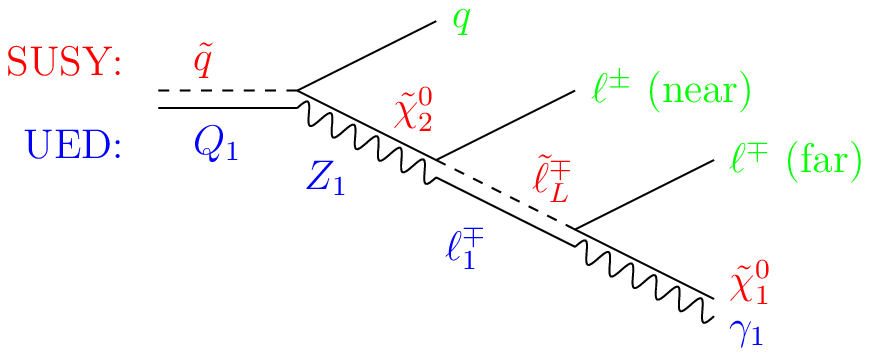}\qquad\includegraphics[width=0.47\textwidth,clip=true]{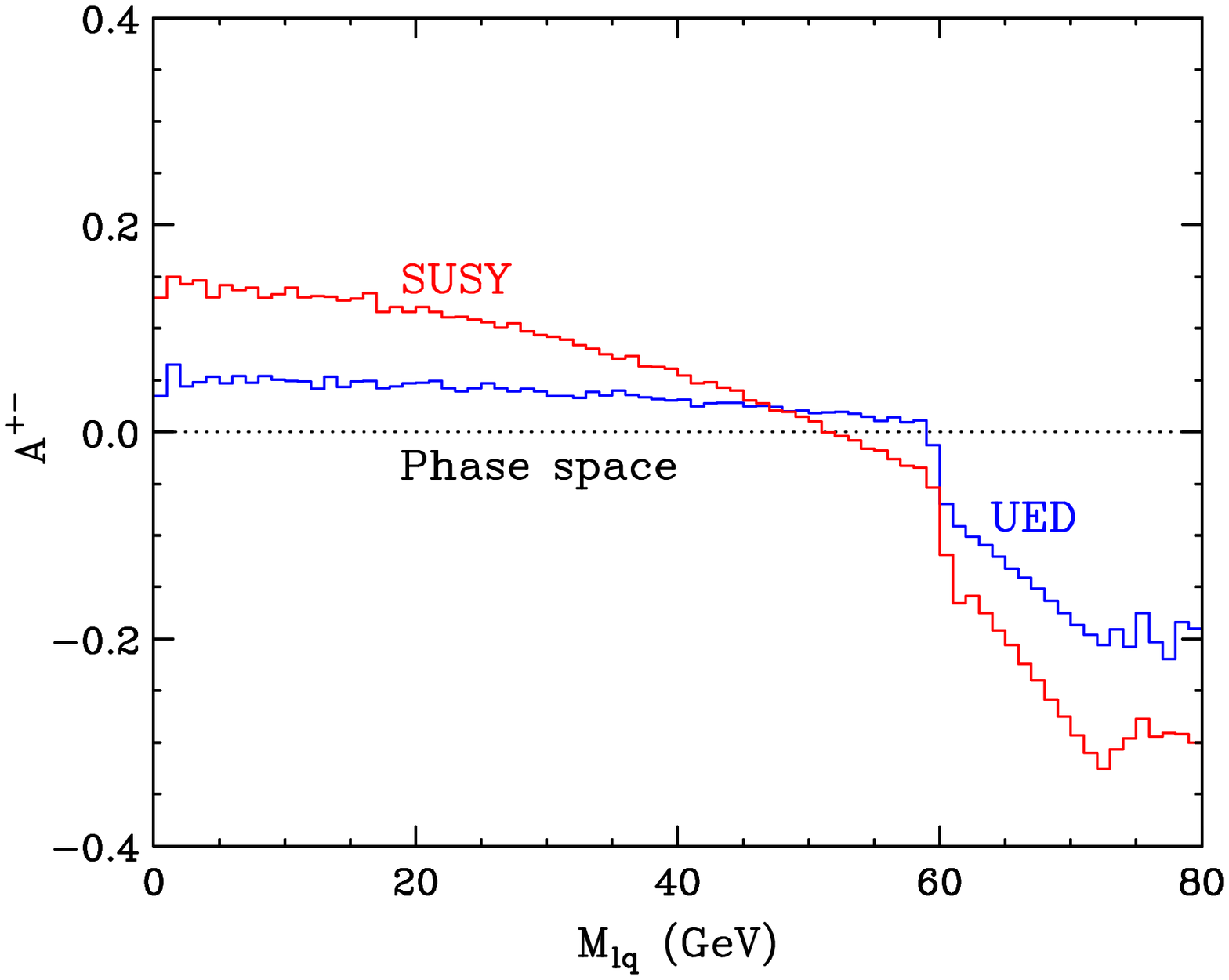}}
\caption{Left: Twin diagrams in supersymmetry and UED. The upper line corresponds to the cascade decay $\tilde q \to q\tilde\chi^0_2 \to q\ell^\pm\tilde\ell^\mp_L \to q\ell^+\ell^-\tilde\chi^0_1$ in supersymmetry. The lower line corresponds to the cascade decay $Q^{(1)} \to q Z^{(1)}\to q\ell^\pm\ell^\mp_1 \to q\ell^+\ell^-\gamma_1$ in UED. In either case the observable final state is the same: $q\ell^+\ell^-\met$. Right: Comparison of the charge asymmetry $A^{+-}$ defined in eq.~(\ref{asymmetry}) as computed in the case of UED with $R^{-1}=500$ GeV and the case of supersymmetry with a matching sparticle spectrum. Adapted from Ref.~\cite{Datta:2005zs}.}
\label{fig:spincorr}
\end{figure}

Ref.~\cite{Barr:2004ze} suggested that a charge asymmetry in the lepton-jet invariant mass distributions from particular cascade decays can be used to discriminate supersymmetry from the case of pure phase space decays and is an indirect indication of the superparticle spins. This technique was applied in Refs.~\cite{Smillie:2005ar,Datta:2005zs} to the case of the discrimination between UED and supersymmetry. Ref.~\cite{Datta:2005zs} compared the case of UED with $R^{-1}=500$ GeV with a supersymmetric model featuring a matching particle spectrum and studied the two cascade decays shown in the left panel of Fig.~\ref{fig:spincorr}. The spin of the intermediate particle ($Z^{(1)}$ in UED or $\tilde\chi^0_2$ in supersymmetry) governs the shape of the lepton-quark invariant mass distributions, $M_{\ell q}$, for the near lepton. In practice, however, one cannot distinguish the near and far lepton, and is forced to include the invariant mass combinations with both leptons~\cite{Barr:2004ze}. This tends to wash out the spin correlations but, due to the different quark and anti-quark content of the proton, a residual effect remains which, in turn, leads to a difference in the production cross sections for squarks and anti-squarks~\cite{Barr:2004ze}. The spin correlations are encoded in the charge asymmetry~\cite{Barr:2004ze}
\begin{equation}
A^{+-} \equiv
\left(\frac{dN(q\ell^+)}{dM_{ql}}-\frac{dN(q\ell^-)}{dM_{ql}}\right)\Biggr/
\left(\frac{dN(q\ell^+)}{dM_{ql}}+\frac{dN(q\ell^-)}{dM_{ql}}\right)\ ,
\label{asymmetry}
\end{equation}
where $q$ stands for both a quark and an anti-quark. $N(q\ell^+)$ and $N(q\ell^-)$ are the numbers of entries with a positively and negatively charged lepton, respectively. The results of Ref.~\cite{Datta:2005zs} on $A^{+-}$ are shown for the UED and supersymmetry cases in the right panel of Fig.~\ref{fig:spincorr}. Evidently, a clean discrimination of the two cases is challenging and probably not completely unambiguous. Further effects (poor statistics near the two ends of the plot, detector effects, backgrounds, etc.) can further dilute the cleanness of this discrimination technique.

Ref.~\cite{Smillie:2005ar} studied, for the same decay chain shown in the left frame of Fig.~\ref{fig:spincorr} other spin correlations, such as the quark-near lepton mass distribution, the dilepton mass distribution and the quark-far lepton mass distribution, and their observable counterparts, arriving at conclusions similar to those of Ref.~\cite{Datta:2005zs}. In particular, it turns out that it would be much easier to rule out a UED setup given a supersymmetry spectrum than vice-versa, as the particle mass degeneracy tends to blur the $A^{+-}$ difference in the two cases.

Recently, Ref.~\cite{Alves:2006df} pointed out that a further handle to disentangle UED from supersymmetry comes from the determination of the spin of the ``gluino'' (or KK gluon) from long decay chains involving the gluino decay into a sbottom, which decays into the second lightest neutralino, followed by a slepton, and finally to the LSP (or the corresponding chain of KK modes in the case of UED). Using, as an example, the supersymmetric benchmark model SPS1a, Ref.~\cite{Alves:2006df} showed that, using a list of asymmetries constructed from lepton-bottom correlations or from pure bottom-bottom correlations, it is possible to distinguish between supersymmetry and UED cascade interpretations, and thus possible to identify the fermionic (bosonic) nature of the gluino (KK gluon).

In a scenario where the compactification radius $R^{-1}$ is very small (or, correspondingly, in the case of supersymmetry, where the mass of sleptons is small) an observable that could help unraveling the spin of KK leptons (sleptons) was proposed in \cite{Barr:2005dz}. The analysis considers the processes
\begin{eqnarray}
q\bar q\rightarrow Z^0/\gamma\rightarrow & l^{(1)+} l^{(1)+} \rightarrow & B^{(1)} \ l^+\ B^{(1)}\ l^-\\
q\bar q\rightarrow Z^0/\gamma\rightarrow & \widetilde l^+ \widetilde l^- \rightarrow & \widetilde\chi^0_1\ l^+\ \widetilde\chi^0_1\ l^-
\end{eqnarray}
and pointed out that since the leptons ($e,\mu$) are highly relativistic, their pseudo-rapidities are to a very good approximation close to their true rapidities. In particular, making use of a function of the pseudopodia difference between the final state leptons, $\Delta\eta_{l^+l^-}$, one no longer has to face the problem of determining the center-of-mass frame along the beam direction, still being sensitive to the KK leptons (sleptons) angular distribution, that, in turn, depends upon the spin of the particles. Ref.~\cite{Barr:2005dz} proposed to use the angular variable
\begin{equation}
\cos\theta_{ll}^*\equiv\cos\left(2\tan^{-1}\exp\left(\Delta\eta_{l^+l^-}/2\right)\right);
\end{equation}
that corresponds to the cosine of the angle between each lepton and the beam axis in the longitudinally boosted frame where the (pseudo-)rapidities of the final state leptons are equal and opposite. $\cos\theta_{ll}^*$ features the virtue of being longitudinally boost invariant, and is on average smaller for SUSY than for UED, providing a potentially spin-sensitive discriminant in KK leptons/sleptons pair production at hadron colliders. Ref.~\cite{Barr:2005dz} showed that with proper cuts it would be possible to statistically distinguish between scalar and fermionic heavy leptons. In particular, the anticipated LHC reach with 100 (300) ${\rm fb}^{-1}$ integrated luminosity extends to 202 (338) GeV for left-handed sleptons, and to 143 (252) GeV for right-handed sleptons \cite{Barr:2005dz}.

\subsection{Discrimination of UED and Supersymmetry at a Multi-TeV $e^+e^-$ Collider}\label{sec:discrlepto}

Assuming the LHC observes signals of new physics consistent with either $n=1$ KK modes in UED or sparticles in supersymmetry, Ref.~\cite{Battaglia:2005zf} studied the discrimination of these scenarios at a post-LHC facility, namely a linear collider tunable over a center-of-mass energy range between 1 TeV and 3 TeV. Ref~\cite{Battaglia:2005zf} concentrated on a minimal UED model featuring $R^{-1}=500$ GeV and $\Lambda R=20$, and adjusted the supersymmetric low-energy scale parameters so that the two smuon masses and the neutralino mass matched the masses of their KK counterparts (the KK muons and the $B^{(1)}$). The most promising process at an ILC with large center-of-mass energy is found to be the pair production of KK muons (smuons), which then decay into a muon pair and two $B^{(1)}$'s (neutralinos). While, in supersymmetry, the production process of smuon pairs proceeds through $\gamma$ and $Z$ s-channel exchanges, in UED, KK muons can be also be pair produced via s-channel $n=2$ KK electroweak gauge boson exchange, a process that can occur close to resonance. 

The signal cross section for a center-of-mass energy of 3 TeV is 14.4 fb for the case of UED and 2.76 fb for supersymmetry. The SM backgrounds come from $\mu^+\mu^-\nu\bar\nu$ final states, mostly due to gauge boson pair production $W^+W^- \to \mu^+\mu^- \nu_\mu\bar{\nu}_\mu$, $Z^0Z^0 \to \mu^+ \mu^- \nu \bar{\nu}$ and from $e^+e^- \to W^+W^- \nu_e \bar{\nu}_e$, $e^+e^- \to Z^0Z^0 \nu_e \bar{\nu}_e$, followed by muonic decays. The total background cross section is $\approx$20~fb at $\sqrt{s}$ = 3~TeV. The background muons are, furthermore, produced at small polar angles, therefore potentially biasing the angular distribution. To reduce this background, Ref~\cite{Battaglia:2005zf} imposed event selection cuts including the requirement of detecting two muons, missing energy in excess to 2.5~TeV, transverse energy below 150~GeV and event sphericity larger than 0.05. The rejection of the $Z^0Z^0$ background was performed discarding events with di-lepton invariant mass compatible with $M_{Z^0}$. A further source of background, $\gamma \gamma \to \mu^+\mu^-$ from underlying $\gamma \gamma$ collisions, can be completely suppressed by a cut on the missing transverse energy, $E_T^{missing} >$ 50~GeV. Finally, events with large beamstrahlung were cut imposing an event sphericity smaller than 0.35 and acolinearity smaller than 0.8. The combination of these criteria provide a factor of $\approx 30$ background suppression in the kinematic region of interest, while not significantly biasing the lepton momentum distribution~\cite{Battaglia:2005zf}. 

\begin{figure}
\centering
\mbox{\includegraphics[width=0.47\textwidth,clip=true]{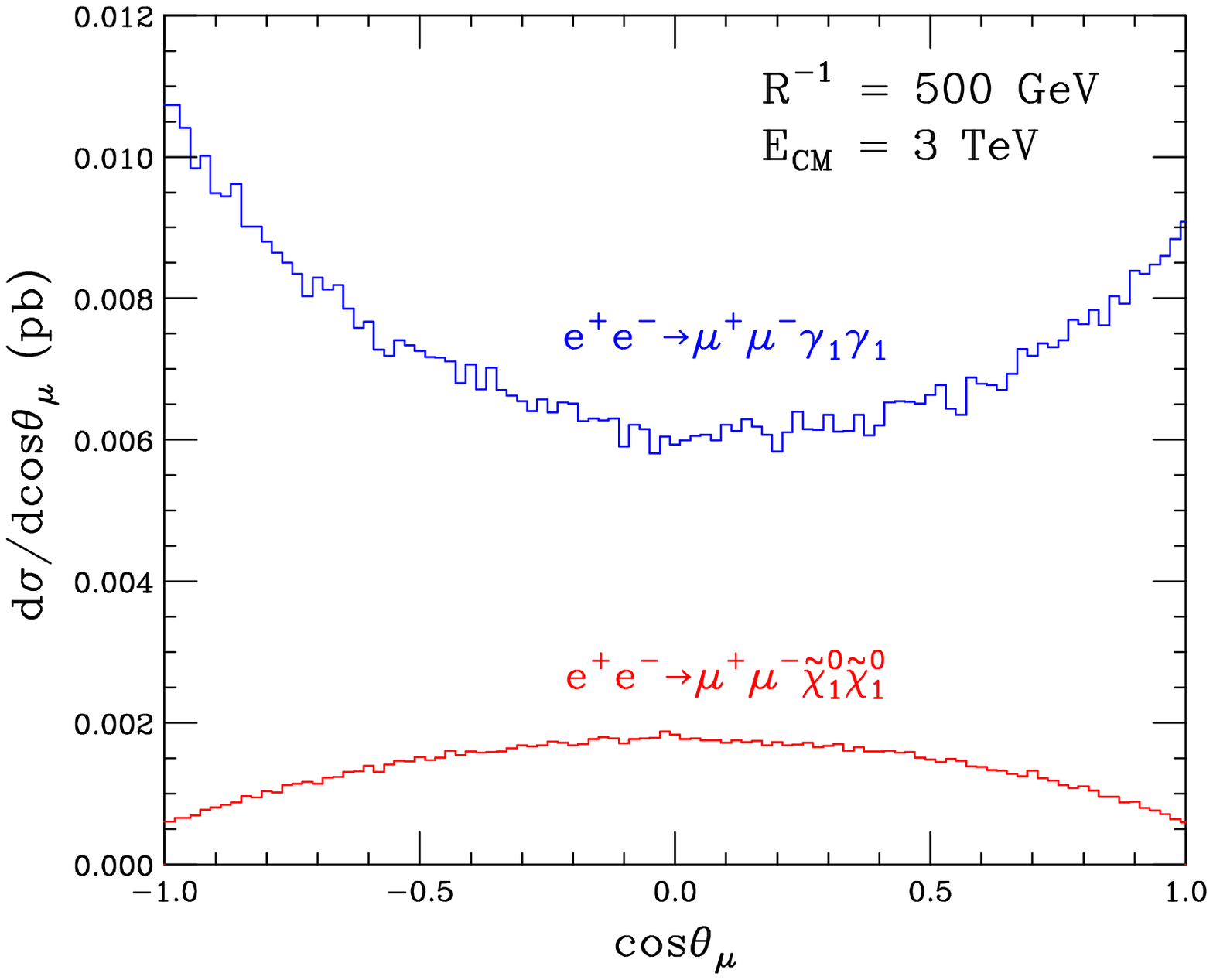}\qquad\includegraphics[width=0.47\textwidth,clip=true]{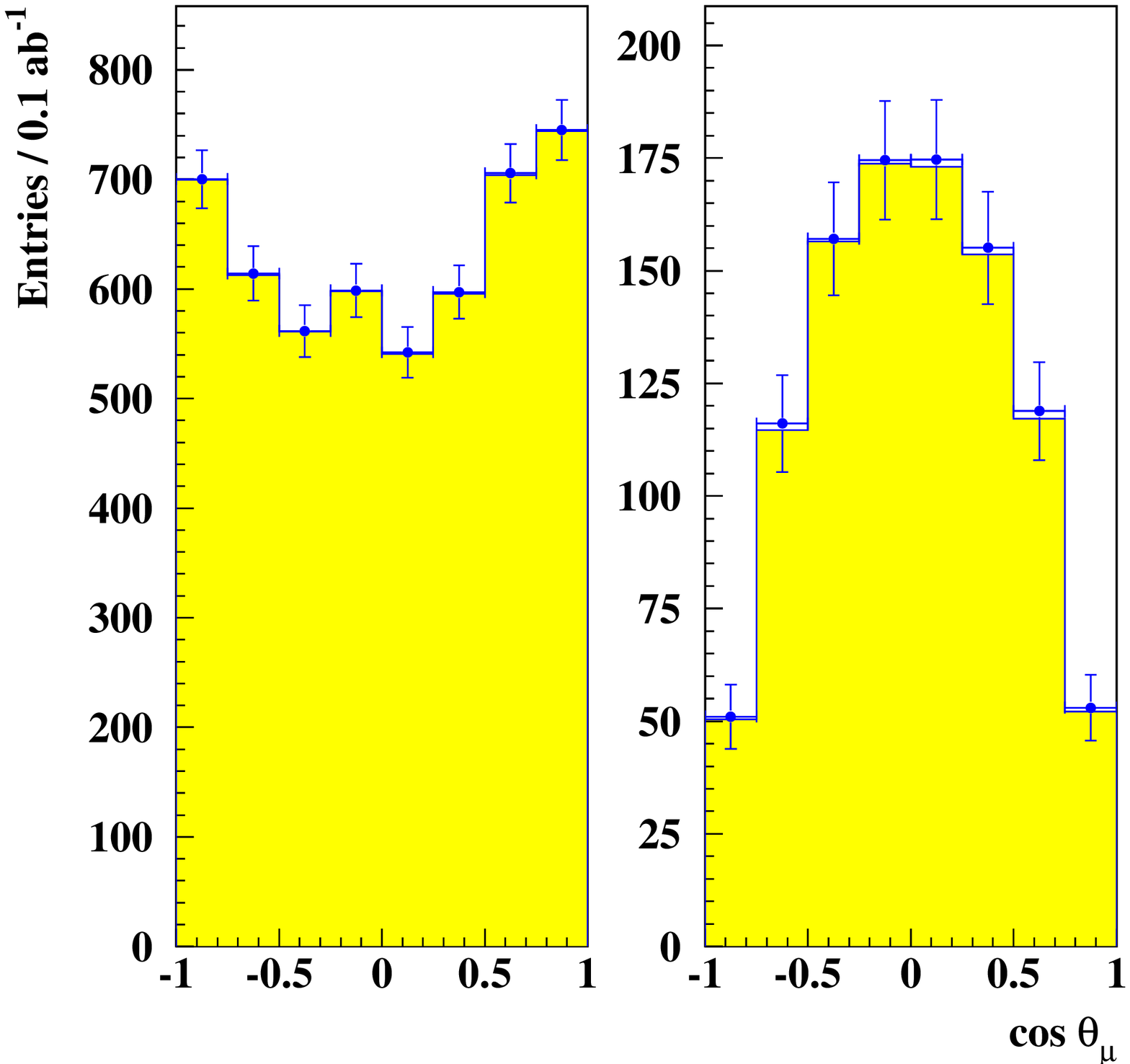}}
\caption{Left: The initial state radiation corrected differential cross-section $d\sigma/d\cos\theta_\mu$ for UED (blue, top) and supersymmetry (red, bottom) as a function of the muon scattering angle $\theta_\mu$. Right: The same, including the effects of event selection, beamstrahlung and detector resolution and acceptance , for UED (left) and supersymmetry (right). The data points are the combined signal and background events, while the yellow-shaded histogram represents the signal only. From Ref.~\cite{Battaglia:2005zf}.}
\label{fig:angdistilc}
\end{figure}
A first distinguishing feature in the KK muons/smuons pair production is the angular distribution of events as a function of the muon scattering angle, $\theta_\mu$. As long as the mass differences $M_{\mu^{(1)}}-M_{B^{(1)}}$ and $M_{\tilde\mu}-M_{\tilde\chi^0_1}$, respectively, remain small, the muon directions are well correlated with those of their parents, for which the angular distributions are given by
\be
\left(\frac{d\sigma}{ d\cos\theta}\right)_{UED} \sim 1+\cos^2\theta.
\ee
in the UED case, and by
\be
\left(\frac{d\sigma}{ d\cos\theta}\right)_{SUSY} \sim 1-\cos^2\theta.
\ee
in the case of supersymmetry, assuming production occurs well above threshold. The two angular distributions can be clearly distinguished even after accounting for initial state radiation, event selection, beamstrahlung and detector resolution and acceptance, as shown in Fig.~\ref{fig:angdistilc}. A $\chi^2$ fit to the normalized polar angle distribution shows that UED and supersymmetry can be distinguished on the sole basis of the distribution shape, with as little as 350~fb$^{-1}$ of data at $\sqrt{s}$ = 3~TeV.

\begin{figure}
\centering
\mbox{\includegraphics[width=0.47\textwidth,clip=true]{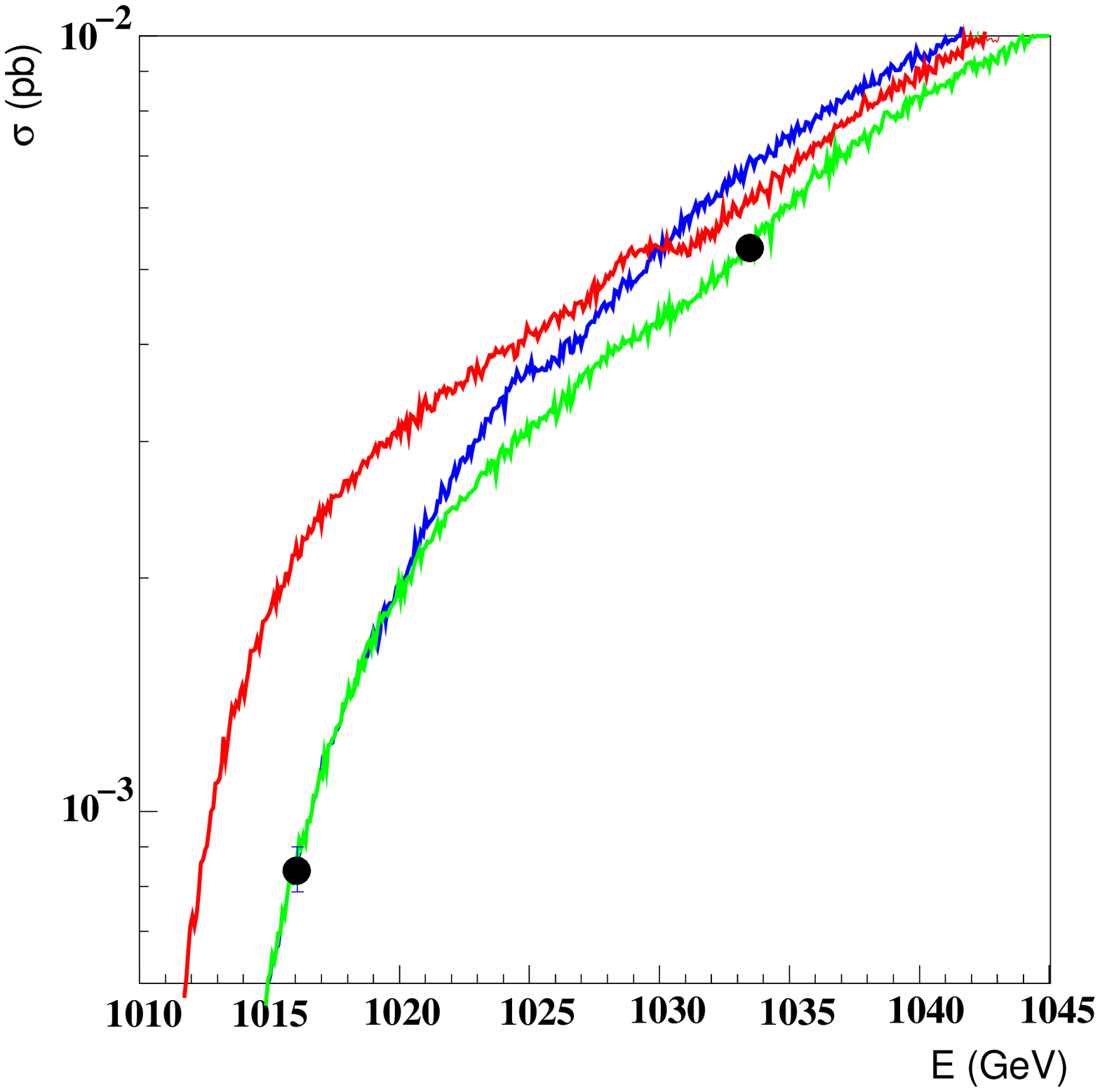}\qquad\includegraphics[width=0.47\textwidth,clip=true]{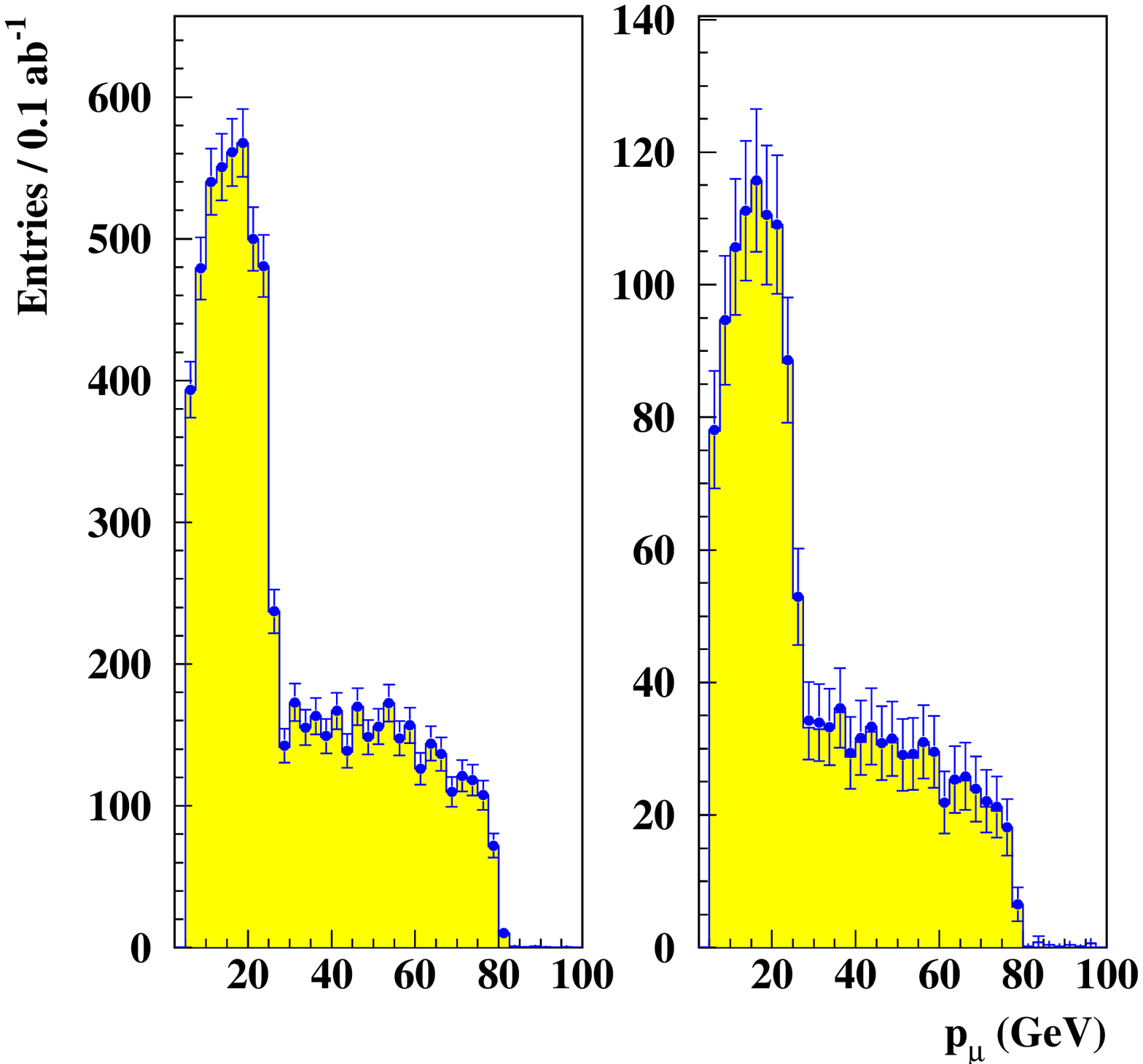}}
\caption{Left: A threshold scan at selected points. The green curve refers to the reference UED parameters while for the red (blue) curve the mass of $\mu_S^{(1)}$ ($\mu_D^{(1)}$) has been lowered by 2.5~GeV. The points indicate the expected statistical accuracy for the cross section determination at the points of maximum mass sensitivity. Right: The muon energy spectrum resulting from KK muon production (left) and smuon production (right), including the effects of event selection, beamstrahlung and detector resolution and acceptance. The data points are the combined signal and background events, while the yellow-shaded histogram is the signal only. From Ref.~\cite{Battaglia:2005zf}.}
\label{fig:ilcvari}
\end{figure}
At an $e^+e^-$ linear collider, the muon excitation masses can be accurately determined through an energy scan across the onset of the pair production threshold. Such a scan, impossible at a machine like the LHC (recall that in a hadron the center of mass energy of the partons involved in the underlying event producing the new physics massive particle pairs is unknown) will also reveal the spin of the particle, as the cross sections for the UED processes rise at threshold $\propto \beta$ while in supersymmetry their threshold onset is $\propto \beta^3$, where $\beta$ is the particle velocity. From the estimated sensitivity, $d \sigma/dM$, and the cross section accuracy, Ref.~\cite{Battaglia:2005zf} claims that the masses of the two UED muon excitations can be determined to $\pm 0.11$~GeV and $\pm 0.23$~GeV for the SU(2) singlet and the doublet states, respectively, with a total luminosity of 1~ab$^{-1}$ shared in three values of the center-of-mass energy, where the particle widths can be disregarded. Fig.~\ref{fig:ilcvari} illustrates the accuracy at which the ILC will be able to tell apart singlet or doublet KK muons with mass differences of 2.5 GeV.

A further handle that will be completely under control at the ILC, unlike the LHC, is the determination of the total production cross section for KK muons or smuons as a function of the center of mass energy. As we stated above, at a center-of-mass energy of 3 TeV, the two cross sections differ by a factor 5, to be contrasted with a projected CLIC absolute luminosity determination that should be measurable to ${\cal{O}}(0.1~\%)$ and the average effective collision energy to ${\cal{O}}(0.01~\%)$.

The mass determination also benefits from the study of the characteristic end-points of the muon energy spectrum (see Fig.~\ref{fig:ilcvari}, right panel). Even though these do not depend on the spin of the decaying parent particles, and will therefore not directly distinguish between UED and supersymmetry, their study can lead to an impressively accurate mass determination, to the level of $\pm$0.19~(stat.)~$\pm$0.21~(syst)~GeV, where the statistical uncertainty is given for 1~ab$^{-1}$ of data and the systematics reflect the uncertainty in the $\mu_1$ masses.

In addition to the possibilities discussed above, the distinction between UED and supersymmetry can proceed through the study of the sharp peak in the photon energy spectrum predicted when one of the mediating s-channel particles is on-shell; in that case, since the decay $Z^{(2)}\to\mu^{(1)}\mu^{(1)}$ is allowed by phase space, there will be a sharp peak in the photon spectrum, due to radiative return to the $Z^{(2)}$. In minimal UED, the peak occurs only for the $Z^{(2)}$ and not for the $B^{(2)}$, since $m_{B^{(2)}}/2<m_{\mu^{(1)}}$, and gives rise to a peak located at a photon energy
\be
E_\gamma = \frac{1}{2}\ E_{CM}\ 
\biggr(1-\frac{M^2_{Z_2}}{ E^2_{CM}} \biggl).
\ee
The same peak cannot occur in supersymmetry, where smuon production is mediated only by the $Z$ and the photon, and both particles are always far from being on-shell (recall that the lower limit on the smuon mass from LEP is roughly $m_{\tilde\mu}\gtrsim 100\ {\rm GeV}\gg m_Z/2$).

Ref.~\cite{Battaglia:2005zf} also considered the prospects for the pair production of KK leptons and quarks. In the case of KK tau pairs, the lower statistics and the inferior jet energy resolution make the channel less competitive than the muon channel. For electrons, instead, the extra t-channel diagram occurring in the pair production of KK electrons enhances the total production cross section and can be up to two orders of magnitude larger than in the case of muons. However, despite the larger event sample, the additional t-channel diagram also distorts the differential angular distributions discussed above, creating a forward peak, which causes the cases of UED and supersymmetry to look very much alike. Finally, in the case of KK quarks, the jet angular distribution will again be indicative of the KK quark spin, and can be used to discriminate against (right-handed) squark production in supersymmetry. Furthermore, the jet energy distribution will again exhibit endpoints which will, in principle, allow for precise mass measurements. The jet angular and energy measurements, however, would not be as clean as in the case of lepton (muon or electron) final states, which would therefore provide the most convincing evidence for UED/supersymmetry discrimination.

\subsection{Discrimination of UED and Supersymmetry With Dark Matter Experiments}\label{sec:discrastro}

In addition to using collider experiments, it may be possible to distinguish UED from supersymmetry with direct and indirect dark matter detection experiments. Following Ref.~\cite{Hooper:2006xe}, we briefly summarize the prospects for this here.

The spin-independent (scalar) elastic scattering cross section for the $B^{(1)}$ with nuclei is quite small compared to that found for neutralinos in many supersymmetric models. In particular, the WIMP-WIMP-Higgs coupling in supersymmetry is typically, although not always, considerably larger than in UED (in part because of the heavy CP-even Higgs in supersymmetry). That being said, the elastic scattering cross section (as determined by the direct detection rate) will, alone, likely be insufficient to discriminate between these models.

In contrast, the spin-dependent (axial-vector) elastic scattering cross section for the $B^{(1)}$ with protons is quite large, $\sigma_{B^{(1)} p, \rm{SD}} \approx 1.8 \times 10^{-6} \, \rm{pb}$ for a 1 TeV mass and a 10\% LKP-KK quark mass splitting (see Sec.~\ref{subsec:spindep}). Neutralinos can possess a similarly large spin-dependent cross section, but only if their higgsino fraction is substantial ($\gsim$ 1\%), which would also imply a fairly large spin-independent cross section~\cite{halzen}.

The discriminating power of these elastic scattering cross sections come from the ratio of the two types. A comparison of the spin-independent cross section, as manifest in direct detection experiments, with the spin-dependent cross section, as manifest in neutrino fluxes from the Sun, can act as a powerful discriminator of supersymmetry and UED.

\begin{figure}
\centering
\includegraphics[width=0.4\textwidth,angle=-90,clip=true]{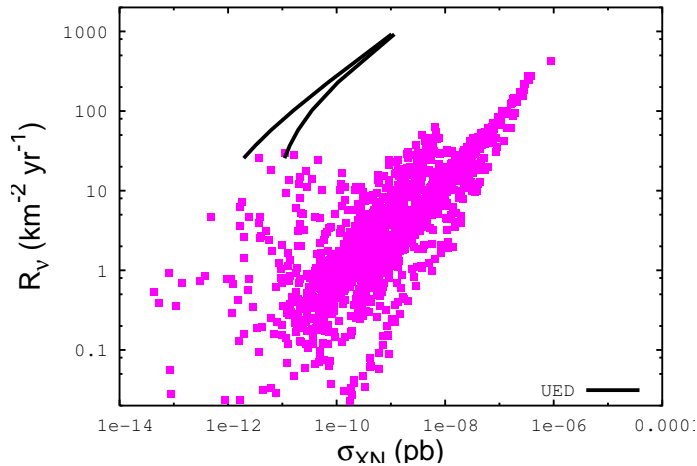}
\caption{A comparison of the neutrino and direct detection rates for neutralinos and Kaluza-Klein dark matter. The mass of the WIMP in each case shown is in the range of 450-550 GeV. The shaded region represents the range of supersymmetric models which satisfy all collider constraints, current direct detection constraints, and do not overproduce the abundance of neutralino dark matter. The thick solid lines denote the UED case varied across a broad range of KK quark masses ($m_{q^{(1)}}=1.05 \, m_{B^{(1)}}$ to \, $1.3 \, m_{B^{(1)}}$). The two UED contours represent the maximum and minimum Higgs masses consistent with electroweak precision observables. From Ref.~\cite{Hooper:2006xe}.}
\label{fig:compare}
\end{figure}

In Fig.~\ref{fig:compare}, a comparison of these two rates is shown for neutralino and $B^{(1)}$ dark matter (see Ref.~\cite{Hooper:2006xe} for details on the adopted scan over supersymmetric model parameters). The range of supersymmetry and UED modes shown demonstrates that, although neither the direct nor the neutrino rate alone will be sufficient to distinguish between these models, the combination of the two rates can be very powerful. For a particular value of the spin-independent elastic scattering, UED generally produces between one and three orders of magnitude more neutrino events. If these rates can each be measured in future experiments, such a comparison would be expected to be very useful in discriminating between UED and supersymmetry.

Other indirect detection channels have the potential to distinguish between UED and supersymmetry by observing distinctive features in the WIMP annihilation spectrum. In particular, the positron and gamma-ray spectra each have considerably larger fluxes at high energies in the case of UED than is predicted for neutralino dark matter. This is a result of the chirality suppression of neutralino annihilations at low velocities to light fermions. $B^{(1)}$s, on the other hand, annihilate a large fraction of the time to light fermion pairs.

\begin{figure}
\centering
\includegraphics[width=0.4\textwidth,angle=0,clip=true]{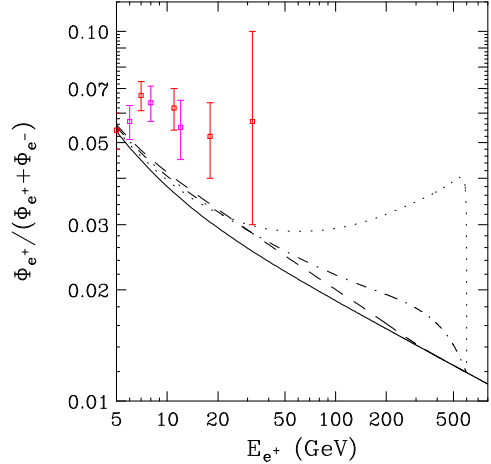}
\caption{The ratio of positrons to positrons+electrons as a function of energy, after diffusion through the galactic halo, for a 600 GeV WIMPs annihilating to $b\bar{b}$ (dashes) to gauge bosons (dot-dash) and to the modes of Kaluza-Klein (UED) dark matter, which are primarily charged leptons (dotted). The errors bars correspond to the measurements of the HEAT experiment \cite{heatanomaly}. To normalize the annihilation rate, a boost factor of approximately 10 has been used. The lower solid line represents the background from the secondary production of cosmic ray positrons.}
\label{fig:poscompare}
\end{figure}

In the case of the cosmic positron spectrum (see Sec.~\ref{sec:ep}), the spectrum is hardened in UED as a result of the large fraction of the annihilations to $e^+ e^-$ (20-25\%) and, to a lesser extent, to $\mu^+ \mu^-$ (20-25\%) and $\tau^+ \tau^-$ (20-25\%). The spectrum resulting from these annihilation modes is shown in Fig.~\ref{fig:poscompare}, where it is compared to the spectra representative of neutralino dark matter (annihilations to $b\bar{b}$ and $W^+ W^-$). As neutralinos annihilate almost entirely to combinations of heavy fermions, gauge and Higgs bosons, positrons are largely generated through relatively soft channels. Even in the case of neutralino annihilations to gauge bosons, the positron spectrum is considerably softer at high energies than is found in UED.

The gamma-ray spectrum produced in dark matter annihilations in UED is also harder at high energies than in the case of supersymmetry. This is a result of the large fraction of annihilations to  $e^+ e^-$ (through final state radiation) and $\tau^+ \tau^-$. The resulting gamma-ray spectrum from $B^{(1)}$ annihilations is shown in Fig.~\ref{gammaspectrum} of Sec.~\ref{sec:gr}. The spectrum predicted for neutralino dark matter, in contrast, is produced through the fragmentation and cascades of heavy quarks and other annihilation products, and is described by the softer dashed line in that figure. If future gamma-ray telescopes were able to measure the dark matter annihilation spectrum with sufficient detail, it would be possible to use this information to distinguish between UED and supersymmetry. In some supersymmetric setups, however, the effects of final state radiation can become non-negligible, leading to a considerably harder spectrum than what expected from fragmentation alone \cite{Bergstrom:2005ss,Bergstrom:2006hk}, blurring, to some extent, the possibility of a clear distinction between UED and supersymmetry.

The prospects for each of the methods described in this section depend strongly on the precise nature of the WIMP being studied and on a number of astrophysical uncertainties. If the $B^{(1)}$ is fairly heavy and it has a fairly large mass splitting between it and the KK quarks, for example, it will be difficult to observe at future direct detection experiments and impossible to detect with planned neutrino telescopes. On the other hand, a lighter UED model spectrum, with a small LKP-KK quark mass splitting, could be effectively studied (and likely distinguished from a neutralino) by using these techniques. The prospects for anti-matter and gamma-ray observations depend on the annihilation rate of dark matter in the galactic halo and other astrophysical environments, making the perspectives of detecting a signature with such techniques difficult to assess. 

The low-energy phenomenology of ``hybrid'' effective theory scenarios where a supersymmetric setup is embedded in an extra-dimensional framework, and where some or all of the SM fields are allowed to propagate in the extra dimensions was also recently addressed from the perspective of dark matter physics in Ref.~\cite{Fucito:2006ch}. As the compactification scale in these models is much larger than the soft-supersymmetry breaking scale, the resulting dark matter candidate is the neutralino, and the related phenomenology is very similar to that of supersymmetry in 4 dimensions~\cite{Fucito:2006ch}.

\clearpage
\newpage

\section{Beyond One Extra Dimension}\label{ch:sixd}

As compared to the numerous papers devoted to five-dimensional UED models, the phenomenology of models with more than one universal extra dimension has been somewhat rarely discussed in the literature so far. In some cases, models with $d\ge2$ have stronger theoretical motivations than the $d=1$ case within the context of UED (see Sec.~\ref{sec:ued}). For instance, the $d=2$ case is motivated by the cancellation of the global SU(2) anomaly \cite{Dobrescu:2004zi} and by the fact that the simplest chiral compactification of two universal extra dimensions, the {\it ``chiral square"}, preserves a discrete symmetry, a subgroup of the six dimensional Lorentz group, which suppresses the proton decay rate even when baryon number is maximally violated at the TeV scale~\cite{Appelquist:2001mj}  (see Secs.~\ref{sec:ued}).

The Feynman rules for gauge theories in $d=2$ UED compactified on the chiral square are given in Ref.~\cite{Burdman:2005sr}. The KK excitations are indicated with two KK numbers ($j,k$) which are manifest in the conservation of the quantity  $(-1)^{j+k}$, somewhat resembling the KK parity of five dimensional UED. The KK gauge bosons in this model feature two scalar fields corresponding to the polarizations along the two extra dimensions. At every KK level, these two scalars mix into a linear combination that is eaten by the spin-1 KK mode, and another which remains as a physical real scalar field, referred to as the {\em``spinless adjoint"}~\cite{Dobrescu:2001ae}.  As a consequence of the presence of the spinless adjoints, the production cross sections and decay modes of other KK states are different than are found in the $d=1$ case. In particular, the (1,1) states, which feature a mass of only $\sqrt{2}/R$ and can be produced in the $s$-channel, are particularly relevant to collider phenomenology~\cite{Allanach:2006fy,Burdman:2005sr,Burdman:2006gy}.

Several aspects of the phenomenology of six dimensional UED models were recently explored. Ref.~\cite{Mohapatra:2002ug} addressed the questions of neutrino masses and nucleon stability, and studied models based on the gauge group $SU(2)_L\times U(1)_{I_{3R}}\times U(1)_{B-L}$ compactified on $T^2/Z_2$ and left-right symmetric extensions featuring a full $SU(2)_L\times SU(2)_R\times U(1)_{B-L}$ gauge group, compactified on the orbifold $T^2/Z_2\times Z^\prime_2$. In both cases, neutrino masses are suppressed by an appropriate orbifold parity assignment for the standard model singlet neutrinos, and the proton decay rate is suppressed due to a residual discrete symmetry, left over from compactification. An interesting phenomenological outcome of these models is that, for low values of the fundamental scale, a dominant decay mode of the neutron is three neutrinos. Ref.~\cite{Mohapatra:2002ug} also pointed out that the model possibly container's a two-component dark matter setup, where the stable relics are KK right handed neutrinos and photons. 

Ref.~\cite{Hsieh:2006qe} elaborated on this latter point, and showed that the left-right symmetric six dimensional UED model of Ref.~\cite{Mohapatra:2002ug} indeed thermally produces the desired amount of two-component cold dark matter, for values of the compactification scale between 400 and 650 GeV, and a mass for the heavy extra gauge boson $Z^\prime$ below 1.5 TeV. Ref.~\cite{Hsieh:2006qe} finds that a generic prediction of this model is a dark matter-nucleon scattering cross section typically in excess of $10^{-44}\ {\rm cm}^2$, accessible to next-generation direct dark matter searches. Furthermore, the parameters of the model giving rise to the appropriate amount of thermal relics are such that the theory can be fully tested at the LHC.
 
In Ref.~\cite{Hsieh:2006fg} the two-component dark matter scenario was further investigated, including the effect of coannihilations. Also, the possibility that the spinless adjoint photon is lighter than the KK photon led the authors to explore this alternate scenario, featuring a the physical scalar gauge boson as one of the two dark matter constituents. The range of parameters is similar to the one outlined in Ref.~\cite{Hsieh:2006qe}, and a discussion of the extra $W_R$ and $Z^\prime$ phenomenology is provided as well.

In Ref.~\cite{Dobrescu:2007ec} dark matter in a six-dimensional UED model was studied. In this scenario, the lightest Kaluza-Klein excitation in this $d=2$ model is a neutral scalar, corresponding to a six-dimensional photon polarized along the extra dimensions. Annihilations of this ``spinless photon'', as it is known, proceed largely through s-channel Higgs exchange to gauge bosons, Higgs bosons and, to a lesser extent, top quarks. Unlike in the $d=1$ case, annihilations of the LKP to heavy fermions are chirality suppressed. 

The thermal relic abundance of spinless photons can be consistent with with measured cosmological dark matter abundance for masses of approximately 500 GeV or less. The elastic scattering cross sections of this dark matter candidate with nuclei are highly suppressed, leading to very small rates in direct detection experiments and in neutrino telescopes.

Many of the phenomenological characteristics of this dark matter candidate in $d=2$ UED resemble those of a supersymmetric neutralino than those of KKDM in $d=1$ models. In particular, the dominant annihilation modes, indirect (gamma ray and antimatter) detection prospects and the preferred mass range are quite similar to those of neutralinos found in typical supersymmetric scenarios.

\clearpage
\newpage

\section{Conclusions}\label{sec:conclusions}

In the opinion of the authors, the universal extra dimensional model reviewed here remains an extremely fertile field for further theoretical and phenomenological investigation. While the UED model possesses several attractive features, numerous aspects of this setup still need to be explored or clarified. Not expecting to be exhaustive, we list below those issues we consider of particular relevance:
\begin{itemize}
\item As the UED model is one of the most promising new physics scenarios accessible to the Tevatron Run-II, we recommend that the CDF and D0 collaborations undertake a careful data analysis to search for signatures of KK states, extending the analysis of Ref.~\cite{Lin:2005ix} not only to a much larger data set, but also to other search channels, and to UED realizations beyond the minimal case. At the same time, it would be useful if dedicated collider simulations exploring non-minimal UED setups and novel search channels became available.
\item In preparation for the beginning of operation at the LHC, we feel that the issue of discriminating among different TeV-scale physics scenarios should be more extensively investigated, possibly contrasting UED and weak-scale supersymmetry with other beyond the Standard Model scenarios. Along these lines, further studies on the spin reconstruction of new particles to be produced at the LHC would certainly be particularly interesting. Furthermore, it would be highly desirable if collider simulation numerical codes were upgraded to include the UED scenario as an option.
\item The realm of UED models beyond one extra dimension certainly deserves the attention of both theorists and phenomenologists. For instance, the collider implications and the dark matter physics of $d\ge2$ UED models have not been so far thoroughly investigated (with the exceptions of the studies mentioned in Sec.~\ref{ch:sixd}), nor was the resulting phenomenology systematically contrasted to the five dimensional case. 
\item The cosmological and phenomenological issues connected to KK gravitons and to radion (gravi-scalar) fields seriously jeopardize the UED model: for instance, the large portion of the parameter space of the minimal UED scenario where the $n=1$ KK graviton is the LKP, as explained in Sec.~\ref{sec:kkgraviton}, is ruled out. We think that further investigations of the physics of KK gravitons in UED and, more in general, of the cosmology of models with UED should be undertaken.
\item It would be highly desirable to encompass the UED framework in a theoretically motivated high-energy completion, and investigate the candidate theories that might lie beyond the UED effective field theory setup.
\item UED triggered a renewed interest in dark matter candidates within the context of extra-dimensional scenarios. We think that this line of investigation should be further pursued starting, for example, with non-minimal or higher dimensional UED constructions.
\end{itemize}

There is a thought that one invariably has after spending many months writing a review of this nature: {\it How long will it be before this article is outdated, or even obsolete?} Strangely enough, we optimistically think that it will not be long at all. 

With the LHC about to begin operation, and numerous dark matter experiments being planned and currently underway, it seems very likely that our understanding of the TeV-scale will change dramatically in the near future. As this transition takes place, there will be a rare and exciting opportunity to witness a new energy scale of physics. The origin of electroweak symmetry breaking, the solution of the hierarchy problem, and many other long shrouded mysteries will soon be uncloaked. 

Whether beneath that cloak lies universal extra dimensions or other new physics, it will be an exciting time to do physics.

\section*{Acknowledgments}
\addcontentsline{toc}{section}{Acknowledgments}
We would like to thank Bogdan Dobrescu for his many helpful comments on this review. We are also grateful to Alan Barr, Torsten Bringmann, Keith Dienes, Emilian Dudas, Pierre Fayet, Tony Gherghetta, Ken Hsieh, Kyoungchul Kong, Savvas Koushiappas, Anirban Kundu, Andrea Lionetto, Konstantin Matchev, Masato Senami, Geraldine Servant, Dejan Stojkovic and Tim Tait for useful correspondence. We also thank the anonymous Referee for carefully reading the manuscript and for providing us with helpful suggestions and insights that improved in various parts our presentation. DH is supported by the Department of Energy and by NASA grant NAG5-10842. SP is supported in part by DoE grants DE-FG03-92-ER40701 and DE-FG02-05ER41361, and NASA grant NNG05GF69G.

\clearpage
\newpage

\appendix
\section{Benchmark Models}\label{ch:benchmarks}

We devote Appendix \ref{ch:benchmarks} to the definition of the four benchmark models used throughout this review, {\bf UED1}-{\bf 4}. For all models we assume vanishing Higgs boundary mass term ($\bar m_h$ in the notation of Ref. \cite{Cheng:2002iz}). All of the models produce a thermal relic abundance of $B^{(1)}$s which is roughly consistent with the cold dark matter abundance as inferred in Ref. \cite{wmap} for a standard $\Lambda$CDM cosmology.

Model {\bf UED1} is motivated by the results of Ref. \cite{Kong:2005hn}, and assumes a particle spectrum given by the radiative mass corrections computed in Ref. \cite{Cheng:2002iz}, and takes into full account coannihilation processes. Second KK level resonances are here irrelevant \cite{Kakizaki:2006dz}. It features $1/R=550$ GeV, $\Lambda R=20$ and $m_h=120$ GeV. In principle, the LKP would be, for this choice, the KK graviton, but we assume that somehow the KK graviton mass is lifted above the $B^{(1)}$ mass and that it doesn't play any role in affecting the relic abundance of LKPs in the Early Universe \cite{Shah:2006gs}.

Model {\bf UED2} is motivated by the results of Ref. \cite{Burnell:2005hm}, where the effect of taking a smaller value for $\Lambda R$ was evaluated, together with full coannihilation processes. We take here $1/R=850$ GeV, $\Lambda R=4$ and again $m_h=120$ GeV.

Model {\bf UED3} is motivated by the results of Ref. \cite{Kakizaki:2006dz}, in particular on the effect of the second KK level resonances when the Higgs mass is large. It features $1/R=1000$ GeV, $\Lambda R=20$ and $m_h=220$ GeV.

Finally, for model {\bf UED4}, we neglect the detailed structure of the UED spectrum including radiative corrections, and simply assume that {\em all} first level KK states, except for the LKP, lie at a mass splitting $\Delta=0.02$ above the mass of the $B^{(1)}\approx1/R=2250$ GeV (i.e. at a mass of 2295 GeV). The choice of the value of $1/R$ is here motivated by the analogous models studied in Ref. \cite{Kong:2005hn}. In principle, this simplified spectrum might occur if suitable boundary kinetic terms occur at the cutoff scale, $\Lambda$. The main motivation for considering this case study is to understand what occurs when the mass range of the KK particles is in the multi-TeV range and, yet, the LKP relic abundance is compatible with the measured density of dark matter.

We collect in Tab.~\ref{tab:spectra} the details of the mass spectra for the models {\bf UED1}, {\bf UED2} and {\bf UED3} (the spectrum of model {\bf UED4} is trivial, as explained above). All masses are given in GeV.

\begin{table}[!t]
\begin{center}
\begin{tabular}{|r |r| r| r|}
\hline
 & {\bf UED1} & {\bf UED2} &{\bf UED3} \\
\hline
$1/R$ & 550 & 850 & 1000 \\
$\Lambda R$ & 20 & 4 & 20 \\
$m_h$ & 120 & 120 & 220 \\
\hline
$m_{g^{(1)}}$    & 705.0 & 967.9 & 1281.8 \\
$m_{W^{\pm(1)}}$ & 587.6 & 876.8 & 1061.4 \\
$m_{Z^{(1)}}$    & 587.8 & 876.9 & 1061.4 \\
$m_{B^{(1)}}$    & 550.7 & 850.0 & 1000.0 \\
\hline
$m_{H_0^{(1)}}$  & 568.0 & 862.1 & 1022.7 \\
$m_{A_0^{(1)}}$  & 562.6 & 858.5 & 1002.9 \\
$m_{H_\pm^{(1)}}$& 560.9 & 857.5 & 1002.0 \\
\hline
$m_{t_1^{(1)}}$  & 637.7 & 912.4 & 1130.6 \\
$m_{t_2^{(1)}}$  & 665.4 & 932.2 & 1180.9 \\
$m_{b_1^{(1)}}$  & 642.5 & 916.1 & 1168.1 \\
$m_{b_2^{(1)}}$  & 643.2 & 916.7 & 1169.4 \\
$m_{Q^{(1)}}$    & 657.6 & 927.0 & 1195.7 \\
$m_{u^{(1)}}$    & 645.2 & 918.1 & 1173.1 \\
$m_{d^{(1)}}$    & 643.2 & 916.6 & 1169.4 \\
$m_{L^{(1)}}$    & 566.5 & 861.8 & 1029.9 \\
$m_{e^{(1)}}$    & 556.0 & 854.3 & 1010.9 \\
\hline
\end{tabular}
\vspace*{1cm}
\caption{\label{tab:spectra} The input parameters and the mass spectrum for models {\bf UED1}, {\bf UED2} and {\bf UED3}. All quantities with the dimensions of a mass are in units of GeV. Model {\bf UED4} features $m_{B^{(1)}}=2250$ GeV, and all the other KK particle masses are set to 2295 GeV.}
\end{center}
\end{table}

\clearpage
\newpage

\section{The UED Feynman Rules and Cross Sections}

\label{ch:appendix}

We collect in this appendix the relevant Feynman rules (Sec.~\ref{sec:feynrules}), all of the annihilation and coannihilation cross sections relevant for the computation of the relic abundance of the lightest Kaluza-Klein particle in the early universe (Sec.~\ref{sec:ann}), and the amplitudes squared for the production of strongly interacting Kaluza-Klein states at hadron colliders (Sec.~\ref{sec:xsechad})

\subsection{The UED Feynman Rules}\label{sec:feynrules}
We indicate here only the interactions between SM particles and level-one KK modes allowed by KK-parity. Following Ref.~\cite{Burnell:2005hm}, we choose to work in the unitary gauge, where the Goldstone bosons appear as the longitudinal polarizations of the massive KK gauge bosons, rather than as external particles.

The mass eigenstates, $f^{(n)}$, of the $n$-th KK fermion modes are in general linear combinations of the weak eigenstates, $f^{\prime(n)}$. For instance, in the up-type quark sector, one has
\be\label{eq:masseigenstates}
\left(\begin{array}{c} u_3^{\prime(n)}\\ Q_t^{\prime(n)} \end{array} \right)=\left( \begin{array}{c c} -\gamma_5\cos\alpha^{(n)} & \sin\alpha^{(n)}\\ \gamma_5\sin\alpha^{(n)} & \cos\alpha^{(n)} \end{array}\right)\left(\begin{array}{c} u_3^{(n)}\\ Q_t^{(n)} \end{array} \right),
\ee
where, neglecting radiative corrections \cite{Cheng:2002iz},
\be
\tan2\alpha^{(n)}=\frac{m_f}{n/R}\ \qquad {\rm for}\ n\ge1.
\ee

\noindent{\bf Fermion-Fermion-Gauge Boson vertices.}
In what follows we consider the vertices involving the {\em mass} eigenstates. For brevity, however, we will suppress the prime to the fermionic fields introduced in Eq.~(\ref{eq:masseigenstates}). Also, we explicitly show only the case of quarks. The Feynman rules for leptons can be trivially obtained replacing the appropriate charges. We indicate with $s_{u,d}$ and $c_{u,d}$ the sine and cosine of the level 1 fermion mixing angles, defined in the previous paragraph.

\begin{figure}[!h]
\begin{center}
\epsfig{file=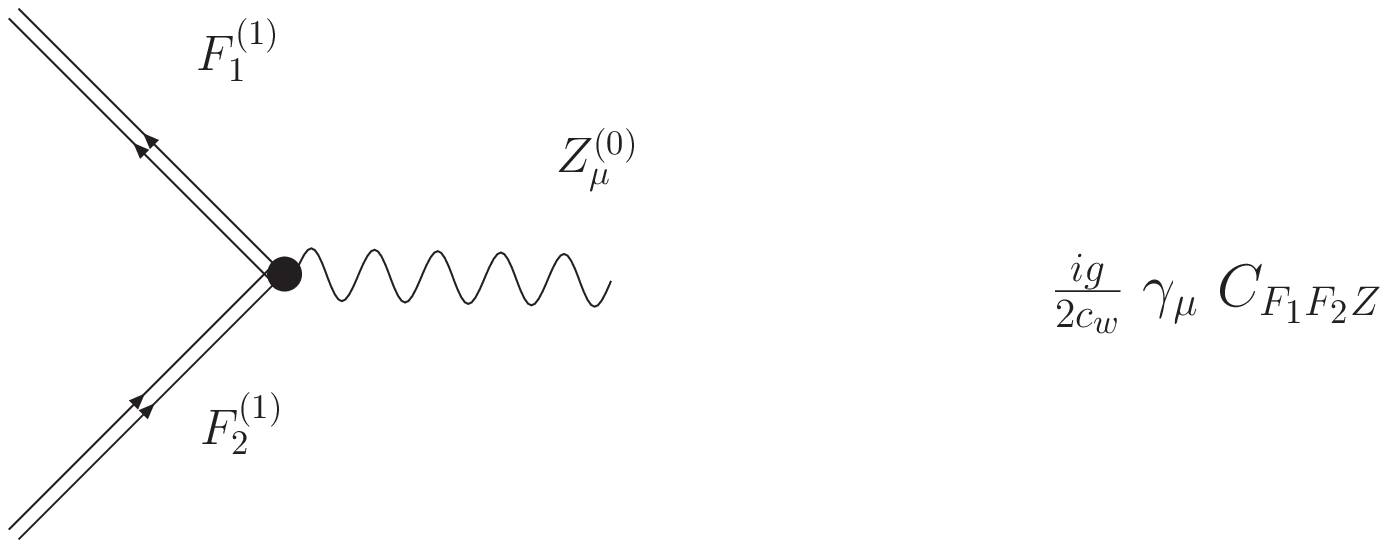,height=3.3cm}
\end{center}
\end{figure}
\begin{table}[!h]
\begin{center}
\begin{tabular}{|c|c|c|c|c|c|c|}
\hline &&&&& \\[-0.25cm]
$\overline{Q^{(1)}_u} Q^{(1)}_u$ &  $\overline{Q^{(1)}_d} Q^{(1)}_d$ & $\overline{Q^{(1)}_u} u^{(1)}$ & $\overline{Q^{(1)}_d} d^{(1)}$ & $\overline{u^{(1)}}u^{(1)}$ & $\overline{d^{(1)}}d^{(1)}$ \\[0.2cm]
\hline &&&&& \\[-0.25cm]
$\displaystyle c_u^2-\frac{4s_w^2}{3}$ & $\displaystyle -c_d^2+\frac{2s_w^2}{3}$& $\displaystyle -s_uc_u\gamma_5$ & $\displaystyle s_dc_d\gamma_5$& $\displaystyle s_u^2-\frac{4s_w^2}{3}$ & $\displaystyle -s_d^2+\frac{2s_w^2}{3}$ \\[0.45cm]
\hline
\end{tabular}\\[0.3cm]
{\em The coefficients $C_{F_1F_2Z}$.}
\end{center}
\end{table}

\begin{figure}[!h]
\begin{center}
\epsfig{file=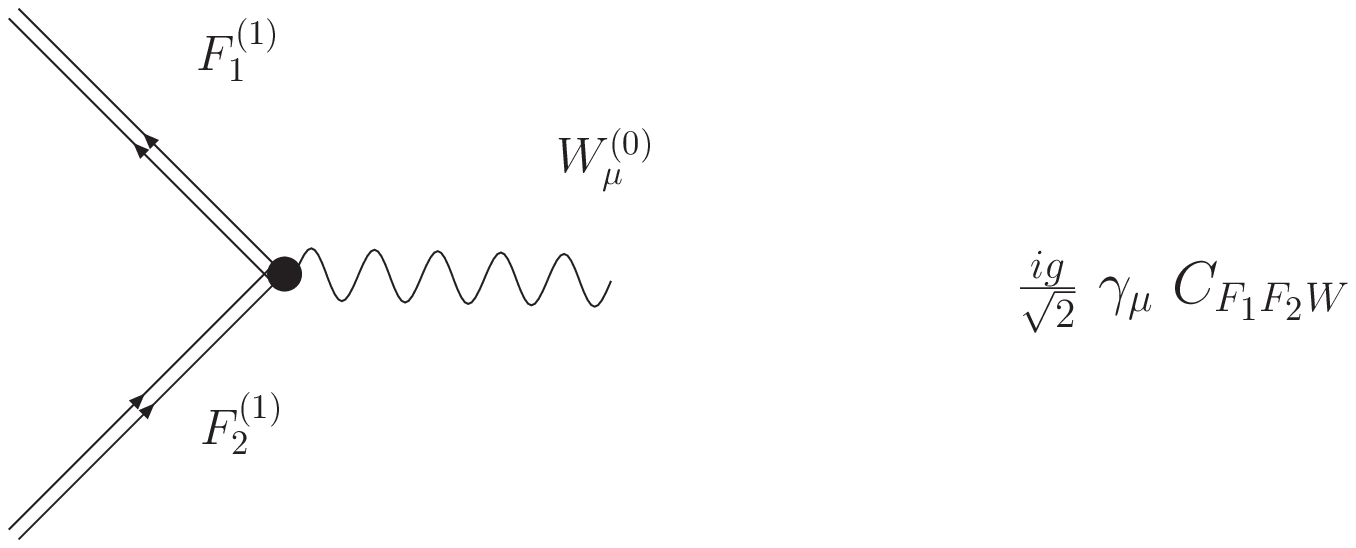,height=3.3cm}
\end{center}
\end{figure}
\begin{table}[!h]
\begin{center}
\begin{tabular}{|c|c|c|c|}
\hline &&& \\[-0.25cm]
$\overline{Q^{(1)}_u} Q^{(1)}_d$ &  $\overline{Q^{(1)}_u} d^{(1)}$ & $\overline{Q^{(1)}_d}u^{(1)}$ & $u^{(1)} u^{(1)}$ \\[0.2cm]
\hline &&& \\[-0.25cm]
 $c_uc_d$ & $c_us_d\gamma_5$& $s_uc_d\gamma_5$ & $s_us_d$ \\[0.2cm]
\hline
\end{tabular}\\[0.3cm]
{\em The coefficients $C_{F_1F_2W}$.}
\end{center}
\end{table}

The other vertices involving two fermions and a gauge boson follow directly from the SM vertices in the weak eigenstate basis. To get the right vertices it is therefore sufficient to perform the rotation from the weak to the mass eigenstate basis. The same applies for all of the vertices involving a gluon or a KK gluon.

\vspace{0.5cm}

\noindent{\bf Fermion-Fermion-Scalar vertices.} We include here the non-trivial vertices involving the mass eigenstates resulting from the superposition of the charged and neutral KK Higgs bosons (or rather, of the zero-level Goldstone bosons which give masses to the $Z$ and the $W$) and of the gauge bosons. We indicate the physical charged and neutral states by $H^{(1)_{\pm}}$ and $A^{(1)}$, respectively. We neglect here the effect of radiative corrections to the masses. The vertex involving the charged KK ``Higgs'' can be represented as in the figures. The vertices involving a scalar KK ``Higgs'', together with the relevant $C^A_{L,R}$ coefficients are also given below. Again, the leptonic vertices follow from those reported above via a simple substitution of the relevant charges. The vertices involving the KK first-level Higgs are the same as those in the SM once the proper rotation of the weak eigenstates into the mass eigenstates is performed.

\begin{figure}[!h]
\begin{center}
\epsfig{file=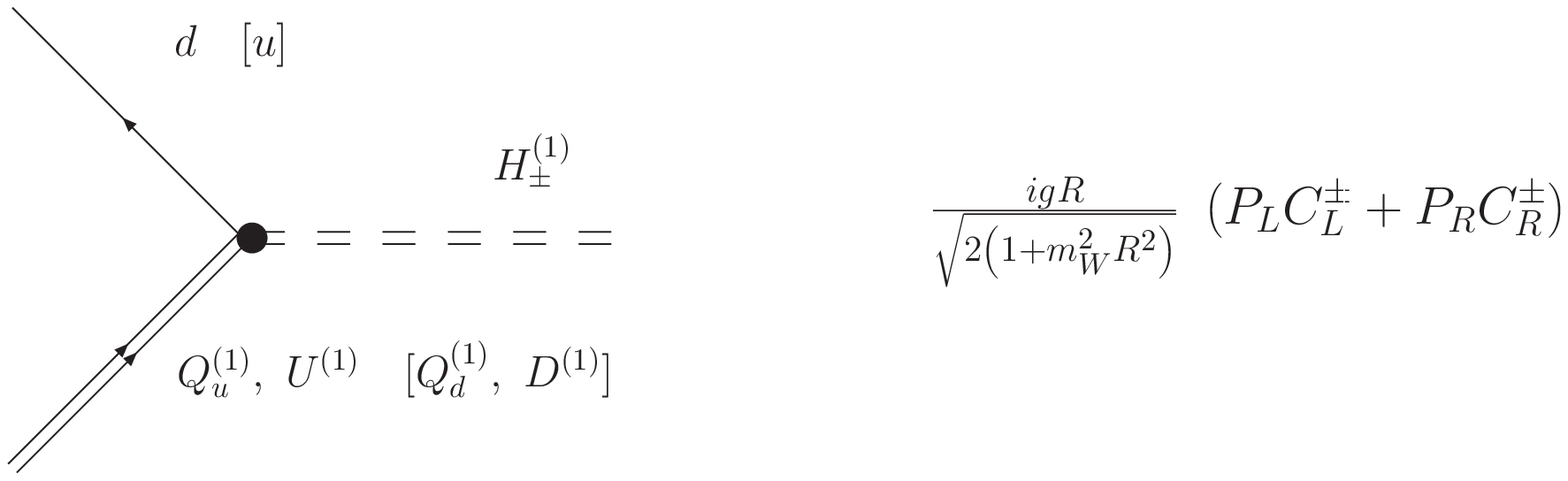,height=3.3cm}
\end{center}
\end{figure}
%
\begin{table}[!h]
\begin{center}
\begin{tabular}{|c|c|c|c|c|}
\hline &&&& \\[-0.25cm]
&$\overline{Q^{(1)}_u} d$ &  $\overline{u^{(1)}} d$ & $\overline{Q^{(1)}_d} u$ &  $\overline{d^{(1)}} u$ \\[0.2cm]
\hline &&&& \\[-0.25cm]
$C^{\pm}_{L}$ &$\displaystyle m_Wc_u-\frac{m_us_u}{Rm_W}$&$\displaystyle m_Ws_u+\frac{m_uc_u}{Rm_W}$&$\displaystyle m_Wc_d-\frac{m_ds_d}{Rm_W}$&$\displaystyle m_Ws_d+\frac{m_dc_d}{Rm_W}$\\[0.2cm]
\hline &&&& \\[-0.25cm]
$C^{\pm}_{R}$ &$\displaystyle \frac{m_uc_u}{Rm_W}$&$\displaystyle -\frac{m_us_u}{Rm_W}$&$\displaystyle \frac{m_dc_d}{Rm_W}$&$\displaystyle -\frac{m_ds_d}{Rm_W}$ \\[0.2cm]\hline
\end{tabular}\\[0.3cm]
{\em The coefficients $C^{\pm}_{L,R}$.}
\end{center}
\end{table}

\begin{figure}[!h]
\begin{center}
\epsfig{file=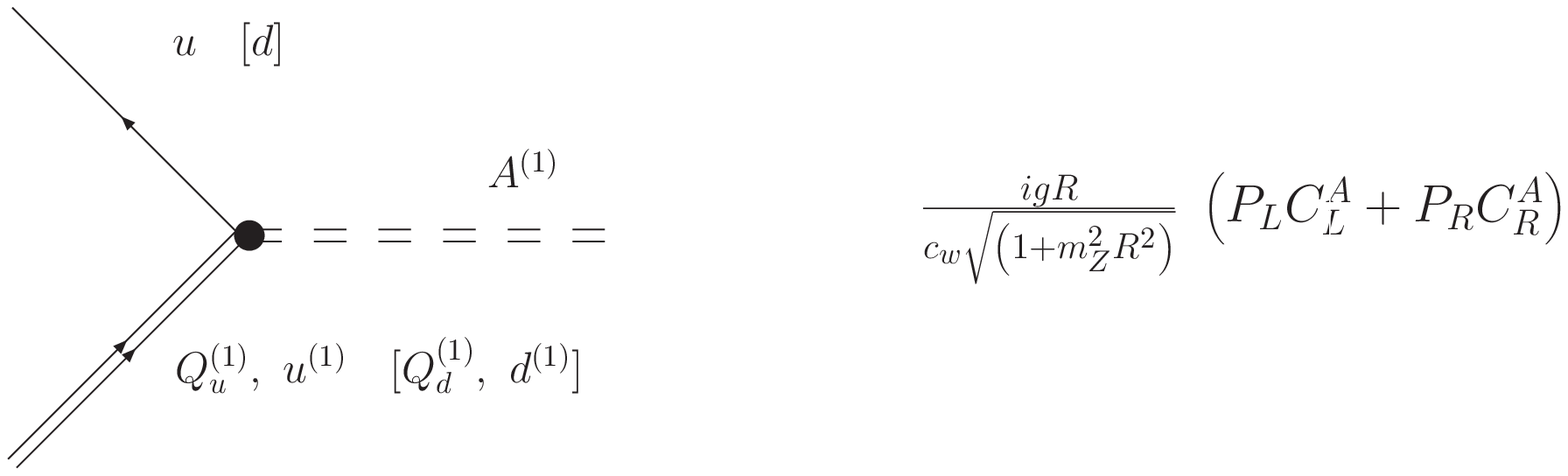,height=3.3cm}
\end{center}
\end{figure}
\begin{table}[!h]
\begin{center}
\begin{tabular}{|c|c|c|}
\hline && \\[-0.25cm]
&$\overline{Q^{(1)}_u} u$ &  $\overline{u^{(1)}} u$ \\[0.2cm]
\hline && \\[-0.25cm]
$C^{A}_{L}$ &$\displaystyle m_Zc_u\left(\frac{1}{2}-\frac{2s_w^2}{3}\right)-\frac{m_us_u}{Rm_W}$&$\displaystyle m_Zs_u\left(\frac{1}{2}-\frac{2s_w^2}{3}\right)+\frac{m_uc_u}{Rm_W}$\\[0.2cm]
\hline && \\[-0.25cm]
$C^{A}_{R}$ &$\displaystyle -m_Zs_u\frac{2s_w^2}{3}+\frac{m_uc_u}{Rm_W}$ &$\displaystyle -m_Zc_u\frac{2s_w^2}{3}-\frac{m_us_u}{Rm_W}$\\[0.2cm]

\hline\hline && \\[-0.25cm]
& $\overline{Q^{(1)}_d} d$ &  $\overline{d^{(1)}} d$ \\[0.2cm]
\hline && \\[-0.25cm]
$C^{A}_{L}$ &$\displaystyle m_Zc_d\left(-\frac{1}{2}+\frac{s_w^2}{3}\right)-\frac{m_ds_d}{Rm_W}$&$\displaystyle m_Zs_d\left(-\frac{1}{2}+\frac{s_w^2}{3}\right)+\frac{m_dc_d}{Rm_W}$\\[0.2cm]
\hline && \\[-0.25cm]
$C^{A}_{R}$ &$\displaystyle m_Zs_d\frac{s_w^2}{3}+\frac{m_dc_d}{Rm_W}$&$\displaystyle m_Zc_d\frac{s_w^2}{3}-\frac{m_ds_d}{Rm_W}$ \\[0.2cm]\hline
\end{tabular}\\[0.3cm]
{\em The coefficients $C^A_{L,R}$.}
\end{center}
\end{table}

\clearpage
\newpage

\noindent{\bf Gauge Boson-Gauge Boson-Scalar vertices.} The table below shows the vertices involving two gauge bosons and one scalar. We use the sign convention of Ref.~\cite{Burnell:2005hm} and adopt the factor, $(-1)^\beta$, which accounts for the $\epsilon^{abc}$ from the commutators of the $SU(2)_W$ generators. $\beta$ is even for vertices where the electric charge entering the vertex increases in the counterclockwise direction, and is odd otherwise. We do not include explicitly $H^{(1)}A^{(0,1)}_\mu A^{(1,0)}_\nu$, where $A=W^\pm,B,W^3$, as they follow exactly from those in the Standard Model. Also, for notational ease (to avoid double superscripts), though with a slight abuse of notation, in this appendix we indicate the neutral component of the $SU(2)_W$ gauge bosons, $W^3$, with the symbol $Z$.
\begin{table}[!h]
\begin{center}
\begin{tabular}{|l c|l|}
\hline && \\[-0.25cm]
$A^{(1)} W^{+(1)}_\mu W^{-(0)}_\nu$ &\qquad& $\displaystyle (-1)^\beta\frac{-igc_wm_Z\ g_{\mu\nu}}{\sqrt{1+m_Z^2R^2}}$\\[0.45cm]
\hline && \\[-0.25cm]
$A^{(1)} W^{-(1)}_\mu W^{+(0)}_\nu$ & & $\displaystyle (-1)^\beta\frac{igc_wm_Z\ g_{\mu\nu}}{\sqrt{1+m_Z^2R^2}}$\\[0.45cm]
\hline && \\[-0.25cm]
$H^{(1)}_- W^{+(1)}_\mu Z^{(0)}_\nu$ & & $\displaystyle \frac{igm_Z(s_w^2+(-1)^\beta c_w^2)\ g_{\mu\nu}}{\sqrt{1+m_W^2R^2}}$\\[0.45cm]
\hline && \\[-0.25cm]
$H^{(1)}_+ W^{-(1)}_\mu Z^{(0)}_\nu$ & & $\displaystyle -\frac{igm_Z(s_w^2+(-1)^\beta c_w^2)\ g_{\mu\nu}}{\sqrt{1+m_W^2R^2}}$\\[0.45cm]
\hline && \\[-0.25cm]
$H^{(1)}_+ Z^{(1)}_\mu W^{-(0)}_\nu$ & & $\displaystyle -\frac{igm_Z(s_w^2-(-1)^\beta c_w^2)\ g_{\mu\nu}}{\sqrt{1+m_W^2R^2}}$\\[0.45cm]
\hline && \\[-0.25cm]
$H^{(1)}_- Z^{(1)}_\mu W^{+(0)}_\nu$ & & $\displaystyle \frac{igm_Z(s_w^2-(-1)^\beta c_w^2)\ g_{\mu\nu}}{\sqrt{1+m_W^2R^2}}$\\[0.45cm]
\hline && \\[-0.25cm]
$H^{(1)}_- W^{+(1)}_\mu B^{(0)}_\nu$ & & $\displaystyle \frac{igm_Zs_wc_w((-1)^\beta-1)\ g_{\mu\nu}}{\sqrt{1+m_W^2R^2}}$\\[0.45cm]
\hline && \\[-0.25cm]
$H^{(1)}_+ W^{-(1)}_\mu B^{(0)}_\nu$ & & $\displaystyle -\frac{igm_Zs_wc_w((-1)^\beta-1)\ g_{\mu\nu}}{\sqrt{1+m_W^2R^2}}$\\[0.45cm]
\hline && \\[-0.25cm]
$H^{(1)}_+ B^{(1)}_\mu W^{-(0)}_\nu$ & & $\displaystyle \frac{igm_Zs_wc_w((-1)^\beta+1)\ g_{\mu\nu}}{\sqrt{1+m_W^2R^2}}$\\[0.45cm]
\hline && \\[-0.25cm]
$H^{(1)}_- B^{(1)}_\mu W^{+(0)}_\nu$ & & $\displaystyle -\frac{igm_Zs_wc_w((-1)^\beta+1)\ g_{\mu\nu}}{\sqrt{1+m_W^2R^2}}$\\[0.45cm]
\hline
\end{tabular}
\end{center}
\end{table}
\clearpage

\noindent{\bf Gauge Boson-Scalar-Scalar vertices.} The table below shows the vertices involving two first KK level mode scalars and a gauge boson.
\begin{table}[!h]
\begin{center}
\begin{tabular}{|l c|l|}
\hline && \\[-0.25cm]
$H^{(1)}_+H^{(1)}_- B^{(0)}_\mu$ &\qquad& $\displaystyle g s_w\left((p_{H^{(1)}_+})_\mu-(p_{H^{(1)}_-})_\mu\right)$\\[0.45cm]
\hline && \\[-0.25cm]
$H^{(1)}_+H^{(1)}_- Z^{(0)}_\mu$ & & $\displaystyle \frac{g}{1+m_W^2R^2}\left(c_w^2m_W^2R^2-\frac{c_w^2-s_w^2}{2c_w}\left((p_{H^{(1)}_+})_\mu-(p_{H^{(1)}_-})_\mu\right)\right)$\\[0.45cm]
\hline && \\[-0.25cm]
$A^{(1)}H^{(1)}_-W^{+(0)}_\mu$ & & $\displaystyle \frac{g (m_W^2R^2+1/2)}{\sqrt{(1+m_W^2R^2)(1+m_Z^2R^2)}}\left((p_{H^{(1)}_-})_\mu-(p_{A^{(1)}})_\mu\right)$\\[0.45cm]
\hline && \\[-0.25cm]
$A^{(1)}H^{(1)}_+W^{-(0)}_\mu$ & & $\displaystyle -\frac{g (m_W^2R^2+1/2)}{\sqrt{(1+m_W^2R^2)(1+m_Z^2R^2)}}\left((p_{H^{(1)}_+})_\mu-(p_{A^{(1)}})_\mu\right)$\\[0.45cm]
\hline && \\[-0.25cm]
$A^{(1)}H^{(1)} Z^{(0)}_\mu$ & & $\displaystyle \frac{g m_Z}{c_w\sqrt{1+m_Z^2R^2}}\left((p_{H^{(1)}})_\mu-(p_{A^{(1)}})_\mu\right)$\\[0.45cm]
\hline && \\[-0.25cm]
$H^{(1}_\pm H^{(1)} W^{\mp(0)}_\mu$ & & $\displaystyle \frac{g m_W}{\sqrt{1+m_W^2R^2}}\left((p_{H^{(1)}})_\mu-(p_{H^{(1}_\pm})_\mu\right)$\\[0.45cm]
\hline
\end{tabular}
\end{center}
\end{table}

\noindent{\bf Gauge Boson-Gauge Boson-Scalar-Scalar vertices.} We show in the following table the vertices involving a pair of zero-mode gauge bosons and a pair of first KK level mode scalars.
\begin{table}[!h]
\begin{center}
\begin{tabular}{|l c|l|}
\hline && \\[-0.25cm]
$H^{(1)}_+H^{(1)}_- W^{+(0)}_\mu W^{-(0)}_\nu$ &\qquad& $\displaystyle \frac{g^2\ g_{\mu\nu}}{2(1+m_W^2R^2)}\left(1+2m_W^2R^2\right)$\\[0.45cm]
\hline && \\[-0.25cm]
$A^{(1)}A^{(1)}W^{+(0)}_\mu W^{-(0)}_\nu$ & & $\displaystyle \frac{g^2\ g_{\mu\nu}}{2(1+m_Z^2R^2)}\left(1+4c_w^2m_Z^2R^2\right)$\\[0.45cm]
\hline && \\[-0.25cm]
$H^{(1)}_+H^{(1)}_-B^{(0)}_\mu B^{(0)}_\nu$ & & $\displaystyle 2e^2\ g_{\mu\nu}$\\[0.25cm]
\hline && \\[-0.25cm]
$H^{(1)}_+H^{(1)}_-Z^{(0)}_\mu Z^{(0)}_\nu$ & & $\displaystyle \frac{g^2\ g_{\mu\nu}}{1+m_W^2R^2} \left(\frac{(c_w^2-s_w^2)^2}{2c_w^2}+2c_w^4m_Z^2R^2\right)$\\[0.45cm]
\hline && \\[-0.25cm]
$H^{(1)}_+H^{(1)}_-Z^{(0)}_\mu B^{(0)}_\nu$ & & $\displaystyle \frac{g e\ g_{\mu\nu}}{1+m_W^2R^2} \left(\frac{c_w^2-s_w^2}{c_w}+2c_w^3m_Z^2R^2\right)$\\[0.45cm]
\hline && \\[-0.25cm]
$A^{(1)}H^{(1)}_\pm Z^{(0)}_\mu W^{\mp(0)}_\nu$ & & $\displaystyle \frac{g^2\ g_{\mu\nu}}{2\sqrt{(1+m_W^2R^2)(1+m_Z^2R^2)}} \left(\frac{s_w^2}{c_w}-2c_w^3m_Z^2R^2\right)$\\[0.65cm]
\hline && \\[-0.25cm]
$A^{(1)}H^{(1)}_\pm B^{(0)}_\mu W^{\mp(0)}_\nu$ & & $\displaystyle -\frac{g e\ g_{\mu\nu}}{2\sqrt{(1+m_W^2R^2)(1+m_Z^2R^2)}} \left(1+c_w^2m_Z^2R^2\right)$\\[0.65cm]
\hline
\end{tabular}
\end{center}
\end{table}
\clearpage
\noindent{\bf Three-Scalar vertices.} The following table collects the vertices featuring three scalars, of which two are KK first level modes and one is a zero-mode.
\begin{table}[!h]
\begin{center}
\begin{tabular}{|l c|l|}
\hline && \\[-0.25cm]
$A^{(1)}A^{(1)}H$ &\qquad& $\displaystyle -\frac{g}{c_wm_Z(1+m_Z^2R^2)}\left(m_Z^2(1+m_Z^2R^2)+m_h^2/2\right)$\\[0.45cm]
\hline && \\[-0.25cm]
$H^{(1)}_+H^{(1)}_-H$ & & $\displaystyle -\frac{g}{c_wm_Z(1+m_W^2R^2)}\left(m_W^2(1+m_W^2R^2)+m_h^2/2\right)$\\[0.45cm]
\hline
\end{tabular}
\end{center}
\end{table}

\noindent{\bf Four-Scalar vertices.} We collect in the following table the vertices involving four scalars, of which two are first KK level modes and two are zero-modes.

\begin{table}[!h]
\begin{center}
\begin{tabular}{|l c|l|}\hline
&& \\[-0.25cm]
$H^{(1)}_+H^{(1)}_-HH$ &\qquad& $\displaystyle-\frac{g^2}{4(1+m_W^2R^2)}\left(\frac{m_h^2}{m_W^2}+2m_W^2R^2\right)$\\[0.45cm]
\hline
&& \\[-0.25cm]
$A^{(1)}A^{(1)}HH$ & & $\displaystyle-\frac{g^2}{4c_w^2(1+m_Z^2R^2)}\left(\frac{m_h^2}{m_Z^2}+2m_Z^2R^2\right)$\\[0.45cm]
\hline
\end{tabular}
\end{center}
\end{table}

\subsection{Annihilation and Coannihilation Cross Sections}\label{sec:ann}

We review here the annihilation and coannihilation cross sections relevant for the computation of the LKP relic abundance \cite{Servant:2002aq,Kakizaki:2005en,Kakizaki:2005uy,Burnell:2005hm,Kong:2005hn,Kakizaki:2006dz}. We collect the various (co-)annihilation processes below by the type of (co-)annihilating particles (gauge bosons, fermions and Higgs bosons). We neglect in what follows the mass splitting between particles belonging to the same KK level, and call their common mass $m\equiv1/R$, as well as EWSB. $\phi$ indicates generically the SM charged and neutral Higgs bosons, and $G$ indicates the zero mode Goldstone bosons. Following Ref.~\cite{Kong:2005hn}, we indicate $s$ as the center-of-mass energy squared, with
\be
\beta=\sqrt{1-\frac{4m^2}{s}} 
\ee
and with
\be
L=\log\left(\frac{1-\beta}{1+\beta}\right)=-2{\rm tanh}^{-1}\beta.
\ee
The other symbols we employ in the cross sections will be defined later. We also list the relevant resonant annihilation cross sections as computed in Refs.~\cite{Kakizaki:2005en,Kakizaki:2005uy,Kakizaki:2006dz}.\\

\small
\noindent {\bf\normalsize I. Gauge Boson-Gauge Boson}\\
\be
\sigma\left(B^{(1)}B^{(1)}\rightarrow f\bar f\right)=\frac{N_c(gt_w)^4\left(Y_{f_L}^4+Y_{f_R}^4\right)}{72\pi s^2\beta^2}\left(-5s(2m^2+s)L-7s\beta\right)&&
\ee
\be
&&\sigma\left(B^{(1)}B^{(1)}\rightarrow \phi\phi^*\right)=\frac{(gt_wY_\phi)^4}{12\pi s\beta}
\ee
\be
&&\sigma\left(B^{(1)}B^{(1)}\rightarrow H^{(2)}\rightarrow t\bar t\right)=\frac{g^2t_w^4m_w^2}{36\beta m}\frac{\Gamma^{H^{(2)}}_{t\bar t}}{(s-m^2_{H^{(2)}})^2+4-m^2_{H^{(2)}}\Gamma^2_{H^{(2)}}}\left(3+\frac{s(s-4m^2)}{4m^4}\right),
\ee
where
\be
&&\Gamma_{H^{(2)}}=\Gamma^{H^{(2)}}_{t\bar t}+\Gamma^{H^{(2)}}_{HH}+\Gamma^{H^{(2)}}_{AA},
\ee
\be
\Gamma^{H^{(2)}}_{t\bar t}=\frac{y_t \alpha_s^2 m}{12\pi^3}\log^2(\Lambda^2R^2),
\ee
\be
\Gamma^{H^{(2)}}_{HH}=\frac{\lambda^2m_W^2}{32\pi g^2 m^2}\sqrt{m_h^2-4 m_W^2}\quad {\rm if}\ m_h>2m_W,\ {\rm otherwise\ }\Gamma^{H^{(2)}}_{HH}=0,
\ee
\be
\Gamma^{H^{(2)}}_{AA}=\frac{\lambda^2m_W^2}{64\pi g^2 m^2}\sqrt{m_h^2-4 m_Z^2}\quad {\rm if}\ m_h>2m_Z,\ {\rm otherwise\ }\Gamma^{H^{(2)}}_{AA}=0.
\ee
\be
\sigma\left(Z^{(1)}Z^{(1)}\rightarrow f\bar f\right)=\frac{N_cg^4}{1152\pi s^2\beta^2}\left(-5s(2m^2+s)L-7s\beta\right)&&
\ee
\be
&&\sigma\left(Z^{(1)}Z^{(1)}\rightarrow \phi\phi^*\right)=\frac{g^4}{192\pi s\beta}
\ee
\be
&&\sigma\left(Z^{(1)}Z^{(1)}\rightarrow W^+W^-\right)=\frac{g^4}{18\pi m^2s^3\beta^2}\left(12m^2(s-2m^2)L+s\beta(12m^4+3sm^2+4s^2)\right)
\ee
\be
&&\sigma\left(W^{+(1)}W^{-(1)}\rightarrow W^+W^-\right)=\frac{g^4}{36\pi m^2s^3\beta^2}\left(12m^2(s-2m^2)L+s\beta(12m^4+3sm^2+4s^2)\right)
\ee
\be
&&\sigma\left(W^{+(1)}W^{-(1)}\rightarrow \gamma\gamma\right)=\frac{g^4s_w^4}{36\pi m^2s^3\beta^2}\left(12m^2(s-2m^2)L+s\beta(12m^4+3sm^2+4s^2)\right)
\ee
\be
&&\sigma\left(W^{+(1)}W^{-(1)}\rightarrow \gamma Z\right)=\frac{g^4s_w^2c_w^2}{18\pi m^2s^3\beta^2}\left(12m^2(s-2m^2)L+s\beta(12m^4+3sm^2+4s^2)\right)
\ee
\be
&&\sigma\left(W^{+(1)}W^{-(1)}\rightarrow Z Z\right)=\frac{g^4c_w^4}{36\pi m^2s^3\beta^2}\left(12m^2(s-2m^2)L+s\beta(12m^4+3sm^2+4s^2)\right)
\ee
\be
&&\sigma\left(W^{+(1)}W^{-(1)}\rightarrow f\bar f\right)=\frac{g^2}{576 \pi s^2\beta^2}\left((12m^2+5s)L+2\beta(4m^2+5s)\right)
\ee
\be
&&\sigma\left(W^{+(1)}W^{-(1)}\rightarrow W^+W^-\right)=\frac{g^4}{18\pi m^2s^2\beta^2}\left(2m^2(3m^2+2s)L+\beta(11m^4+5sm^2+2s^2)\right)
\ee
\be
&&\sigma\left(W^{+(1)}W^{-(1)}\rightarrow \phi\phi^*\right)=\frac{g^4(s-m^2)}{144\pi s^2\beta}
\ee
\be
&&\sigma\left(B^{(1)}Z^{(1)}\rightarrow f\bar f\right)=\frac{-N_cg^4t_w^2T_f^2}{288 \pi s^2\beta^2}\left(5(2m^2+s)L+7s\beta\right)
\ee
\be
&&\sigma\left(B^{(1)}W^{-(1)}\rightarrow f\bar f^\prime\right)=\frac{-N_cg^4t_w^2Y_f^2}{144 \pi s^2\beta^2}\left(5(2m^2+s)L+7s\beta\right)
\ee
\be
&&\sigma\left(B^{(1)}Z^{(1)}\rightarrow \phi\phi^*\right)=\frac{g^4t_w^2}{192\pi s\beta}
\ee
\be
&&\sigma\left(B^{(1)}W^{-(1)}\rightarrow \phi_d\phi_u^*\right)=\frac{g^4t_w^2}{96\pi s\beta}
\ee
\be
&&\sigma\left(Z^{(1)}W^{-(1)}\rightarrow \phi_d\phi_u^*\right)=\frac{g^4\beta}{288\pi s}
\ee
\be
&&\sigma\left(Z^{(1)}W^{-(1)}\rightarrow f\bar f^\prime\right)=\frac{-N_cg^4}{576 \pi s^2\beta^2}\left((14m^2+5s)L+\beta(16m^2+13s)\right)
\ee
\be
&&\sigma\left(Z^{(1)}W^{-(1)}\rightarrow ZW^-\right)=\frac{g^4c_w^2}{18 \pi m^2s^2\beta^2}\left(2m^2(3m^2+2s)L+\beta(11m^4+5sm^2+2s^2)\right)
\ee
\be
&&\sigma\left(Z^{(1)}W^{-(1)}\rightarrow \gamma W^-\right)=\frac{g^4t_w^2}{18 \pi m^2s^2\beta^2}\left(2m^2(3m^2+2s)L+\beta(11m^4+5sm^2+2s^2)\right)
\ee
\be
&&\sigma\left(g^{(1)}g^{(1)}\rightarrow gg\right)=\frac{g_3^4}{64 \pi m^2s^3\beta^2}\left(8m^2(s^2+3sm^2-3m^4)L+s\beta(34m^4+13sm^2+8s^2)\right)
\ee
\be
&&\sigma\left(g^{(1)}g^{(1)}\rightarrow q\bar q\right)=\frac{-g_3^4}{3456 \pi s^2\beta^2}\left(2(20s+49m^2)L+\beta(72m^2+83s)\right)
\ee
\be
&&\sigma\left(g^{(1)}B^{(1)}\rightarrow q\bar q\right)=\frac{g_3^2g^2t_w^2}{144 \pi s^2\beta^2}\left(-5(2m^2+s)L-7s\beta\right)
\ee
\be
&&\sigma\left(g^{(1)}Z^{(1)}\rightarrow q\bar q\right)=\frac{g_3^2g^2}{576 \pi s^2\beta^2}\left(-5(2m^2+s)L-7s\beta\right)
\ee\\ \vspace*{0.5cm}

\noindent{\bf\normalsize II. Gauge Boson and Fermion}\\
\be
&&\sigma\left(Z^{(1)}f_R^{(1)}\rightarrow {\rm any}\ {\rm SM}\ {\rm final}\ {\rm state}\right)=\sigma\left(W^{\pm(1)}f_R^{(1)}\rightarrow {\rm any\ SM\ final\ state}\right)=0
\ee
\be
&&\sigma\left(B^{(1)}l^{\pm(1)}\rightarrow \gamma l^\pm\right)=\frac{g^4t_w^4(-6L-\beta)}{96 \pi s\beta^2}
\ee
\be
&&\sigma\left(B^{(1)}q^{(1)}\rightarrow g q^\pm\right)=\frac{g^2t_w^2g_3^2(-6L-\beta)}{72 \pi s\beta^2}
\ee
\be
&&\sigma\left(Z^{(1)}q_L^{(1)}\rightarrow g q\right)=\frac{g^2g_3^2(-6L-\beta)}{288 \pi s\beta^2}
\ee
\be
&&\sigma\left(Z^{(1)}q_L^{(1)}\rightarrow g q^\prime\right)=\frac{g^2g_3^2(-6L-\beta)}{144 \pi s\beta^2}
\ee
\be
&&\sigma\left(B^{(1)}l_L^{(1)}\rightarrow \gamma l\right)=\sigma\left(B^{(1)}l_L^{(1)}\rightarrow Z l\right)=\frac{g^4t_w^4(-6L-\beta)}{1536 \pi s\beta^2 s_w^2}
\ee
\be
&&\sigma\left(Z^{(1)}l_L^{(1)}\rightarrow \gamma l\right)=\sigma\left(Z^{(1)}l_L^{(1)}\rightarrow Z l\right)=\frac{g^4(-6L-\beta)}{1536 \pi s\beta^2 c_w^2}
\ee
\be
&&\sigma\left(B^{(1)}l_L^{(1)}\rightarrow W\nu_l\right)=\frac{g^4t_w^2(-6L-\beta)}{768 \pi s\beta^2}
\ee
\be
&&\sigma\left(Z^{(1)}l_L^{(1)}\rightarrow W\nu_l\right)=\frac{g^4}{768 \pi m^2 s\beta^2}\left(26m^2L+\beta(23m^2+32s)\right)
\ee
\be
&&\sigma\left(W^{(1)}l_L^{(1)}\rightarrow (\gamma+Z)\nu_{l^\prime}\right)=\frac{g^4}{768 \pi m^2 s\beta^2}\left(m^2(32c_w^2-6)L+\beta m(24c_w^2-1)+32s\beta c_w^2\right)
\ee
\be
&&\sigma\left(W^{-(1)}l_L^{(1)}\rightarrow W^- l_L\right)=\sigma\left(W^{+(1)}\nu_l^{(1)}\rightarrow W^+ \nu_l\right)=\frac{g^4}{192 \pi m^2 s\beta^2}\left(-3m^2 L+4s\beta\right)
\ee
\be\nonumber
&&\sigma\left(W^{+(1)}l_L^{(1)}\rightarrow W^+ l_L\right)=\sigma\left(W^{-(1)}\nu_l^{(1)}\rightarrow W^- \nu_l\right)=\frac{g^4}{384 \pi m^2 s\beta^2}\left(16m^2 L+\beta(11m^2+8s)\right)\\
&&
\ee
\be
&&\sigma\left(g^{(1)}q^{(1)}\rightarrow gq\right)=\frac{g_3^4}{846 \pi m^2 s\beta^2}\left(24m^2 L+\beta(25m^2+36s)\right)
\ee\\ \vspace*{0.5cm}

\noindent{\bf\normalsize III. Gauge Boson and Higgs}\\
\be\nonumber
&&\sigma\left(g^{(1)}H^{(1)}\rightarrow t\bar t\right)=\sigma\left(g^{(1)}A^{(1)}\rightarrow t\bar t\right)=\sigma\left(g^{(1)}H^{+(1)}\rightarrow t\bar b\right)=\frac{g_3^2y_t^2}{48 \pi m^2 s^2 \beta^2}\left(2m^2(s-m^2)L+s\beta(2s-5m^2)\right)\\
&&
\ee
\be\nonumber
&&\sigma\left(Z^{(1)}H^{(1)}\rightarrow ZH\right)=\sigma\left(Z^{(1)}A^{(1)}\rightarrow ZH\right)=\frac{g^4(L+4\beta)}{96 \pi s \beta^2 c_w^2}\left(2m^2(s-m^2)L+s\beta(2s-5m^2)\right)\\
&&
\ee
\be\nonumber
&&\sigma\left(Z^{(1)}H^{(1)}\rightarrow W^-G^+\right)=\sigma\left(Z^{(1)}A^{(1)}\rightarrow W^-G^+\right)=\sigma\left(Z^{(1)}H^{(1)}_+\rightarrow W^+G\right)=\sigma\left(Z^{(1)}H^{(1)}_+\rightarrow W^+H\right)=\\
&&
=\sigma\left(W^{+(1)}H^{(1)}\rightarrow W^+G\right)=\sigma\left(W^{+(1)}H^{(1)}\rightarrow W^+H\right)=\frac{g^4(4m^2L+\beta(4s+m^2))}{96 \pi m^2 s \beta^2}
\ee
\be
&&\sigma\left(Z^{(1)}H^{(1)}\rightarrow t\bar t\right)=\sigma\left(Z^{(1)}A^{(1)}\rightarrow t\bar t\right)=\frac{g^2y_t^2}{64 \pi m^2 s \beta^2}\left(2m^2L+\beta(4s-11m^2)\right)
\ee
\be
&&\sigma\left(Z^{(1)}H_+^{(1)}\rightarrow ZG^+\right)=\frac{g^4(1-2s_w^2)^2}{96 \pi s \beta^2 c_w^2}\left(L+4\beta\right)
\ee
\be
&&\sigma\left(Z^{(1)}H_+^{(1)}\rightarrow \gamma G^+\right)=\frac{g^4s_w^2}{24 \pi s \beta^2}\left(L+4\beta\right)
\ee
\be\nonumber
&&\sigma\left(Z^{(1)}H_+^{(1)}\rightarrow t\bar b\right)=\sigma\left(W^{+(1)}H^{(1)}\rightarrow t\bar b\right)=\sigma\left(W^{+(1)}A^{(1)}\rightarrow t\bar b\right)=\frac{g^2y_t^2}{64 \pi m^2 s \beta^2}\left(2m^2L+\beta(4s-11m^2)\right)\\
&&
\ee
\be
&&\sigma\left(B^{(1)}H^{(1)}\rightarrow ZH\right)=\sigma\left(B^{(1)}A^{(1)}\rightarrow ZG\right)=\frac{g^4t_w^2}{96 \pi s \beta^2 c_w^2}\left(L+4\beta\right)
\ee
\be
&&\sigma\left(B^{(1)}A^{(1)}\rightarrow W^-G^+\right)=\sigma\left(B^{(1)}H^{(1)}\rightarrow W^-G^+\right)=\sigma\left(B^{(1)}H^{(1)}_+\rightarrow W^+G\right)=\\
&&
=\sigma\left(B^{(1)}H^{(1)}_+\rightarrow W^+H\right)=\frac{g^4t_w^2}{96 \pi s \beta^2}\left(L+4\beta\right)
\ee
\be
&&\sigma\left(B^{(1)}H^{(1)}\rightarrow t\bar t\right)=\sigma\left(B^{(1)}A^{(1)}\rightarrow t\bar t\right)=\sigma\left(B^{(1)}H^{(1)}_+\rightarrow t\bar b\right)=\\
&&
=\frac{g^2t_w^2y_t^2}{576 \pi m^2 s^2 \beta^2c_w^2}\left(-2m^2(7s+8m^2)L+s\beta(4s-43m^2)\right)
\ee
\be
&&\sigma\left(B^{(1)}H^{(1)}_+\rightarrow ZG^+\right)=\sigma\left(B^{(1)}A^{(1)}_+\rightarrow ZG^+\right)=\frac{g^4t_w^2(1-2s_w^2)^2}{96 \pi s \beta^2 c_w^2}\left(L+4\beta\right)
\ee
\be
&&\sigma\left(B^{(1)}H^{(1)}_+\rightarrow \gamma G^+\right)=\frac{g^4s_w^2}{24 \pi s \beta^2}\left(L+4\beta\right)
\ee
\be
&&\sigma\left(W^{+(1)}H^{(1)}_+\rightarrow W^+G^+\right)=\frac{g^4}{96 \pi m^2 s \beta^2}\left(12m^2L+\beta(6m^2+5s)\right)
\ee
\be\nonumber
&&\sigma\left(W^{+(1)}H^{(1)}\rightarrow ZG^+\right)=\frac{g^4}{96 \pi m^2 s \beta^2 c_w^2}\left(m^2(2-s_w^2-2s_w^4)L-\beta[m^2(4s_w^4-s_w^2+1)+s(3s_w^4-7s_w^2+4)]\right)\\
&&
\ee
\be
&&\sigma\left(W^{+(1)}H^{(1)}\rightarrow \gamma G^+\right)=\sigma\left(W^{+(1)}A^{(1)}\rightarrow \gamma G^+\right)=\frac{g^4t_w^2s_w^2}{96 \pi m^2 s \beta^2}\left(-2m^2L+\beta(4m^2+3s)\right)
\ee
\be
&&\sigma\left(W^{+(1)}H^{(1)}\rightarrow W^+H\right)=\sigma\left(W^{+(1)}A^{(1)}\rightarrow W^+G\right)\frac{g^4}{96 \pi s \beta^2}\left(L+4\beta\right)
\ee
\be
&&\sigma\left(W^{+(1)}H^{(1)}_-\rightarrow W^+G^-\right)=\frac{g^4}{96 \pi s \beta^2}\left(-2m^2L+\beta(4m^2+3s)\right)
\ee
\be
&&\sigma\left(W^{+(1)}H^{(1)}_-\rightarrow ZG\right)=\sigma\left(W^{+(1)}A^{(1)}_-\rightarrow ZH\right)=\\
&&
=\frac{g^4}{96 \pi m^2 s \beta^2 c_w^2}\left(m^2(12s_w^4-15s_w^2+4)L+\beta[m^2(6s_w^4-3s_w^2+1)+s(5s_w^4-9s_w^2+4)]\right)
\ee
\be
&&\sigma\left(W^{+(1)}H^{(1)}_-\rightarrow \gamma G\right)=\sigma\left(W^{+(1)}A^{(1)}_-\rightarrow \gamma H\right)=\frac{g^4s_w^2}{96 \pi m^2 s \beta^2}\left(12m^2L+\beta(6m^2+5s)\right)
\ee
\be
&&\sigma\left(W^{+(1)}H^{(1)}_-\rightarrow t\bar t\right)=\frac{g^2}{64 \pi m^2 s \beta^2}\left(-4m^2Ly_t^2+\beta[g^2s_w^2m^2+2y_t^2(4s^2-11m^2)]\right)
\ee\\ \vspace*{0.5cm}

\noindent{\bf\normalsize IV. Fermion-Fermion}\\
\be
&&\sigma\left(l^{+(1)}_Rl^{-(1)}_R\rightarrow f\bar f\right)=\frac{N_cg^4t_w^4Y_l^2(Y^2_{f_L}+Y^2_{f_R})(s+2m^2)}{24 \pi s^2 \beta}
\ee
where the fermion $f\ne l$.
\be\nonumber
&&\sigma\left(l^{+(1)}_Rl^{-(1)}_R\rightarrow l^+l^-\right)=\frac{g^4t_w^4Y_{l_R}^4(5\beta s+2(2s+3m^2)L}{32 \pi s^2 \beta^2}+\\
&&
+\frac{g^4t_w^4Y_{l_R}^4(\beta(4s+9m^2)+8m^2L}{64 \pi m^2 s \beta^2}+\frac{g^4t_w^4Y_{l_R}^2(Y^2_{l_L}+Y^2_{l_R})(s+2m^2)}{24 \pi s^2 \beta}
\ee
\be
&&\sigma\left(l^{\pm(1)}_Rl^{\pm(1)}_R\rightarrow l^\pm l^\pm\right)=\frac{g^4t_w^4Y_{l}^4\left(-m^2(4s-5m^2)L-\beta s(2s-m^2)\right)}{32 \pi m^2 s^2 \beta^2}
\ee
\be
&&\sigma\left(l^{\pm(1)}_Rl^{\prime\pm(1)}_R\rightarrow l^\pm l^{\prime\pm}\right)=\frac{g^4t_w^4Y_{l}^4(4s-3m^2)}{64 \pi m^2 s \beta}
\ee
\be
&&\sigma\left(l^{\pm(1)}_Rl^{\prime\mp(1)}_R\rightarrow l^\pm l^{\prime\mp}\right)=\frac{g^4t_w^4Y_{l}^4\left(\beta(4s+9m^2)+8m^2L\right)}{64 \pi m^2 s \beta^2}
\ee
where $l$ and $l^\prime$ belong to different lepton families.
\be
&&\sigma\left(l^{+(1)}_Rl^{-(1)}_R\rightarrow \phi\phi^*\right)=\frac{g^4t_w^4Y_{l}^2Y_\phi^2(s+2m^2)}{48 \pi s^2 \beta}
\ee
\be
&&\sigma\left(l^{+(1)}_Rl^{-(1)}_R\rightarrow (ZZ+Z\gamma+\gamma\gamma)\right)=\frac{g^4t_w^4Y_{l}^4\left(2(s^2+4sm^2-8m^4)-\beta s(s+4m^2)\right)}{8 \pi s^3 \beta^2}
\ee
\be
&&\sigma\left(\nu_l^{(1)}\overline{\nu_l^{(1)}}\rightarrow f\bar f\right)=\frac{N_c g^4(\tilde Y_{L}^2+\tilde Y_{R}^2)\left(s+2m^2\right)}{96 \pi s^2 \beta c_w^4}
\ee
where $\tilde Y\equiv T^3-Q_fs_w^2$.
\be
\sigma\left(\nu_l^{(1)}\overline{\nu_l^{(1)}}\rightarrow Z^{(2)}\rightarrow q\bar q\right)=\sigma\left(l_L^{+(1)}l^{-(1)}_L\rightarrow Z^{(2)}\rightarrow q\bar q\right)=\frac{g^2 s\beta}{24m}\frac{\Gamma_{q\bar q}}{\left(s-m^2_{Z^{(2)}}\right)^2+m^2_{Z^{(2)}}\Gamma^2_{Z^{(2)}}},
\ee
where
\be\label{eq:gammaqq}
\Gamma_{q\bar q}=\frac{27 g^2 g_3^4 m}{1024\pi^5}\ln^2(\Lambda^2R^2),
\ee
and
\be
\Gamma_{Z^{(2)}}=\Gamma_{q\bar q}+\Gamma^{Z^{(2)}}_{HH,HA}+\frac{g^2 m}{8\pi}\left(1-\frac{4m^2_{l_L^{(1)}}}{m^2_{Z^{(2)}}}\right)^{3/2}+\frac{3g^2 m}{8\pi}\left(1-\frac{m^2_{l_L^{(2)}}}{m^2_{Z^{(2)}}}\right)^{2},
\ee
where
\be
\Gamma^{Z^{(2)}}_{HH,HA}=\frac{g^2 m}{96\pi}\left(1-4\frac{m^2_{H^{(1)}_\pm}}{m^2_{Z^{(2)}}}\right)^{3/2}+\frac{g^2 m}{96\pi}\left(1-2\frac{m^2_{H^{(1)}}+m^2_{A^{(1)}}}{m^2_{Z^{(2)}}}\right)^{3/2}.
\ee
\be
\sigma\left(\nu_l^{(1)}\overline{\nu_l^{(1)}}\rightarrow \phi\phi^* \right)=\sigma\left(l^{+(1)}_Ll^{-(1)}_L\rightarrow \phi\phi^* \right)=\frac{g^4\tilde Y_{\phi}^2\left(s+2m^2\right)}{192 \pi s^2 \beta c_w^4}
\ee
\be\nonumber
\sigma\left(\nu_l^{(1)}\overline{\nu_l^{(1)}}\rightarrow ZZ \right)=\sigma\left(l^{+(1)}_Ll^{-(1)}_L\rightarrow (ZZ+Z\gamma+\gamma\gamma) \right)=\frac{g^4\left((8m^4-4m^2s-s^2)L-\beta s(s+4m^2)\right)}{128 \pi s^3 \beta^2 c_w^4}\\
&&
\ee
\be\nonumber
\sigma\left(\nu_l^{(1)}\overline{\nu_l^{(1)}}\rightarrow W^+W^- \right)=\sigma\left(l^{+(1)}_Ll^{-(1)}_L\rightarrow W^+W^- \right)&=&-\frac{g^4(s+2m^2)}{96 \pi s^2 \beta}+\\
&&\nonumber
+\frac{g^4(\beta s-2m^2L)}{32 \pi s^2 \beta^2}-\frac{g^4\left(\beta(s+4m^2)+(s+2m^2)L\right)}{32 \pi s^2 \beta^2}\\
&&
\ee
\be
\sigma\left(\nu_l^{(1)}\nu_l^{(1)}\rightarrow \nu_l\nu_l \right)=\sigma\left(l^{\pm(1)}_Ll^{\pm(1)}_L\rightarrow l^\pm l^\pm \right)=\frac{g^4\left(\beta s(2s-m^2)-m^2(4s-5m^2)L\right)}{512 \pi m^2 s^2 \beta^2 c_w^4}
\ee
\be
\sigma\left(\nu_l^{(1)}\nu_{l^\prime}^{(1)}\rightarrow \nu_l\nu_{l^\prime}\right)=\sigma\left(l^{\pm(1)}_Ll^{\prime\pm(1)}_L\rightarrow l^\pm l^{\prime \pm} \right)=\sigma\left(\nu_l^{(1)}l^{\prime(1)}_L\rightarrow \nu_l l^\prime \right)=\frac{g^4(4s-3m^2)}{1024 \pi m^2 s \beta c_w^4}
\ee
\be
&&\sigma\left(\nu_l^{(1)}\overline{\nu_{l^\prime}^{(1)}}\rightarrow \nu_l\overline{\nu_{l^\prime}} \right)=\sigma\left(l^{\pm(1)}_Ll^{\prime\mp(1)}_L\rightarrow l^\pm l^{\prime \mp} \right)=\frac{g^4\left(\beta(4s+9m^2)+8m^2L\right)}{1024 \pi m^2 s \beta^2 c_w^4}
\ee
\be
&&\sigma\left(\nu_l^{(1)}\overline{\nu_{l^\prime}^{(1)}}\rightarrow l^-l^{\prime +} \right)=\frac{g^4\left(\beta(4s+9m^2)+8m^2L\right)}{256 \pi m^2 s \beta^2}
\ee
\be\nonumber
\sigma\left(\nu_l^{(1)}\overline{\nu_{l}^{(1)}}\rightarrow l^-l^{+} \right)&=&\frac{g^4\tilde Y_L \hat g_L^2\left(5\beta s+2(2s+3m^2)L\right)}{64 \pi s^2 \beta^2 c_w^4}+\frac{g^4(\tilde Y^4_L+\tilde Y^4_R)(s+2m^2)}{96 \pi s^2 \beta c_w^4}+
\\&&+\frac{\hat g_L^4\left(\beta(4s+9m^2)+8m^2L\right)}{64 \pi m^2 s \beta^2}
\ee
where $\hat g_L=g/(2c_w)$ for neutrinos and $\hat g_L=g/\sqrt{2}$ for charged leptons.
\be
&&\sigma\left(l^{+(1)}_Ll^{-(1)}_L\rightarrow f\bar f\ {\rm or}\ l^+_Rl^-_R  \right)=\frac{N_c\tilde g^4(s+2m^2)}{24 \pi s^2 \beta}
\ee
where $\tilde g^2\equiv g^2(Y_fY_{l_R}t_w^2+T^3_fT^3_{l_L})$
\be\nonumber
\sigma\left(l^{+(1)}_Ll^{-(1)}_L\rightarrow \nu_l\overline{\nu_l}\ {\rm or}\ l^+_Ll^-_L  \right)&=&\frac{\hat g^2_L \tilde g^2\left(5\beta s+2(2s+3m^2)L\right)}{32 \pi s^2 \beta^2}+
\\&&+\frac{\hat g^4_L \left(\beta(4s+9m^2)+8m^2L\right)}{64 \pi m^2 s \beta^2}+\frac{\tilde g^4(s+2m^2)}{24 \pi s^2 \beta}
\ee
\be
&&\sigma\left(l^{-(1)}_L\overline{\nu_l^{(1)}}\rightarrow f\bar f^\prime\right)=\frac{N_c g^4(s+2m^2)}{96 \pi s^2 \beta}
\ee
\be
\sigma\left(l^{-(1)}_L\overline{\nu_l^{(1)}}\rightarrow W^{-(2)}\rightarrow q\bar q^\prime\right)=\sigma\left(l_L^{+(1)}l^{-(1)}_L\rightarrow Z^{(2)}\rightarrow q\bar q\right)=\frac{g^2 s\beta}{24m}\frac{\Gamma_{q\bar q}}{\left(s-m^2_{W^{(2)}}\right)^2+4m^2\Gamma^2_{W^{(2)}}},
\ee
where $\Gamma_{q\bar q}$ is given in Eq.~(\ref{eq:gammaqq}), and where 
\be
\Gamma_{W^{(2)}}=\Gamma_{q\bar q}+\Gamma^{W^{(2)}}_{HH,HA}+\frac{g^2 m}{8\pi}\left(1-\frac{4m^2_{l_L^{(1)}}}{m^2_{W^{(2)}}}\right)^{3/2}+\frac{3g^2 m}{8\pi}\left(1-\frac{m^2_{l_L^{(2)}}}{m^2_{W^{(2)}}}\right)^{2},
\ee
where
\be
\Gamma^{W^{(2)}}_{HH,HA}=\frac{g^2 m}{96\pi}
\ee
\be
&&\sigma\left(l^{-(1)}_L\overline{\nu_l^{(1)}}\rightarrow \phi^*_u\phi_d\right)=\frac{g^4(s+2m^2)}{192 \pi s^2 \beta}
\ee
\be\nonumber
\sigma\left(l^{-(1)}_L\overline{\nu_l^{(1)}}\rightarrow W^-(Z+\gamma)\right)&=&\frac{g^4t_w^2\left((8m^4-4m^2s-s^2)L\beta s(s+4m^2)\right)}{32 \pi s^3 \beta^2}-\frac{5g^4(s+2m^2)}{48 \pi s^2 \beta}+\\
&&\nonumber
+\frac{g^4(\beta s-2m^2L)}{32 \pi s^2 \beta^2}+\frac{g^4\left(\beta(s+4m^2)+(s+2m^2)L\right)}{64 \pi s^2 \beta}+\frac{g^4m^2L}{16 \pi s^2}\\
&&
\ee
\be\nonumber
&&\sigma\left(l^{-(1)}_L\overline{\nu_l^{(1)}}\rightarrow l^-\overline{\nu_l}\right)=\frac{g^4t_w\left(5\beta s+2(2s+3m^2)L\right)}{64 \pi s^2 \beta^2(2s_w^2-1)}+\frac{g^4t^2_w\left(\beta(4s+9m^2)+8m^2L\right)}{64 \pi m^2 s \beta^2(2s_w^2-1)^2}+\frac{g^4(s+2m^2)}{96 \pi s^2 \beta}\\
&&
\ee
\be\nonumber
&&\sigma\left(l^{-(1)}_L\nu_l^{(1)}\rightarrow l^-\nu_l\right)=\frac{g^4t_w\left(-2m^2(4s-5m^2)L+m^2\beta s\right)}{64 \pi m^2 s^2 \beta^2(2s_w^2-1)}+\frac{\beta s(4s-3m^2)}{64 \pi m^2 s^2 \beta^2}\left(\frac{g^4t_w^2}{(2s_w^2-1)^2}+\frac{g^4}{4}\right)\\
&&
\ee
\be
&&\sigma\left(l^{-\prime(1)}_L\nu_l^{(1)}\rightarrow l^-\nu_{l^\prime}\right)=\frac{g^4(4s-3m^2)}{256 \pi m^2 s \beta}
\ee
\be
&&\sigma\left(l^{-\prime(1)}_L\nu_l^{(1)}\rightarrow l^{^-\prime}\nu_{l}\right)=\frac{g^4t_w^2(4s-3m^2)}{64 \pi m^2 s \beta(2s_w^2-1)^2}
\ee
\be
&&\sigma\left(l^{-\prime(1)}_L\overline{\nu_l^{(1)}}\rightarrow l^{^-\prime}\overline{\nu_{l}}\right)=\frac{g^4t_w^2\left(\beta(4s+9m^2)+8m^2L\right)}{64 \pi m^2 s \beta^2(2s_w^2-1)^2}
\ee
\be
&&\sigma\left(l^{(1)}_L\overline{l^{\prime(1)}}_L\rightarrow \nu_{l}\overline{\nu_{l^\prime}}\right)=\frac{g^4\left(\beta(4s+9m^2)+8m^2L\right)}{256 \pi m^2 s \beta^2}
\ee
\be
&&\sigma\left(l^{(1)}_Ll^{(1)}_R\rightarrow ll\right)=\frac{g^4t_w^4Y^2_{L}Y^2_{R}\left(8m^2L+\beta(9m^2+4s)\right)}{64 \pi m^2 s \beta^2}
\ee
\be
&&\sigma\left(l^{(1)}_L\overline{l^{(1)}}_R\rightarrow l\overline{l}\right)=\frac{g^4t_w^4Y^2_{L}Y^2_{R}\left(4s-3m^2\right)}{64 \pi m^2 s \beta^2}
\ee
\noindent Below we collect the KK quark-KK lepton pair annihilation cross sections:
\be
&&\sigma\left(l^{(1)}_L u^{(1)}_L\rightarrow lu\right)=\frac{\left(4g^2t_w^2Y_{l_L}Y_{u_L}-g^2\right)^2(4s-3m^2)}{1024 \pi m^2 s \beta}
\ee
\be
&&\sigma\left(\nu^{(1)}_l u^{(1)}_L\rightarrow \nu_l u\right)=\frac{\left(4g^2t_w^2Y_{l_L}Y_{u_L}+g^2\right)^2(4s-3m^2)}{1024 \pi m^2 s \beta}
\ee
\be
&&\sigma\left(l^{(1)}_L u^{(1)}_L\rightarrow \nu d\right)=\frac{g^4(4s-3m^2)}{256 \pi m^2 s \beta}
\ee
\be
&&\sigma\left(\nu^{(1)}_l \overline{u^{(1)}_L}\rightarrow l \overline{d}\right)=\frac{g^4\left(8m^2L+\beta(9m^2+4s)\right)}{256 \pi m^2 s \beta^2}
\ee
\be
&&\sigma\left(l^{(1)}_L \overline{u^{(1)}_L}\rightarrow l \overline{u}\right)=\frac{\left(4g^2t_w^2Y_{l_L}Y_{u_L}+g^2\right)^2\left(8m^2L+\beta(9m^2+4s)\right)}{1024 \pi m^2 s \beta^2}
\ee
\be
&&\sigma\left(\nu_l^{(1)} \overline{u^{(1)}_L}\rightarrow \nu_l \overline{u}\right)=\frac{\left(4g^2t_w^2Y_{l_L}Y_{u_L}+g^2\right)^2\left(8m^2L+\beta(9m^2+4s)\right)}{1024 \pi m^2 s \beta^2}
\ee
\be
\sigma\left(l^{(1)}_R u^{(1)}_L\rightarrow lu\right)=\sigma\left(l^{(1)}_L u^{(1)}_R\rightarrow lu\right)=\sigma\left(l^{(1)}_L \overline{u^{(1)}_R}\rightarrow l\overline{u}\right)=\sigma\left(l^{(1)}_Ll^{(1)}_R\rightarrow ll\right)
\ee
\be
\sigma\left(l^{(1)}_R \overline{u^{(1)}_L}\rightarrow l\overline{u}\right)=\sigma\left(l^{(1)}_L\overline{u^{(1)}_R}\rightarrow l\overline{u}\right)=\sigma\left(l^{(1)}_R u^{(1)}_R\rightarrow lu\right)=\sigma\left(l^{(1)}_L\overline{l^{(1)}}_R\rightarrow l\overline{l}\right)
\ee
\noindent The annihilation cross section involving the annihilation of KK quarks through strong interactions are collected below:
\be
&&\sigma\left(q^{(1)} \overline{q^{(1)}}\rightarrow q^\prime\overline{q}^\prime\right)=\frac{g_3^4(s+2m^2)}{54 \pi s^2 \beta}
\ee
\be
&&\sigma\left(q^{(1)} q^{(1)}\rightarrow qq\right)=\frac{g_3^4\left(2m^2(4s-5m^2)L+s\beta(6s-5m^2)\right)}{432 \pi m^2 s^2 \beta^2}
\ee
\be
&&\sigma\left(q^{(1)} \overline{q^{(1)}}\rightarrow q\overline{q}\right)=\frac{g_3^4\left(4m^2(4s-3m^2)L+\beta(32m^4+33sm^2+12s^2)\right)}{864 \pi m^2 s^2 \beta^2}
\ee
\be
&&\sigma\left(q^{(1)} \overline{q^{(1)}}\rightarrow gg\right)=\frac{-g_3^4\left(4(m^4+4sm^2+s^2)L+s\beta(31m^2+7s)\right)}{54 \pi s^3 \beta^2}
\ee
\be\nonumber
\sigma\left(q^{(1)} q^{\prime(1)}\rightarrow qq^\prime\right)=\sigma\left(u^{(1)}_R \overline{d^{(1)}_L}\rightarrow u\overline{d}\right)&=&\sigma\left(u^{(1)}_R d^{(1)}_R\rightarrow ud\right)=\sigma\left(u^{(1)}_R \overline{u^{(1)}_L}\rightarrow u\overline{u}\right)=\\
&&=\frac{g_3^4(4s-3m^2)}{288 \pi m^2 s \beta}
\ee
\be\nonumber
\sigma\left(q^{(1)} \overline{q^{\prime(1)}}\rightarrow q\overline{q}^\prime\right)=\sigma\left(u^{(1)}_R \overline{d^{(1)}_R}\rightarrow u\overline{d}\right)&=&\sigma\left(u^{(1)}_R d^{(1)}_L\rightarrow ud\right)=\sigma\left(u^{(1)}_R u^{(1)}_L\rightarrow uu\right)=\\
&=&\frac{g_3^4\left(8m^2L+\beta(9m^2+4s)\right)}{288 \pi m^2s \beta^2}
\ee\\ \vspace*{0.5cm}

\noindent{\bf\normalsize V. Fermion-Higgs}\\
\be
\sigma\left(f_R^{(1)}H^{(1)}\rightarrow fG\right)=\sigma\left(f_R^{(1)}A^{(1)}\rightarrow fH\right)=\sigma\left(f_R^{(1)}H_\pm^{(1)}\rightarrow fG^\pm\right)=\frac{g^4t_w^4Y_f^2(m^2L+s\beta)}{32 \pi m^2 s \beta^2}
\ee
\be\nonumber
&&\sigma\left(t_R^{(1)}H^{(1)}\rightarrow tg\right)=\sigma\left(t_L^{(1)}H^{(1)}\rightarrow tg\right)=\sigma\left(t_R^{(1)}A^{(1)}\rightarrow tg\right)=\sigma\left(t_R^{(1)}H_-^{(1)}\rightarrow bg\right)=\\
&&
=\frac{1}{2}\sigma\left(b_L^{(1)}H_+^{(1)}\rightarrow tg\right)=\frac{-g_3^2y_t^2(2L+3\beta)}{48 \pi s \beta^2}
\ee
\be\nonumber
\sigma\left(t_R^{(1)}H_+^{(1)}\rightarrow tG^+\right)&=&\frac{1}{288 \pi m^2 s^2 \beta^2 c_w^2}\Big(12 g^2c_w^2s_w^2m^2y_t^2+m^2L[12c_w^2s_w^2(m^2-s)y_t^2g^2+\\
&&
+9c_w^4(m^2-s)y_t^2+4sg^4s_w^4]+s\beta[4sg^4s_w^4-9c_w^4m^2y_t^4]\Big)
\ee
\be\nonumber
\sigma\left(f_L^{(1)}H^{(1)}\rightarrow fG\right)&=&\sigma\left(f_L^{(1)}A^{(1)}\rightarrow fH\right)=\sigma\left(f_L^{(1)}H_\pm^{(1)}\rightarrow fG^\pm\right)=\\
&&=\frac{g^4(T_f^3c_w-2s_wt_wY_f)^2(m^2L+s\beta)}{128 \pi m^2 s \beta^2 c_w^2}
\ee
\be\nonumber
&&\sigma\left(f_+^{(1)}H^{(1)}\rightarrow f_-G^+\right)=\sigma\left(f_-^{(1)}H^{(1)}\rightarrow f_+G^-\right)=\sigma\left(f_+^{(1)}A^{(1)}\rightarrow f_-G^+\right)=\sigma\left(f_-^{(1)}A^{(1)}\rightarrow f_+G^-\right)=\\
&&
=\sigma\left(f_-^{(1)}H_+^{(1)}\rightarrow f_+G\right)=\sigma\left(f_+^{(1)}H_-^{(1)}\rightarrow f_-G\right)=\frac{g^4(m^2L+s\beta)}{64 \pi m^2 s \beta^2}
\ee
\noindent where $f$ is any lepton or quark except $t_L^{(1)}$ and $b_L^{(1)}$, and $f_{+}$ ($f_{-}$) denotes isospin $+1/2$ (isospin $-1/2$) fermions.
\be
\sigma\left(t_L^{(1)}H_+^{(1)}\rightarrow tG^+\right)=\frac{g^4(c_w-2s_wt_wY_f)^2(m^2L+s\beta)}{128\pi m^2 s \beta^2 c_w^2}+\frac{y_t^4(m^2L+s\beta)}{32\pi s^2 \beta^2}
\ee
\be
\sigma\left(t_L^{(1)}H_+^{(1)}\rightarrow tW^+\right)=2\sigma\left(b_L^{(1)}H^{(1)}\rightarrow Gb\right)=\frac{-g^2y_t^2L}{32\pi s \beta^2}
\ee
\be\nonumber
\sigma\left(b_L^{(1)}H_-^{(1)}\rightarrow bG^-\right)=\frac{g^4(c_w-2s_wt_wY_f)^2(m^2L+s\beta)}{128\pi m^2 s \beta^2 c_w^2}+\frac{y_t^2L[sg^2(c_w^2-2s_w^2y_b)+c_w^2(s-m^2)y_t^2-c_w^2s\beta^2y_t^2]}{32\pi s^2\beta^2 c_w^2}\\
&&
\ee\\ \vspace*{0.5cm}

\noindent{\bf\normalsize VI. Higgs-Higgs}\\
\be\nonumber
\sigma\left(H^{(1)}H^{(1)}\rightarrow G^+G^-\right)=\sigma\left(A^{(1)}A^{(1)}\rightarrow G^+G^-\right)=\frac{8m^2\lambda g^2L+\beta[(2s+m^2)g^4+4\lambda g^2 m^2+4\lambda^2 m^2]}{64 \pi m^2 s \beta^2}\\
&&
\ee
\be
\sigma\left(H^{(1)}H^{(1)}\rightarrow HH\right)=\sigma\left(A^{(1)}A^{(1)}\rightarrow GG\right)=\frac{9\lambda^2}{32 \pi s \beta}
\ee
\be\nonumber
\sigma\left(H^{(1)}H^{(1)}\rightarrow GG\right)=\sigma\left(A^{(1)}A^{(1)}\rightarrow HH\right)=\frac{8m^2\lambda g^2c_w^2L+\beta[(2s+m^2)g^4+4\lambda g^2 c_w^2 m^2+4\lambda^2 m^2 c_w^4]}{64 \pi m^2 s \beta^2 c_w^4}\\
&&
\ee
\be
\sigma\left(H^{(1)}H^{(1)}\rightarrow ZZ\right)=\sigma\left(A^{(1)}A^{(1)}\rightarrow ZZ\right)=\frac{g^4\left(s\beta(s+4m^2)+4m^2(s-2m^2)L\right)}{64 \pi s^3 \beta^2 c_w^4}
\ee
\be
\sigma\left(H^{(1)}H^{(1)}\rightarrow W^+W^-\right)=\sigma\left(A^{(1)}A^{(1)}\rightarrow W^+W^-\right)=\frac{g^4\left(s\beta(s+4m^2)+4m^2(s-2m^2)L\right)}{32 \pi s^3 \beta^2 c_w^4}
\ee
\be
\sigma\left(H^{(1)}H^{(1)}\rightarrow t\bar t\right)=\sigma\left(A^{(1)}A^{(1)}\rightarrow t\bar t\right)=\frac{3y_t^4\left(-(s+2m^2)L-2s\beta\right)}{16 \pi s^2 \beta^2}
\ee
\be
\sigma\left(A^{(1)}A^{(1)}\rightarrow H^{(2)}\rightarrow t\bar t\right)=\sigma\left(H_+^{(1)}H_-^{(1)}\rightarrow H^{(2)}\rightarrow t\bar t\right)=\frac{\lambda^2 m_W^2}{g^2 m\beta}\frac{\Gamma^{H_{t\bar t}^{(2)}}}{(s-m^2_{H^{(2)}})^2+m^2_{H^{(2)}}\Gamma^2_{H^{(2)}}}
\ee
\be
\sigma\left(H_+^{(1)}H_+^{(1)}\rightarrow G^+G^+\right)=\frac{-16 g^2m^2\lambda c_w^2L+\beta\left((2s+m^2)g^4-8\lambda m^2 g^2 c_w^2+16\lambda^2 m^2 c_w^4\right)}{128 \pi m^2 s \beta^2 c_w^4}
\ee
\be\nonumber
\sigma\left(H_+^{(1)}H_-^{(1)}\rightarrow t\bar t\right)=\frac{72 c_w^2 y_t^2(-3sc_w^2y_t^2-4m^2 g^2s_w^2)L-432s\beta c_w^4 y_t^4+s\beta^3(20 s_w^4-12 s_w^2+9)g^4-144s\beta c_w^2y_t^2 g^2s_w^2}{1152 \pi s^2 \beta^2 c_w^4}\\
&&
\ee
\be\nonumber
\sigma\left(H_+^{(1)}H_-^{(1)}\rightarrow b\bar b\right)&=&\frac{1}{1152 \pi s^2 \beta^2 c_w^4}\Big(72 c_w^2 y_t^2(-3sc_w^2y_t^2+g^2m^2(4s_w^2-3)L)+\\
&&\nonumber
-432s\beta c_w^4 y_t^4+s\beta^3(20 s_w^4-24 s_w^2+9)g^4-36s\beta c_w^2y_t^2 g^2(s_w^4-7s_w^2+3)\Big)\\
&&
\ee
\be\nonumber
\sigma\left(H_+^{(1)}H_-^{(1)}\rightarrow G^+G^-\right)=-\frac{6m^2 g^2 s(g^2+4\lambda c_w^2)L-48\lambda^2 m^2 s c_w^4+\beta\left(g^4(m^4-7m^2 s-3s^2)-12\lambda m^2s c_w^2 g^2\right)}{192 \pi m^2 s^2 \beta^2 c_w^4}\\
&&
\ee
\be\nonumber
&&\sigma\left(H_+^{(1)}H_-^{(1)}\rightarrow HH\right)=\sigma\left(H_+^{(1)}H_-^{(1)}\rightarrow GG\right)=\frac{1}{2}\sigma\left(H^{(1)}H^{(1)}\rightarrow G^+G^-\right)=\\
&&\nonumber
=\frac{8 m^2 g^2\lambda L+\beta\left((2s+m^2)g^4+4\lambda m^2 g^2+4\lambda^2 m^2\right)}{128 \pi m^2 s \beta^2}
\ee
\be
\sigma\left(H_+^{(1)}H_-^{(1)}\rightarrow GH\right)=\frac{g^4\left(24m^2c_w^2 s L-\beta\big[4(1-2s_w^2)^2m^4+s(92s_w^4-140s_w^2+47)-24s^2c_w^4\big]\right)}{768 \pi m^2 s^2 \beta^2 c_w^4}
\ee
\be
\sigma\left(H_+^{(1)}H_-^{(1)}\rightarrow ZZ\right)=\frac{g^4(1-2s_w^2)^4\left(4m^2(s-2m^2)L+s\beta(s+4m^2)\right)}{64 \pi s^3 \beta^2 c_w^4}
\ee
\be
\sigma\left(H_+^{(1)}H_-^{(1)}\rightarrow \gamma Z\right)=\frac{g^4s_w^2(1-2s_w^2)^2\left(4m^2(s-2m^2)L+s\beta(s+4m^2)\right)}{8 \pi s^3 \beta^2 c_w^2}
\ee
\be
\sigma\left(H_+^{(1)}H_-^{(1)}\rightarrow \gamma \gamma\right)=\frac{g^4s_w^4\left(4m^2(s-2m^2)L+s\beta(s+4m^2)\right)}{4 \pi s^3 \beta^2}
\ee
\be
\sigma\left(H_+^{(1)}H_-^{(1)}\rightarrow W^+W^-\right)=\frac{g^4\left(6m^2L+\beta(s+11m^2)\right)}{24 \pi s^2 \beta^2}
\ee
\be
\sigma\left(H_+^{(1)}H_-^{(1)}\rightarrow f\bar f\right)=\frac{N_c g^4\beta\left(Q_f(t_w^2(1-2s_w^2)-2s_w^2)-\frac{1-2s_w^2}{c_w^2}T^3_f\right)}{96 \pi s c_w^2}
\ee
\be\nonumber
\sigma\left(H^{(1)}A^{(1)}\rightarrow HG\right)=\frac{6m^2g^2s(g^2-2\lambda c_w^2)L-\beta\left((m^4-7sm^2-3s^2)g^4+6\lambda m^2 s s_w^2 c_w^2 g^4-12\lambda^2 m^2 s c_w^4\right)}{192 \pi m^2 s^2 \beta^2 c_w^4}\\
&&
\ee
\be\nonumber
\sigma\left(H^{(1)}A^{(1)}\rightarrow G^+G^-\right)=\frac{g^4\left(24m^2c_w^2 sL+\beta\big[-4(1-2s_w^2)^2m^4+sm^2(-92s_w^4+140s_w^2-47)+24s^2c_w^4\big]\right)}{192 \pi m^2 s^2 \beta^2 c_w^4}\\
&&
\ee
\be
\sigma\left(H^{(1)}A^{(1)}\rightarrow W^+W^-\right)=\frac{g^4\left(12m^2(2m^2+s)L+s\beta(32m^2+s)\right)}{96 \pi s^3 \beta^2}
\ee
\be\nonumber
\sigma\left(H^{(1)}A^{(1)}\rightarrow t\bar t\right)=\frac{-54y_t^2\left(g^2m^2+c_w^2(s-2m^2)y_t^2\right)L+\beta\big[-54sy_t^4c_w^2-27 g^2 y_t^2 s+g^4 s\beta^2(32s_w^4-24s_w^2+9)/4c_w^2\big]}{288 \pi s^2 \beta^2 c_w^2}\\
&&
\ee
\be
\sigma\left(H^{(1)}A^{(1)}\rightarrow t\bar t\right)=\frac{N_c\beta g^4\left(\tilde Y^2_L+\tilde Y^2_R\right)}{96 \pi s c_w^4}
\ee
\be\nonumber
\sigma\left(H^{(1)}H_+^{(1)}\rightarrow HG^+\right)&=&\sigma\left(A^{(1)}H_+^{(1)}\rightarrow GG^+\right)=\\
\nonumber
&=&\frac{6m^2 g^2 s(g^2-2\lambda)L-12\lambda^2m^2s-\beta\big[(m^4-7sm^2-3s^2)g^4+6\lambda m^2 s s_w^2g^4\big]}{192 \pi m^2 s^2 \beta^2}\\
&&
\ee
\be\nonumber
&&\sigma\left(H^{(1)}H_+^{(1)}\rightarrow GG^+\right)=\sigma\left(A^{(1)}H_+^{(1)}\rightarrow HG^+\right)=\frac{g^4}{768 \pi m^2 s^2 \beta^2 c_w^4}\times\\
&&\nonumber
\times\left(12m^2(1-2s_w^2)(2-3s_w^2)L-\beta\big[(4m^4+47sm^2-24s^2)c_w^4-24(m^2-s)s s_w^2c_w^2-3s(m^2+4s)s_w^4\big]\right)\\
&&
\ee
\be
&&\sigma\left(H^{(1)}H_+^{(1)}\rightarrow ZW^+\right)=\sigma\left(A^{(1)}H_+^{(1)}\rightarrow ZW^+\right)=\frac{g^4}{96 \pi s^3 \beta^2 c_w^4}\times\\
&&
\times\left(6m^2L\big[(1-2s_w^2)(4m^2-2ss_w^2+s)+s\big]+s\beta[4s_w^4(s+11m^2)+(1-2s_w^2)(s+32m^2)\big]\right)
\ee
\be
\sigma\left(H^{(1)}H_+^{(1)}\rightarrow \gamma W^+\right)=\sigma\left(A^{(1)}H_+^{(1)}\rightarrow \gamma W^+\right)=\frac{g^4s_w^2\left(6m^2L+\beta(11m^2+s)\right)}{24 \pi s^2 \beta^2}
\ee
\be
\sigma\left(H^{(1)}H_+^{(1)}\rightarrow t\overline{b}\right)=\frac{-6y_t^2(2g^2m^2+sy_t^2)L+\beta\big[s\beta^2g^4-6sy_t^2s_w^2g^2-12sy_t^4\big]}{64 \pi s^2 \beta^2}
\ee
\be
\sigma\left(H^{(1)}H_+^{(1)}\rightarrow f\overline{f}^\prime\right)=\sigma\left(A^{(1)}H_+^{(1)}\rightarrow f\overline{f}^\prime\right)=\frac{N_c g^4\beta}{192 \pi s}
\ee


\subsection{Production Cross Sections for strongly interacting KK modes at Hadron Colliders}\label{sec:xsechad}

We list below the spin-averaged amplitudes squared, summed over initial for the production of strongly interacting KK particles at hadron colliders~\cite{Smillie:2005ar}. The symbols $q_L$ and $q_R$ generically indicate up and down-type quarks of any generation.

\be
\left|\mathcal{M}(q \overline{q} \rightarrow q^{(1)} \overline{q^{(1)}}\right|^2=\frac{4 g_3^4}{9} \left[ \frac{2 m^2}{s} + \frac{(t-m^2)^2+(u-m^2)^2}{s^2} \right]
\ee

\be\nonumber 
\left|\mathcal{M}(q\overline{q} \rightarrow q^{(1)} \overline{q^{(1)}})\right|^2&=&\frac{g_3^4}{9}\left[ 2 m^2 \left( \frac{4}{s}+\frac{s}{(t-m^2)^2}-\frac{1}{t-m^2} \right) \right. \\ 
&& \left. + \frac{23}{6} + \frac{2s^2}{(t-m^2)^2}+\frac{8s}{3 (t-m^2)} + \frac{6 (t-m^2)}{s} + \frac{8 (t-m^2)^2}{s^2} \right]
\ee

\be\nonumber 
\left|\mathcal{M}(qq \rightarrow q^{(1)} q^{(1)})\right|^2&=&\frac{g_3^4}{27} \left[ m^2 \left( 6\frac{t-m^2}{(u-m^2)^2}+6\frac{u-m^2}{(t-m^2)^2}-\frac{s}{(t-m^2)(u-m^2)}\right)\right. \\ 
&& \quad \left. +2\left( 3\frac{(t-m^2)^2}{(u-m^2)^2}+3\frac{(u-m^2)^2}{(t-m^2)^2} + 4\frac{s^2}{(t-m^2) (u-m^2)}-5 \right) \right]
\ee

\be\nonumber 
\left|\mathcal{M}(gg \rightarrow q^{(1)} \overline{q^{(1)}})\right|^2&=&g_3^4 \left[m^4 \frac{-4}{(t-m^2)(u-m^2)} \left( \frac{s^2}{6 (t-m^2) (u-m^2)}-\frac{3}{8} \right) + \right. \\ \nonumber 
&& \qquad \left. + m^2 \frac{4}{s} \left(\frac{s^2}{6(t-m^2) (u-m^2)}-\frac{3}{8} \right) + \right. \\ 
&& \qquad \left. + \frac{s^2}{6 (t-m^2) (u-m^2)} -\frac{17}{24}+\frac{3 (t-m^2) (u-m^2)}{4s^2}\right]
\ee

\be\nonumber 
\left|\mathcal{M}(gq \rightarrow g^{(1)} q^{(1)})\right|^2&=&\frac{-g_3^4}{3} \left[ \frac{5s^2}{12 (t-m^2)^2}+\frac{s^3}{(t-m^2)^2 (u-m^2)}+\right. \\ 
&& \qquad \left. + \frac{11 s (u-m^2)}{6 (t-m^2)^2}+\frac{5(u-m^2)^2}{12 (t-m^2)^2}+\frac{(u-m^2)^3}{s (t-m^2)^2}\right]
\ee

\be
\left|\mathcal{M}(q \overline{q}^{\prime} \rightarrow q^{(1)} \overline{q^{\prime(1)}})\right|^2=\frac{g_3^4}{18} \left[ 4 m^2 \frac{s}{(t-m^2)^2}+5+4\frac{s^2}{(t-m^2)^2}+8\frac{s}{t-m^2} \right]
\ee

\be
\left|\mathcal{M}(qq^{\prime} \rightarrow q^{(1)} q^{\prime(1)})\right|^2=\frac{2 g_3^4}{9} \left[ - m^2 \frac{s}{(t-m^2)^2}+\frac{1}{4}+\frac{s^2}{(t-m^2)^2} \right]
\ee

\be\nonumber 
\left|\mathcal{M}(qq \rightarrow q^{(1)}_L q^{(1)}_R)\right|^2&=&\frac{g_3^4}{9}\left[ m^2 \left( \frac{2s^3}{(t-m^2)^2 (u-m^2)^2} - \frac{4s}{(t-m^2) (u-m^2)} \right)+\right. \\ 
&& \qquad \left.+ 2\frac{s^4}{(t-m^2)^2 (u-m^2)^2} - 8\frac{s^2}{(t-m^2) (u-m^2)}+5 \right]
\ee
\be\nonumber 
&&\left|\mathcal{M}(q \overline{q}^{\prime} \rightarrow q_L^{(1)}\overline{q_R^{(1)\prime}})\right|^2=\left|\mathcal{M}(q \overline{q} \rightarrow q_L^{(1)}\overline{q_R^{(1)}}, q_R^{(1)}\overline{q_L^{(1)}})\right|^2=\\ \nonumber 
&& =\left|\mathcal{M}(q q^{\prime} \rightarrow q_L^{(1)}q_R^{(1)\prime}, q_R^{(1)}q_L^{(1)\prime})\right|^2=\left|\mathcal{M}(q \overline{q}^{\prime} \rightarrow q_R^{(1)}\overline{q_L^{(1)\prime}}\right|^2=\\ 
&& =\frac{g_3^4}{9} \left[ 2 m^2 \left( \frac{1}{t-m^2}+\frac{u-m^2}{(t-m^2)^2} \right) + \frac{5}{2} + \frac{4 (u-m^2)}{t-m^2}+\frac{2 (u-m^2)^2}{(t-m^2)^2} \right]
\ee

\be\nonumber 
\left|\mathcal{M}(g g \rightarrow g^{(1)} g^{(1)})\right|^2&=&\frac{9 g_3^4}{4} \left[ 3 m^4 \frac{s^2+(t-m^2)^2+(u-m^2)^2}{(t-m^2)^2 (u-m^2)^2}- 3 m^2 \frac{s^2+(t-m^2)^2+(u-m^2)^2}{s (t-m^2) (u-m^2)}+ \right. \\ 
&&\qquad \quad \left. +1+\frac{(s^2+(t-m^2)^2+(u-m^2)^2)^3}{4s^2(t-m^2)^2(u-m^2)^2}-\frac{(t-m^2) (u-m^2)}{s^2} \right]
\ee

\be\nonumber 
\left|\mathcal{M}(q \overline{q} \rightarrow g^{(1)} g^{(1)})\right|^2&=&\frac{2 g_3^4}{27} \left[m^2 \left( -\frac{4s^3}{(t-m^2)^2 (u-m^2)^2}+\frac{57s}{(t-m^2)(u-m^2)}-\frac{108}{s} \right) \right. \\ 
&& \qquad \quad \left. + \frac{20s^2}{(t-m^2)(u-m^2)}-93+\frac{108(t-m^2)^2 (u-m^2)}{s^2} \right]
\ee

\normalsize


\end{document}